\documentclass[prd,floats,floatfix,showpacs,nofootinbib,dvipdfm]{revtex4}
\usepackage{graphicx}
\usepackage{epsf}
\usepackage{dcolumn}
\usepackage{bm}
\usepackage{slashed}
\usepackage{amssymb}
\usepackage{natbib}
\usepackage{amsmath}
\usepackage{amsthm}
\usepackage{url}
\usepackage{color}
\newcommand{\fracc}[2]{\frac{\textstyle{#1}}{\textstyle{#2}}}
\newlength{\espaco}
\newtheorem*{theorem}{Lemma (Lichnerowicz)}

\begin{document}

\title{The Quasi-Maxwellian Equations of General Relativity: Applications to the Perturbation Theory}

\author{M. Novello\footnote{M. Novello is Cesare Lattes ICRANet Professor}} \email{novello@cbpf.br}
\author{E. Bittencourt}\email{eduhsb@cbpf.br}
\author{J. M. Salim} \email{jsalim@cbpf.br}

\affiliation{Instituto de Cosmologia Relatividade Astrofisica ICRA -
CBPF\\ Rua Dr. Xavier Sigaud, 150, CEP 22290-180, Rio de Janeiro,
Brazil}

\date{\today}

\begin{abstract}
A comprehensive review of the equations of general relativity in the quasi-Maxwellian (QM) formalism introduced by Jordan, Ehlers and Kundt is made. Our main interest concerns its applications to the analysis of the perturbation of standard cosmology in the Friedman-Lema\^itre-Robertson-Walker framework. The major achievement of the QM scheme is its use of completely gauge independent quantities.  We shall see that in the QM-scheme we deal directly with observable quantities. This reveals its advantage over the old method introduced by Lifshitz et al that deals with perturbation in the standard Einstein framework. For completeness, we compare the QM-scheme to the gauge-independent method of Bardeen, a procedure consisting on particular choices of the perturbed variables as a combination of gauge dependent quantities.
\end{abstract}

\maketitle

\addcontentsline{toc}{section}{Contents}
\tableofcontents
\clearpage

\section{Introduction}

There are two formal ways to deal with the dynamics of General Relativity which we call the Einstein frame and the Jordan-Ehlers-Kundt frame (JEK frame, for short):
\begin{itemize}
\item{The Einstein frame (1915) corresponds to a second order differential equations relating the curvature tensor to the energy-momentum tensor;}
\item{The JEK frame (1960) relates the derivatives of the conformal Weyl tensor to the derivatives of the energy-momentum tensor using Bianchi's identities.}
\end{itemize}

Although the demonstration of the equivalence of JEK formulation to general relativity (GR) have been realized by Lichnerowicz in the early 60, its role in the development of applications have been less active than one could expect. A possible cause of this is the fact that almost all introductory books on GR do not present the JEK frame as an alternative formulation of gravitation. Indeed, only few advanced books--- cf. Zakharov (1973), Choquet-Bruhat (2009), Hawking-Ellis (1977)---show an overview on this. In particular, as a direct consequence of this is the fact that the great majority of analysis of perturbation theory is made ignoring completely the possibility of use JEK frame to develop a consistent and worldwide method for this.

The main goal of the present review is to make a little more popular the use of the JEK frame in the realm of perturbation theory of Friedman universes. Indeed, the Lifshitz-Bardeen method and the JEK frame, under the same initial conditions, give the same results for the perturbations in the linear regime, as we shall see in Sec.\ [\ref{previous}]. The main interest on JEK rests on its unambiguous way to deal with perturbation within the standard cosmological FLRW scenario.

In this paper we concentrate our attention to the (cosmological) perturbation scheme presented in Sec.\ [\ref{pert_th}], although we will describe some examples of well-known solutions---Schwarzschild, Kasner, Friedman (singular and nonsingular)---according to the JEK frame (cf. Section\ [\ref{part_sol}]) to show how this method could be used to obtain new solutions of general relativity (details of this discussion in Sec.\ [\ref{init_cond}]).

Finally, the JEK-frame is called alternatively the Quasi-Maxwellian version of general relativity. The reason for this name is manifest from its strike similitude to Maxwell's equations of electrodynamics. In the appendix, this similarity is used to exhibit an example of modification of general relativity by extending further such analogy to the case of the electrodynamics inside a dielectric medium.

\subsection{Definitions, Notations and a Brief Mathematical Compendium}

We shall list below all the definitions that should be used in this review:

\begin{itemize}
\item{The structure of the space-time is represented by a Riemannian geometry $g_{\mu\nu}(x^{\alpha})$, with Lorentzian signature $(+,-,-,-)$;}
\item{The Levi-Civita tensor $\eta_{\mu\nu\alpha\beta} = \sqrt{-g}\, \epsilon_{\mu\nu\alpha\beta}$, where $g$ is the determinant of $g_{\mu\nu}$ and $\epsilon_{\mu\nu\alpha\beta}$ is the completely antisymmetric pseudo-tensor, $\epsilon_{0123}=1$;}
\item{The Christoffel symbols are defined by $
\Gamma^{\alpha}_{\beta\mu} = \fracc{1}{2} g^{\alpha\lambda} (g_{\beta\lambda,\mu}+g_{\mu\lambda,\beta}-g_{\beta\mu,\lambda})$;}
\item{The geodesic equation is
\begin{equation}
\nonumber
\fracc{d^2x^{\mu}}{d\lambda^2}+\Gamma^{\mu}_{\alpha\beta}\fracc{dx^{\alpha}}{d\lambda}\fracc{dx^{\beta}}{d\lambda}=0;
\end{equation}}
\item{The Riemann tensor is defined by
\begin{equation}
\nonumber
R^{\alpha}{}_{\beta\mu\nu}=\Gamma^{\alpha}_{\beta\mu,\nu}-\Gamma^{\alpha}_{\beta\nu,\mu}+\Gamma^{\alpha}_{\nu\tau}\Gamma^{\tau}_{\beta\mu}
-\Gamma^{\alpha}_{\mu\tau}\Gamma^{\tau}_{\beta\nu};
\end{equation}
The traces of Riemann tensor define the Ricci tensor $R_{\mu\nu}=R^{\alpha}{}_{\mu\alpha\nu}$ and the curvature scalar $R=R^{\alpha}{}_{\alpha}$;}
\item{The decomposition of the energy-momentum tensor into irreducible parts, with respect to a normalized observer field $V^{\alpha}$, is given by
\begin{equation}
\nonumber
T_{\mu\nu}=\rho V_{\mu}V_{\nu}-ph_{\mu\nu}+V_{(\mu}q_{\nu)}+\pi_{\mu\nu},
\end{equation}
where $\rho$ is the energy density, $p$ is the isotropic pressure, $q_{\mu}$ is the heat flux and $\pi_{\mu\nu}$ is the anisotropic pressure. We use parentheses `$(\,)$' for symmetrization and brackets `$[\,]$' for skew-symmetrization;}
\item{The Weyl tensor is defined by
\begin{equation}
\nonumber
W_{\alpha\beta\mu\nu}=R_{\alpha\beta\mu\nu}-M_{\alpha\beta\mu\nu}+\fracc{1}{6}Rg_{\alpha\beta\mu\nu},
\end{equation}
where $$2M_{\alpha\beta\mu\nu} = R_{\alpha\mu}g_{\beta\nu} + R_{\beta\nu}g_{\alpha\mu} - R_{\alpha\nu}g_{\beta\mu} - R_{\beta\mu}g_{\alpha\nu}$$ and $$g_{\alpha\beta\mu\nu}=g_{\alpha\mu}g_{\beta\nu}-g_{\alpha\nu}g_{\beta\mu}.$$ The dual is denoted by
$$W^{\,*}_{\alpha\beta\mu\nu} = \frac{1}{2} \eta_{\alpha\beta}{}^{\rho\sigma} W_{\rho\sigma\mu\nu}.$$ Remark that $W^{\,*}_{\alpha\beta\mu\nu} = W^{\,\,\,\,\,\,\,\,*}_{\alpha\beta\mu\nu}$;}
\item{The electric and magnetic parts of the Weyl tensor are, respectively, $$E_{\alpha\beta}\equiv-W_{\alpha\mu\beta\nu}V^{\mu}V^{\nu}$$ and $$H_{\alpha\beta}\equiv-^{*}W_{\alpha\mu\beta\nu}V^{\mu}V^{\nu}$$;}
\item{The tensor defined by $h_{\mu\nu}\equiv g_{\mu\nu}-V_{\mu}V_{\nu}$ projects tensorial quantities in the rest space $\Sigma$ of the observers. Note that $h_{\mu\nu}V^{\nu}=0$ and $h_{\mu\nu}h^{\nu}{}_{\lambda}=h_{\mu\lambda}$.}
\item{Einstein's equations (EE) are given by
$$R_{\mu\nu}-\frac{1}{2}Rg_{\mu\nu}+\Lambda g_{\mu\nu}=-kT_{\mu\nu},$$
where $\Lambda$ is the cosmological constant and $k\equiv8\pi G_N/c^4$ which we shall set equal to 1, unless stated otherwise;}
\item{The covariant derivative of $V_{\mu}$ can be decomposed into its irreducible parts, that is, $$V_{\mu;\nu} = \sigma_{\mu\nu} + \omega_{\mu\nu} + \fracc{1}{3}\theta h_{\mu\nu} + a_{\mu}V_{\nu},$$ where $\theta=V^{\alpha}{}_{;\alpha}$ is the expansion coefficient, $$\sigma_{\mu\nu}\equiv \fracc{1}{2} h_{\mu}^{\alpha} h_{\nu}^{\beta} V_{(\alpha;\beta)} - \frac{\theta}{3} h_{\mu\nu}$$ is the shear tensor, $$\omega_{\mu\nu}\equiv \fracc{1}{2} h_{\mu}^{\alpha} h_{\nu}^{\beta} V_{[\alpha;\beta]}$$ is the vorticity and $a^{\mu}$ is the acceleration;}
\end{itemize}

One can use the quantities defined above to obtain the evolution equations of the kinematical quantities:

\begin{itemize}
\item{Raychaudhuri equation
\begin{equation}
\label{raych}
\dot\theta+\fracc{\theta^2}{3}+2\sigma^2+2\omega^2-a^{\mu}{}_{;\mu}=-\fracc{1}{2}(\rho+3p),
\end{equation}
where $2\sigma^2\equiv\sigma^{\mu\nu}\sigma_{\mu\nu}$ and $2\omega^2\equiv\omega^{\mu\nu}\omega_{\mu\nu}$ and $\dot X\equiv X_{,\alpha}V^{\alpha}$ (this last definition will be used throughout the text);}
\item{The evolution of the shear tensor is
\begin{equation}
\begin{array}{l}
\label{sigma_ev}
h_{\alpha}{}^{\mu}h_{\beta}{}^{\nu}\dot\sigma_{\mu\nu}+\fracc{1}{3}h_{\alpha\beta}(a^{\lambda}{}_{;\lambda}-2\sigma^2-2\omega^2)+a_{\alpha}a_{\beta}-
\fracc{1}{2}h_{\alpha}{}^{\mu}h_{\beta}{}^{\nu}(a_{\mu;\nu}+a_{\nu;\mu})+\\[2ex]
+\fracc{2}{3}\theta\sigma_{\alpha\beta}+\sigma_{\alpha\mu}\sigma^{\mu}{}_{\beta}+\omega_{\alpha\mu}\omega^{\mu}{}_{\beta}=-E_{\alpha\beta}-\fracc{1}{2}\pi_{\alpha\beta};
\end{array}
\end{equation}}
\item{The evolution equation for the vorticity tensor is given by
\begin{equation}
\label{vort_ev}
h_{\alpha}{}^{\mu}h_{\beta}{}^{\nu}\dot\omega_{\mu\nu}-\fracc{1}{2}h_{\alpha}{}^{\mu}h_{\beta}{}^{\nu}(a_{\mu;\nu}-a_{\nu;\mu})+
\fracc{2}{3}\theta\omega_{\alpha\beta}-\sigma_{\beta\mu}\omega^{\mu}{}_{\alpha}+\sigma_{\alpha\mu}\omega^{\mu}{}_{\beta}=0;
\end{equation}}
\item{These kinematical quantities must satisfy three constraint equations:
\begin{equation}
\label{eq_vinc_1}
\fracc{2}{3}\theta_{,\mu}h^{\mu}{}_{\lambda}-(\sigma^{\alpha}{}_{\gamma}+\omega^{\alpha}{}_{\gamma})_{;\alpha}h^{\gamma}{}_{\lambda}-
a^{\nu}(\sigma_{\lambda\nu}+\omega_{\lambda\nu})=-q_{\lambda};
\end{equation}
\begin{equation}
\label{eq_vinc_2}
\omega^{\alpha}{}_{;\alpha}+2\omega^{\alpha}a_{\alpha}=0,
\end{equation}
where $\omega^{\alpha} = -\fracc{1}{2} \eta^{\alpha\beta\gamma\delta} \omega_{\beta\gamma} V_{\delta}$ and
\begin{equation}
\label{eq_vin_3}
-\fracc{1}{2}h_{(\tau}{}^{\epsilon}h_{\lambda)}{}^{\alpha}\eta_{\epsilon}{}^{\beta\gamma\nu}V_{\nu}(\sigma_{\alpha\beta}+\omega_{\alpha\beta})_{;\gamma}+a_{(\tau}\omega_{\lambda)}=H_{\tau\lambda};
\end{equation}}
\item{The conservation law of the energy-momentum tensor expressed in terms of its components is the conservation equation
\begin{equation}
\label{proj_cons_law_en_mom_1}
\dot\rho+(\rho+p)\theta+\dot q^{\mu}V_{\mu}+q^{\alpha}{}_{;\alpha}-\pi^{\mu\nu}\sigma_{\mu\nu}=0,
\end{equation}
and the generalized Euler equation
\begin{equation}
\label{proj_cons_law_en_mom_2}
(\rho+p)a_{\alpha}-p_{,\mu}h^{\mu}{}_{\alpha}+\dot q_{\mu}h^{\mu}{}_{\alpha}+\theta q_{\alpha}+q^{\nu}\theta_{\alpha\nu}+q^{\nu}\omega_{\alpha\nu}+\pi_{\alpha}{}^{\nu}{}_{;\nu}\\
+\pi^{\mu\nu}\sigma_{\mu\nu}V_{\alpha}=0.
\end{equation}}
\end{itemize}
These formulas summarize the kinematical part of the QM-approach. Now we shall focus on the dynamical equations in terms of the Weyl tensor.

\subsection{Quasi-Maxwellian Equations}

The quasi-Maxwellian equations \cite{salim,no_sa_ma_re_se,jek_eq,hawking_66} are obtained from Bianchi's identities written in terms of the Weyl tensor, i.e.

\begin{equation}
\label{bianchi_R}
{W^{\alpha\beta\mu\nu}}_{;\nu}=\fracc{1}{2}R^{\mu[\alpha ;\beta]}-\fracc{1}{12}g^{\mu[\alpha}R^{,\beta]}.
\end{equation}
Substituting the Einstein equations, we get
\begin{equation}
\label{bianchi_T}
W^{\alpha\beta\mu\nu}{}_{;\nu}=-\fracc{1}{2}T^{\mu[\alpha;\beta]}+\fracc{1}{6}g^{\mu[\alpha}T^{,\beta]}.
\end{equation}
From practical analysis, it is worth to project these equations with respect to a vector field $V^{\alpha}$ and its orthogonal hyper-surface of spatial metric $h_{\mu\nu}$. There are four possibilities to do such decomposition and, therefore, four linearly independent equations:

\begin{itemize}
\item{Projection $V_{\beta}V_{\mu}h_{\alpha}{}^{\sigma}$ gives
\begin{equation}
\label{quase_max1}
\begin{array}{l}
h^{\epsilon\alpha}h^{\lambda\gamma}E_{\alpha\lambda;\gamma}+\eta^{\epsilon}{}_{\beta\mu\nu}V^{\beta}H^{\nu\lambda}\sigma^{\mu}{}_{\lambda}+3H^{\epsilon\nu}\omega_{\nu}=\fracc{1}{3}h^{\epsilon\alpha}\rho_{,\alpha}+\fracc{\theta}{3}q^{\epsilon}+\\[1ex]
-\fracc{1}{2}(\sigma^{\epsilon}{}_{\nu}-3\omega^{\epsilon}{}_{\nu})q^{\nu}+\fracc{1}{2}\pi^{\epsilon\mu}a_{\mu}+\fracc{1}{2}h^{\epsilon\alpha}\pi_{\alpha}{}^{\nu}{}_{;\nu};
\end{array}
\end{equation}}
\item{Projection $\eta^{\sigma\lambda}{}_{\alpha\beta}V_{\mu}V_{\lambda}$ yields
\begin{equation}
\label{quase_max2}
\begin{array}{l}
h^{\epsilon\alpha}h^{\lambda\gamma}H_{\alpha\lambda;\gamma}-\eta^{\epsilon}{}_{\beta\mu\nu}V^{\beta}E^{\nu\lambda}\sigma^{\mu}{}_{\lambda}-3E^{\epsilon\nu}\omega_{\nu}=(\rho+p)\omega^{\epsilon}-\fracc{1}{2}\eta^{\epsilon\alpha\beta\lambda}V_{\lambda}q_{\alpha;\beta}+\\[1ex]
+\fracc{1}{2}\eta^{\epsilon\alpha\beta\lambda}(\sigma_{\mu\beta}+\omega_{\mu\beta})\pi^{\mu}{}_{\alpha}V_{\lambda};
\end{array}
\end{equation}}
\item{Projection $h_{\mu}{}^{(\sigma}\eta^{\tau)\lambda}{}_{\alpha\beta}V_{\lambda}$ gives
\begin{equation}
\label{quase_max3}
\begin{array}{l}
h_{\mu}{}^{\epsilon}h_{\nu}{}^{\lambda}\dot H^{\mu\nu}+\theta H^{\epsilon\lambda}-\fracc{1}{2}H_{\nu}{}^{(\epsilon}h^{\lambda)}{}_{\mu}V^{\mu;\nu}+\eta^{\lambda\nu\mu\gamma}\eta^{\epsilon\beta\tau\alpha}V_{\mu}V_{\tau}H_{\alpha\gamma}\theta_{\nu\beta}+\\[1ex]
-a_{\alpha}E_{\beta}{}^{(\lambda}\eta^{\epsilon)\gamma\alpha\beta}V_{\gamma}+\fracc{1}{2}E_{\beta}{}^{\mu}{}_{;\alpha}h_{\mu}^{(\epsilon}\eta^{\lambda)\gamma\alpha\beta}V_{\gamma}=-\fracc{3}{4}q^{(\epsilon}\omega^{\lambda)}+\fracc{1}{2}h^{\epsilon\lambda}q^{\mu}\omega_{\mu}+\\[1ex]
+\fracc{1}{4}\sigma_{\beta}{}^{(\epsilon}\eta^{\lambda)\alpha\beta\mu}V_{\mu}q_{\alpha}
+\fracc{1}{4}h^{\nu(\epsilon}\eta^{\lambda)\alpha\beta\mu}V_{\mu}\pi_{\nu\alpha;\beta};
\end{array}
\end{equation}}
\item{Projection $V_{\beta}h_{\mu(\tau}h_{\sigma)\alpha}$ yields
\begin{equation}
\label{quase_max4}
\begin{array}{l}
h_{\mu}{}^{\epsilon}h_{\nu}{}^{\lambda}\dot E^{\mu\nu}+\theta E^{\epsilon\lambda}-\fracc{1}{2}E_{\nu}{}^{(\epsilon}h^{\lambda)}{}_{\mu}V^{\mu;\nu}+\eta^{\lambda\nu\mu\gamma}\eta^{\epsilon\beta\tau\alpha}V_{\mu}V_{\tau}E_{\alpha\gamma}\theta_{\nu\beta}+\\[1ex]
+a_{\alpha}H_{\beta}{}^{(\lambda}\eta^{\epsilon)\gamma\alpha\beta}V_{\gamma}-\fracc{1}{2}H_{\beta}{}^{\mu}{}_{;\alpha}h_{\mu}^{(\epsilon}\eta^{\lambda)\gamma\alpha\beta}V_{\gamma}=\fracc{1}{6}h^{\epsilon\lambda}(q^{\mu}{}_{;\mu}-q^{\mu}a_{\mu}-\pi^{\mu\nu}\sigma_{\mu\nu})+\\[1ex]
-\fracc{1}{2}(\rho+p)\sigma^{\epsilon\lambda}+\fracc{1}{2}q^{(\epsilon}a^{\lambda)}-\fracc{1}{4}h^{\mu(\epsilon}h^{\lambda)\alpha}q_{\mu;\alpha}+\fracc{1}{2}h_{\alpha}{}^{\epsilon}h_{\mu}{}^{\lambda}\dot\pi^{\alpha\mu}+\fracc{1}{4}\pi_{\beta}{}^{(\epsilon}\sigma^{\lambda)\beta}+\\[1ex]
-\fracc{1}{4}\pi_{\beta}{}^{(\epsilon}\omega^{\lambda)\beta}+\fracc{1}{6}\theta\pi^{\epsilon\lambda}.
\end{array}
\end{equation}}
\end{itemize}
Eqs.\ (\ref{quase_max1})-(\ref{quase_max4}) are the QM equations. Now let us show the consistence of the QM-formalism and its equivalence to the dynamics of general relativity.

\subsection{Equivalence between QM Equations and GR}\label{equiv_qm_gr}

The QM formalism is supported by the theorems that we shall mention in this section. Although our quoted references are mostly interested into its formal aspects, we shall focus here on the physical results.

Following the same steps given by Lichnerowicz (1960) in order to prove the equivalence between QM equations and GR, we start considering a manifold ${\cal M}$ with $n+1$ dimensions endowed with a hyperbolic metric $g_{\mu\nu}$ satisfying Einstein's equations. Suppose the existence of a hyper-surface $\Sigma$ with local equation $f(x^{\alpha})=0$. One assumes that the discontinuity of the derivatives of $g_{\mu\nu}$ when it crosses $\Sigma$ is provided by Hadamard's conditions, i.e.

\begin{equation}
\label{had_cond}
[g_{\mu\nu,\alpha,\beta}]_{\Sigma}=a_{\mu\nu}f_{,\alpha}f_{,\beta},
\end{equation}
where the amplitude of the discontinuities $a_{\mu\nu}$, under local coordinates, transforms as

\begin{equation}
\label{coor_trans_a_mu_nu}
a'_{\alpha\beta}=J^{\mu}_{\alpha}J^{\nu}_{\beta}(a_{\mu\nu}+t_{(\mu}l_{\nu)}),
\end{equation}
where $l_{\mu}\equiv f_{,\mu}$. The discontinuity of the Jacobian matrix $J^{\mu}_{\alpha}$ is defined by

\begin{equation}
\label{had_cond_jac_mat}
[J^{\mu}_{\alpha,\beta,\gamma}]_{\Sigma}=t^{\mu}t_{\alpha}l_{\beta}l_{\gamma},
\end{equation}
where $t_{\alpha}$ is an arbitrary vector.

As a consequence, it follows that the discontinuity relations for the Riemann tensor is

\begin{equation}
\label{had_cond_rie}
[R_{\alpha\beta\mu\nu}]_{\Sigma}=\fracc{1}{2}(a_{\alpha\mu}l_{\beta}l_{\lambda}+a_{\beta\lambda}l_{\alpha}l_{\mu}-a_{\alpha\lambda}l_{\beta}l_{\mu}-a_{\beta\mu}l_{\alpha}l_{\lambda}),
\end{equation}
and for the Ricci tensor is

\begin{equation}
\label{had_cond_ricci}
[R_{\alpha\beta}]_{\Sigma}=\fracc{1}{2}g^{\rho\sigma}(a_{\alpha\rho}l_{\beta}l_{\sigma}+a_{\beta\sigma}l_{\alpha}l_{\rho}-a_{\rho\sigma}l_{\beta}l_{\alpha}-a_{\alpha\beta}l_{\rho}l_{\sigma}).
\end{equation}

The validity of Einstein's equations for an empty space-time on $\Sigma$ implies that $[R_{\mu\nu}]_{\Sigma}=0$, if and only if, the null vector $l^{\alpha}$ is an eigenvector of the matrix $a_{\mu\nu}$, that is

\begin{equation}
\label{rel_a_l}
a_{\alpha\beta}l^{\beta}=\fracc{a}{2}l_{\alpha},
\end{equation}
where $a\equiv g^{\mu\nu}a_{\mu\nu}$. This result shows that the coefficients of discontinuity are not arbitrary. Now let us analyze the discontinuity relation imposed by the Bianchi identity for the Riemann tensor. First of all, consider $f(x^{\alpha})=0$ as being the local equation of the hyper-surface $\Sigma$. From the definition of $l_{\alpha}$, it has null vorticity, i.e., $l_{[\alpha;\beta]}=0$. The cyclic identity applied to the discontinuity of the Riemann tensor implies that

\begin{equation}
\label{cyc_dis_rie1}
l_{\rho}[R_{\alpha\beta\lambda\mu}]+l_{\lambda}[R_{\alpha\beta\mu\rho}]+l_{\mu}[R_{\alpha\beta\rho\lambda}]=0,
\end{equation}
and

\begin{equation}
\label{cyc_dis_rie2}
l_{\rho}[R^{\rho}{}_{\mu\alpha\beta}]=0.
\end{equation}
The covariant derivative of Eq.\ (\ref{cyc_dis_rie1}) yields

\begin{equation}
\label{cov_der_cyc_dis_rie}
(l_{\rho}[R_{\alpha\beta\lambda\mu}]+l_{\lambda}[R_{\alpha\beta\mu\rho}]+l_{\mu}[R_{\alpha\beta\rho\lambda}])_{;\nu}=0.
\end{equation}
Making the contraction between the indices $\rho$ and $\nu$, and then, considering that the Einstein equations for an empty space-time ($R_{\mu\nu}=0$) hold on $\Sigma$, we obtain

\begin{equation}
\label{rel_diff_rie}
2l^{\rho}[R_{\alpha\beta\lambda\mu}]_{;\rho}+l^{\rho}{}_{;\rho}[R_{\alpha\beta\lambda\mu}]=0.
\end{equation}
It means that if $[R_{\alpha\beta\lambda\mu}]$ vanishes at some point $x$ of $\Sigma$, then it vanishes along the whole isotropic geodesic passing by $x$.

In the general case, where $l^{\alpha}$ is considered either space-type or null-like, we have the following result: Let $\Omega$ be an oriented hyper-surface in the space-time intersecting $\Sigma(x^0=0)$ and defining a 2-surface $U$. If one gives Cauchy's data $g_{\mu\nu}$ and $g_{\mu\nu,\lambda}$ on $\Omega$, such that crossing on $\Sigma$ the second derivatives admit the discontinuities $[g_{\alpha\beta,00}]=(a_{\alpha\beta})_{\Sigma}$, then $a_{\alpha\beta}$ on $U$ must satisfy the condition

\begin{equation}
\label{con_a_u}
\left(a_{\mu\nu}-\fracc{a}{2}g_{\mu\nu}\right)l^{\nu}\Big|_{U}=0.
\end{equation}
Equivalently, if one gives for all points $x$ of $U$ the tensor $[R_{\alpha\beta\mu\nu}]_{U}$ admitting as fundamental vector $(l^{\mu})_{U}$, which contracted to the Riemann tensor is zero, then the Cauchy data considered correspond to the solution of Einstein's equations such that the curvature tensor, when crossing $\Sigma$, admits a discontinuity $[R_{\alpha\beta\mu\nu}]$. The tensor $[R_{\alpha\beta\mu\nu}]$ is necessarily the solution of Eq.\ (\ref{rel_diff_rie}) corresponding to the initial data $[R_{\alpha\beta\mu\nu}]_{U}$.

The results above for non-empty Einstein's equations $G_{\mu\nu}=-kT_{\mu\nu}$ are easily proven due to the continuity of $T_{\mu\nu}$ through $\Sigma$ and can be seen in Lichnerowicz, 1960 or Novello \& Salim, 1985. From these considerations we can state the following lemma:

\begin{theorem}
Bianchi's identity together with the convenient Cauchy data represented by Eq.\ (\ref{con_a_u}) are equivalent to Einstein's equations.
\end{theorem}

This is the main result that will be used in this review. Next we apply this method of dealing with Quasi-Maxwellian formalism by analyzing some special solutions of the equations of general relativity.

\section{Particular Solutions of GR from QM Equations}\label{part_sol}

In order to present specific examples of how QM equations work to, we reproduce some known solutions of the general relativity theory using the quasi-Maxwellian framework. Our task is simplified if we use Gaussian coordinates, where $g_{0\mu}=\delta_{\mu}^0$ and, moreover, a foliation described by the observer field $V^{\mu}\equiv\delta^{\mu}_0$. This coordinate system will be used to deduce all solutions presented here.

\subsection{Schwarzschild Solution}\label{schwarz_sol}

The Schwarzschild metric in a Gaussian coordinate system takes the form:

\begin{equation}
\label{schwarz_gc_qm}
ds^2=dT^2-B(T,R)dR^2-r^2(T,R)d\Omega^2.
\end{equation}
A geodesic observer in this metric is $V^{\mu}=\delta^{\mu}_0$. The expansion coefficient $\theta$ is given by

\begin{equation}
\label{exp_diag_max}
\theta=\fracc{1}{2}\left(\fracc{\dot B}{B}+\fracc{4\dot
r}{r}\right),
\end{equation}
where $\dot Y(T,R)\equiv\partial Y/\partial T$. Using the metric\ (\ref{schwarz_gc_qm}), we write the corresponding shear tensor $\sigma^{\mu}{}_{\nu}$ and the electric part of Weyl tensor $E^{\mu}{}_{\nu}$ in the matrix form:

\begin{equation}
\label{shear_diag_max}
[\sigma^i{}_j]=f(T,R)
\left(
\begin{array}{ccc}
1&0&0\\
0&-\fracc{1}{2}&0\\
0&0&-\fracc{1}{2}
\end{array}
\right),
\end{equation}

\begin{equation}
\label{elec_diag_max}
[E^i{}_j]=g(T,R)
\left(
\begin{array}{ccc}
1&0&0\\
0&-\fracc{1}{2}&0\\
0&0&-\fracc{1}{2}
\end{array}
\right),
\end{equation}
where

\begin{equation}
\label{f_diag_max}
f(T,R)=\fracc{1}{3}\left(\fracc{\dot B}{B}-\fracc{2\dot r}{r}\right).
\end{equation}
and

\begin{equation}
\label{expl_g_schwarz}
12g(T,R)=-2\fracc{\ddot B}{B}+\fracc{\dot B^2}{B^2}-4\fracc{r''}{rB}+2\fracc{\dot r}{r}\fracc{\dot B}{B}+2\fracc{r'}{r}\fracc{B'}{B^2}+4\fracc{\ddot r}{r}-\fracc{4}{r^2}-4\fracc{\dot r^2}{r^2}+4\fracc{r'^2}{r^2B},
\end{equation}
where $Y'(T,R)\equiv\partial Y/\partial R$. The magnetic part of Weyl tensor $H_{\alpha\beta}$, the vorticity $\omega_{\alpha\beta}$ and the acceleration $a^{\mu}$ are identically zero. The set of quasi-Maxwellian equations\ (\ref{quase_max1})-(\ref{quase_max4}) reduces to the form

\begin{subequations}
\label{qm_schwarz}
\begin{eqnarray}
&&g'+3\fracc{r'}{r}g=0,\label{qm_schwarz1}\\[1ex]
&&\dot g+3\fracc{\dot r}{r}g=0\label{qm_schwarz2}
\end{eqnarray}
\end{subequations}

The evolution equations of the remaining kinematical quantities are provided by the Raychaudhuri equation and the shear evolution

\begin{subequations}
\label{kin_evol_schwarz}
\begin{eqnarray}
&&\dot\theta+\fracc{\theta^2}{3}+\fracc{3}{2}f^2=0,\label{kin_evol_schwarz1}\\
&&\dot f+\fracc{f^2}{2}+\fracc{2}{3}\theta f=-g.\label{kin_evol_schwarz2}
\end{eqnarray}
\end{subequations}
Finally, the only non-trivial remaining constraint equation is

\begin{equation}
\label{vin_schwarz}
f'+3\fracc{r'}{r}f-\fracc{2}{3}\theta'=0.
\end{equation}

This is nothing but the Schwarzschild solution in Gaussian coordinates. Indeed, functions $B(T,R)$ and $r(T,R)$ are obtained by imposing on the Cauchy surface $T=T_0$ the Lichnerowicz condition
\begin{equation}
\label{cauchy_schw_cond}
\Big(R_{\mu\nu}=0\Big)\,\Big|_{T_0},
\end{equation}
Then it follows
\begin{subequations}
\label{eins_schwarz}
\begin{eqnarray}
&&B_E(T,R)=\fracc{r'^2}{1+F(R)},\label{eins_schwarz1}\\
&&\dot r_E(T,R)=-\sqrt{F+r_H/r},\label{eins_schwarz2}\\
&&r'_E(T,R)=w'(R)\sqrt{F+r_H/r},\label{eins_schwarz3}
\end{eqnarray}
\end{subequations}
where $F(R)$ and $w(R)$ are arbitrary functions and $r_H$ is an arbitrary constant\footnote{We introduce the subscript indexes $E$ and $QM$ to distinguish the solution of the Einstein equations valid only on the Cauchy surface and the solution propagated by the Quasi-Maxwellian equations, respectively.}. These expressions play the role of initial conditions on the Cauchy surface such that the functions $B_{QM}$ and $r_{QM}$ must be equal to $B_{E}$ and $r_{E}$ on $T_0$, and then, they are evolved to the whole space-time.

From Eqs.\ (\ref{qm_schwarz}), it follows
\begin{equation}
\label{g_qm_schwarz}
g=-\fracc{k}{r^3},
\end{equation}
where $k$ is another arbitrary constant. From Eq.\ (\ref{vin_schwarz}), we have

\begin{equation}
\label{b_vinc_schwarz}
B_{QM}=\fracc{r'^2}{1+h(R)},
\end{equation}
where $h(R)$ is an arbitrary function. One can write Eqs.\ (\ref{kin_evol_schwarz}) in terms of the functions $B$ and $r$, as follows

\begin{subequations}
\label{expl_kin_evol_schwarz}
\begin{eqnarray}
&&\fracc{\ddot B}{B}-\fracc{1}{2}\fracc{\dot B^2}{B^2}-2\fracc{\ddot r}{r}=-3g,\label{expl_kin_evol_schwarz1}\\
&&\fracc{\ddot B}{B}-\fracc{1}{2}\fracc{\dot B^2}{B^2}+4\fracc{\ddot r}{r}=0.\label{expl_kin_evol_schwarz2}
\end{eqnarray}
\end{subequations}

Substituting Eq.\ (\ref{b_vinc_schwarz}) in\ (\ref{expl_kin_evol_schwarz2}) yields

\begin{equation}
\label{pre_eq_inter_dot_r_schwarz}
\fracc{\ddot r'}{\ddot r}+2\fracc{r'}{r}=0
\end{equation}
which can be integrated with respect to $R$ resulting in

\begin{equation}
\label{eq_inter_dot_r_schwarz}
\ddot r=\fracc{a(T)}{r^2},
\end{equation}
where $a(T)$ is an arbitrary function of $T$. Using the subtraction between\ (\ref{expl_kin_evol_schwarz1}) and\ (\ref{expl_kin_evol_schwarz2}), and then, manipulating Eqs.\ (\ref{b_vinc_schwarz}) and (\ref{g_qm_schwarz}), it follows that $a(T)$ is a constant ($a(T)=-k/2$). From Eq.\ (\ref{eq_inter_dot_r_schwarz}), we get

\begin{equation}
\label{eq_dot_r_schwarz}
\dot r^2_{QM}=2y(R)+\fracc{k}{r},
\end{equation}
where $y(R)$ is an arbitrary function. Substituting the constraint (\ref{b_vinc_schwarz}) in the definition of $g$---Eq.\ (\ref{expl_g_schwarz})---and again taking the subtraction between Eqs.\ (\ref{expl_kin_evol_schwarz}), we obtain

\begin{equation}
\label{eq_inter_hy}
3\fracc{\ddot r}{r}-\fracc{\dot r}{r}\fracc{\dot r'}{r'}+\fracc{1}{2}\fracc{h'}{rr'}+\fracc{\dot r^2}{r^2}-\fracc{h}{r^2}=0.
\end{equation}
Substituting\ (\ref{eq_dot_r_schwarz}) into (\ref{eq_inter_hy}) yields

\begin{equation}
\label{eq_hy}
\fracc{h}{r^2}-\fracc{2y}{r^2}=x(T),
\end{equation}
where $x(T)$ is an arbitrary function of $T$. Eq.\ (\ref{eq_dot_r_schwarz}) gives

\begin{equation}
\label{before_int_eq_dot_r_schwarz}
\int\fracc{dr}{\sqrt{h+k/r}}=\int dT,
\end{equation}
which we integrate to get

\begin{equation}
\label{int_eq_dot_r_schwarz}
\fracc{\sqrt{(hr+k)r}}{h}-\fracc{k}{2h^{3/2}}\ln{(k+2hr+2\sqrt{(hr+k)hr})}=T+b(R).
\end{equation}
Partially deriving this equation with respect to $R$, yields

\begin{equation}
\label{eq_line_r_schwarz}
r'_{QM}=b'\sqrt{h+\fracc{k}{r}}.
\end{equation}

We still must determine the arbitrary functions in\ (\ref{eins_schwarz}). In order to do this, we set

\begin{equation}
\label{crit_schwarz_qm}
F\equiv\alpha^2-1,\hspace{.5cm}w(R)\equiv\alpha R,
\end{equation}
where $\alpha$ is a constant parameter. Then, considering the Jacobian matrix $J^{\alpha}{}_{\beta}$

\begin{equation}
[J^{\alpha}{}_{\beta}]\equiv\left[\fracc{\partial x^{\alpha}}{\partial\bar x^{\beta}}\right]=
\left(
\begin{array}{cccc}
\alpha/A&-(\alpha^2-A)/A&0&0\\
-\sqrt{\alpha^2-A}&\alpha\sqrt{\alpha^2-A}&0&0\\
0&0&1&0\\
0&0&0&1
\end{array}
\right),
\end{equation}
where $x^{\alpha}=(t,r,\theta,\phi)$, $\bar x^{\beta}=(T,R,\theta,\phi)$ and $A=1-r_H/r$, we map the metric given in Gaussian coordinates by\ (\ref{schwarz_gc_qm}) into the well-known Schwarzschild solution in the usual Schwarzschild coordinates

\begin{equation}
\label{schwarz}
ds^2=\left(1-\fracc{r_H}{r}\right)dt^2-\left(1-\fracc{r_H}{r}\right)^{-1}dr^2-r^2d\Omega^2,
\end{equation}
if we make the identification $r_H=2MG/c^2$. The parameter $\alpha$ is interpreted as the mechanical energy of a test particle obtained by integration of the geodesic equations.

Due to the similar functional form of Eqs.\ (\ref{b_vinc_schwarz}), (\ref{eq_dot_r_schwarz}) and (\ref{eq_line_r_schwarz}) in comparison to Eqs.\ (\ref{eins_schwarz}), the determination of the arbitrary functions on the hyper-surface $T_0$ is trivial:

\begin{subequations}
\label{jun_fun_eins_qm_schwarz}
\begin{eqnarray}
&&B_E(T_0,R)=B_{QM}(T_0,R)\Longrightarrow h(R)=F(R),\label{jun_fun_eins_qm_schwarz1}\\
&&\dot r_E(T_0,R)=\dot r_{QM}(T_0,R)\Longrightarrow k=k_1,\label{jun_fun_eins_qm_schwarz2}\\
&&r'_E(T_0,R)=r'_{QM}(T_0,R)\Longrightarrow b(R)=w(R).\label{jun_fun_eins_qm_schwarz3}
\end{eqnarray}
\end{subequations}
Therefore, the metric\ (\ref{schwarz_gc_qm}) becomes

\begin{equation}
\label{fin_schwarz_gc_qm}
ds^2=dT^2-\left(\alpha^2-1+\fracc{2M}{r(T,R)}\right)dR^2-r^2(T,R)d\Omega^2.
\end{equation}

The Schwarzschild internal solution is more involved to be obtained from the above steps, because the Cauchy surface $T_0$ for this case is different from the spherically symmetric surface $r=r_0$---used in Schwarzschild coordinates to apply the match conditions between the internal and the external parts. Besides, Gaussian observers do not decompose the energy-momentum tensor as a perfect fluid contrary to the Schwarzschild observers $V^{\mu}=\sqrt{g_{00}}\delta^{\mu}_0$, which turn calculations more cumbersome. Indeed, the energy-momentum tensor associated to the Gaussian observers $\delta^{\mu}_0$ is

\begin{equation}
\label{ener_mom_tens_schwarz_int_gauss}
T^{(G)}_{\mu\nu}=(\rho_{G}+p_{G})V^{\mu}V^{\nu}-p_{G}g_{\mu\nu}+V_{(\mu}q_{\nu)}+\pi_{\mu\nu}.
\end{equation}
The energy-momentum tensor decomposed in terms of the observer field $u_{\mu}=e^{-\nu(T,R)}(\alpha,1,0,0)$ is given by

\begin{equation}
\label{ener_mom_tens_schwarz_int}
T_{\mu\nu}=(\rho+p)u^{\mu}u^{\nu}-pg_{\mu\nu},
\end{equation}
where $\nu=\nu(T,R)$ must satisfy the Tolman-Oppenheimer-Volkov equation and $\alpha$ is the external parameter associated to the co-moving test particle. The physical properties of the fluid in both representations are linked by the following expressions

\begin{subequations}
\label{ener_mom_gauss}
\begin{eqnarray}
&&\rho_G=(\rho+p)\alpha^2e^{-\nu}-p,\label{ener_mom_gauss1}\\[2ex]
&&p_G=-\fracc{1}{3}[(\rho+p)(1-\alpha^2e^{-\nu})-3p],\label{ener_mom_gauss2}\\[2ex]
&&q^{i}=(\rho+p)\alpha e^{-\nu}\,\delta^{i}_1,\label{ener_mom_gauss3}\\[2ex]
&&\pi^{i}{}_{j}=\fracc{2}{3}(1-\alpha^2e^{-\nu})\,\mbox{diag}(1,-1/2,-1/2).\label{ener_mom_gauss4}
\end{eqnarray}
\end{subequations}
Substituting these equations in the quasi-Maxwellian equations, one exactly find the Schwarzschild internal solution containing some arbitrary functions. Match conditions must be used in order that this solution is joined to Einstein's equations on the hyper-surface. Choosing the Cauchy surface $T=T_0$, one fixes all arbitrary functions obtaining the Schwarzschild stellar solution  with such procedure.

\subsection{Kasner Solution}

Different from the case of Schwarzschild metric, we shall use the Hadamard method in order to get the Kasner solution. We set for the Bianchi-I anisotropic metric the form

\begin{equation}
\label{an_met}
ds^2=dt^2-a^2(t)dx^2-b^2(t)dy^2-c^2(t)dz^2.
\end{equation}
In this case there is solely one non-trivial QM-equation: the ``time" evolution of the electric part of the Weyl tensor that reads

\begin{equation}
\label{el_weyl_kasn}
\dot E^{\epsilon\lambda} + \theta \, E^{\epsilon\lambda} - \fracc{3}{2} \, \sigma^{\mu(\epsilon}E^{\lambda)}{}_{\mu} + h^{\epsilon\lambda}\sigma^{\mu}{}_{\nu} \, E^{\nu}{}_{\mu}=0,
\end{equation}
where we consider an observer field $V^{\mu}\equiv\delta^{\mu}_0$ and $T_{\mu\nu}=0$.
The non-trivial equations of the kinematical quantities are

\begin{equation}
\label{raych_sig_kasn}
 \dot\theta+\fracc{\theta^2}{3}+2\sigma^2=0,
\end{equation}
and
\begin{equation}
\label{raych_sig_kasn2}
\dot\sigma_{\alpha\beta}+E_{\alpha\beta}-\fracc{2}{3}h_{\alpha\beta}\sigma^2 + \fracc{2}{3}\theta\sigma_{\alpha\beta}+\sigma_{\alpha\mu}\sigma^{\mu}{}_{\beta}=0.
\end{equation}

To obtain the Kasner solution, we set

\begin{equation}
\label{kasn_ans}
\begin{array}{l}
a(t)=t^{p_1},\hspace{.5cm} b(t)=t^{p_2},\hspace{.5cm}c(t)=t^{p_3},
\end{array}
\end{equation}
where $p_1$, $p_2$ and $p_3$ must satisfy

\begin{equation}
\label{kasn_cons}
\begin{array}{l}
p_1+p_2+p_3=1,\\[2ex]
(p_1)^2+(p_2)^2+(p_3)^2=1.
\end{array}
\end{equation}
It is necessary to consider the Kasner solution on a Cauchy hyper-surface $t=t_0$ as initial condition for the Eqs.\ (\ref{el_weyl_kasn})-(\ref{raych_sig_kasn2}). For this case, one can reduce time evolution of $E_{\mu\nu}$, $\sigma{\mu\nu}$ and $\theta$ in some algebraic constraints, that is:

\begin{equation}
\label{alg_rel_kasn}
\begin{array}{l}
\dot E^{\mu}{}_{\nu}=-2\theta E^{\mu}{}_{\nu},\\[2ex]
\dot{\sigma}^{\mu}{}_{\nu}=-\theta\sigma^{\mu}{}_{\nu},\\[2ex]
\dot{\theta}=-2\theta^2.
\end{array}
\end{equation}
Note that these expressions are valid only on the Cauchy surface $t=t_0$. Using relations\ (\ref{alg_rel_kasn}) in Eqs.\ (\ref{el_weyl_kasn}), (\ref{raych_sig_kasn}) and (\ref{raych_sig_kasn2}), these equations become just algebraic expressions for the Kasner background. As a consequence, one can define three special variables for the Kasner background:

\begin{equation}
\label{null_kasn_var}
\begin{array}{l}
X_{\alpha\beta}\equiv E_{\alpha\beta}-\fracc{2}{3}h_{\alpha\beta}\sigma^2 - \fracc{1}{3}\theta\sigma_{\alpha\beta}+\sigma_{\alpha\mu}\sigma^{\mu}{}_{\beta},\\[2ex]
Y_{\alpha\beta}\equiv\theta E_{\alpha\beta} + \fracc{3}{2}\sigma^{\mu}{}_{(\alpha}E_{\beta)}{}_{\mu} - h_{\alpha\beta}\sigma^{\mu}{}_{\nu}E^{\nu}{}_{\mu},\\[2ex]
W\equiv2\sigma^2-\fracc{2}{3}\theta^2.
\end{array}
\end{equation}
These quantities are identically zero for the Kasner solution (on $t_0$) and they represent the minimal set of variables which contains all the information about such metric, because they come from the nontrivial equations of the QM-formalism. Once $X_{\alpha\beta}$, $Y_{\alpha\beta}$ and $W$ are zero on $t_0$, the QM-equation will propagate these quantities to the hyper-surface in the vicinity of $t_0$ retaining their null values, due to their tensorial features. In other words, we obtain the validity of Kasner solution for the whole space-time, according to the theorems of Sec.\ [\ref{equiv_qm_gr}].

\subsection{Friedman Solution}\label{fried_sol}

Consider the isotropic metric given in the Gaussian coordinate system:

\begin{equation}
\label{fried_met}
ds^2\equiv dt^{2} + g_{ij} dx^{i} dx^{j}=dt^2-a^2(t)[d\chi^2-\sigma^2(\chi)d\Omega^2],
\end{equation}
in which $g_{ij} = - a^{2}(t) \gamma_{ij}(x^{k})$. The material content of this universe is represented by a perfect fluid, with energy density $\rho$, pressure $p$ and equation of state $p=\lambda\rho$, where $\lambda$ is a constant.

A direct inspection of Eqs.\ (\ref{quase_max1}) -- (\ref{quase_max4}) shows that the quasi-Maxwellian equations for this metric are identically zero, because the metric is conformally flat (we use an observer field $V^{\mu}=\delta^{\mu}_0$). The only kinematical surviving equation is the Raychaudhuri equation

\begin{equation}
\label{theta_ev_fried}
\dot\theta+\fracc{1}{3}\theta^2=-\fracc{1}{2}(1+3\lambda)\rho.
\end{equation}
The energy-momentum conservation reduces to

\begin{subequations}
\label{cons_en_mom_fried}
\begin{eqnarray}
&&\dot\rho+(\rho+p)\theta=0,\label{cons_en_mom_fried1}\\[2ex]
&&h_{\alpha}{}^{\mu}p_{,\mu}=0,\Longrightarrow p=p(t).\label{cons_en_mom_fried2}
\end{eqnarray}
\end{subequations}
Thus,

\begin{equation}
\label{theta_ev_fried_expl}
3\fracc{\ddot a}{a}=-\fracc{1}{2}(1+3\lambda)\rho,
\end{equation}
and
\begin{equation}
\label{cons_en_mom_fried_expl}
\dot\rho+3(1+\lambda)\rho\fracc{\dot a}{a}=0.
\end{equation}
Eq.\ (\ref{cons_en_mom_fried_expl}) can be integrated to yield

\begin{equation}
\label{cons_en_mom_fried_expl_int}
\rho=\rho_0a^{-3(1+\lambda)},
\end{equation}
where $\rho_0$ is a constant. Substituing\ (\ref{cons_en_mom_fried_expl_int}) in\ (\ref{theta_ev_fried}), we obtain the Friedman equation

\begin{equation}
\label{fried_eq}
\fracc{\dot a^2}{a^2}-\fracc{\epsilon}{a^2}=\fracc{1}{3}\rho,
\end{equation}
where $\epsilon$ is a constant of integration.

Finally, we can exhibit the scale factor $a(t)$ in terms of a quadrature equation

\begin{equation}
\label{scale_factor}
\int\fracc{da}{\sqrt{\rho_0 a^{-(1+3\lambda)}+3\epsilon}}=\fracc{1}{\sqrt{3}}(t-t_0),
\end{equation}
where $t_0$ is a constant of integration.

The initial conditions necessary to solve this problem are $a(t)$, $\dot a(t)$, $\ddot a(t)$ and $\sigma(r)$ on the Cauchy surface. On the other hand, we have $\lambda$, $\epsilon$, $\rho_0$ and $t_0$. Instead of specifying each initial condition of the Cauchy problem, one can equivalently fix each free parameter. It can be done if we write the Riemann, Ricci and curvature tensors in terms of the 3-geometry of the background $h_{\mu\nu}$, as:

\begin{equation}
\nonumber
\begin{array}{l}
\hat{R}_{\alpha\beta\mu\nu} = -\fracc{\epsilon}{a^{2}}\,(h_{\alpha\mu}\, h_{\beta\nu} - h_{\alpha\nu}\, h_{\beta\mu}), \\
\hat{R}_{\beta\nu} = -\fracc{2\epsilon}{a^{2}}\,h_{\beta\nu}, \\
\hat{R} = -\fracc{6\epsilon}{a^{2}},
\end{array}
\end{equation}
where we use the following relation that holds for the 3-geometry:

\begin{equation}
\nonumber
\hat{\nabla}_{\alpha}\, \hat{\nabla}_{\beta}\, \hat{X}_{\gamma} - \hat{\nabla}_{\beta}\, \hat{\nabla}_{\alpha}\, \hat{X}_{\gamma} = - {\hat{R}^{\lambda}}\mbox{}_{\gamma\beta\alpha}\, \hat{X}_{\lambda}, \\
\end{equation}
The symbol (\,$\hat{}$\,) means a projection on the hyper-surface defined by $h_{\mu\nu}$.

The explicit expression of $\hat{R} $ is obtained from the Friedman metric as follows

\begin{equation}
\label{tri_curv}
-6\epsilon\equiv \hat R a^2=-4\fracc{\sigma''}{\sigma}+\fracc{2}{\sigma^2}-2\fracc{\sigma'^2}{\sigma^2}.
\end{equation}

The only three possible solutions for this equation are listed in Table \ref{tab}, which joins the solutions of Friedmann equation for different values of $\lambda$ and $\epsilon$. The constant $a_0$, which is written in terms of $\rho_0$ and $t_0$, has different values for each solution in the table and it is commonly interpreted as the ``current size of the Universe".

\begin{table}[htb]
\centering
\begin{tabular}{|c|c|c|c|c|c|}
\hline
& & & & & \\[1ex]
$\rho$ & $\lambda$ & $\epsilon$ & $\theta$ & $a(\eta)$ & $t(\eta)$\\[1ex]
& & & & & \\[1ex]
\hline \hline
& & & & & \\[1ex]
$\fracc{4}{3}t^{-2}$ & $0$ & $0$ & $2t^{-1}$ & $a_0\eta^{2/3}$ & $\eta$\\[1ex] \hline
& & & & & \\[1ex]
$\fracc{3}{4}t^{-2}$ & $1/3$ & $0$ & $\fracc{3}{2}t^{-1}$ & $a_0\eta^{1/3}$ & $\eta$\\[1ex] \hline
& & & & & \\[1ex]
$\fracc{6}{a_0^2}(1-\cos\eta)^{-3}$ & $0$ & $1$ & $\fracc{3}{a_0}\fracc{\sin\eta}{(1-\cos\eta)^2}$ & $a_0(1-\cos\eta)$ & $a_0(\eta-\sin\eta)$\\[1ex] \hline
& & & & & \\[1ex]
$\fracc{3}{a^2_0}\fracc{1}{\sin^4\eta}$ & $1/3$ & $1$ & $\fracc{3}{a^2_0}\fracc{\cos\eta}{\sin^2\eta}$ & $a_0\sin\eta$ & $a_0(1-\cos\eta)$\\[1ex] \hline
& & & & & \\[1ex]
$\hspace{.3cm}\fracc{6}{a_0^2}(\cosh\eta-1)^{-3}$ \hspace{.3cm} & \hspace{.3cm} $0$ \hspace{.3cm} & \hspace{.3cm} $-1$ \hspace{.3cm} & \hspace{.3cm} $\fracc{3}{a_0}\fracc{\sinh\eta}{(\cosh\eta-1)^2}$ \hspace{.3cm} & \hspace{.3cm} $a_0(\cosh\eta-1) $ \hspace{.3cm} & \hspace{.3cm} $a_0(\sinh\eta-\eta)$\hspace{.3cm} \\ [1ex] \hline
& & & & & \\[1ex]
$\fracc{3}{a^2_0}\fracc{1}{\sinh^4\eta}$ & $1/3$ & $-1$ & $\fracc{3}{a^2_0}\fracc{\cosh\eta}{\sinh^2\eta}$ & $a_0\sinh\eta$ & $a_0(\cosh\eta-1)$\\[1ex]
& & & & & \\\hline
\end{tabular}
\caption{Fundamental quantities of Friedman Universe. (Units system k=c=1.)}
\label{tab}
\end{table}

\subsection{Nonsingular Solutions}

There are many proposals of cosmological solutions without a primordial singularity. Such models are based on a variety of distinct mechanisms, such as cosmological constant, non-minimal couplings, nonlinear Lagrangians involving quadratic terms in the curvature, modifications of the geometric structure of space-time, non-equilibrium thermodynamics, among others---cf. de Sitter (1917), Murphy (1973), Novello and Salim (1979), Salim and de Olivera (1988), Mukhanov and Brandenberger (1992), Mukhanov and Sornborger (1993), Novello et al. (1993), Moessner and Trodden (1995), Saa et al. (2000). Recently, an inhomogeneous and anisotropic nonsingular model for the early universe filled with a Born-Infeld-type nonlinear electromagnetic field was presented by Garcia-Salcedo and Breton (2000). Further investigations on regular cosmological solutions can be found in Klippert et al. (2000), Veneziano (2000) or Ac\'acio de Barros et al. (1998). A complete listing of nonsingular solutions can be seen at Novello and Bergliaffa (2008). Here we shall analyze some of these examples presented in the literature--- cf. Refs. \cite{art_marcela} and \cite{lorenci}, for instance---using the Quasi-Maxwellian formalism.

\subsubsection{A WIST Model}

In the Weyl integrable space–time model (WIST)---cf. Novello et al. (1993), Salim and Sautu (1996) and Fabris et al. (1998)---as well as in string theory (Gasperini, 2003), there are models that describe the geometry $g_{\mu\nu}$ coupled to a scalar field. In those models, there are nonsingular solutions for an FLRW geometry. In order to search for a simple bounce scenario in cosmology, described by an analytical exact solution, we fix our attention on the background discussed by Novello et al. (1993), Salim and Sautu (1996) and Oliveira et al. (1997). Basically, the model concerns a modified Riemannian geometry with metric $g_{\mu\nu}$ and an extra Weyl affinity given by

\begin{equation}
\nonumber
\Gamma^{\mu}_{\alpha\beta}=\left\{{}^{\mu}_{\alpha\beta}\right\} + \fracc{1}{2}(\delta^{\mu}_{\alpha}\omega_{,\beta}+\delta^{\mu}_{\beta}\omega_{,\alpha}-g_{\alpha\beta}\omega^{,\mu}).
\end{equation}

In the Weyl manifold, the vacuum field equations can be rewritten in terms of a Riemannian geometry plus a term dependent of the Weyl field $\omega$ included in the non-metric part of the affinity. At this level, the field equations can be represented by a perfect fluid with the $4$-velocity given by $V_{\mu} = \partial_{\mu}\omega/ \omega^2$, where $\omega^2\equiv g^{\alpha\beta}\omega_{,\alpha}\omega_{,\beta}$ and equation of state $p = \rho$. Originally, the scalar field is part of the affinity. However, it is transposed to the right side of the field equations and it can be interpreted as a perfect fluid. In this case, its effective energy density appears as a negative quantity. The quasi-Maxwellian equations of motion of the background written in the conformal time are

\begin{equation}
\label{weyl_eq1}
(a')^2+\epsilon a^2 + \fracc{\lambda^2}{6}(w'a)^2 = 0 ,
\end{equation}
and
\begin{equation}
\label{weyl_eq2}
w'= \gamma a^{-2},
\end{equation}
where $\gamma$ is a constant and $\lambda^2$ is the coupling constant between the scalar field and the metric tensor. It follows from these equations that

\begin{equation}
\label{res_weyl12}
(a')^2 = -\epsilon a^2 - \fracc{a_{0}^2}{a^2}\,,
\end{equation}
where we defined $a_{0}^2=\lambda\gamma/\sqrt{6}$. Only solutions with three curvature $\epsilon = -1$ are possible. The scale factor, solution to\ (\ref{weyl_eq1}), is given by

\begin{equation}
\label{sol_weyl_eq1}
a(\eta) = a_o\sqrt{\cosh(2\eta + \delta)},
\end{equation}
where $\delta$ is a constant of integration. The scalar factor displays a bounce produced by the scalar field that was introduced as the Weyl part of the affinity, due to the nonmetricity condition.

\subsubsection{Nonsingular Solution From Nonlinear Electrodynamics}\label{non_sol_non_elec}

The standard cosmological model, based on Friedman-Lema\^itre-Robertson-Walker (FLRW) geometry with Maxwell electrodynamics as its source, leads to a cosmological singularity at a finite time in the past as seen in Sec.\ [\ref{fried_sol}]. Such a mathematical singularity itself shows that, around the point of maximum condensation, the curvature and the energy density are arbitrarily large, thus being beyond the domain of applicability of the model. This difficulty raises also secondary problems, such as the horizon problem: the Universe seems to be very homogeneous over scales which approach its causally correlated region, as pointed out by Brandenberger (1996). These secondary problems are usually solved by introducing geometric scalar fields (for a review on this approach see Kofman et al. (1997) and references therein).

This section shall present that homogeneous and isotropic nonsingular FLRW solutions that are obtained by considering a toy model generalization of Maxwell electrodynamics. Here it is presented as a local covariant and gauge-invariant Lagrangian which depends on the field invariants up to the second order, as a source of classical Einstein equations. This modification is expected to be relevant when the fields reach high values, as occurs in the primeval era of the Universe. Consequences of the inevitability of the singularity through the singularity theorems (see. Hawking and Ellis, 1973) are circumvented by the appearance of a high (nevertheless finite) negative pressure in the early phase of FLRW geometry. The influence of other kinds of matter on the evolution of the universe were also taken into account. It is shown that standard matter, even in its ultra-relativistic state, is unable to modify the regularity of the obtained solution.

Heaviside non-rationalized units are used. The volumetric spatial average of an arbitrary quantity $X$ for a given instant of time $t$ is defined as

\begin{equation}
\label{spa_aver}
\big<X\big>\equiv\lim_{V\rightarrow V_o}\fracc{1}{V}\int \sqrt{-g}\,d^3x^i\,X ,
\end{equation}
where $V=\int\sqrt{-g}\,d^3x^i$, and $V_o$ stands for the time dependent volume of the whole space. An extended discussion about averages in cosmological models can be seen in Refs. \cite{gasp12,gasp11,wilt11,wilt12}.

\underline{{\bf Average Process:}} since the spatial sections of FLRW geometry are isotropic, electromagnetic fields can generate such a universe only if an averaging procedure is performed---cf. Tolman and Ehrenfest (1930), Hindmarsh and Everett (1998). The standard way to do this is just to set the following mean values for the electric $E_i$ and magnetic $H_i$ fields:

\begin{equation}
\label{aver_fie_1}
\big<E_i\big>=0,\hspace{1cm} \big<H_i\big>=0, \hspace{1cm} \big<E_iH_j\big>=0,
\end{equation}

\begin{equation}
\label{aver_fie_2}
\big<E_iE_j\big>=-\fracc{1}{3}E^2g_{ij},
\end{equation}

\begin{equation}
\label{aver_fie_3}
\big<H_iH_j\big>=-\fracc{1}{3}H^2g_{ij}.
\end{equation}

The energy-momentum tensor associated with Maxwell Lagrangian is given by

\begin{equation}
\label{ener_mon_tens_max}
T_{\mu\nu}=F_{\mu}{}^{\alpha}F_{\alpha\nu}+\fracc{1}{4}Fg_{\mu\nu},
\end{equation}
in which $F\equiv F_{\mu\nu}F^{\mu\nu}=2(H^2-E^2)$. Using the above average values, it follows that Eq.\ (\ref{ener_mon_tens_max}) reduces to a perfect fluid configuration with energy density $\rho_{\gamma}$ and pressure $p_{\gamma}$ as

\begin{equation}
\label{aver_ener_mon_tens_max}
\big<T_{\mu\nu}\big>=(\rho_{\gamma}+p_{\gamma})V_{\mu}V_{\nu}-p_{\gamma}g_{\mu\nu},
\end{equation}
where

\begin{equation}
\label{rho_p_rel_max}
\rho_{\gamma}=3p_{\gamma}=\fracc{1}{2}(E^2+H^2).
\end{equation}
From the Raychaudhuri equation, we can see that the singular nature of FLRW universes comes from the fact that both the energy density and the pressure are positive definite for all time. Thus Einstein equations for the above energy-momentum configuration yield

\begin{equation}
\label{sc_fac_rad}
a(t)=\sqrt{a^2_ot-\epsilon t^2},
\end{equation}
where $a_o$ is an arbitrary constant.

\underline{{\bf Nonsingular FLRW Universes:}} Nonlinear generalization of Maxwell electromagnetic Lagrangian will be considered up to second order terms in the field invariants $F$ and $G\equiv\frac{1}{2}\eta_{\alpha\beta\mu\nu}F^{\alpha\beta}F^{\mu\nu}=-4(\vec E\cdot \vec H)$ as

\begin{equation}
\label{gen_lag_max}
L=-\fracc{1}{4}F+\alpha F^2+\beta G^2,
\end{equation}
where $\alpha$ and $\beta$ are arbitrary constants\footnote{If we consider that the origin of these corrections comes from quantum fluctuations, then the value of the constants $\alpha$ and $\beta$ are fixed---see Heisenberg and Euler (1936).}. Maxwell electrodynamics can be formally obtained from Eq.\ (\ref{gen_lag_max}) by setting $\alpha=\beta=0$. Alternatively, it can also be dynamically obtained from the nonlinear theory in the limit of small fields. The energy-momentum tensor for arbitrary nonlinear electromagnetic theories reads

\begin{equation}
\label{gen_ener_mom_tens_max}
T_{\mu\nu}=-4L_FF_{\mu}{}^{\alpha}F_{\alpha\nu}+(GL_G- L)g_{\mu\nu},
\end{equation}
in which $L_F$ represents the partial derivative of the Lagrangian with respect to the invariant $F$ and similarly for the invariant $G$. In the linear case, expression\ (\ref{gen_ener_mom_tens_max}) reduces to the usual form\ (\ref{ener_mon_tens_max}).

Since we are mainly interested in the analysis of the behavior of this system in the early universe, where matter should be identified with a primordial plasma -- for instance, Tajima et al. (1992), Giovannini and Shaposhnikov (1998), and Campos and Hu (1998)---we are led to limit our considerations to the case in which only the average of the squared magnetic field $H^2$ survives as was also done by Tajima et al. (1992), Dunne (1997), Joyce and Shaposhnikov (1997), Giovannini and Shaposhnikov (1998), Dunne and Hall (1998). This is formally equivalent to put $E^2=0$ in Eq.\ (\ref{aver_fie_2}), and physically means to neglect bulk viscosity terms in the electric conductivity of the primordial plasma.

The homogeneous Lagrangian\ (\ref{gen_lag_max}) requires some spatial averages over large scales, as given by Eqs.\ (\ref{aver_fie_1}) – (\ref{aver_fie_3}). If one intends to make similar calculations on smaller scales, then either more involved non homogeneous Lagrangians should be used or some additional magneto-hydrodynamical effect introduced, as were done by Thompson and Blaes (1998), and Subramanian and Barrow (1998), should be devised in order to achieve correlation at the desired scale (see Jedamzik et al., 1998). Since the average procedure is independent from the equations of the electromagnetic field, we can use the above formulas\ (\ref{aver_fie_1})–(\ref{aver_fie_3}) to arrive at a counterpart of expression\ (\ref{aver_ener_mon_tens_max}) for the non-Maxwellian case. The average energy-momentum tensor is identified as a perfect fluid\ (\ref{aver_ener_mon_tens_max}) with modified expressions for the energy density $\rho_{\gamma}$ and pressure $p_{\gamma}$ as

\begin{equation}
\label{rho_h0}
\rho_{\gamma}=\fracc{1}{2}H^2(1-8\alpha H^2),
\end{equation}
and
\begin{equation}
\label{p_h0}
p_{\gamma}=\fracc{1}{6}H^2(1-40\alpha H^2).
\end{equation}
Inserting expressions\ (\ref{rho_h0}) and (\ref{p_h0}) in the continuity equation for a Friedman model\ (\ref{cons_en_mom_fried1}) yields

\begin{equation}
\label{h_sca_fac}
H=\fracc{H_o}{a^2},
\end{equation}
where $H_o$ is a constant. With this result, a similar procedure applied to Eq.\ (\ref{fried_eq}) leads to

\begin{equation}
\label{gen_fried_eq}
\dot a^2=\fracc{kH^2_o}{6a^2}\left(1-\fracc{8\alpha H^2_o}{a^4}\right)-\epsilon,
\end{equation}
where $k$ is the Einstein constant. As far as the right-hand side of Eq.\ (\ref{gen_fried_eq}) must not be negative, it follows that, regardless of the value of $\epsilon$, for $\alpha>0$ the scale factor $a(t)$ cannot be arbitrarily small.

The solution of Eq.\ (\ref{gen_fried_eq}) is implicitly given as

\begin{equation}
\label{sol_gen_fried_eq}
ct=\pm \int^{a(t)}_{a_o}\fracc{dz}{\sqrt{\fracc{kH^2_o}{6z^2}-\fracc{8\alpha kH^4_o}{6z^6}-\epsilon}},
\end{equation}
where $a(0)=a_o$. The linear case\ (\ref{sc_fac_rad}) can be achieved from Eq.\ (\ref{sol_gen_fried_eq}) by setting $\alpha=0$.

A closed form of Eq.\ (\ref{sol_gen_fried_eq}) for $\epsilon=\pm1$ can be derived as

\begin{figure}[!htb]
\centering
\includegraphics[width=8cm,height=8cm]{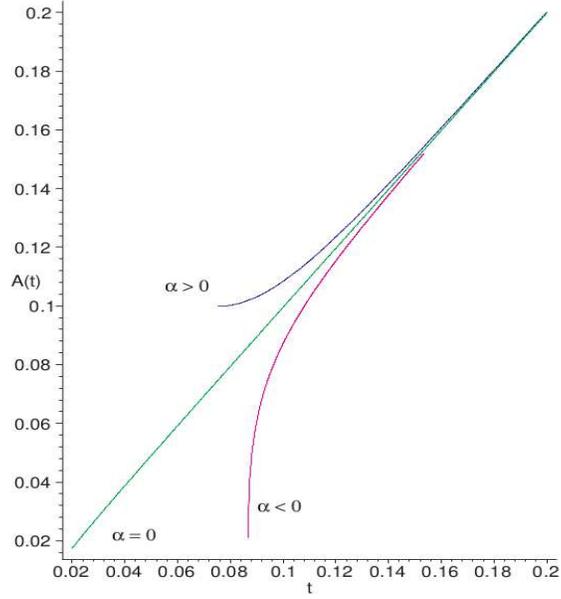}
\caption{Plots from Eq.\ (\ref{sol_gen_fried_eq}). We set $a(1)=1$, $kH^2_o=12$, and $\alpha H^2_o=(0;\pm1,25.10^{-4})$ as illustrative values.}
\label{fig1}
\end{figure}

{\small \begin{equation}
\label{clos_sol_gen_fried_eq}
ct=\pm\left[\fracc{(x_1-x_3){\cal E}\left(\arcsin{\sqrt{\fracc{z-x_1}{x_2-x_1}}},\sqrt{\fracc{x_1-x_2}{x_1-x_3}}\right)+x_3{\cal F}\left(\arcsin{\sqrt{\fracc{z-x_1}{x_2-x_1}}},\sqrt{\fracc{x_1-x_2}{x_1-x_3}}\right)}{\sqrt{x_1-x_3}}\right]\Big|^{z=a^2(t)}_{z=a^2_o},
\end{equation}}
where $x_1$, $x_2$, $x_3$ are the three roots of the equation $8\alpha kH^4_o-kH^2_ox+3\epsilon x^3=0$, and

\begin{equation}
\label{ellyp_func}
{\cal F}(x,\kappa)\equiv\int^{\sin x}_{0}\fracc{dz}{\sqrt{(1-z^2)(1-\kappa^2z^2)}},\hspace{1cm} {\cal E}(x,\kappa)\equiv\int^{\sin x}_{0}\sqrt{\fracc{1-\kappa^2z^2}{1-z^2}}dz,
\end{equation}
are the elliptic functions of the first and second kinds, respectively (see expressions $8.111.2$ and $8.111.3$ in Gradshteyn and Ryzhik, 1965). The behavior of $a(t)$ for $\epsilon=\pm1$ is displayed in the Fig.\ (\ref{fig1}).

For the Euclidean section, by suitably choosing the origin of time, expression\ (\ref{sol_gen_fried_eq}) can be solved as

\begin{equation}
\label{eucl_sc_fac}
a^2=H_o\sqrt{\fracc{2}{3}(kc^2t^2+12\alpha)}.
\end{equation}
From Eq.\ (\ref{h_sca_fac}), the average strength of the magnetic field $H$ evolves in time as

\begin{equation}
\label{mag_ev_eucl}
H^2=\fracc{3}{2}\fracc{1}{kc^2t^2+12\alpha}.
\end{equation}

Expression\ (\ref{eucl_sc_fac}) is singular for $\alpha<0$, as there is a time $t=\sqrt{-12\alpha/kc^2}$ for which $a(t)$ is arbitrarily small. Otherwise, for $\alpha>0$ we recognize that at $t=0$ the radius of the universe attains a minimum value $a_{min}$, which is given by

\begin{equation}
\label{min_sc_fac}
a^2_{min}=H_o\sqrt{8\alpha}.
\end{equation}
Therefore, the actual value of $a_{min}$ depends on $H_o$, which turns out to be the only free parameter of the present model. The energy density $\rho_{\gamma}$ given by Eq.\ (\ref{rho_h0}) reaches its maximum value $\rho_{max}=1/64\alpha$ at the instant $t=t_c$, where

\begin{equation}
\label{rho_max}
t_c=\fracc{1}{c}\sqrt{\fracc{12\alpha}{k}}.
\end{equation}
For smaller values of $t$ the energy density decreases, vanishing at $t=0$, while the pressure becomes negative. Only for times $t\lesssim10\sqrt{\alpha/kc^2}$ the nonlinear effects are relevant for cosmological solution of the normalized scale factor, as shown in Fig. 2. Indeed, solution\ (\ref{eucl_sc_fac}) fits the standard expression\ (\ref{sc_fac_rad}) of the Maxwell case at the limit of large times.

\begin{figure}[!htb]
\centering
\includegraphics[width=15cm,height=10cm]{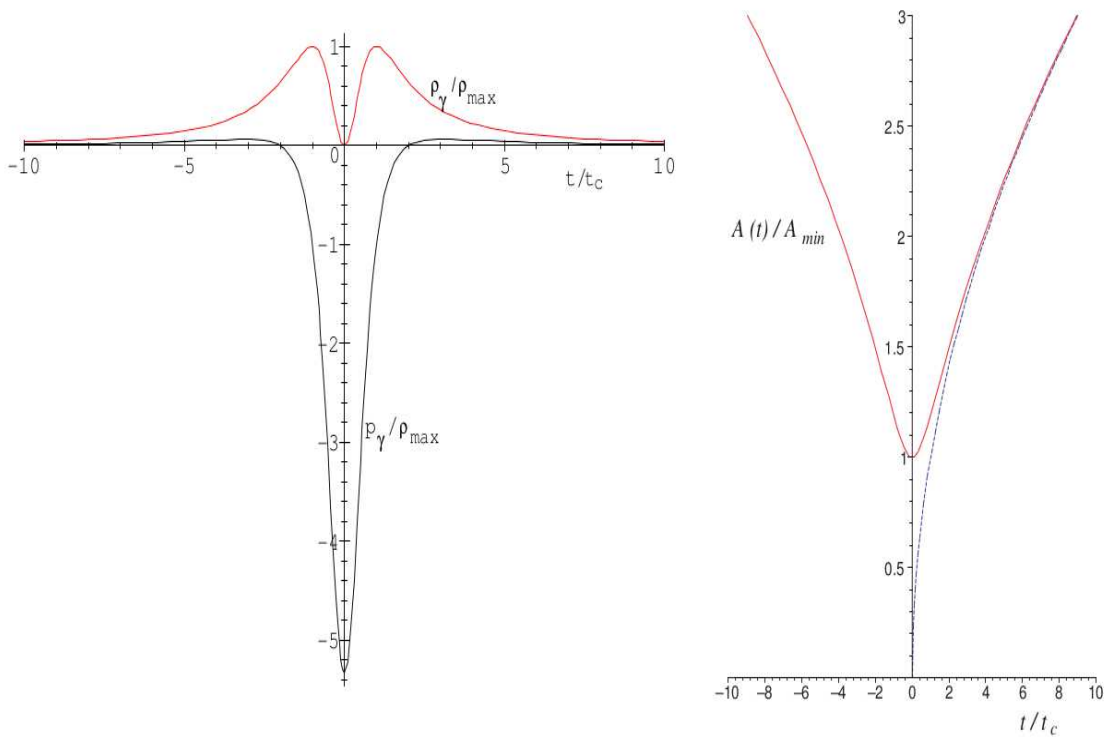}
\caption{On the left panel: time dependence of the electromagnetic energy density $\rho_{\gamma}$ and pressure $p_{\gamma}$. $\rho_{max}=1/64\alpha$ and $t_c$ is given by Eq.\ (\ref{rho_max}). On the right panel: nonsingular behavior of the scale factor $a(t)$. $a_{min}$ and $t_c$ are given from Eqs.\ (\ref{min_sc_fac}), (\ref{rho_max}). The corresponding classical expression\ (\ref{sc_fac_rad}) is shown (dashed line) for comparison, with $a_o= a_{min}$.}
\label{fig2}
\end{figure}

The energy-momentum tensor\ (\ref{gen_ener_mom_tens_max}) is not trace-free for $\alpha\neq0$. Thus, the equation of state $p_{\gamma}=p_{\gamma}(\rho_{\gamma})$ is no longer given by the Maxwellian value; it has instead a quintessential-like term---see Caldwell et al. (1998)---which is proportional to the constant $\alpha$. That is

\begin{equation}
\label{p_rho_fin}
p_{\gamma}=\fracc{1}{3}\rho_{\gamma}-\fracc{16}{3}\alpha H^4.
\end{equation}
Equation\ (\ref{p_rho_fin}) can also be written in the form

\begin{equation}
\label{rew_p_rho_fin}
p_{\gamma}=\fracc{1}{3}\rho_{\gamma}-\fracc{1}{24\alpha}\{(1 - 32\alpha\rho_{\gamma})+[1-2\Theta(t-t_c)]\sqrt{1-64\alpha\rho_{\gamma}}\},
\end{equation}
where $\Theta(z)$ is the Heaviside step function. The right-hand side of Eq.\ (\ref{rew_p_rho_fin}) behaves as $(1-64\alpha\rho_{\gamma})\rho_{\gamma}/3$ for $t>t_c$ in the Maxwell limit $\alpha\rho_{\gamma}\ll1$.

The maximum temperature corresponding to $t=t_c$ is given by

\begin{equation}
\label{t_max}
T_{max}=\left(\fracc{c}{24\alpha\sigma}\right)^{1/4},
\end{equation}
where $\sigma$ is the Stefan-Boltzmann constant.

Therefore, the consequences of the minimal coupling of gravity with second order nonlinear electrodynamics, from the cosmological point of view, propose relevant modification only in the primeval era of the universe. Indeed, the class of theories $\alpha>0$ leads to nonsingular solutions for which the scale factor $a(t)$ attains a minimum value. The regularity of this cosmological solution is to be attributed to the fact that, for some interval of time, the quantity $\rho+3p$ becomes negative.

\section{Perturbation Theory in QM Formalism}\label{pert_th}

Since the original paper of Lifshitz and Khalatnikov (1963), it has been a common practice to start the examination of the perturbation theory of Einstein's equations of general relativity by considering variations of non observable quantities such as $\delta g_{\mu\nu}$. However, the main drawback of this procedure is that it mixes true perturbations and arbitrary coordinate transformations. A solution for this difficulty was found by looking for gauge-independent combinations which are written in terms of the metric tensor and its derivatives by many authors (cf. Bardeen 1980, Hawking 1966, Jones 1976, Olson 1976, Brandenberger 1983, Vishniac 1990, and Mukhanov 1992).

The fundamental element of the gauge problem in the perturbation theory of RG was clear and geometrically detailed by Stewart's Lemma (1974, 1990): gauge invariant variables (scalars or not) are those which are identically null on the background. After that, in Stewart sense, Hawking (1966) used the QM equations to argue that the applicability of such alternative formalism of RG is restricted to the standard cosmology problem---the problem of a homogenous, isotropic and conformally flat case.

Although it is correct in the strict sense, this argument led to a disadvantage of the QM formalism compared to the other methods based on the Lifshitz program. It justifies the wide use of the complex Newmann-Penrose formalism (1962). Despite of it, we can prove that some objects of this formalism are physically unobservable.

Here we will follow a simpler and more direct path, which corresponds to choosing, from the beginning, as the basis of our analysis, the gauge-invariant physically observable quantities. The dynamics for these fundamental quantities will then be analyzed and any remaining gauge-dependent objects which one can deal with will be obtained from this fundamental set.

There are basically two fundamental approaches by which the perturbation theory can be elaborated: one of them makes use of Einstein's standard equations (Lifshitz 1963) and the other is based on the equivalent quasi-Maxwellian description (cf. Jordan 1961, Hawking 1966 and Novello 1983). In this paper, we will focus on the second approach.

\subsection{Perturbed Quasi-Maxwellian Equations}

We state here the perturbed linearized quasi-Maxwellian equations for gravity. They shall be used in the following sections to treat the dynamics of the perturbed quantities, writing all these dynamical variables in the form

\begin{equation}
\nonumber
A_{(perturbed)}=A_{(background)}+(\delta A).
\end{equation}
After straightforward manipulations, we obtain the perturbed QM equations as follows

\begin{equation}
\begin{array}{l}
{h_{\mu}}^{\alpha}{h_{\nu}}^{\beta}(\delta E^{\mu\nu})^{\bullet}+ \theta (\delta E^{\alpha\beta}) - \fracc{1}{2} (\delta{E_{\nu}}^{(\alpha}) {h^{\beta )}}_{\mu}V^{\mu ;\nu} + \fracc{\theta}{3}\eta^{\beta\nu\mu\varepsilon} \eta^{\alpha\gamma\tau\lambda} V_{\mu}V_{\tau}(\delta E_{\varepsilon\lambda}) h_{\gamma\nu}+\\[2ex]
- \fracc{1}{2} (\delta {{H_{\lambda}}^{\mu}})_{;\gamma} {h_{\mu}}^{(\alpha}\eta^{\beta )\tau\gamma\lambda}V_{\tau} = - \fracc{1}{2}(\rho + p)(\delta \sigma^{\alpha\beta}) +  \fracc{1}{6}h^{\alpha\beta} (\delta q^{\mu})_{;\mu} - \fracc{1}{4} h^{\mu (\alpha} h^{\beta )\nu}
(\delta q_{\mu})_{;\nu}+\\[2ex]
+ \fracc{1}{2} h^{\mu\alpha} h^{\beta \nu} (\delta\Pi_{\mu\nu})^{\bullet} + \fracc{1}{6}\theta (\delta\Pi^{\alpha\beta})\,,
\label{apb13}
\end{array}
\end{equation}

\begin{equation}
\begin{array}{l}
{h_{\mu}}^{\alpha}{h_{\nu}}^{\beta}(\delta H^{\mu\nu})^{\bullet} + \theta(\delta H^{\alpha\beta}) - \fracc{1}{2} (\delta {H_{\nu}}^{(\alpha})
{h^{\beta )}}_{\mu} V^{\mu ;\nu} + \fracc{\theta}{3} \eta^{\beta\nu\mu\varepsilon} \eta^{\alpha\lambda\tau\gamma} V_{\mu}V_{\tau} (\delta H_{\varepsilon\gamma}) h_{\lambda\nu}+ \\[2ex]
- \fracc{1}{2} (\delta {{E_{\lambda}}^{\mu}})_{;\tau} {h_{\mu}}^{(\alpha}\eta^{\beta )\tau\gamma\lambda} V_{\gamma} = \fracc{1}{4} h^{\nu (\alpha}
\eta^{\beta )\varepsilon\tau\mu} V_{\mu} (\delta\Pi_{\nu\varepsilon})_{;\tau}\,,
\label{apb14}
\end{array}
\end{equation}

\begin{equation}
(\delta H_{\alpha\mu})_{;\nu} h^{\alpha\varepsilon} h^{\mu\nu} = (\rho + p) (\delta\omega^{\varepsilon}) - \frac{1}{2} \hspace{0.1cm}
\eta^{\varepsilon\alpha\beta\mu} V_{\mu} \hspace{0.1cm}(\delta q_{\alpha})_{;\beta}\,,
\label{apb15}
\end{equation}
and
\begin{equation}
(\delta E_{\alpha\mu})_{;\nu} h^{\alpha\varepsilon}h^{\mu\nu} = \fracc{1}{3} (\delta\rho)_{,\alpha} h^{\alpha\varepsilon} - \fracc{1}{3}
\dot{\rho} (\delta V^{\varepsilon}) - \fracc{1}{3} \rho_{,0} (\delta V^{0}) V^{\varepsilon} + \frac{1}{2} {h^{\varepsilon}}_{\alpha} (\delta\Pi^{\alpha\mu})_{;\mu} + \frac{\theta}{3} \hspace{0.1cm}(\delta q^{\varepsilon})\,.
\label{apb16}
\end{equation}
The perturbed equations for the kinematical quantities are
\begin{equation}
(\delta\theta)^{\bullet} + \dot{\theta} (\delta V^{0}) +\frac{2}{3}\theta (\delta\theta) - (\delta {a^{\alpha}})_{;\alpha} = - \frac{(1 + 3\lambda)}{2} (\delta\rho)\,,
\label{apb17}
\end{equation}

\begin{equation}
(\delta\sigma_{\mu\nu})^{\bullet} + \frac{1}{3}h_{\mu\nu} (\delta {a^{\alpha}})_{;\alpha} - \frac{1}{2}(\delta a_{(\alpha})_{;\beta)} {h_{\mu}}^{\alpha} {h_{\nu}}^{\beta} + \frac{2}{3}\theta (\delta\sigma_{\mu\nu}) = - (\delta E_{\mu\nu}) - \frac{1}{2} (\delta\Pi_{\mu\nu})\,,
\label{apb18}
\end{equation}

\begin{equation}
(\delta\omega^{\mu})^{\bullet} + \frac{2}{3}\theta (\delta\omega^{\mu}) = \frac{1}{2} \eta^{\alpha\mu\beta\gamma} (\delta a_{\beta})_{ ;\gamma} V_{\alpha}\,,
\label{apb19}
\end{equation}

\begin{equation}
\frac{2}{3}(\delta\theta)_{,\lambda} {h^{\lambda}}_{\mu} -  \frac{2}{3}\dot{\theta} (\delta V_{\mu}) + \frac{2}{3} \dot{\theta} (\delta V^{0}) {\delta_{\mu}}^{0} - {(\delta {\sigma^{\alpha}}_{\beta} + \delta {\omega^{\alpha}}_{\beta})}_{;\alpha} {h^{\beta}}_{\mu} = - (\delta q_{\mu})\,,
\label{apb20}
\end{equation}

\begin{equation}
(\delta {\omega^{\alpha}})_{;\alpha} = 0
\label{apb21}
\end{equation}
and
\begin{equation}
(\delta H_{\mu\nu}) = - \frac{1}{2} {h^{\alpha}}_{(\mu} {h^{\beta}}_{\nu )} ((\delta\sigma_{\alpha\gamma})_{;\lambda} + (\delta\omega_{\alpha\gamma})_{;\lambda}) {\eta_{\beta}}^{\varepsilon\gamma\lambda} V_{\varepsilon}\,.
\label{apb22}
\end{equation}
The perturbed equations for the conservation of the energy-momentum tensor yields
\begin{equation}
(\delta\rho)^{\bullet} + \dot{\rho} (\delta V^{0}) + \theta (\delta\rho + \delta p) + (\rho + p) (\delta\theta) + (\delta q^{\alpha})_{;\alpha} = 0\,,
\label{apb23}
\end{equation}
and
\begin{equation}
\dot{p} (\delta V_{\mu}) + p_{,0} (\delta V^{0}) {\delta_{\mu}}^{0} - (\delta p)_{,\beta} {h^{\beta}}_{\mu} + (\rho + p) (\delta a_{\mu})+ h_{\mu\alpha} (\delta q^{\alpha})^{\bullet} +  \frac{4}{3} \theta (\delta q_{\mu}) + h_{\mu\alpha} (\delta \pi^{\alpha\beta})_{;\beta} = 0.
\label{apb24}
\end{equation}
Let us give some examples of how this method of perturbation works by considering the solutions derived last section.
\subsection{Schwarzschild Solution}
The analysis of the stability of the Schwarzschild geometry in the work of Regge and Wheeler (1957) uses the standard Lifshitz method. These authors analyze the spectral decomposition of linear perturbations in terms of two fundamental modes, called ``evens" and ``odds", and conclude affirmatively about the stability of this geometry. The main difficulties presented\footnote{At those times, everybody knows whether the divergence presented by the perturbation at the event horizon was real or only an effect caused by choice of coordinates.} were solved by Vishveshwara (1970) by means of a convenient coordinate transformation.

An important characteristic of these works, which are not conveniently emphasized in the literature, consists on the impossibility of application of the harmonic decomposition. In effect, ``even" and ``odd" modes are both obtained from the scalar derivation, where only the degrees of freedom of spin $1$ are present. Klippert (1998) has shown that it is possible to develop a systematic way to appropriately apply the QM formalism to the Schwarzschild solution, even in Stewart sense---cf. Stewart (1974). Let us review briefly this approach constructing a \textit{manifold's basis}.

\subsubsection{Construction of the Schwarzschild Basis}
The complete set of eigenfunctions of the Laplace-Beltrami operator ($\hat \nabla^2$). The manifold is a sub-manifold orthogonal to $V^{\mu}=\delta^{\mu}_0$ and we deal with the Schwarzschild geometry in Gaussian coordinates. The Laplacian is formally written as

\begin{equation}
\nonumber
\hat\nabla^2\equiv \hat\nabla^{\alpha}\hat\nabla_{\alpha}
\end{equation}
where $\hat\nabla_{\alpha}\equiv h_{\alpha}{}^{\beta}\nabla_{\beta}$ and $h_{\alpha}{}^{\beta}=\delta_{\alpha}{}^{\beta}-V_{\alpha}V^{\beta}$. The scalar component of the Schwarzschild base is a function $Q(x^{\alpha})$ such that

\begin{equation}
\label{sc_lap_schw}
\begin{array}{l}
\hat\nabla^2Q=-\lambda_sQ
\end{array}
\end{equation}
explicitly in terms of the metric\ (\ref{fin_schwarz_gc_qm}) yields

\begin{equation}
\label{exp_sc_lap_schw}
\begin{array}{l}
\fracc{1}{r^2\sqrt{\alpha^2-A}}\fracc{\partial}{\partial R}\left(\fracc{r^2}{\sqrt{\alpha^2-A}}\fracc{\partial Q}{\partial R}\right)+\fracc{1}{r^2}\left[\fracc{\partial^2Q}{\partial\theta^2}+\cot\theta\fracc{\partial Q}{\partial\theta}+\fracc{1}{\sin^2\theta}\fracc{\partial^2Q}{\partial\phi^2}\right]=\lambda Q,
\end{array}
\end{equation}
where $A(T,R)=1-2M/r(T,R)$. The angular variables will give origin to the spherical harmonics. Afterwards, we consider the particular case in which $\alpha^2=1$ to get the easier differential equation below for the basis $Q$ in terms of $T$ and $R$ coordinates

\begin{equation}
\label{exp_sc_lap_schw_a1}
\begin{array}{l}
\fracc{3}{2}(T+R)\fracc{\partial^2F}{\partial R^2}+\fracc{5}{2}\fracc{\partial F}{\partial R}+\left[\fracc{4}{3}\fracc{l(l+1)}{T+R}+\lambda\sqrt{2M}\left(-\fracc{3}{2}\sqrt{2M}(T+R)\right)^{1/3}\right]F=0,
\end{array}
\end{equation}
where it was assumed that $Q(T,R,\theta,\phi)\equiv F(T,R)Y_m^l(\theta,\phi)$ and $Y_m^l(\theta,\phi)$ are the spherical harmonics. We can completely integrate the equation above to obtain the general solution in terms of Bessel functions, as follows

\begin{equation}
\label{sol_exp_sc_lap_schw_a1}
F(T,R)=\fracc{1}{(T+R)^{1/3}}\left[a\,J_{\alpha}(x) + b\,Y_{\alpha}(x)\right],
\end{equation}
where we defined

\begin{equation}
\nonumber
\alpha\equiv\fracc{\sqrt{1-8l-8l^2}}{2},\hspace{1cm} x\equiv\fracc{\sqrt{6c}\,(T+R)^{2/3}}{2},
\end{equation}
being $c\equiv-\lambda\sqrt{2M}[(3/2)\sqrt{2M}]^{1/3}$ and, $a$ and $b$ constants of integration. $J_{\alpha}(x)$ and $Y_{\alpha}(x)$ are Bessel functions of first and second kind, respectively.

\subsubsection{Gauge-Invariant Variables}

From the subsection\ [\ref{schwarz_sol}], the decomposition induced by $V^{\mu}=\delta^{\mu}_0$ leads to a degenerated shear tensor (two identical eigenvalues) proportional to the electric part of the Weyl tensor. We thus introduce the following geometrical objects

\begin{equation}
\label{gau_inv_var_sch}
\begin{array}{lcl}
X_{\mu\nu}\equiv\sigma_{\mu\nu}-\fracc{2\sigma^2}{\sigma^3}\sigma^{\alpha}{}_{\mu}\sigma_{\alpha\nu}+\fracc{2\sigma^2}{\sigma^3}\fracc{2\sigma^2}{3}h_{\mu\nu},\\[2ex]
Y_{\mu\nu}\equiv E_{\mu\nu}-\fracc{E^{\alpha}{}_{\beta}\sigma^{\beta}{}_{\alpha}}{2\sigma^2}\sigma_{\mu\nu},\\[2ex]
Z_{\mu\nu}\equiv H_{\mu\nu}-\fracc{H^{\alpha}{}_{\beta}\sigma^{\beta}{}_{\alpha}}{2\sigma^2}\sigma_{\mu\nu}.
\end{array}
\end{equation}
We remark that these tensors present some particular algebraic features: they are symmetric, traceless, orthogonal to shear ($X^{\alpha}{}_{\beta} \sigma^{\beta}{}_{\alpha} = Y^{\alpha}{}_{\beta} \sigma^{\beta}{}_{\alpha} = Z^{\alpha}{}_{\beta} \sigma^{\beta}{}_{\alpha}=0$) and the most important characteristic is that they are null on the background. Therefore, they constitute a set of ``good" variables (in Stewart's sense, 1974) to do perturbation theory for the Schwarzschild case.

\subsubsection{Dynamics}

Using the QM-equations, we can calculate the propagation equations of $X_{\mu\nu}$, $Y_{\mu\nu}$ and $Z_{\mu\nu}$ along to the geodesics represented by the vector field $V^{\mu}$. It is useful to rewrite the outcome in terms of these objects, to get a closed dynamical system. We restrict ourselves to the exhibition of the propagation equations for the perturbations associated to the gauge invariant variables as follows

\begin{equation}
\label{pert_x_ev_schwa}
\begin{array}{ll}
\delta\dot X_{\mu\nu}=&-\left(\fracc{4}{3}\theta+\fracc{\sigma^3}{2\sigma^2}+2\fracc{\sigma:E}{2\sigma^2}\right)\delta X_{\mu\nu}-2\sigma^{\lambda}{}_{(\mu}\delta X_{\nu)\lambda}+\delta Y_{\mu\nu}+2\fracc{2\sigma^2}{\sigma^3}\sigma^{\lambda}{}_{(\mu}\delta Y_{\nu)\lambda} +\\[2ex]
&+\left[h^{\alpha}_{\mu}h^{\beta}_{\nu} - 2\fracc{2\sigma^2}{\sigma^3}h^{\alpha}{}_{(\mu}\sigma_{\nu)}{}^{\beta} + \fracc{1}{2\sigma^2}(\sigma^{\alpha\beta} - X^{\alpha\beta})(\sigma_{\mu\nu} - X_{\mu\nu} + 2\fracc{2\sigma^2}{\sigma^3}\fracc{2\sigma^2}{3}h_{\mu\nu})\right].\\[2ex]
&.\left(\delta a_{(\alpha;\beta)} + \fracc{1}{2}\delta\Pi_{\alpha\beta}\right) + \fracc{1}{3}\delta a^{\lambda}{}_{;\lambda}\left[2\fracc{2\sigma^2}{\sigma^3}\sigma_{\mu\nu}-h_{\mu\nu}\right],
\end{array}
\end{equation}

\begin{equation}
\label{pert_y_ev_schwa}
\begin{array}{ll}
\delta\dot Y_{\mu\nu}=&-4\fracc{\sigma^3}{2\sigma^2}\fracc{\sigma:E}{2\sigma^2}X_{\mu\nu}+\left(\fracc{\sigma:E}{2\sigma^2}-\theta\right)\delta Y_{\mu\nu}+3\sigma^{\lambda}{}_{(\mu}\delta Y_{\nu)\lambda}+\left(h^{\alpha}_{\mu}h^{\beta}_{\nu} -\fracc{1}{2\sigma^2} \sigma^{\alpha\beta}\sigma_{\mu\nu}\right).\\[2ex]
&.\left[h^{\lambda}{}_{(\alpha}\eta_{\beta)}{}^{\epsilon\gamma\tau}V_{\tau}\delta Z_{\lambda\epsilon;\gamma} - \fracc{1}{2}\delta\dot\Pi_{\alpha\beta} + \left(\fracc{\theta}{3} - \fracc{\sigma:E}{2\sigma^2}\right)\delta\Pi_{\alpha\beta}+ \sigma^{\lambda}{}_{(\alpha}\delta\Pi_{\beta)\lambda}-\fracc{\sigma:E}{2\sigma^2}\delta a_{(\alpha;\beta)}\right] +\\[2ex]
&- \sigma^{\lambda}{}_{(\mu}\delta\omega_{\nu)\lambda} + \fracc{1}{6}\left(\delta q^{\lambda}_{;\lambda}+\sigma^{\alpha\beta}\delta\Pi_{\alpha\beta}+2\fracc{\sigma:E}{2\sigma^2}\delta a^{\lambda}_{;\lambda}\right)h_{\mu\nu},
\end{array}
\end{equation}
and

\begin{equation}
\label{pert_z_ev_schwa}
\begin{array}{ll}
\delta\dot Z_{\mu\nu}=&\left(h^{\alpha}_{\mu}h^{\beta}_{\nu} -\fracc{\sigma^{\alpha\beta}}{2\sigma^2}\sigma_{\mu\nu}\right)\left\{h^{\lambda}{}_{(\alpha}\eta_{\beta)}{}^{\epsilon\gamma\tau}V_{\tau}\left(\delta Y_{\lambda\epsilon;\gamma}-\fracc{1}{2}\delta\Pi_{\lambda\epsilon;\gamma} - \fracc{\sigma:E}{2\sigma^2}\delta\omega_{\lambda\epsilon;\gamma}\right)+\right.\\[2ex]
&\left.- \sigma^{\lambda}{}_{(\alpha}\eta_{\beta)\lambda\epsilon\gamma}V^{\epsilon}\left(2\fracc{\sigma:E}{2\sigma^2}\delta a^{\gamma} -\fracc{1}{2}\delta q ^{\gamma}\right) - \delta\left[\sigma_{\lambda(\alpha}\eta_{\beta)}{}^{\lambda\gamma\tau}V_{\tau}\left(\fracc{\sigma:E}{2\sigma^2}\right){}_{;\gamma}\right]\right\}+\\[2ex]
&- \left(\theta - \fracc{\sigma:E}{2\sigma^2}\right)\delta Z_{\mu\nu}+3\sigma^{\lambda}{}_{(\mu}\delta Z_{\nu)\lambda},
\end{array}
\end{equation}

These equations\ (\ref{pert_x_ev_schwa}) (\ref{pert_y_ev_schwa}) and (\ref{pert_z_ev_schwa}) are not completely independent. They must satisfy constraint equations, which rewritten in terms of the geometrical objects described above are

\begin{equation}
\label{x_div_schwa}
\begin{array}{ll}
h^{\alpha}_{\epsilon}h^{\nu}_{\mu} X^{\mu}{}_{\alpha;\nu} = & -\fracc{\sigma^3}{2\sigma^2}h^{\alpha}{}_{\epsilon}\sigma^{\mu\nu}\sigma_{\alpha\mu;\nu} + \left[\sigma^{\alpha}{}_{\epsilon} - X^{\alpha}{}_{\epsilon} + \fracc{2\sigma^2}{\sigma^3}\fracc{2\sigma^2}{3}h^{\alpha}{}_{\epsilon}\right]\fracc{(\sigma^3/2\sigma^2)_{;\alpha}}{(\sigma^3/2\sigma^2)} + h^{\alpha}{}_{\epsilon}\left(\fracc{2\sigma^2}{\sigma^3}\fracc{2\sigma^2}{3}\right){}_{;\alpha}\\[2ex]
&+\left(h^{\alpha}_{\epsilon}- 2\fracc{2\sigma^2}{\sigma^3}\sigma^{\alpha}{}_{\epsilon}\right)\left[h^{\mu\nu}\omega_{\alpha\mu;\nu} + \fracc{2}{3}\theta_{;\alpha} + (\sigma_{\alpha\lambda} + \omega_{\alpha\lambda})a^{\lambda}-q_{\alpha}\right],
\end{array}
\end{equation}

\begin{equation}
\label{y_ev_schwa}
\begin{array}{ll}
h^{\alpha}_{\epsilon}h^{\nu}_{\mu}Y^{\mu}{}_{\alpha;\nu} = & \eta_{\epsilon}{}^{\alpha\beta\gamma}V_{\gamma}\sigma^{\lambda}{}_{\alpha}Z_{\beta\lambda} - 3\left(Z_{\epsilon\lambda} + \fracc{\sigma:H}{2\sigma^2}\sigma_{\epsilon\lambda}\right)\omega^{\lambda} - \sigma^{\lambda}{}_{\epsilon}\left(\fracc{\sigma:E}{2\sigma^2}\right){}_{;\lambda}+\fracc{1}{3}h^{\alpha}_{\epsilon}\rho_{;\alpha}+\\[2ex]
& - \fracc{\sigma:E}{2\sigma^2}\left[h^{\alpha}_{\epsilon}h^{\mu\nu}\omega_{\alpha\mu;\nu} + \fracc{2}{3}h^{\alpha}_{\epsilon}\theta_{;\alpha} +(\sigma_{\epsilon\lambda} + \omega_{\epsilon\lambda})a^{\lambda}-q_{\epsilon}\right] + \\[2ex] &+\fracc{1}{2}\left[(\sigma_{\epsilon\lambda}-3\omega_{\epsilon\lambda})q^{\lambda} - h_{\alpha\epsilon}\Pi^{\alpha\beta}{}_{;\beta} + \Pi_{\epsilon\lambda}a^{\lambda}\right] - \fracc{\theta}{3}q_{\epsilon},
\end{array}
\end{equation}
and

\begin{equation}
\label{z_ev_schwa}
\begin{array}{ll}
h^{\alpha}_{\epsilon}h^{\nu}_{\mu}Z^{\mu}{}_{\alpha;\nu} = & \eta_{\epsilon}{}^{\alpha\beta\gamma}V_{\gamma}\sigma^{\lambda}{}_{\alpha}Y_{\beta\lambda} + 3\left(Y_{\epsilon\lambda} + \fracc{\sigma:E}{2\sigma^2}\sigma_{\epsilon\lambda}\right)\omega^{\lambda} + (\rho+p)\omega_{\epsilon} + \\[2ex] &+\fracc{1}{2}\eta_{\epsilon}{}^{\alpha\beta\gamma}V_{\gamma}\left[q_{\alpha;\beta} + \Pi^{\lambda}{}_{\alpha}(\sigma_{\beta\lambda}-\omega_{\beta\lambda})\right],
\end{array}
\end{equation}

Applying the Lifshitz method, we can decompose the gauge invariant variables set in terms of the basis $Q(x^{\alpha})$ and analyze the stability of the dynamical system for the scalar perturbation, but such analysis will not be pursued in this work. More details can be seen in Klippert (1998).

\subsection{Kasner Solution}

Kasner universes constitute a paradigm of the Bianchi-type I anisotropic space-times. We shall follow the procedure used last section and present a minimal closed set of gauge-independent observables for an adequate basis built for this specific background, and subsequently employed in a dynamical system that is written in the framework of the quasi-Maxwellian equations. It is then found that the method can be carried out to its end and a closed dynamical system obtained. All three types of perturbation---scalar, vectorial and tensorial---are presented and discussed in the same way, but we limit here our analysis to the scalar perturbation case.

From the seminal work of Belinsky, Khalatnikov and Lifshitz (1970) it has been shown that---for any kind of regular matter satisfying the usual energy conditions (cf. Hawking and Ellis, 1977) in the neighborhood of singularity---the Bianchi-type I Kasner solution works as an attractor for all the other solutions. In this sense these geometries are good paradigms for anisotropic models, which have been extensively analyzed in scientific literature (see Novello 1990, 1998).

The problem of stability of anisotropic cosmological models and the analysis of perturbations has also been extensively studied in the literature using the method based on the perturbations of the metric tensor and the dynamics determined by Einstein's equations by Mataresse (1993-5), Miedema (1994), Noh (1995), Ib\'a\~nez (1995) and Mutoh (1997).

In order to apply the linear perturbation theory, we will obtain basis which are analogous to the spherical harmonics bases. The dynamical system is then studied in the framework of the quasi-Maxwellian (QM) equations. However, a slight change in the method must be effected at this point, due to the existence of non-null tensorial quantities such as the shear and the Weyl tensor in the background, in analogy to the Bardeen (1980) variables. The dynamical equations for these extra quantities are obtained from QM equations.

\subsubsection{The Anisotropic Basis}

In order to make the temporal dependence of the perturbations explicit in the quasi-Maxwellian (QM) equations, it is necessary to obtain a basis in terms of which all perturbed quantities can be written. Since we are dealing here with an anisotropic background, we shall not writte in terms of the spherical harmonics. A new basis must then be constructed.

In this section, the three types of basis -- scalar, vectorial, and tensorial -- are exhibited. Since we are considering a background free of any matter, an apparent difficulty appears regarding the adequacy of analyzing matter-related perturbations. However, it is possible to write a generalized solution for the specific Kasner model, which presents matter-related terms that are of a lower order than the geometrical ones. Therefore, although it has been argued that the matter-related terms do not contribute to the unperturbed background, they might have an important contribution after perturbation.

\underline{{\bf The scalar basis:}} In order to be able to obtain a scalar basis $\{Q(x,y,z)\}$ for the Kasner background, we will impose that it satisfies the equation

\begin{equation}
\label{klgor_eq}
\hat\nabla^2 Q =n^2 Q,
\end{equation}

where $n^2$ is a function of time. In Cartesian coordinates, this equation is integrated to give

\begin{equation}
\label{solklgor_eq}
Q(x,y,z)=N \exp[-i(n_1x+n_2y+n_3z)]\equiv Ne^{-in_jx^j},
\end{equation}
where $N$ and $n_j$ are arbitrary constants (a choice made to simplify our calculations), and the following relation:

\begin{equation}
n^2=-h_{\alpha\beta}n^{\alpha}n^{\beta},
\end{equation}
is inferred.

Using results\ (\ref{klgor_eq}) and (\ref{solklgor_eq}), we proceed to write vector and symmetric traceless second-order tensor bases, which will be used to define the corresponding perturbed quantities

\begin{equation}
\label{ves_ten_from_scal}
\begin{array}{lcl}
\hat Q_{\alpha}&\equiv&\hat\nabla_{\alpha} Q=-in_{\alpha} Q,\\
\hat Q_{\alpha\beta}&\equiv&\hat\nabla_{\beta} Q_{\alpha}-\fracc{1}{3}n^2h_{\alpha\beta} Q,
\end{array}
\end{equation}
where the properties of symmetry and tracelessness of the tensor can be directly shown. From the vector definition, $\hat Q_{\alpha}$ has only spatial components. The tensor $\hat Q_{\alpha\beta}$ written in terms of the scalar $Q$ is

\begin{equation}
\label{ten_in_terms_scal}
\hat Q_{\alpha\beta}=-\left(n_{\alpha}n_{\beta}+\fracc{n^2}{3}h_{\alpha\beta}\right) Q.
\end{equation}

The following properties are then obtained:
\begin{equation}
\label{rel_svt_scal_kasner}
\begin{array}{l}
\hat\nabla^{\alpha}\hat Q_{\alpha}=n^2Q,\\
\hat\nabla^{\mu}\hat Q_{\mu\nu}=\fracc{2}{3}n^2\hat Q_{\nu},\\
\hat\nabla^2\hat Q_{\alpha}=n^2\hat Q_{\alpha},\\
\hat\nabla^2\hat Q_{\alpha\beta}=n^2\hat Q_{\alpha\beta}\\
\dot Q=0,\\
(\hat Q_{\alpha})\dot{}=(\sigma_{\alpha}{}^{\beta}+\fracc{\theta}{3}h_{\alpha\beta})\hat Q_{\beta},\\
(\hat Q_{\alpha\beta})\dot{}=-\fracc{2}{3}\theta\hat Q_{\alpha\beta}-\sigma_{(\alpha}{}^{\mu}\hat Q_{\beta)m}-\fracc{2}{3}n^2\sigma_{\alpha\beta} Q+\fracc{2}{3}h_{\alpha\beta}\sigma^{\mu\nu}\hat Q_{\mu\nu}.
\end{array}
\end{equation}
Let us choose a specific direction of propagation for the scalar basis.\footnote{This procedure was adopted in several instances in the literature. See, for example, Sagnotti (1981).} Therefore, we set

\begin{equation}
\label{simpl_scal}
n_1=n_2=0\Rightarrow n^2=t^{-2p_3}(n_3)^2,
\end{equation}
Thus, the scalar basis and its correlated quantities take the very simple form:
\begin{equation}
\label{final_sc_basis}
\begin{array}{lcl}
Q=Ne^{-in_3z},\\[2ex]
\hat Q_{\alpha}=-in_3(0,0,Q),\\[2ex]
\hat Q_{\alpha\beta}=\fracc{n^2}{3} Q
\left(\begin{array}{ccc}
t^{2p_1}&0&0\\
0&t^{2p_2}&0\\
0&0&-2t^{2p_3}
\end{array}\right).
\end{array}
\end{equation}

\underline{{\bf The vectorial basis:}} Similar to the scalar case, let us impose for the vector basis $\{\hat P_{\alpha} \}$ the equation

\begin{equation}
\label{klgor_eqvec}
\hat\nabla^2\hat P_{\alpha}=m^2\hat P_{\alpha},
\end{equation}
where $m^2$ is a function of $t$. Integrating this equation, one finds

\begin{equation}
\label{solklgor_eqvec}
\hat P_{\alpha}={\cal P}^0{}_{\alpha}e^{-im_jx^j}.
\end{equation}
We choose ${\cal P}^0{}_{\alpha}$ and $m_j$ as constants.\footnote{This choice is necessary to avoid that complex terms or explicit dependences on spatial coordinates occur when the simple derivatives of the vector basis are calculated.} From Eqs.\ (\ref{klgor_eqvec}) and\ (\ref{solklgor_eqvec}), it follows:

\begin{equation}
\label{vec_rel}
m^2=-h^{\alpha\beta}m_{\alpha}m_{\beta}.
\end{equation}

In order that $\{\hat P_{\alpha}\}$ constitute a basis, two properties must be valid. The first one concerns the fact that $\hat P_{\alpha}$ must be spatial quantities. This is immediately satisfied, upon the choice ${\cal P}^0{}_0=0$, or

\begin{equation}
\label{vec_rel_print}
V^{\alpha}\hat P_{\alpha}=0.
\end{equation}
This property must also be preserved in time, i.e.,

\begin{equation}
\label{ev_vec_rel_print}
(V^{\alpha}\hat P_{\alpha})\dot{}=0.
\end{equation}
which is identically valid for ${\cal P}_{\mu}{}^0=$const.

The second property, namely that no scalar quantities can be obtained from the vector $\hat P_{\alpha}$, implies that

\begin{equation}
\label{div_p_kasner}
\hat\nabla^{\alpha}\hat P_{\alpha}=0,
\end{equation}

and can be written as

\begin{equation}
\label{proj_div_p_kasner}
h^{\alpha\beta}m_{\alpha}\hat P_{\beta}=0.
\end{equation}

This property must also be conserved in time; hence it follows

\begin{equation}
\label{imp_div_p_kasner}
\hat\nabla^{\alpha}\hat P_{\alpha}=0\Longrightarrow \sigma^{\alpha\beta}m_{\alpha}\hat P_{\beta}=0.
\end{equation}

The conditions\ (\ref{proj_div_p_kasner}) and (\ref{imp_div_p_kasner}) can also be written in terms of the ${\cal P}^0{}_{\alpha}$:

\begin{equation}
\label{cond_ab_rew_kasner}
t^{-2p_1}m_1{\cal P}^0_{x}+t^{-2p_2}m_2{\cal P}^0_{y}+t^{-2p_3}m_3{\cal P}^0_{z}=0
\end{equation}

and

\begin{equation}
\label{der_cond_ab_rew_kasner}
p_1t^{-2p_1}m_1{\cal P}^0_{x}+p_2t^{-2p_2}m_2{\cal P}^0_{y}+p_3t^{-2p_3}m_3{\cal P}^0_{z}=0.
\end{equation}

From ${\cal P}_{\alpha}$, it is possible to construct three quantities: a symmetric, traceless second-order tensor (which we will denote by $\hat P_{\alpha\beta}$), a
pseudo-vector denoted by $\hat P^*_{\alpha}$, and finally the corresponding pseudotensor $\hat P^*_{\alpha\beta}$. The respective definitions for these quantities are the following:

\begin{equation}
\label{rel_pstar}
\begin{array}{lcl}
\hat P_{\alpha\beta}&\equiv&\hat\nabla_{(\alpha}\hat P_{\beta)},\\
\hat P^*_{\alpha}&\equiv&\eta_{\alpha}{}^{\beta\mu\nu}V_{\beta}(\hat\nabla_{\nu}\hat P_{\mu}),\\
\hat P^*_{\alpha\beta}&\equiv&\hat\nabla_{(\alpha}\hat P^*{}_{\beta)}.
\end{array}
\end{equation}

The first of the definitions\ (\ref{rel_pstar}) is immediately rewritten as

\begin{equation}
\label{re_1st_rel_pstar}
\hat P_{\alpha\beta}=-im_{(\alpha}\hat P_{\beta)}
\end{equation}
and its corresponding Laplacian and time-projected derivative are proven to be given by

\begin{equation}
\label{lapl_re_1st_rel_pstar}
\hat\nabla^2\hat P_{\alpha\beta}=m^2\hat P_{\alpha\beta},
\end{equation}

\begin{equation}
\label{dot_re_1st_rel_pstar}
(\hat P_{\alpha\beta})\dot{}=-\fracc{2}{3}\theta\hat P_{\alpha\beta}-\sigma^{\gamma}{}_{(\alpha}\hat P_{\beta)\gamma}.
\end{equation}
The pseudovector $\hat P^*_{\alpha}$, from the second definition in Eq.\ (\ref{rel_pstar}), is

\begin{equation}
\label{re_2nd_rel_pstar}
\hat P^*_{\alpha}=-i\eta_{\alpha}{}^{\beta\mu\nu}V_{\beta}m_{\nu}\hat P_{\mu}
\end{equation}
It follows---from the fact that all these quantities describe the perturbations---that the same properties that define $\hat P_{\alpha}$ should also be valid for $\hat P^*_{\alpha}$. Therefore, the pseudovector should be both a spatial and a divergenceless quantity, which is satisfied. These conditions must be preserved in time. This is identically valid for the first property. The preservation condition for the null divergence property is given by

\begin{equation}
\label{div_con_p_star}
(\hat\nabla^{\alpha}\hat P^*_{\alpha})\dot{}=0\Longrightarrow\sigma^{\alpha\beta}m_{\alpha}\hat P^*_{\beta}=0,
\end{equation}
which is then rewritten in terms of the ${\cal P}^0_{\alpha}$ as

\begin{equation}
\label{re_div_con_p_star}
(p_2-p_3)m_2m_3{\cal P}^0_x+(p_3-p_1)m_1m_3{\cal P}^0_y+(p_1-p_2)m_1m_2{\cal P}^0_z=0
\end{equation}
In addition to the above condition, the following useful results are also obtained:

\begin{equation}
\label{lapl_rel_pstar}
\hat\nabla^2\hat P^*_{\alpha}=m^2\hat P^*_{\alpha},
\end{equation}
and
\begin{equation}
\label{dot_rel_pstar}
(\hat P^*_{\alpha})\dot{}=-\fracc{1}{3}\theta\hat P^*_{\alpha}-\sigma^{\beta}{}_{\alpha}\hat P^*_{\beta}.
\end{equation}

The last quantity to be considered is the symmetric, traceless pseudotensor $\hat P^*_{\alpha\beta}$. From the third definition\ (\ref{rel_pstar}) we get

\begin{equation}
\label{ten_rel_kasn}
\hat P^*_{\alpha\beta}=-im_{(\alpha}\hat P^*_{\beta)},
\end{equation}
and the relations below follow immediately:

\begin{equation}
\label{div_rel_pstar_tens}
\hat\nabla^{\alpha}\hat P^*_{\alpha\beta}=m^2\hat P^*_{\alpha\beta},
\end{equation}

\begin{equation}
\label{lapl_rel_pstar_tens}
\hat\nabla^2\hat P^*_{\alpha\beta}=m^2\hat P^*_{\alpha\beta},
\end{equation}
and
\begin{equation}
\label{dot_rel_pstar_tens}
(\hat P^*_{\alpha\beta})\dot{}=-\fracc{2}{3}\theta\hat P^*_{\alpha\beta}-\sigma^{\gamma}{}_{(\alpha}\hat P^*_{\beta)\gamma}.
\end{equation}

The most general form for the vectorial basis implies obtaining suitable ${\cal P}^0_{\alpha}$ and $m_j$, and replacing them in the basis expression\ (\ref{solklgor_eqvec}). These quantities are defined from the conditions\ (\ref{cond_ab_rew_kasner}),\ (\ref{der_cond_ab_rew_kasner}), and\ (\ref{re_div_con_p_star}) as\footnote{The condition\ (\ref{cond_vec_basis}) above implies making a choice of a specific direction for the basis and has been made a number of times in literature. See, for example, Ref. \cite{sagnotti}.}

\begin{equation}
\label{cond_vec_basis}
\begin{array}{l}
m_1\equiv m_2=0,\\
{\cal P}^0_z=0,
\end{array}
\end{equation}
and, using Eq.\ (\ref{cond_vec_basis}) above in Eq.\ (\ref{solklgor_eqvec}), we get

\begin{equation}
\label{p_vec_kasn}
\hat P_{\alpha}=e^{-im_3z}({\cal P}^0_x,{\cal P}^0_y,0),
\end{equation}
where $m_3$, ${\cal P}^0_x$ and ${\cal P}^0_y$ are arbitrary constants. In addition, the quantities $\hat P_{\alpha\beta}$, $\hat P^*_{\alpha}$, and $\hat P^*_{\alpha\beta}$ are written, in this case, as

\begin{equation}
\label{final_p_tens_kasner}
\hat P_{\alpha\beta}=-im_3e^{-im_3z}
\left(\begin{array}{ccc}
0&0&{\cal P}^0_x\\
0&0&{\cal P}^0_y\\
{\cal P}^0_x&{\cal P}^0_y&0
\end{array}\right),
\end{equation}

\begin{equation}
\label{final_pstar_vec_kasner}
\hat P^*_{\alpha}=i\eta^{0123}m_3e^{-im_3z}(-t^{2p_1}{\cal P}^0_y,t^{2p_2}{\cal P}^0_x,0),
\end{equation}

\begin{equation}
\label{final_pstar_tens_kasner}
\hat P^*_{\alpha\beta}=\eta^{0123}(m_3)^2e^{-im_3z}
\left(\begin{array}{ccc}
0&0&-t^{2p_1}{\cal P}^0_y\\
0&0&t^{2p_2}{\cal P}^0_x\\
-t^{2p_1}{\cal P}^0_y&t^{2p_2}{\cal P}^0_x&0
\end{array}\right).
\end{equation}

\underline{{\bf The tensorial basis:}} We will begin by defining the tensor $\hat U^{\mu}{}_{\nu}(t,x,y,z)$, which is written in matrix form as

\begin{equation}
\label{tens_basis_kasner}
\hat U^{\mu}{}_{\nu}=
\left(\begin{array}{cccc}
0&0&0&0\\
0&\alpha&\psi&\phi\\
0&\eta&\beta&\epsilon\\
0&\chi&\zeta&\gamma
\end{array}\right),
\end{equation}
where $(\alpha,\psi,\phi,\eta,\beta,\epsilon,\chi,\zeta,\gamma)$ are functions of all four coordinates. The choice we made for our tensorial basis enables us to simplify somewhat the future calculations. It is easy to write the totally contravariant tensor as:

\begin{equation}
\label{contr_tens_basis_kasner}
\hat U^{\mu}{}_{\nu}=h^{\mu\alpha}\hat U_{\alpha\nu}
\end{equation}
with $\hat U_{\alpha\nu}=\hat U_{\nu\alpha}$.

In order that $\hat U^{\mu}{}_{\nu}$ be, in fact, a basis, it has to be a solution of the equation

\begin{equation}
\label{tens_kg_eq_kasn}
\hat\nabla^2\hat U^{\mu}{}_{\nu}=k^2\hat U^{\mu}{}_{\nu},
\end{equation}
where $k^2$ is a function of time. Solving this equation, we obtain an explicit form for the tensorial basis

\begin{equation}
\label{sol_tens_kg_eq_kasn}
\hat U^{\mu}{}_{\nu}={\cal U}^{\mu}{}_{\nu}e^{-ik_jx^j},
\end{equation}
where the ${\cal U}^{\mu}{}_{\nu}$ will be taken as covariantly constant tensors and $k_j$ are arbitrary constants which are related to the wave number $k^2$ and the components of the metric tensor through the following relation:\footnote{This choice was made in order to avoid terms with spatial coordinate dependence when calculating the derivative $(\hat U^{\mu}{}_{\nu})\dot{}$}

\begin{equation}
\label{wave_numb_rel_kasn}
k^2=-h^{jl}k_jk_l=\left[\left(\fracc{k_1}{a(t)}\right)^2+\left(\fracc{k_2}{b(t)}\right)^2+\left(\fracc{k_3}{c(t)}\right)^2\right].
\end{equation}
As in the previous cases, the tensor basis must obey the following properties:

$(I)$ The tensor basis should be orthogonal to $V_{\alpha}$:

\begin{equation}
\label{rel_v_tbas_kasn}
V_{\mu}\hat U^{\mu}{}_{\nu}=0.
\end{equation}

$(II)$ Scalars cannot be obtained from the tensor basis:

\begin{equation}
\label{rel_traceless_tbas_kasn}
h^{\mu\nu}\hat U_{\mu\nu}=\hat U^{\mu}{}_{\mu}=0,
\end{equation}
or, using Eq.\ (\ref{tens_basis_kasner}),

\begin{equation}
\label{re_rel_traceless_tbas_kasn}
\alpha+\beta+\gamma=0.
\end{equation}

$(III)$ Vectors cannot be obtained from the tensor basis:

\begin{equation}
\label{rel_divless_tbas_kasn}
\hat\nabla^{\mu}\hat U_{\mu\nu}=0,
\end{equation}
which gives

\begin{equation}
\label{re_rel_divless_tbas_kasn}
k_{\mu}\hat U^{\mu}{}_{\nu}=0,
\end{equation}

Equation\ (\ref{re_rel_divless_tbas_kasn}) above can be rewritten, using Eq.\ (\ref{sol_tens_kg_eq_kasn}), as

\begin{equation}
\label{sub_solkgt_div_tbas_kasn}
\begin{array}{l}
k_1\alpha+k_2\eta+k_3\chi=0,\\[2ex]
k_1\psi+k_2\beta+k_3\zeta=0,\\[2ex]
k_1\phi+k_2\epsilon+k_3\gamma=0,
\end{array}
\end{equation}
It is easily seen that all the properties above are preserved in time.

At this point it becomes necessary to define a quantity that enables us to write pseudo-tensorial perturbations. Therefore, we define the dual $\hat U^{*\mu}{}_{\nu}$ as

\begin{equation}
\label{dual_tens_basis_kasn}
\hat U^*_{\mu\nu}\equiv\fracc{1}{2}h^{\alpha}_{(\mu}h^{\beta}_{\nu)}\eta_{\beta}{}^{\lambda\epsilon\gamma}V_{\lambda}(\hat\nabla_{\epsilon}\hat U_{\gamma\alpha}),
\end{equation}
which can be then rewritten as

\begin{equation}
\label{re_dual_tens_basis_kasn}
\hat U^{*\mu}{}_{\nu}\equiv-\fracc{i}{2}k_{\epsilon}(\eta_{\gamma}{}^{\lambda\epsilon\mu}V_{\lambda}\hat U^{\gamma}{}_{\mu}+\eta^{\gamma\lambda\epsilon}{}_{\nu}V_{\lambda}\hat U^{\mu}{}_{\gamma}).
\end{equation}
It follows that all the properties of the tensorial basis $\hat U^{\mu}{}_{\nu}$ are equally valid for the dual tensor $\hat U^{*\mu}{}_{\nu}$ and that they are preserved in time.

We can obtain an explicit form for $\hat U^{\mu}{}_{\nu}$. From Eqs.\ (\ref{kasn_ans}) and\ (\ref{tens_basis_kasner}) we have that

\begin{equation}
\label{expl_u_basis_kasn}
\begin{array}{l}
\hat U^2{}_1\equiv\eta=g_{11}g^{22}\hat U^1_2\equiv t^{2(p_1-p_2)}\psi,\\[2ex]
\hat U^3{}_1\equiv\chi=g_{11}g^{33}\hat U^1_3\equiv t^{2(p_1-p_3)}\phi,\\[2ex]
\hat U^3{}_2\equiv\zeta=g_{22}g^{33}\hat U^2_3\equiv t^{2(p_2-p_3)}\epsilon.
\end{array}
\end{equation}
Using the above results and the null trace condition\ (\ref{re_rel_traceless_tbas_kasn}) in condition\ (\ref{sub_solkgt_div_tbas_kasn}), we find that

\begin{equation}
\label{re_sub_solkgt_div_tbas_kasn}
\begin{array}{l}
k_1\alpha+k_2t^{2(p_1-p_2)}\psi+k_3t^{2(p_1-p_3)}\phi=0,\\
k_1\psi+k_2\beta+k_3t^{2(p_2-p_3)}\epsilon=0,\\
k_1\phi+k_2\epsilon-k_3(\alpha+\beta)=0.
\end{array}
\end{equation}

However, since both the $k_j$ and the ${\cal U}^{\mu}{}_{\nu}$ are constant, we can see that each term in the three above relations must also be constants. We then choose a specific direction for the basis (Sagnotti, 1981), taking

\begin{equation}
\label{rest_wave_numb_kasn}
\begin{array}{l}
k_1=k_2=0,\\
k_3\neq0.
\end{array}
\end{equation}
Then, equations\ (\ref{re_sub_solkgt_div_tbas_kasn}) simplify to

\begin{equation}
\label{sim_67}
\begin{array}{l}
t^{2(p_1-p_3)}\phi=0,\\
t^{2(p_2-p_3)}\epsilon=0,\\
\alpha+\beta=0,
\end{array}
\end{equation}
and give, as a consequence,

\begin{equation}
\label{cons_sim_67}
\begin{array}{l}
\phi=\epsilon=0,\\
\beta=-\alpha\Rightarrow\gamma=0.
\end{array}
\end{equation}
The tensorial basis and its dual are, then, written in matrix form as

\begin{equation}
\label{final_u_basis}
\hat U^{\mu}{}_{\nu}={\cal U}^{\mu}{}_{\nu} e^{-ik_3z}=
\left(\begin{array}{ccc}
\alpha&\psi&0\\
\eta&-\alpha&0\\
\chi&\zeta&0
\end{array}\right),
\end{equation}

\begin{equation}
\label{final_dual_u_basis}
\hat U^{*\mu}{}_{\nu}=-\fracc{i}{2}\eta^{0123}k_3
\left(\begin{array}{ccc}
(t^{2p_1}\psi+t^{2p_2}\eta)&-2t^{2p_2}\alpha&0\\
-2t^{2p_1}\alpha&-(t^{2p_1}\psi+t^{2p_2}\eta)&0\\
t^{2p_1}\chi&-t^{2p_2}\zeta&0
\end{array}\right).
\end{equation}

\subsubsection{The Gauge-Invariant Variables and Their Dynamics}

The dynamical system for an anisotropic background will be obtained in the framework of the quasi-Maxwellian (QM) equations. In the Kasner background, the QM equations are reduced to the set

\begin{equation}
\label{shear_ev_kasner}
(\sigma_{ij})\dot{}+E_{ij}+\fracc{2}{3}\theta\sigma_{ij}-\fracc{1}{3}(2\sigma^2)h_{ij}+\sigma_{ik}\sigma^{k}{}_{j}=0,
\end{equation}

\begin{equation}
\label{eij_ev_kasner}
(E_{ij})\dot{}+3\theta E_{ij}+\fracc{3}{2}\sigma^{\mu}{}_{(i}E_{j)\mu}-h_{ij}\sigma^{\mu\nu}E_{\mu\nu}=0,
\end{equation}

\begin{equation}
\label{exp_ev_kasner}
\dot\theta+\fracc{1}{3}\theta^2+2\sigma^2=0,
\end{equation}
The natural step is, then, to write the perturbed QM equations, by making the usual choice: $A_{(pert.)}=A_{(back.)}+(\delta A)$. However, a modification in this method becomes necessary at this point: the three non-null quantities in the Kasner background ($\sigma_{\mu\nu}$, $E_{\mu\nu}$ and $\theta$) should be replaced by ``artificial" quantities in order to eliminate all gauge-dependent terms from the dynamical system equations. These new variables are written in terms of the original, gauge-dependent variables. Nevertheless they constitute ``good" quantities, in the sense that they are zero in the Kasner background, as per the Stewart lemma (1974). This procedure is analogous to the one implemented by Bardeen (1980), but the variables obtained in the present case are much simpler, as we shall see in the next section.

\underline{Minimal closed set of variables for the Kasner background}. The starting point to obtain the new variables is the set of QM equations for the Kasner background, Eqs.\ (\ref{shear_ev_kasner}) -- (\ref{exp_ev_kasner}). If we employ the following relations (easily demonstrated and specifically valid for the Kasner background):

\begin{equation}
\label{back_kasn_rel}
\begin{array}{l}
(\sigma^{\alpha}{}_{\beta})\dot{}=-\theta\,\sigma^{\alpha}{}_{\beta},\\
(E^{\alpha}{}_{\beta})\dot{}=-2\,\theta\,E^{\alpha}{}_{\beta},\\
\dot\theta=-\theta^2,
\end{array}
\end{equation}
we are able to define the new variables that are to replace the original ones as

\begin{equation}
\label{new_x_back__kasn}
X^{\alpha}{}_{\beta}\equiv E^{\alpha}{}_{\beta}-\fracc{1}{3}\theta\sigma^{\alpha}{}_{\beta}-\fracc{2\sigma^2}{3}h^{\alpha}{}_{\beta}+\sigma^{\alpha}{}_{\mu}\sigma^{\mu}{}_{\beta},
\end{equation}

\begin{equation}
\label{new_y_back__kasn}
Y^{\alpha}{}_{\beta}\equiv\theta E^{\alpha}{}_{\beta}+\fracc{3}{2}\sigma^{\alpha}{}_{\mu}E^{\mu}{}_{\beta}+\fracc{3}{2}\sigma^{\mu}{}_{\beta}E^{\alpha}{}_{\mu}-h^{\alpha}{}_{\beta}\sigma^{\mu\nu}
E_{\mu\nu},
\end{equation}

\begin{equation}
\label{new_w_back__kasn}
W\equiv2\sigma^2-\fracc{2}{3}\theta^2.
\end{equation}
These three variables are easily proven to be zero for the Kasner background and, therefore, ``good" ones to be perturbed. They may, then, replace the shear, electric part of the Weyl tensor and expansion as the new variables in the dynamic system. An additional simplifying choice will be made, namely the relation between the energy density $\rho$ and the pressure $p$ is

\begin{equation}
\label{kasne_eq_state}
p=\lambda\rho,\hspace{1cm}\lambda\equiv{\mbox const.},
\end{equation}
even after the background is perturbed. This choice---also made for the FLRW case (see Novello, 1995a,b and 1996)---will be considered as valid throughout this analysis. The complete minimal closed set of variables to appear in the dynamical system is therefore\footnote{Since the acceleration -- the variable $a_{\alpha}$ -- does not have a dynamical equation of its own, it must be eliminated from the dynamical system in order to be really closed. This will be achieved by fixing a value for the function $(\delta a_{\alpha})$.}

\begin{equation}
\label{min_set_kasn}
{\cal M}=\{X_{\alpha\beta}, Y_{\alpha\beta}, H_{\alpha\beta}, \pi_{\alpha\beta}, q_{\alpha}, a_{\alpha},\omega_{\alpha}, W, \rho\}.
\end{equation}
The next step in this analysis is to obtain the complete dynamics for the new variables, $X_{\alpha\beta}$, $Y_{\alpha\beta}$, $W$. The resulting set of equations and the remaining QM equations must, then, be rewritten in terms of the new gauge-independent variables in {\cal M}. This constitutes the dynamical system of equations which is used to study the perturbations of Kasner model. Such complete dynamical system as well as the steps for its derivation can be seen in Novello et al. (2000). The next sections will deal with the three perturbation cases and the results obtained for each one.

\subsubsection{Scalar Perturbations}

In this case, the minimal closed set of observables ${\cal M}$ involves practically all the original variables of the system, which we proceed to present here in terms of the scalar basis $Q$:

\begin{equation}
\label{min_set_terms_q}
\begin{array}{ll}
(\delta X_{\alpha\beta})=X(t)\hat Q_{\alpha\beta}, \hspace{.5 cm}&(\delta Y_{\alpha\beta})=Y(t)\hat Q_{\alpha\beta},\\
(\delta\pi_{\alpha\beta})=\pi(t)\hat Q_{\alpha\beta}, \hspace{.5 cm}&(\delta q_{\alpha})=q(t)\hat Q_{\alpha},\\
(\delta a_{\alpha})=\psi(t)\hat Q_{\alpha}, \hspace{.5 cm}&(\delta \rho)=R(t) Q,\\
(\delta W)=W(t) Q, \hspace{.5 cm}&(\delta p)=p(t) Q,
\end{array}
\end{equation}
where the spatial part of the velocity, $(\delta V_k)$, is also zero in the background (but not an adequate variable, since its value in the background depends on the choice of an observer), and it is written as

\begin{equation}
\label{dv_sc_basis}
(\delta V_k)=V(t)\hat Q_k.
\end{equation}

Since we are dealing with scalar perturbations, the vorticity and related perturbations, for instance the magnetic part of Weyl tensor, are not defined---see details in Novello (1995a) and Goode (1989). The relation between the shear and the anisotropic pressure is still valid, but in this case the viscosity $\xi$ is also a ``good" variable (i.e., it is a gauge-independent variable for it is zero in the Kasner background), written as

\begin{equation}
\nonumber
(\delta \xi)=\xi(t)\ Q
\end{equation}
and must be considered. From Eqs.\ (\ref{min_set_terms_q}) and\ (\ref{ten_in_terms_scal}) it is possible to obtain $(\delta\xi)$ in terms of $(\delta\pi_{\alpha\beta})$ as

\begin{equation}
\label{xi_term_pi}
\xi(t)=-\fracc{1}{(2\sigma^2)}(\sigma^{\mu\nu}n_{\mu}n_{\nu})\pi(t).
\end{equation}
Once again -- as in the previous cases -- we take the perturbed anisotropic pressure $(\delta\pi_{\alpha\beta})$ as zero, in order to simplify the dynamical system to be solved. Thus,

\begin{equation}
\label{sim_xi_pi}
\pi(t)=\xi(t)=0.
\end{equation}
Using the results obtained for the scalar basis, we then obtain the dynamical system for the scalar perturbations case

\begin{equation}
\label{sc_x_ev}
\begin{array}{l}
\left(\dot X+\theta X-Y+\fracc{1}{2}q+\fracc{1}{3}\theta\psi\right)\hat Q_{\mu\nu}+\left(-\fracc{2}{3}n^2X+\fracc{1}{3}n^2\psi+\fracc{1}{3}(W+R)\right)[\sigma_{\mu\nu} Q ]+\\[2ex]
-\psi\left[\sigma^{\alpha}{}_{(\mu}\hat Q_{\nu)\alpha}-\fracc{2}{3}h_{\mu\nu}\sigma^{\alpha\beta}\hat Q_{\alpha\beta}\right]=0,
\end{array}
\end{equation}

\begin{equation}
\label{sc_y_ev}
\begin{array}{l}
\left(\dot Y+\fracc{4}{3}\theta Y+\fracc{1}{2}\theta q\right)\hat Q_{\mu\nu} + \left[-\fracc{2}{3}n^2Y+\theta(1+\lambda)R\right][\sigma_{\mu\nu} Q]+(n^2q-2n^2\psi+W-R)[E_{\mu\nu} Q]+\\[2ex]
+\fracc{1}{2}\left(-5Y+\fracc{3}{2}q\right)\left(\sigma^{\alpha}{}_{(\mu}\hat Q_{\nu)\alpha}-\fracc{2}{3}h_{\mu\nu}\sigma^{\alpha\beta}\hat Q_{\alpha\beta}\right)+\fracc{3}{2}(X-\psi)\left[E^{\alpha}{}_{(\mu}\hat Q_{\nu)\alpha}-\fracc{2}{3}h_{\mu\nu}E^{\alpha\beta}\hat Q_{\alpha\beta}\right]=0,
\end{array}
\end{equation}

\begin{equation}
\label{sc_w_ev}
\dot W+\fracc{2}{3}\theta W-2(\sigma^{\alpha\beta}n_{\alpha}n_{\beta})X+2\left[(\sigma^{\alpha\beta}n_{\alpha}n_{\beta})+\fracc{2}{3}\theta n^2\right]\psi-\fracc{2}{3}\theta(1+3\lambda)R=0,
\end{equation}

\begin{equation}
\label{sc_psi_constr}
\left\{[p_1(1-p_1)-p_2(1-p_2)]+\fracc{1}{12}(1-3p_3)(p_1-p_2)\right\}\psi=0,
\end{equation}

\begin{equation}
\label{sc_r_ev}
\dot R+\theta(1+\lambda)R+n^2q=0,
\end{equation}

\begin{equation}
\label{sc_q_ev}
\dot q+\theta q-\lambda t^{-8/3}R=0,
\end{equation}

\begin{equation}
\label{sc_constr2}
\left[\fracc{2}{3}n^2X-\theta q+\fracc{1}{3}(W-R)\right]\hat Q_{\alpha}+\fracc{1}{2}\psi\left[\left(\sigma^{\alpha}{}_{\gamma}\sigma^{\gamma\beta}+\fracc{4}{3}\theta\sigma_{\alpha}{}^{\beta}\right)\hat Q_{\beta}\right]=0.
\end{equation}
Equation\ (\ref{sc_psi_constr}) is satisfied in two cases: (1) $p_1\neq p_2\Rightarrow\psi=0$ and (2) $p_1=p_2\Rightarrow$ isotropy plane, and the simplest choice here is (1), with $\psi=0$.

The system becomes thus

\begin{equation}
\label{simpl_x_ev}
\left(\dot X+\theta X-Y+\fracc{1}{2}q\right)\hat Q_{\mu\nu}+\left(-\fracc{2}{3}n^2X+\fracc{1}{3}(W+R)\right)[\sigma_{\mu\nu} Q ]=0,
\end{equation}

\begin{equation}
\label{simpl_y_ev}
\begin{array}{l}
\left(\dot Y+\fracc{4}{3}\theta Y+\fracc{1}{2}\theta q\right)\hat Q_{\mu\nu} + \left[-\fracc{2}{3}n^2Y+\theta(1+\lambda)R\right][\sigma_{\mu\nu} Q]+(n^2q+W-R)[E_{\mu\nu} Q]+\\[2ex]
+\fracc{1}{2}\left(-5Y+\fracc{3}{2}q\right)\left(\sigma^{\alpha}{}_{(\mu}\hat Q_{\nu)\alpha}-\fracc{2}{3}h_{\mu\nu}\sigma^{\alpha\beta}\hat Q_{\alpha\beta}\right)+\fracc{3}{2}X\left[E^{\alpha}{}_{(\mu}\hat Q_{\nu)\alpha}-\fracc{2}{3}h_{\mu\nu}E^{\alpha\beta}\hat Q_{\alpha\beta}\right]=0,
\end{array}
\end{equation}

\begin{equation}
\label{simpl_w_ev}
\dot W+\fracc{2}{3}\theta W-2(\sigma^{\alpha\beta}n_{\alpha}n_{\beta})X-\fracc{2}{3}\theta(1+3\lambda)R=0,
\end{equation}

\begin{equation}
\label{simpl_r_ev}
\dot R+\theta(1+\lambda)R+n^2q=0,
\end{equation}

\begin{equation}
\label{simpl_q_ev}
\dot q+\theta q-\lambda t^{-8/3}R=0,
\end{equation}

\begin{equation}
\label{simpl_constr2}
\left[\fracc{2}{3}n^2X-\theta q+\fracc{1}{3}(W-R)\right]\hat Q_{\alpha}=0.
\end{equation}

Although Eqs.\ (\ref{simpl_x_ev}) and\ (\ref{simpl_y_ev}) cannot be factored out in the scalar basis, the remaining equations in the system, Eqs.\ (\ref{simpl_w_ev}),\ (\ref{simpl_constr2}), can be separately integrated. The constraint\ (\ref{simpl_constr2}) eliminates the variable $X$ from the reduced system

\begin{equation}
\label{simpl_x}
X=\fracc{3}{2}\fracc{\theta}{n^2}t^{8/3}q+\fracc{1}{2n^2}R-\fracc{1}{2n^2}W,
\end{equation}
so that the dynamics of $W$, Eq.\ (\ref{simpl_w_ev}), is written as

\begin{equation}
\label{simpl2_w_ev}
\dot W+\left[\fracc{2}{3}\theta W + \fracc{1}{n^2} (\sigma^{\alpha\beta}n_{\alpha}n_{\beta}) \right]W-\left[\fracc{2}{3}\theta(1+3\lambda) + (\sigma^{\alpha\beta}n_{\alpha}n_{\beta})\right]R-3\fracc{\theta}{n^2}t^{8/3}(\sigma^{\alpha\beta}n_{\alpha}n_{\beta})q=0,
\end{equation}
and, with the relation

\begin{equation}
\label{rel_sigma_theta}
\fracc{1}{n^2}(\sigma^{\alpha\beta}n_{\alpha}n_{\beta})=\fracc{1}{3}\theta(1-p_3),
\end{equation}
the final dynamics for $W$ thus becomes:

\begin{equation}
\label{final_w_ev}
\dot W+(1-p_3)\theta W-(1+2\lambda-p_3)\theta R+(3p_3-1)\theta t^{8/3}q=0,
\end{equation}
and the reduced dynamical system---which is closed in the variables $W$, $R$, and $q$---is given by Eqs.\ (\ref{simpl_r_ev}),\ (\ref{simpl_q_ev}), and\ (\ref{final_w_ev}). As in the previous cases, this reduced system can be solved for the following Ansatz:

\begin{equation}
\label{sc_ansatz_kasn}
\begin{array}{l}
q(t)=q_0t^x,\\
R(t)=R_0t^y,\\
W(t)=W_0t^w,
\end{array}
\end{equation}
where $q_0$, $R_0$, $W_0$ and $x,y,w$ are constants to be determined. An immediate inspection of the powers of t in the three equations gives the relation between $x$, $y$, and $w$

\begin{equation}
\label{rel_xyw}
y=w=x+\fracc{5}{3},
\end{equation}
while the rest of the equations gives the following results:

\begin{equation}
\label{other_rel1}
R_0=\fracc{1}{\lambda}(x+1)q_0,
\end{equation}
and
\begin{equation}
\label{other_rel2}
W_0=[(x+1)(1+2\lambda-p_3)+\lambda(1-3p_3)](w+1-p_3)^{-1}q_0 ,
\end{equation}
where $w\neq(p_3-1)$ must be valid, and

\begin{equation}
\label{n3_rel}
(n_3)^2=\fracc{1}{\lambda}(x+1)\left(x+\fracc{8}{3}+\lambda\right).
\end{equation}
The results above are not applicable to the case $\lambda=0$; the details of this specific solution can be seen in \cite{no_sa_ma_re_se4}.

In order to analyze the stability of the above solution, we will impose that $(n_3)^2$ be positive. From Eq.\ (\ref{n3_rel}), we obtain -- considering $\lambda$ as positive\footnote{The case of $\lambda<0$ gives $(n_3)^2<0$ for all values of $x$, and therefore will not be considered here.} -- two possibilities: (1) $x>-1$; (2) $x<-(8/3-\lambda)$, and we have the following results for each case, exhibited in Table \ref{tab4}.

\begin{table}[htb]
\centering
\begin{tabular}{|c|cc|cc|}
\hline
\hline
& & & & \\[1ex]
Value for $x$       & & $q(t)$   & & $R(t)$ and $W(t)$ \\[1ex] \hline
& & & & \\[1ex]
\hspace{.3cm}$x<-(8/3+\lambda)$\hspace{.3cm} & & stable   & & stable \\[1ex]
& & & & \\[1ex]
$-1<x<0$            & & stable   & & \hspace{.3cm}unstable (faster than $t^{2/3}$)\hspace{.3cm}\\[1ex]
& & & & \\[1ex]
$x=0$               & & \hspace{.3cm} constant \hspace{.3cm}& & \hspace{.3cm}unstable (faster than $t^{5/3}$)\hspace{.3cm}\\[1ex]
& & & & \\[1ex]
$x>0$               & & unstable & & \hspace{.3cm}unstable (faster than $t^{5/3}$)\hspace{.3cm}\\[1ex]
& & & & \\[1ex]
\hline
\hline
\end{tabular}
\caption{Stability analysis results for scalar perturbations.}
\label{tab4}
\end{table}

We see that, although unstable solutions exist, the instability is---as we will see in the case of tensorial perturbations---not catastrophic. Another point which should be mentioned is that the matter density is unstable for any choice of the constant exponent $x$ that implies $(n_3)^2$ positive.

This concludes the analysis for the scalar perturbations case $(\lambda\neq0)$, and the case with zero $\lambda$ must now be studied. For $\lambda=0$, the reduced dynamical system is given by

\begin{equation}
\label{sc_w_ev_l0}
\dot W+(1-p_3)\theta W-(1-p_3)\theta R+(3p_3-1)\theta^2t^{8/3}q=0,
\end{equation}

\begin{equation}
\label{sc_r_ev_l}
\dot R+\theta R+n^2q=0,
\end{equation}

\begin{equation}
\label{c_q_ev_l0}
\dot q+\theta q=0.
\end{equation}

Equation\ (\ref{c_q_ev_l0}) can be directly integrated as

\begin{equation}
\label{sol_c_q_ev_l0}
q(t)=q_0t^{-1},
\end{equation}
where $q_0$ is a constant, while the ansatz

\begin{equation}
\label{ans_r_w_l0}
\begin{array}{l}
R(t)=R_0t^y,\\[2ex]
W(t)=W_0t^w,
\end{array}
\end{equation}
can be used in Eqs.\ (\ref{sc_w_ev_l0}) and\ (\ref{sc_r_ev_l}) to give

\begin{equation}
\nonumber
\begin{array}{l}
y=w=\fracc{2}{3},\\[2ex]
W_0=\fracc{(1-p_3)}{(w+1-p_3)}R_0,\\[2ex]
q_0=\fracc{5}{3(n_3)^2}R_0.
\end{array}
\end{equation}
The above results give a partially unstable solution for $\lambda=0$, with $R$ and $W$ divergent in $t$, while $q$ tends to zero for $t\rightarrow\infty$.

\subsubsection{Vectorial Perturbations}

The perturbations associated to the state of motion of a fluid (without taking energy density perturbations into account) are described, in principle, by the following minimal closed set of variables:\footnote{The same observation regarding the variable $a_{\alpha}$ that was stated for the tensorial case holds here as well.}

\begin{equation}
\label{cev_min_set}
{\cal M}=\{X_{\alpha\beta}, Y_{\alpha\beta}, H_{\alpha\beta}, q_{\alpha},a_{\alpha}, \omega_{\alpha}\}.
\end{equation}
The perturbed quantities can then be written in terms of the vectorial basis $\hat P_{\alpha}$ as

\begin{equation}
\label{Min_set_terms_p}
\begin{array}{ll}
(\delta X_{\alpha\beta})=X(t)\hat P_{\alpha\beta}, \hspace{.5 cm}&(\delta Y_{\alpha\beta})=Y(t)\hat P_{\alpha\beta},\\
(\delta H_{\alpha\beta})=H(t)\hat P^*_{\alpha\beta}, \hspace{.5 cm}&(\delta q_{\alpha})=q(t)\hat P_{\alpha},\\
(\delta a_{\alpha})=\psi(t)\hat P_{\alpha}, \hspace{.5 cm}&(\delta \omega_{\alpha})=\Omega(t)\hat P^*_{\alpha},\\
(\delta V_k)=V(t)\hat P_k,&
\end{array}
\end{equation}
where the variable $V(t)$ is, again, not an adequate one, since it depends on the initial choice of an observer. In order that the basis can be successfully factored out from the dynamical system equations, we are forced to choose between two possibilities: $(1)$ Eliminate one of the components of the basis, say ${\cal P}^0_y$, and $(2)$ analyze solely the case of a background with an isotropy plane (namely, the Kasner solution).

We will consider here the choice $(1)$ above, since it gives a more general result (and also contains the specific case $(2)$, as it shall be shortly seen). The dynamical system for the vectorial case can then be obtained by factoring out the basis, as

\begin{equation}
\label{vec_x_ev}
\dot X+\theta X-Y+ \fracc{m^2}{2}H+\fracc{1}{4}q-\fracc{1}{2}\theta p_2\psi=0,
\end{equation}

\begin{equation}
\label{vec_y_ev}
\dot Y-\fracc{1}{2}\theta(5p_2+1)Y+\fracc{3}{2}\theta^2p_2(p_2-1)X+\fracc{3}{4}\theta m^2(p_2-1)H-\fracc{3}{8}\theta(p_2-1)q-\fracc{3}{4}\theta^2p_2(p_2-1)\psi=0,
\end{equation}

\begin{equation}
\label{vec_h_ev}
\dot H-\fracc{1}{2}\theta (5p_1+1/3)H-\fracc{1}{2}X+\fracc{1}{4}\theta(3(p_3-p_2)+2/3)\Omega+\fracc{1}{4}\fracc{\theta^2}{m^2}(p_2-p_3)(3 p_1+1/3)\psi=0,
\end{equation}

\begin{equation}
\label{vec_q_ev}
\dot q+\theta q=0,
\end{equation}

\begin{equation}
\label{vec_omega_ev}
\dot\Omega+\fracc{1}{3}\theta(6p_2+1)\Omega-\fracc{1}{2}t^{-8/3}\psi=0,
\end{equation}

\begin{equation}
\label{vec_constr1_kasn}
m^2t^{-8/3}X+\theta m^2(p_2-1/3)H+\fracc{1}{2}\theta m^2(p_1+1)\Omega-\theta q+\fracc{1}{2}\theta^2(p_1-1/3)(p_1+1)\psi=0,
\end{equation}

\begin{equation}
\label{vec_constr2_kasn}
m^2H-\theta(p_1-p_3)X+3\theta^2t^{8/3}p_2(p_2-1)\Omega+\fracc{1}{2}q=0.
\end{equation}

The system\ (\ref{vec_x_ev}) --\ (\ref{vec_constr2_kasn}) is not closed because there is no dynamics for the acceleration $\psi(t)$. However, the constraint\ (\ref{vec_constr1_kasn}) can be employed to eliminate this variable in terms of the other variables in ${\cal M}$, closing therefore the remaining dynamical system for a given background, specified by the triad $(p_1, p_2, p_3)$. In order to obtain a specific solution for the dynamical system, we must make a choice for $(p_1, p_2, p_3)$. Here we solve the specific case of the Kasner solution with an isotropy plane, which is algebraically simpler to solve. This choice, applied to Eqs.\ (\ref{vec_x_ev}) --\ (\ref{vec_constr2_kasn}), gives the following dynamical system:

\begin{equation}
\label{simpl_vec_x_ev}
\dot X+\theta X-Y+ \fracc{m^2}{2}H+\fracc{1}{4}q+\fracc{1}{3}\theta\psi=0,
\end{equation}

\begin{equation}
\label{simpl_vec_y_ev}
\dot Y+\fracc{13}{6}\theta Y-\fracc{1}{3}\theta^2X+\fracc{1}{4}\theta m^2H+\fracc{1}{8}\theta q+\fracc{1}{6}\theta^2\psi=0,
\end{equation}

\begin{equation}
\label{simpl_vec_h_ev}
\dot H+\fracc{11}{6}\theta H-\fracc{1}{2}X-\fracc{1}{12}\theta\Omega+\fracc{7}{12}\fracc{\theta^2}{m^2}\psi=0,
\end{equation}

\begin{equation}
\label{simpl_vec_q_ev}
\dot q+\theta q=0,
\end{equation}

\begin{equation}
\label{simpl_vec_omega_ev}
\dot\Omega+\fracc{5}{3}\theta\Omega-\fracc{1}{2}t^{-8/3}\psi=0,
\end{equation}

\begin{equation}
\label{simpl_vec_constr1_kasn}
m^2t^{-8/3}X+\fracc{1}{3}\theta m^2H+\fracc{5}{6}\theta m^2\Omega-\theta q+\fracc{5}{18}\theta^2\psi=0,
\end{equation}

\begin{equation}
\label{simpl_vec_constr2_kasn}
m^2H-\theta X-\fracc{2}{3}\theta^2t^{8/3}\Omega+\fracc{1}{2}q=0.
\end{equation}

The acceleration $\psi$ is given in terms of the other variables by\ (\ref{simpl_vec_constr1_kasn}), which then closes the system

\begin{equation}
\label{vec_psi_kasn}
\psi=\fracc{18}{5\theta}q-\fracc{18}{5}\fracc{m^2}{\theta^2}t^{-8/3}X-\fracc{6}{5}\fracc{m^2}{\theta}H-3\fracc{m^2}{\theta}\Omega,
\end{equation}
and the closed dynamical system is then written as

\begin{equation}
\label{clos_vec_x_ev}
\dot X+\fracc{1}{5\theta}(5\theta^2-6t^{-8/3}m^2)X-Y+\fracc{1}{10}m^2H-m^2\Omega+\fracc{29}{20}q=0,
\end{equation}

\begin{equation}
\label{clos_vec_y_ev}
\dot Y+\fracc{13}{6}\theta Y-\fracc{1}{15}(5\theta^2+9t^{-8/3}m^2)X+\fracc{1}{20}\theta m^2H-\fracc{1}{2}\theta m^2\Omega+\fracc{29}{40}\theta q=0,
\end{equation}

\begin{equation}
\label{clos_vec_h_ev}
\dot H+\fracc{17}{15}\theta H-\fracc{26}{10}t^{-8/3}X-\fracc{11}{6}\theta\Omega+\fracc{21}{10}\fracc{\theta}{m^2}q=0,
\end{equation}

\begin{equation}
\label{clos_vec_omega_ev}
\dot\Omega+\fracc{1}{6\theta}(10\theta^2-9t^{-8/3}m^2)\Omega-\fracc{9}{5}\fracc{m^2}{\theta^2}t^{-16/3}X- \fracc{3}{5}\fracc{m^2}{\theta}t^{-8/3}H - \fracc{9}{5\theta}t^{-8/3}q=0,
\end{equation}

\begin{equation}
\label{clos_vec_q_ev}
\dot q+\theta q=0,
\end{equation}

\begin{equation}
\label{clos_vec_constr1_kasn}
\theta X-m^2H+\fracc{2}{3}\theta^2t^{8/3}\Omega-\fracc{1}{2}q=0.
\end{equation}

Equation\ (\ref{clos_vec_q_ev}) can be directly integrated in $q(t)$, giving

\begin{equation}
\label{sol_clos_vec_q_ev}
q(t)=q_0t^{-1},
\end{equation}
where $q_0$ is an integration constant. The rest of the system can be solved for the following ansatz:

\begin{equation}
\label{ansatz_vec_kasn}
\begin{array}{l}
X(t)=X_0t^x,\\
Y(t)=Y_0t^y,\\
H(t)=H_0t^z,\\
\Omega(t)=\Omega_0t^w,
\end{array}
\end{equation}
where $X_0, Y_0, H_0, \Omega_0, x, y, z,$ and $w$ are constants to be determined from Eqs.\ (\ref{clos_vec_x_ev}) --\ (\ref{clos_vec_constr1_kasn}), as well as the constant $m_3$. The exponents are easily obtained as

\begin{equation}
\label{exp_vec_ans}
\begin{array}{l}
x=0,\\
y=-1,\\
z=w=-5/3,
\end{array}
\end{equation}
and the remaining constants must satisfy the following conditions:

\begin{equation}
\label{eq_rem_cons}
\begin{array}{l}
8[5-3(m_3)^2]X_0-20Y_0+2(m_3)^2H_0-20(m_3)^2\Omega_0+29q_0=0,\\[2ex]
8[5+9(m_3)^2]X_0-380Y_0-6(m_3)^2H_0+60(m_3)^2\Omega_0-87q_0=0,\\[2ex]
78(m_3)^2X_0-68(m_3)^2H_0+55(m_3)^2\Omega_0-63q_0=0,\\[2ex]
54(m_3)^2X_0+18(m_3)^2H_0-5[11-9(m_3)^2]\Omega_0+54q_0=0,\\[2ex]
6X_0-6(m_3)^2H_0+4\Omega_0-3q_0=0.
\end{array}
\end{equation}
A simple but rather tedious manipulation gives four of the constants (say $X_0, Y_0 , H_0,$ and $\Omega_0$) as proportional to the fifth ($q_0$), as well as polynomials of $(m_3)^2$:

\begin{equation}
\label{cons_fun_q0m3}
\begin{array}{l}
X_0=-\fracc{1}{4}P_1q_0,\\[2ex]
H_0=-\fracc{3}{4(m_3)^2}P_2q_0,\\[2ex]
Y_0=\fracc{1}{19}\left[\fracc{2}{5}[5+9(m_3)^2]X_0-\fracc{3}{10}(m_3)^2H_0+3(m_3)^2\Omega_0-\fracc{87}{20}q_0\right],
\end{array}
\end{equation}
where

\begin{equation}
\nonumber
\begin{array}{l}
P_1\equiv\fracc{M_1[1+3(m_3)^2]^{-1}}{[450(m_3)^4+1434(m_3)^2-935]},\\[2ex]
P_2\equiv\fracc{M_2[1+3(m_3)^2]^{-1}}{[450(m_3)^4+1434(m_3)^2-935]},\\[2ex]
P_3\equiv\fracc{[39(m_3)^6+5(m_3)^4-5(m_3)^2+8]}{[450(m_3)^4+1434(m_3)^2-935]}\\[2ex]
M_1\equiv4455(m_3)^8+5328(m_3)^6+1524(m_3)^4-3583(m_3)^2+598,\\[2ex]
M_2\equiv900(m_3)^8+2523(m_3)^6+3591(m_3)^4-3465(m_3)^2-240.
\end{array}
\end{equation}
The condition over $(m_3)^2$ is then obtained as

\begin{equation}
\nonumber
24[15-11(m_3)^2]X_0+22(m_3)^2H_0-220(m_3)^2\Omega_0+319q_0=0,
\end{equation}
which---upon substitution of $X_0, H_0, \Omega_0$ and $q_0$---gives a fifth-order equation on $(m_3)^2$ that is satisfied for at least one positive value of $(m_3)^2$ namely, $(m_3)^2\approx2.9252$, and the consistency of the ansatz is then proven. The solution for the vectorial perturbation case, for a Kasner background with an isotropy plane is then written as

\begin{equation}
\label{fin_vec_var}
\begin{array}{l}
q(t)=q_0t^{-1},\\
X(t)=X_0,\\
Y(t)=Y_0t^{-1},\\
H(t)=H_0t^{-5/3},\\
\Omega(t)=\Omega_0t^{-5/3},
\end{array}
\end{equation}
with constants $X_0, Y_0, H_0,$ and $\Omega_0$ given in terms of $q_0$ by means of Eq.\ (\ref{cons_fun_q0m3}). Differently from the tensorial case, as we shall see, it is seen from Eq.\ (\ref{fin_vec_var}) above that the only existing solution is stable, although the decrease is not fast.

\subsubsection{Tensorial Perturbations}

We will proceed in this section to analyze the case of gravitational waves in the Kasner background. Here, the minimal closed set ${\cal M}$ will reduce to the four tensorial variables:

\begin{equation}
\label{tens_var_kasn}
{\cal M}=\{X_{\alpha\beta}, Y_{\alpha\beta}, H_{\alpha\beta}, \pi_{\alpha\beta}\}.
\end{equation}
However, from causal thermodynamics, a relation between the shear and the anisotropic pressure can be obtained as

\begin{equation}
\label{an_pres_rel}
\tau(\pi_{ij})\dot{}+\pi_{ij}=\xi\sigma_{ij},
\end{equation}
where $\tau$ is the relaxation parameter and $\xi$ the viscosity parameter. We will make here the same choice made for the FLRW case and consider $\tau$ as negligible; the viscosity will also be taken roughly as a constant,\footnote{In the nonequilibrium thermodynamics, both $\tau$ and $\xi$ are functions of the variables of equilibrium of the system, such as the density $\rho$ and the temperature $T$.} which makes Eq.\ (\ref{an_pres_rel})

\begin{equation}
\label{an_pres_rel_sim}
\pi_{ij}=\xi\sigma_{ij}.
\end{equation}
This result poses a problem after perturbation, for it would then imply that the anisotropic pressure---a ``good" variable in the sense of Stewart (cf. Novello, 1983), one that is zero in the background and therefore gauge-independent, can be written in terms of the shear---which, in our case, is nonzero and (in the sense of Stewart) coordinate dependent; therefore, a ``bad" variable. This apparent dilemma can be solved by taking the viscosity as a ``good" variable itself (being zero in the background and therefore gauge independent as well). However, this is a scalar quantity and, as such, not defined for tensorial perturbations. The solution to this problem is, then, to consider the viscosity itself as zero, thus writing

\begin{equation}
\label{delta_pi}
(\delta\pi_{ij})=\xi(\delta\sigma_{ij})=0,
\end{equation}
after perturbation, so that the consistency of the dynamical system is maintained. This further reduces the set ${\cal M}$:

\begin{equation}
\label{min_ten_set_red_kasn}
{\cal M}=\{X_{\alpha\beta}, Y_{\alpha\beta}, H_{\alpha\beta}\}.
\end{equation}.

We can, at this point, expand the perturbed quantities in ${\cal M}$ in terms of the tensorial basis $\hat U^{\mu}{}_{\nu}$ as

\begin{equation}
\label{exp_tens_min_set}
\begin{array}{lcl}
(\delta X^i{}_j)=\sum_{(n)}X(t)^{(n)}\hat U_{(n)}{}^i{}_j,\\
(\delta Y^i{}_j)=\sum_{(n)}Y(t)^{(n)}\hat U_{(n)}{}^i{}_j,\\
(\delta H^i{}_j)=\sum_{(n)}H(t)^{(n)}\hat U^*{}_{(n)}{}^i{}_j.
\end{array}
\end{equation}

We will deal here with linear equations so that we are able to suppress the summation and the extra indices from now on, since we can deal with each component $(n)$ separately. The same reasoning will be applied to the vectorial and the scalar cases as well.

With Eq.\ (\ref{exp_tens_min_set}) above, a perturbed dynamical system can then be written. Starting from the original QM equations, and rewriting them in terms of the new variables above, we can exhibit the perturbed dynamical system for the tensorial case as follows:

\begin{equation}
\label{d_x_dot}
\begin{array}{l}
h^{\mu}_{\alpha} h^{\beta}_{\nu} (\delta X^{\alpha}{}_{\beta})\dot{} + \fracc{5}{3}\theta(\delta X^{\mu}{}_{\nu}) + \sigma^{\mu}_{\alpha}(\delta X^{\alpha}{}_{\nu}) + \sigma^{\alpha}_{\nu}(\delta X^{\mu}{}_{\alpha}) - \fracc{2}{3}h^{\mu}_{\nu}\sigma^{\alpha}_{\beta}(\delta X^{\beta}{}_{\alpha}) - (\delta Y^{\mu}{}_{\nu}) +\\[2ex]
+\fracc{1}{2}\eta^{\mu\gamma}{}_{\alpha}{}{\beta}V_{\gamma}h_{\nu}^{\lambda}(\delta H^{\alpha}_{\lambda})_{;\beta} + \fracc{1}{2}\eta_{\nu}{}^{\gamma}{}_{\alpha}{}^{\beta}V_{\gamma}h^{\mu\lambda}(\delta H^{\alpha}_{\lambda})_{;\beta} = 0,
\end{array}
\end{equation}

\begin{equation}
\label{d_y_dot}
\begin{array}{l}
h_{\alpha}^{\mu}h_{\nu}^{\beta}(\delta Y^{\alpha}{}_{\beta})\dot{} + 2\theta(\delta Y^{\mu}{}_{\nu}) - \fracc{3}{2}\sigma^{\mu}_{\alpha}(\delta Y^{\alpha}{}_{\nu}) - \fracc{3}{2}\sigma^{\alpha}_{\nu}(\delta Y^{\mu}{}_{\alpha}) + h^{\mu}_{\nu}\sigma^{\alpha}_{\beta}(\delta Y^{\beta}{}_{\alpha}) + \fracc{3}{2}E^{\mu}_{\alpha}(\delta X^{\alpha}{}_{\nu}) +\\[2ex]
+\fracc{3}{2}E^{\alpha}_{\nu}(\delta X^{\mu}{}_{\alpha}) - h^{\mu}_{\nu} E^{\alpha}_{\beta}(\delta X^{\beta}{}_{\alpha}) + \fracc{1}{2}\theta\eta^{\mu\gamma\alpha\beta}V_{\gamma}h_{\nu\lambda}(\delta H^{\lambda}_{\alpha})_{;\beta} + \fracc{1}{2}\theta\eta_{\nu}{}^{\gamma\alpha\beta}V_{\gamma}h^{\mu}_{\lambda}(\delta H^{\lambda}_{\alpha})_{;\beta} +\\[2ex] +\fracc{3}{4}\eta^{\mu\gamma\alpha\beta}V_{\gamma}\sigma_{\nu\lambda}(\delta H^{\lambda}_{\alpha})_{;\beta} + \fracc{3}{4}\eta_{\nu}{}^{\gamma\alpha\beta}V_{\gamma}\sigma^{\mu}_{\lambda}(\delta H^{\lambda}_{\alpha})_{;\beta}
+\fracc{3}{4}\eta^{\gamma\lambda\alpha\beta}V_{\lambda}\sigma_{\gamma}^{\mu}h_{\nu\tau}(\delta H^{\tau}_{\alpha})_{;\beta} +\\[2ex] +\fracc{3}{4}\eta^{\gamma\lambda\alpha\beta}V_{\lambda}\sigma_{\gamma\nu}h^{\mu}_{\tau}(\delta H^{\tau}_{\alpha})_{;\beta} - h^{\mu}_{\nu}\eta^{\gamma\lambda\alpha\beta}V_{\lambda}\sigma_{\gamma\tau}(\delta H^{\tau}_{\alpha})_{;\beta} = 0,
\end{array}
\end{equation}

\begin{equation}
\label{sig_d_x}
2\sigma^{\alpha}_{\beta}(\delta X^{\beta}_{\alpha})=0,
\end{equation}

\begin{equation}
\begin{array}{l}
\label{d_h_dot}
h_{\alpha}^{\mu}h_{\nu}^{beta}(\delta H^{\alpha}{}_{\beta})\dot{} + \fracc{4}{3}\theta(\delta H^{\mu}{}_{\nu}) - \fracc{1}{2}\sigma^{\mu}_{\alpha}(\delta H^{\alpha}{}_{\nu}) - \fracc{1}{2}\sigma^{\alpha}_{\nu}(\delta H^{\mu}{}_{\alpha})+ \eta^{\mu\alpha\gamma}{}_{\epsilon}\eta_{\nu}{}^{\beta\lambda\tau}V_{\gamma}V_{\lambda}\sigma_{\alpha\beta}(\delta H^{\epsilon}_{\tau}) +\\[2ex] \fracc{1}{3}\theta\eta^{\mu\alpha\gamma}{}_{\epsilon}\eta_{\nu}{}^{\beta\lambda\tau}V_{\gamma}V_{\lambda}\sigma_{\alpha\beta}(\delta H^{\epsilon}_{\tau}) - \fracc{1}{2}\eta^{\mu\gamma\alpha\beta}V_{\gamma}h_{\nu\lambda}(\delta H^{\lambda}_{\alpha})_{;\beta}- \fracc{1}{2}\eta_{\nu}{}^{\gamma\alpha\beta}V_{\gamma}h^{\mu}_{\lambda}(\delta H^{\lambda}_{\alpha})_{;\beta}
\end{array}
\end{equation}

\begin{equation}
\label{cov_der_d_x}
h^{\alpha}_{\beta}h^{\mu\nu}(\delta X^{\beta}{}_{\mu})_{;\nu}+\eta^{\alpha\beta\mu\nu}V_{\beta}\sigma_{\mu\gamma}(\delta H^{\gamma}_{\nu})=0,
\end{equation}

\begin{equation}
\label{cov_der_d_h}
h^{\alpha}_{\beta}h^{\mu\nu}(\delta H^{\beta}{}_{\mu})_{;\nu}+\eta^{\alpha\beta\mu\nu}V_{\beta}\sigma_{\mu\gamma}(\delta X^{\gamma}_{\nu})=0.
\end{equation}

This dynamic system has to be decomposed in terms of the tensor basis and then solved for the specific Kasner solution. However, it is immediately seen (upon writing the perturbed terms in the tensorial basis) that the following restriction on the background must be accepted in order that the basis can be factored out from the equations

\begin{equation}
p_1=p_2,
\label{sim_kasn_exp}
\end{equation}
which implies the existence of an \textit{isotropy plane} in the Kasner original, non-perturbed background. There are two such solutions, named Kasner and Milne, and defined by

($1$) \textit{Kasner solution}

\begin{equation}
\nonumber
\begin{array}{l}
p_1=p_2=2/3,\\
p_3=-1/3,
\end{array}
\end{equation}

($2$) \textit{Milne solution}

\begin{equation}
\nonumber
\begin{array}{l}
p_1=p_2=0,\\
p_3=1.
\end{array}
\end{equation}

Here we will only consider the Kasner solution, since the Milne case has already been analyzed by Novello (1995b).

Some additional choices will be made over the tensorial basis which, while not constituting a material change on the $\hat U^{\mu}{}_{\nu}$, will simplify the algebraic steps to be taken towards a final closed dynamical system

\begin{equation}
\nonumber
\hat U^3{}_1=\hat U^3{}_2=0,
\end{equation}
whereupon we write

\begin{equation}
\label{u_sim_kasn}
\hat U^{\mu}{}_{\nu}=
\left(\begin{array}{ccc}
\alpha&0&0\\
0&-\alpha&0\\
0&0&0
\end{array}\right),
\end{equation}

\begin{equation}
\label{ustar_sim_kasn}
\hat U^{*\mu}{}_{\nu}=-\fracc{i}{2}\eta^{0123}k_3
\left(\begin{array}{ccc}
0&-2t^{2p_2}\alpha&0\\
-2t^{2p_1}\alpha&0&0\\
0&0&0
\end{array}\right).
\end{equation}

Proceeding to analyze the dynamical system, we find that Eq.\ (\ref{sig_d_x}) is identically satisfied, since
\begin{equation}
\label{u_sig}
(\sigma^{\alpha}{}_{\beta}\hat U^{\beta}{}_{\alpha})=0,
\end{equation}
while Eqs.\ (\ref{cov_der_d_x}) and\ (\ref{cov_der_d_h}) are also identically valid through similar arguments. The remaining equations can then be rewritten, giving

\begin{equation}
\label{re_ten_eq_kasn}
\begin{array}{l}
\dot X+\fracc{7}{3}\theta X-Y+k^2H=0,\\
\dot Y+\theta Y+\fracc{2}{3}\theta^2X+2\theta k^2H=0,\\
\dot H+\fracc{2}{3}\theta H-t^{-4/3}X=0,
\end{array}
\end{equation}
which constitutes a closed dynamical system in the variables $(X, Y, H)$. This result is analogous to the ones obtained in the FLRW case, but in this case the system can be completely solved by using the relations

\begin{equation}
\nonumber
\begin{array}{l}
\theta=t^{-1},\\
k^2=t^{-2/3}(k_3)^2,\hspace{.5 cm} k_3\equiv const,
\end{array}
\end{equation}
and considering a simple form for the desired solution, in terms of powers of $t$,

\begin{equation}
\label{xyh_pow_t}
\begin{array}{l}
X(t)=X_0t^x,\\
Y(t)=Y_0t^y,\\
H(t)=H_0t^w,\\
\end{array}
\end{equation}
with $X_0$, $Y_0$, $H_0$, $x$, $y$ and $w$ as constants to be determined.

\begin{table}[htb]
\centering
\begin{tabular}{|c|c|c|c|c|}
\hline
\hline
& & & & \\ [1ex]
Value for $w$ & & $X(t)$   & $Y(t)$   & $H(t)$ \\[1ex] \hline
& & & & \\ [1ex]
$w<-2/3$      & & stable   & stable   & stable \\[1ex]
& & & & \\ [1ex]
$w=-2/3$      & & null     & null     & stable \\[1ex]
& & & & \\ [1ex]
\hspace{.3cm}$-2/3<w<-1/3$ \hspace{.3cm}& & stable   & stable   & stable \\[1ex]
& & & & \\ [1ex]
$w=-1/3$      & & constant & stable   & stable \\[1ex]
& & & & \\ [1ex]
$-1/3<w<0$    & & unstable & stable   & stable \\[1ex]
& & & & \\ [1ex]
$w=0$         & & unstable & stable   & constant \\[1ex]
& & & & \\ [1ex]
$0<w<2/3$     & & unstable & stable   & unstable \\[1ex]
& & & & \\ [1ex]
$w=2/3$       & & unstable & constant & unstable \\[1ex]
& & & & \\ [1ex]
$w>2/3$       & & \hspace{.3cm}unstable \hspace{.3cm}& \hspace{.3cm}unstable \hspace{.3cm}& \hspace{.3cm}unstable \hspace{.3cm}\\[1ex]
& & & & \\ [1ex]
\hline
\hline
\end{tabular}
\caption{Stability analysis results for tensorial perturbations.}
\label{tab3}
\end{table}

Replacing this ansatz in the dynamical system above, Eq.\ (\ref{re_ten_eq_kasn}), we obtain the following equations:

\begin{equation}
\label{re_ten_eq_kasn}
\begin{array}{l}
(3x+7)X_0t^{(x-1)}-3Y_0t^y+3(k_3)^2H_0t^{(w-2/3)}=0,\\[2ex]
2X_0t^{(x-1)}-3(y+1)Y_0t^y+6(k_3)^2H_0t^{(w-2/3)}=0,\\[2ex]
3X_0t^{(x-1)}-(3w+2)H_0t^{(w-2/3)}=0.
\end{array}
\end{equation}
It is immediately seen that the only non-trivial solutions satisfy the conditions

\begin{equation}
\label{cond_res_kasn_tens}
y=x-1=w-\fracc{2}{3}\Rightarrow w=x-\fracc{1}{3},
\end{equation}
and the dynamical system reduces to the following conditions on the triad $(X_0, Y_0, H_0)$ and the constant $(k_3)^2$:

\begin{equation}
\label{sub_cond_above_re_ten_eq_kasn}
\begin{array}{l}
(3x+7)X_0-3Y_0+3(k_3)^2H_0=0,\\
2X_0-3(y+1)Y_0+6(k_3)^2H_0=0,\\
3X_0-(3w+2)H_0=0.
\end{array}
\end{equation}
These conditions then give $X_0$ and $Y_0$ in terms of $H_0$ and the exponents $x, y, w$:

\begin{equation}
\label{x0_y0_func_h0}
\begin{array}{l}
X_0=\fracc{(3w+2)}{3}H_0,\\[2ex]
Y_0=\fracc{2}{3}\fracc{(x+2)(3w+2)}{(y+3)}H_0,
\end{array}
\end{equation}
which, upon employing Eqs.\ (\ref{cond_res_kasn_tens}) and\ (\ref{x0_y0_func_h0}) in the ansatz\ (\ref{xyh_pow_t}), gives

\begin{equation}
\label{fin_xyh_pow_t}
\begin{array}{l}
X(t)=\fracc{(3w+2)}{3}t^{1/3}H(t),\\[2ex]
Y(t)=\fracc{2}{3}\fracc{(3w+2)^2}{(3w+7)}t^{-2/3}H(t),\\[2ex]
H(t)=H_0t^w.
\end{array}
\end{equation}
We also determine the constant $(k_3)^2$ in terms of $x, y, w$ as

\begin{equation}
\label{k_3_func_xyw}
(k_3)^2=-\fracc{1}{9}(3w+2)^2.
\end{equation}
If we take the arbitrary constant $H_0$ as positive, an analysis of the stability of the solutions above, Eq.\ (\ref{fin_xyh_pow_t}), gives the results exhibited in Table \ref{tab3}.

From this table, it is immediately seen that tensorial perturbations of a Kasner background may present---upon a choice of the exponent $w$---the same kind of instability which will be present in Friedman-Lemaître-Robertson-Walker spacetime. This instability is not of a catastrophic kind (as in the Einstein
model), but is rather gradual. This is a reasonable development, since we are not interested in eliminating the background anisotropy upon perturbation.

\subsection{Friedman Universe: Scalar Perturbations}\protect\label{scalarpert}

In the case of the spatially homogeneous and isotropic FLRW cosmological model, the vanishing of Weyl conformal tensor suggests that the QM approach is more useful. Therefore, the variation of Weyl conformal tensor $\delta W_{\alpha\beta\mu\nu}$ is the basic quantity to be considered, once there is certainly no doubt that $\delta W_{\alpha\beta\mu\nu}$ is a true perturbation, which can not be achieved by a coordinate transformation. This solves {\it ab initio} the gauge problem that was pointed out before.

From a technical standpoint, instead of considering tensorial quantities, one should restrain oneself to scalar ones. There are two ways to implement this:

\begin{itemize}
\item{Expand the relevant quantities in terms of a complete basis of functions (e.g. the spherical harmonics basis);}
\item{Analyze the invariant geometric quantities one can construct from $g_{\mu\nu}$ and its derivatives in the Riemannian background structure, that is, examine the 14 Debever invariants.}
\end{itemize}

Either way, we shall see that the net result is that there is a set of perturbed quantities which can be divided into ``good" quantities (i.e., the ones whose unperturbed counterparts have zero value in the background and, consequently, Stewart\rq s lemma (cf. Stewart 1974) guarantees that the associated perturbed quantity is really a gauge-independent one) and ``bad" ones (whose corresponding values in the background are nonzero). One should limit therefore the analysis only to the ``good" ones.

This same kind of behavior occurs for the geometrical structure of the model for both the kinematic and dynamic quantities of matter. Therefore the
``good" quantities which constitute the set of variables with which we work should then be chosen from these particular scalars that come from these three structures: geometric, kinematic and dynamic. Does that mean that the present approach effectively avoids the gauge problem?

To answer this question affirmatively one should be able to exhibit a set of ``good" variables in such a way that its corresponding dynamics is closed. That is, if we call ${\cal M}_{[A]}$ the set of these variables, Einstein's equations should provide the dynamics of each element of ${\cal M}_{[A]}$, depending only on the background evolution quantities (and, eventually, on other elements of ${\cal M}_{[A]}$). This would exhaust the perturbation problem and we shall show that this is indeed the case.

What should be learned from this discussion is that one should then understand the gauge problem not as a basic difficulty on the perturbation theory but just as a simple matter of asking a bad question\footnote{Let us point out that some of the gauge dependent terms are particularly relevant, $\delta\rho$ including.}. One could imagine---which has been used a number of times in the literature (Hawking 1966, Olson 1976 and Mukhanov 1992)---that for FRW cosmology the perturbations of its main characteristics (the energy density, $\delta\rho$ the scalar of curvature $\delta R$ and the Hubble expansion factor $\delta\theta$) would be natural quantities to be considered as basic for the perturbation scheme. However, these are not ``good" scalars, since they are not zero in the background \footnote{However, as we shall see soon, we can construct associated ``good" quantities in terms of these scalars.}. We shall see in the next sections which scalars replace these ones.

Consider FRW geometry written in the standard Gaussian coordinate system as in Eq.\ (\ref{fried_met}). The 3-dimensional geometry has constant curvature and thus the corresponding Riemannian tensor $\hat R_{ijkl}$ can be written as

\begin{equation}
\nonumber
\hat R_{ijkl} = -\epsilon\gamma_{ijkl}.
\end{equation}
where $\gamma_{ijkl}\equiv\gamma_{ik}\gamma_{jl}-\gamma_{il}\gamma_{jk}$ was defined in Eq. (\ref{tri_curv}). For a while, it is necessary to distinguish covariant derivative in the 4-dimensional space-time by the symbol (;) and the 3-dimensional derivative by (${}_{\|}$).

Since the original Lifshitz paper, it has shown to be useful to develop all perturbed quantities in the spherical harmonics basis. Once we are limiting ourselves to irrotational perturbations, it suffices to our purposes to take into account only the scalar $Q(x^{k})$ (with $\dot{Q} = 0$) and its derived vector and tensor quantities. We have thus

\begin{equation}
\begin{array}{l}
Q_{i} \equiv Q_{,i}, \\
Q_{ij} \equiv Q_{,i;j},
\end{array}
\protect\label{d12}
\end{equation}
where the scalar $Q$ obeys the eigenvalue equation defined in the 3-dimensional background space by:

\begin{equation}
\hat\nabla^{2}Q = mQ,
\protect\label{d13}
\end{equation}
where $m$ is the wave number of the scalar eigenfunction, with

\begin{equation}
\protect\label{mQ}
m=\left\{
\begin{array}{lll}
q^2+1, & 0<q<\infty,     & \epsilon=1   \mbox{\rm\ (open)},\\
q,     & 0<q<\infty,     & \epsilon=\ \ 0\mbox{\rm\ (plane) },\\
n^2-1, & n=1,\,2,\ldots, & \epsilon=-1   \mbox{\rm\ (closed)},
\end{array}
\right.
\end{equation}
and

\begin{equation}
\hat\nabla^{2}Q \equiv \gamma^{ik} Q_{,i\| k} = \gamma^{ik} Q_{,i;k},
\protect\label{d14}
\end{equation}
where the symbol $\hat\nabla^{2}$ denotes the 3-dimensional Laplace-Beltrami operator. The traceless operator $\bar{Q}_{ij}$ is defined as

\begin{equation}
\bar{Q}_{ij} = \frac{1}{m} Q_{ij} - \frac{1}{3}Q \gamma_{ij},
\protect\label{d15}
\end{equation}
and the divergence of $\bar{Q}_{ij}$ is given by

\begin{equation}
{\bar{Q}^{ik}}_{  \| k} = 2 \left(\frac{1}{3} + \frac{\epsilon}{m}\right)
\hspace{0.1cm}Q^{i}.
\protect\label{d16}
\end{equation}
We remark that $Q$ is a 3-dimensional object; therefore indices are raised with $\gamma^{ij}$, the 3-space metric.

In Debever (1964), the complete 14 algebraically independent invariants constructed with the curvature tensor were presented. Considering that we are using an adimensional metric tensor, we can classify them with respect to
dimensionality as follows:

\vspace{0.5cm}
\begin{center}
\begin{tabular}{||c|c||}
\hline \hline
& \\[1ex]
\hspace{.3cm}Dimensionality \hspace{.3cm}& Invariants \\ \hline
& \\[1ex]
$L^{-2}$ & $I_{5}$ \\ \hline
& \\[1ex]
$L^{-4}$ & $I_{1}$, $I_{3}$, $I_{6}$ \\ \hline
& \\[1ex]
$L^{-6}$ & \hspace{.3cm}$I_{2}$, $I_{4}$, $I_{7}$, $I_{9}$, $I_{12}$ \hspace{.3cm}\\ \hline
& \\[1ex]
$L^{-8}$ & $I_{8}$, $I_{10}$, $I_{13}$ \\ \hline
& \\[1ex]
$L^{-10}$ & $I_{11}$, $I_{14}$ \\
\hline \hline
\end{tabular}
\end{center}
\vspace{0.5cm}

The expressions for these invariants are:

\begin{equation}
\nonumber
\begin{array}{lll}
I_{1} = W_{\alpha\beta\mu\nu} W^{\alpha\beta\mu\nu}, & & I_{8} = C_{\alpha\beta} C^{\beta\mu} C_{\mu\lambda} C^{\alpha\lambda}, \\
I_{2} = {W_{\alpha\beta}}^{\rho\sigma} {W_{\rho\sigma}}^{  \mu\nu}{W_{\mu\nu}}^{  \alpha\beta}, & & I_{9} = C_{\mu\nu} D^{\mu\nu}, \\
I_{3} = W^{\alpha\beta\mu\nu}\,{}^{*}W_{\alpha\beta\mu\nu}, & & I_{10} = D_{\mu\nu} D^{\mu\nu}, \\
I_{4} = W^{\alpha\beta\rho\sigma} {W_{\rho\sigma}}^{  \mu\nu}\,{}^{*}W_{\mu\nu\alpha\beta}, & & I_{11} = C_{\alpha\beta} D^{\beta\mu} {D_{\mu}}^{ \alpha}, \\
I_{5} = R, & & I_{12} = \tilde{D}_{\mu\nu} C^{\mu\nu}, \\
I_{6} = C_{\mu\nu} C^{\mu\nu}, & & I_{13} = \tilde{D}_{\mu\nu} D^{\mu\nu}, \\
I_{7} = C_{\alpha\beta} C^{\beta\mu} {C_{\mu}}^{\alpha}, & & I_{14} = \tilde{D}_{\mu\nu} \tilde{D}^{\nu\alpha} {C^{\mu}}_{ \alpha}.
\end{array}
\end{equation}
in which we used the following definitions:

\begin{equation}
\begin{array}{l}
C_{\mu\nu} \equiv R_{\mu\nu} - \frac{1}{4} R g_{\mu\nu}, \\[2ex]
D_{\mu\nu} \equiv W_{\mu\alpha\nu\beta} C^{\alpha\beta}, \\[2ex]
\tilde{D}_{\mu\nu} \equiv {}^{*}W_{\mu\alpha\nu\beta}
C^{\alpha\beta}.
\end{array}
\protect\label{d17}
\end{equation}

\subsubsection{Fundamental Perturbations of FLRW Universe}\protect\label{fundamental}

As we observed previously, a complete examination of the perturbation theory should naturally include the analysis of the evolution of the Debever metric invariants associated to FLRW geometry.

The only non identically zero invariants of FLRW geometry are given by

\begin{equation}
\nonumber
\begin{array}{l}
I_{5} = (1 - 3\lambda) \rho, \\[2ex]
I_{6} = \frac{3}{4} (1 + \lambda)^{2} \rho^{2}, \\[2ex]
I_{7} = - \frac{3}{8} (1 + \lambda)^{3} \rho^{3}, \\[2ex]
I_{8} = \frac{21}{64} (1 + \lambda)^{4} \rho^{4},
\end{array}
\end{equation}
in which we used Einstein's equations, and the stress-energy tensor is that of a perfect fluid.

If we restrict ourselves to the linear perturbation theory, the only invariants which have non identically zero linear perturbation terms are $I_{5}$, $I_{6}$, $I_{7}$, $I_{8}$,
$I_{9}$ and $I_{12}$. Among these the first four are nonzero in the background and the latter two are zero, since the geometry is conformally flat. This could lead to the conclusion that $I_{9}$ and $I_{12}$ are the \mbox{\lq \lq good\rq \rq} scalars to be examined. However, a direct calculation shows that the latter two invariants have zero
linear perturbation. Indeed, it follows from FLRW geometry that the perturbation of $I_{9}$ reduces to

\begin{equation}
\nonumber
\delta I_{9} = C^{\mu\nu} C^{\alpha\beta} \delta W_{\mu\alpha\nu\beta}.
\end{equation}
Then (due to the fact that Weyl tensor is trace-free) the above quantity identically vanishes. This result depends of course on the fact that the source of the background geometry is given by a perfect fluid. In effect we have in this case

\begin{equation}
\nonumber
\delta I_{9} =  (\rho + p)^{2} \left(V^{\mu} V^{\nu} -
\frac{1}{4} g^{\mu\nu} \right)
\left(V^{\alpha} V^{\beta} - \frac{1}{4} g^{\alpha\beta} \right)
\delta W_{\mu\alpha\nu\beta},
\end{equation}
which is zero. Likewise $\delta I_{12}$, given by

\begin{equation}
\nonumber
\delta I_{12} = C^{\mu\nu} C^{\alpha\beta} \delta\,{}^*W_{\mu\alpha\nu\beta},
\end{equation}
also vanishes.

The corresponding perturbations for the remaining invariants are given by

\begin{equation}
\nonumber
\begin{array}{l}
\delta I_{5} = (1 - 3\lambda)\delta\rho, \\[2ex]
\delta I_{6} = \frac{3}{2} (1 + \lambda)^{2}\rho \hspace{0.1cm}\delta\rho, \\[2ex]
\delta I_{7} = - \frac{9}{8} (1 + \lambda)^{3} \rho^{2} \hspace{0.1cm}
\delta\rho, \\[2ex]
\delta I_{8} = \frac{21}{16} (1 + \lambda)^{4} \rho^{3} \hspace{0.1cm}
\delta\rho.
\end{array}
\end{equation}

It follows from these results that the perturbations of these quantities are algebraically related\footnote{One can write these invariants in a pure geometrical way without using Einstein\rq s equations. This does not modify our argument.}. Besides, once all these scalars have a non-zero background value, they do not belong to the minimum set of good quantities that we are searching for.

Corresponding difficulties occur for the standard kinematical and dynamical variables, that is, the expansion parameter $\theta$ and the density of energy $\rho$ suffer from the same disease.

This is thus the bad choice for the basic variables which we should avoid. Let us now turn our attention to which good variables should be considered as the fundamental ones.

\underline{{\bf Geometric Perturbations:}} From the previous section it follows that

\begin{equation}
\nonumber
\sqrt{\delta E_{ij} \delta E^{ij}},
\end{equation}
is the only quantity that characterizes without ambiguity a true perturbation of the Debever invariants\footnote{This is a consequence of the vanishing of the perturbation of the magnetic part of Weyl tensor(cf.above).}. We need thus to only
consider the perturbed $E_{ij}$ since, as we shall see, any other metric quantity does not belong to the ``good" basic nucleus needed for a complete knowledge of the true perturbations. We then set the expansion of this tensor in terms of the spherical harmonic basis

\begin{equation}
\delta E_{ij} = E(t) \hspace{0.1cm}\bar{Q}_{ij}(x^{k}).
\protect\label{g1}
\end{equation}
Thus $E(t)$ is the geometric quantity whose dynamics we are looking for.

\underline{{\bf Kinematical Perturbations:}} We restrict our considerations only to linear perturbation terms. The
normalization of the 4-velocity yields that the variation of the time component of the perturbed velocity is related to the variation of the (0-0) component of the metric tensor, that is:

\begin{equation}
\delta V_{0} = \frac{1}{2} \delta g_{00}.
\protect\label{k1}
\end{equation}
The corresponding contravariant quantities are related as follows:

\begin{equation}
\delta V^{0} = \frac{1}{2} \delta g^{00} = - \delta V_{0}.
\protect\label{k2}
\end{equation}

The expansion of the perturbations of the 4-velocity in terms of the
spherical harmonic basis is\footnote{The vorticity
is of course zero, since we are limiting ourselves to the irrotational case.}

\begin{equation}
\begin{array}{l}
\delta V_{0} = \frac{1}{2} \beta (t) \hspace{0.1cm}Q(x^{i})
+ \frac{1}{2} Y(t), \\[2ex]
\delta V_{k} = V(t) \hspace{0.1cm}Q_{k}(x^{i}).
\end{array}
\protect\label{k3}
\end{equation}
For the acceleration we set

\begin{equation}
\delta a_{k} = \Psi (t) \hspace{0.1cm}Q_{k}(x^{i}).
\protect\label{k4}
\end{equation}
For the shear

\begin{equation}
\delta \sigma_{ij} = \Sigma (t) \hspace{0.1cm}\bar{Q}_{ij}(x^{k}),
\protect\label{k5}
\end{equation}
and for the expansion we set

\begin{equation}
\delta\theta = H(t) \hspace{0.1cm}Q(x^{i}) + Z(t),
\protect\label{k6}
\end{equation}
where $Y(t)$ and $Z(t)$ are homogeneous terms that are not true perturbations.

Let us point out that, once we are limiting ourselves to the analysis of true perturbed quantities, the important kinematical variable whose dynamics we need to examine is only $\Sigma(t)$, since the other gauge-invariant quantity $\Psi$ is a function of $\Sigma$ (and $E$), as we shall see ($\beta$ is just a matter of choosing of the coordinate system).

\underline{{\bf Matter Perturbation:}} Since we are considering a background geometry in which there is a state equation relating the pressure and the energy density, i.e. $p = \lambda\rho$, we will consider the standard procedure that accepts the preservation of this state equation under arbitrary perturbations. Besides, our frame is such that there is no heat flux. Thus the general form of the
perturbed energy-momentum tensor is given by

\begin{equation}
\delta T_{\mu\nu} = (1 + \lambda) \hspace{0.1cm}\delta (\rho V_{\mu} V_{\nu})
- \lambda \delta (\rho g_{\mu\nu}) + \delta \Pi_{\mu\nu}.
\protect\label{m1}
\end{equation}

We write $\delta \rho$ in terms of the scalar basis as:

\begin{equation}
\delta\rho = N(t) \hspace{0.1cm}Q(x^{i}) + \mu (t),
\protect\label{m2}
\end{equation}
in which the homogeneous term $\mu (t)$ is not a true
perturbation\footnote{We will set $Y = Z = \mu = 0$, since
these homogeneous terms are just a matter of choosing of the coordinate system. Nevertheless we are not interested in examining pure gauge quantities such as $Y$, $Z$ and $\mu$.}.

According to causal thermodynamics, the evolution equation of the anisotropic pressure is related to the shear through

\begin{equation}
\tau\dot{\Pi}_{ij} + \Pi_{ij} = \xi\sigma_{ij},
\protect\label{extra1}
\end{equation}
in which $\tau$ is the relaxation parameter and $\xi$ is the viscosity parameter as we saw previous section. For simplicity of this present treatment, we will limit ourselves to the case in which $\tau$ can be neglected and $\xi$ is a constant\footnote{In the general case, $\xi$ and $\tau$ are functions of the
equilibrium variables, for instance $\rho$ and the temperature $T$ and, since both variations $\delta\Pi_{ij}$ and $\delta\sigma_{ij}$ are expanded in terms of the traceless tensor $\hat{Q}_{ij}$, it follows that the above relation does not restrain the kind of fluid we are examining. However, if we consider $\xi$ as time-dependent, the quantity $\delta\Pi_{ij}$ must be included in the fundamental set ${\cal M}_{[A]}$.}; Eq.\ (\ref{extra1}) then gives

\begin{equation}
\Pi_{ij} = \xi\sigma_{ij},
\protect\label{extra2}
\end{equation}
and the associated perturbed equation is:
\begin{equation}
\delta\Pi_{ij} = \xi \hspace{0.1cm}\delta\sigma_{ij}.
\protect\label{m3}
\end{equation}

Following the same reasoning as before, $\delta \Pi_{ij}$ is the matter quantity that should enter in the complete system of differential equations which describes the perturbation evolution. One should also be interested in the dynamics of $\delta\rho$ although it is not a fundamental part of the
basic system of equations. We will examine its evolution later on.

The ``good" set ${\cal M}_{[A]}$ has therefore three elements: $\delta E_{ij}$, $\delta\sigma_{ij}$ and $\delta\Pi_{ij}$. But, since $\delta\Pi_{ij}$ is written in terms of $\delta\sigma_{ij}$, the set ${\cal M}_{[A]}$ which will be considered reduces to:

\begin{equation}
\nonumber
{\cal M}_{[A]} = \{\delta E_{ij}, \delta\sigma_{ij}\}.
\end{equation}
So much for definitions. Let us then turn to the analysis of the dynamics.

\subsubsection{Dynamics}\protect\label{dynamics}

In this section we will show that $E(t)$ and $\Sigma (t)$ constitute the fundamental pair of variables in terms of which all the dynamics for the perturbed FRW geometry is given, that is, \mbox{${\cal M}_{[A]} = \{E(t), \Sigma (t)\}$} is the minimal closed set of observables in the perturbation theory of FRW which characterizes and completely determines the spectrum of perturbations. Indeed, the evolution equations for these two quantities (which come from Einstein's equations) generate a dynamical system only involving $E$ and $\Sigma$ (and background quantities) which, when solved, contains all the necessary information for a complete description of all remaining perturbed quantities of FRW geometry. Such conclusion does not seem to have been noticed in the past.

We remark that we will limit ourselves only to the examination of the perturbed quantities that are relevant for the complete knowledge of the system. These equations are the quasi-Maxwellian equations of gravitation and the evolution equations for the kinematical quantities. In Vishniac (1990) and Novello (1988), this system of equations was presented and analyzed.

\underline{{\bf The Perturbed Equation for the Shear:}} The perturbed equation for the shear Eq.\ (\ref{apb18}) is written as:

\begin{eqnarray}
{h_{\alpha}}^{\mu}{h_{\beta}}^{\nu}\,(\delta\sigma_{\mu\nu})\,\dot{} + \frac{2}{3}\theta\,\delta\sigma_{\alpha\beta} + \frac{1}{3} h_{\alpha\beta}\,\delta
{a^{\lambda}}_{;\lambda} - \frac{1}{2}{h_{\alpha}}^{\mu}{h_{\beta}}^{\nu}[\delta a_{\mu ;\nu} + \delta a_{\nu ;\mu}] = \delta M_{\alpha\beta},
\protect\label{s1}
\end{eqnarray}
where

\begin{equation}
M_{\alpha\beta} \equiv R_{\alpha\mu\beta\nu} \hspace{0.1cm}V^{\mu} V^{\nu} -
\frac{1}{3} R_{\mu\nu} \hspace{0.1cm}V^{\mu}V^{\nu} \hspace{0.1cm}h_{\alpha
\beta}.
\protect\label{s2}
\end{equation}

Using the above spherical harmonics expansion and Eq.\ (\ref{m3}), Eq.\ (\ref{s1}) reduces to:

\begin{equation}
\dot{\Sigma} = -E - \frac{1}{2}\xi \hspace{0.1cm}\Sigma + m \hspace{0.1cm}
\Psi.
\protect\label{s3}
\end{equation}

\underline{{\bf The Perturbed Equation for $E_{ij}$:}} The perturbed equation for the electric part of the Weyl tensor is given by\ (\ref{apb13}). Using the above spherical harmonics expansion and Eq.(\ref{m3}), one obtains:

\begin{eqnarray}
\dot{E} = - \frac{(1 + \lambda)}{2} \rho\,\Sigma - \left(\frac{\Theta}{3} + \frac{\xi}{2}\right) E - \frac{\xi}{2}\left(\frac{\xi}{2} +
\frac{\Theta}{3}\right) \Sigma + \frac{m}{2}\xi\,\Psi.
\protect\label{e2}
\end{eqnarray}
This suggests that $E$ and $\Sigma$ may be considered as canonically
conjugated variables. We shall see later on that this is indeed the case.

Equations (\ref{s3}) and (\ref{e2}) contain three variables: $E$, $\Sigma$ and $\Psi$. We will now show that using the conservation law for the matter we can eliminate $\Psi$ in all cases, except when $(1 + \lambda) = 0$. We will return to this particular (vacuum) case in a later section.

The proof is the following. Projecting the conservation equation of the energy-momentum tensor in the 3-space, that is

\begin{equation}
{T^{\mu\nu}}_{;\nu} {h_{\mu}}^{\lambda} = 0,
\protect\label{e3}
\end{equation}
and using the perturbed quantities this equation gives:

\begin{equation}
(1 + \lambda) \rho \hspace{0.1cm}\delta a_{k} - \lambda (\delta \rho)_{,k} +
\lambda \dot{\rho} \hspace{0.1cm}\delta V_{k} + \delta {{\Pi_{k}}^{i}}_{;i}
= 0.
\protect\label{e4}
\end{equation}

Using the decomposition in the spherical harmonics basis we obtain

\begin{equation}
(1 + \lambda) \rho \Psi = \lambda [N - \dot{\rho} V] +
2\xi \left(\frac{1}{3} + \frac{\epsilon}{m}\right) \hspace{0.1cm}A^{-2}
\hspace{0.1cm}\Sigma.
\protect\label{e5}
\end{equation}

Now comes a remarkable result: the right hand side of Eq.\ (\ref{e5}) can be expressed in terms of the variables $E$ and $\Sigma$ only (since we are analyzing here the case where $(1 + \lambda)$ does not vanish). Indeed, from the equation of divergence of the electric tensor---see Eq.\ (\ref{apb16}---we find

\begin{equation}
N - \dot{\rho} V = \left(1 + \frac{3\epsilon}{m}\right) \hspace{0.1cm}\xi
\hspace{0.1cm}\Sigma A^{-2} - 2 \left(1 + \frac{3\epsilon}{m}\right) \hspace{0.1cm}
A^{-2} \hspace{0.1cm}E.
\protect\label{e6}
\end{equation}

Combining these two equations we find that $\Psi$ is given in terms of the background quantities and the basic perturbed terms $E$ and $\Sigma$:

\begin{equation}
(1 + \lambda) \hspace{0.1cm}\rho \hspace{0.1cm}\Psi = 2
\left(1 + \frac{3\epsilon}{m}\right) \hspace{0.1cm}A^{-2} \hspace{0.1cm}
[ -\lambda E + \frac{1}{2} \hspace{0.1cm}\lambda \hspace{0.1cm}\xi
\hspace{0.1cm}\Sigma + \frac{1}{3} \hspace{0.1cm}\xi \hspace{0.1cm}\Sigma].
\protect\label{e7}
\end{equation}

Thus the whole set of perturbed equations reduces, for the variables $E$ and $\Sigma$, to a time-dependent dynamical system:

\begin{equation}
\begin{array}{l}
\dot{\Sigma} = F_{1}(\Sigma, E), \\[2ex]
\dot{E} = F_{2}(\Sigma, E),
\end{array}
\protect\label{e8}
\end{equation}
with

\begin{equation}
\nonumber
F_{1} \equiv -E - \frac{1}{2}\xi \hspace{0.1cm}\Sigma + m\Psi,
\end{equation}
and

\begin{eqnarray}
\nonumber
F_{2} \equiv - \left(\frac{1}{3}\theta + \frac{1}{2}\xi\right)E - \left(\frac{1}{4}\xi^{2} + \frac{(1 + \lambda )}{2}\rho + \frac{1}{6} \xi\theta\right) \Sigma + \frac{m}{2} \xi\Psi,
\end{eqnarray}
in which $\Psi$ is given in terms of $E$ and $\Sigma$ by Eq.\ (\ref{e7}).

\subsubsection{Comparison with Previous Gauge-Invariant Variables}\protect\label{previous}

FLRW cosmology is characterized by the homogeneity of the fundamental variables that specify its kinematics (the expansion factor $\theta$), its dynamics (the energy density $\rho$) and its associated geometry (the scalar of curvature $R$). This means that these three quantities depend only on the global time $t$, characterized by the hyper-surfaces of homogeneity. We can thus use this fact to define in a trivial way 3-tensor associated quantities, which vanish in this geometry, and look for its corresponding non-identically vanishing perturbation. The simplest way to do this is just to let $U$ be a homogeneous variable (in the present case, it can be any one of the quantities $\rho$,
$\theta$ or $R$), that is $U = U(t)$. Then, we use the 3-gradient operator $\hat\nabla_{\mu}$ defined by

\begin{equation}
\hat\nabla_{\mu} \equiv
{h_{\mu}}^{\lambda} \hspace{0.1cm}\nabla_{\lambda},
\protect\label{prev1}
\end{equation}
to produce the desired associated variable

\begin{equation}
U_{\mu} = \hat\nabla_{\mu} U.
\protect\label{prev2}
\end{equation}

According to Ellis (1989), these quantities were discussed and its associated evolution analyzed. In the present section we will exhibit the relation of these variables to our fundamental ones. We shall see that under the conditions of our analysis\footnote{Remind that here we restrain our examination to irrotational perturbation. The formulas which we obtain are thus simpler. However, the method of our analysis is not restrictive and the study of generic cases can be obtained through the same lines.} these quantities are functionals of our basic variables ($E$ and $\Sigma$) and the background ones.

\underline{{\bf The Matter Variable $\chi_{i}$:}} It seems useful to define the fractional gradient of the energy density $\chi_{\alpha}$ as

\begin{equation}
\chi_{\alpha} \equiv \frac{1}{\rho} \hspace{0.1cm}
\hat\nabla_{\alpha} \hspace{0.1cm}\rho.
\protect\label{1234}
\end{equation}

Such quantity $\chi_{\alpha}$ is nothing but a combination of the acceleration and the divergence of the anisotropic stress. Indeed, from Eq.\ (\ref{e4}) it follows (in the frame in which there is no heat flux)

\begin{equation}
\delta \chi_{i} = \frac{(1 + \lambda)}{\lambda} \hspace{0.1cm}
\delta a_{i} + \frac{1}{\lambda\rho} \hspace{0.1cm}
\delta{{\Pi_{i}}^{\beta}}_{;\beta},
\protect\label{1235}
\end{equation}

From what we have learned above, it follows that this quantity can be reduced to a functional of the basic quantities of perturbation, that is $\Sigma$ and $E$, yielding

\begin{equation}
\delta \chi_{i} = -2 \hspace{0.1cm}\left(1 + \frac{3\epsilon}{m}\right) \hspace{0.1cm}
\frac{1}{\rho a^{2}} \hspace{0.1cm}\left(E - \frac{\xi}{2} \Sigma\right)
\hspace{0.1cm}Q_{i}.
\protect\label{1236}
\end{equation}

\underline{{\bf The Kinematical Variable $\eta_{i}$:}} The only non-vanishing quantity of the kinematics of the cosmic
background fluid is the (Hubble) expansion factor $\theta$. This allows us to define the quantity $\eta_{\alpha}$ as:

\begin{equation}
\eta_{\alpha} = {h_{\alpha}}^{\beta} \hspace{0.1cm}\theta_{,\beta}.
\protect\label{1237}
\end{equation}
Using the constraint relation Eq.\ (\ref{eq_vinc_1}) we can relate this quantity to the basic ones:

\begin{equation}
\delta \eta_{i} = -\frac{\Sigma}{a^{2}} \hspace{0.1cm}\left(1 + \frac{3\epsilon}{m}
\right) \hspace{0.1cm}Q_{i}.
\protect\label{1238}
\end{equation}

\underline{{\bf The Geometrical Variable $\tau$:}} We can choose the scalar of curvature $R$ which depends only on the cosmical time $t$ like $\rho$ and $\theta$ to be the $U$-geometrical variable. However, it seems more appealing to use a combined expression $\tau$ involving $R$, $\rho$ and $\theta$ given by

\begin{equation}
\tau = R + (1 + 3\lambda) \hspace{0.1cm}\rho - \frac{2}{3} \hspace{0.1cm}
\theta^{2}.
\protect\label{1239}
\end{equation}
In the unperturbed FLRW background this quantity is defined in terms of the curvature scalar of the 3-dimensional space and the scale factor $a(t)$:

\begin{equation}
\nonumber
\frac{\hat R}{a^{2}}.
\end{equation}
We define then the new associated variable $\tau_{\alpha}$ as

\begin{equation}
\tau_{\alpha} = {h_{\alpha}}^{\beta} \hspace{0.1cm}\tau_{,\beta}.
\protect\label{1311}
\end{equation}
This quantity $\tau_{\alpha}$ vanishes in the background. Its perturbation can be written in terms of the previous variations, since Einstein's equations give

\begin{equation}
\nonumber
\tau = 2 \hspace{0.1cm}\left(\rho - \frac{1}{3} \hspace{0.1cm}
\theta^{2}\right).
\end{equation}

We can thus, without any information loss, limit all our
analysis to the fundamental variables. Nevertheless, just for completeness, let us exhibit the evolution equations for some gauge-dependent variables.

\underline{{\bf Perturbed Equations for $\rho$ and $\theta$:}} From Eq.(\ref{apb23}) and using the decomposition of the perturbed energy density in the scalar basis (defined by Eq.(\ref{m2})) we obtain the equation of evolution for $\delta\rho$ as:

\begin{equation}
\dot{N} - \frac{1}{2} \hspace{0.1cm}\beta \hspace{0.1cm}\dot{\rho} +
(1 + \lambda) \hspace{0.1cm}\theta \hspace{0.1cm}N + (1 + \lambda)
\hspace{0.1cm}\rho \hspace{0.1cm}H = 0.
\protect\label{rt1}
\end{equation}

Applying the same procedure for the perturbed Raychaudhuri equation\ (\ref{apb20}) and using the decomposition in a scalar basis\ (\ref{k6}), we obtain

\begin{equation}
\dot{H} - \frac{1}{2}\beta\dot{\theta} + \frac{2}{3}\theta H + \frac{m}{a^{2}}\Psi + \frac{(1 + 3\lambda)}{2}N = 0.
\protect\label{rt2}
\end{equation}

To solve these two equations we need to fix the gauge $\beta (t)$ and to use the values for $E$ and $\Sigma$ which were obtained from the fundamental closed system found in the previous section (Eqs.\ (\ref{e8})). All the remaining geometrical and kinematical quantities can be likewise obtained. This completely exhausts our analysis of the irrotational perturbations of FLRW universe.

\subsubsection{The Singular Case $(1 + \lambda) = 0$: The Perturbations of de Sitter Universe}\protect\label{deSitter}

We have seen that all the system of reduction to the variables $\Sigma$ and $E$ was based on the possibility of writing the acceleration in terms of $E$ and $\Sigma$. This was possible in all cases, except in the special one in which $(1 + \lambda) = 0$. Although no known fluid exists with such negative pressure, the fact that the vacuum admits such an interpretation has led to the identification of the cosmological constant with this fluid. It is therefore worthwhile to examine this case in the same way as it was done for the previous sections.

At this point it must be remarked that, contrary to all previously studied cases, perturbations of this fluid must necessarily contain contributions which come from the heat flux or the anisotropic pressure. Indeed, if we take both of these quantities as vanishing, then the set of perturbed equations implies that all equations are trivially satisfied, since all perturbative quantities vanish, except for the cases where $\delta p = \overline{\lambda} \hspace{0.1cm}\delta\rho$, with $\overline{\lambda} = 0$, and $\lambda + 1 = 0$. We will analyze these cases below.

When $\delta p = \overline{\lambda} \hspace{0.1cm}\delta\rho$, for $\overline{\lambda} = 0$, the system is stable. Indeed, we obtain for the electric part of Weyl tensor, in the case that $\theta$ is constant in the background, the following expression:

\begin{equation}
\nonumber
E(t) = E_{0} e^{- \frac{\theta}{3} t}.
\end{equation}

The other case of interest is the one in which the condition \mbox{$(1 + \lambda) = 0$} is preserved throughout the perturbation. Looking at Eq.\ (\ref{rt1}) it follows that, from the fact that $\dot{\rho} = 0$ and reminding the reader that $(1 + \lambda) = 0$, temporal variation of the energy density exists only if we take into account the perturbed fluid with heat flux. We then write

\begin{equation}
\nonumber
q_{i} = q(t) \hspace{0.1cm} Q_{i}(x^{k}).
\end{equation}
Equation (\ref{apb23}) gives

\begin{equation}
\dot{N} = \frac{m}{a^{2}} \hspace{0.1cm}q.
\protect\label{dS1}
\end{equation}

The projected conserved equation gives (see Eq.\ (\ref{apb24})):

\begin{equation}
\dot{q} + \theta q + N = \frac{2\xi}{3a^{2}}\left(1 + \frac{3\epsilon}{m}\right)\Sigma.
\protect\label{dS2}
\end{equation}

The evolution equation for the electric part of Weyl tensor gives:

\begin{equation}
\dot{{\cal S}} + \frac{\theta}{3} {\cal S} = - \frac{m}{2}q,
\protect\label{dS3}
\end{equation}
in which we used the definition

\begin{equation}
\nonumber
{\cal S} \equiv E - \frac{1}{2} \hspace{0.1cm}\xi \hspace{0.1cm}\Sigma.
\end{equation}

Finally, from the equation that gives the divergence of $E_{ij}$, we have the constraint

\begin{equation}
\frac{2}{a^{2}}\left(1 + \frac{3\epsilon}{m}\right){\cal S} = - \left(N + \theta q\right).
\protect\label{dS4}
\end{equation}

The evolution equation for the shear provides the value of the acceleration $\Psi$. Equations (\ref{dS2})-(\ref{dS4}) constitute thus a complete system for the variables $E$, $\Sigma$ and $q$. This completes the general explicitly gauge-invariant scheme that we presented here even in the
singular case $(1 + \lambda) = 0$. Notwithstanding, just as an additional comment, it would be interesting to consider the perturbation scheme in the framework of Lanczos potential. This will be done in a later section.

\subsubsection{Hamiltonian Treatment of the Scalar Solution}
\protect\label{Hamiltonian}

The examination of the perturbations in FLRW cosmology, which we analyzed above, admits a Hamiltonian formulation that is worth considering here (cf. Grishchuk, 1990). In this vein, the variables $E$ and $\Sigma$, analyzed in the previous section, are the ones that must be employed to obtain such a formulation. From the evolution equations for $\Sigma$ and $E$ (Eq.\ (\ref{e8})) it follows that they are not canonically conjugated for arbitrary geometries of the background.

The natural step would be to define canonically conjugated variables $Q$ and $P$ as a linear functional of $\Sigma$ and $E$ as\footnote{The attentive reader should notice that in this subsection the quantity $Q$ shall not be confused with the previous scalar basis.}:

\begin{equation}
\left[\begin{array}{cc}
Q \\
P
\end{array}\right] = \left[\begin{array}{ccc}
\alpha & \eta \\
\delta & \beta
\end{array}\right] \hspace{0.2cm}\left[\begin{array}{cc}
\Sigma \\
E
\end{array}\right].
\protect\label{h1A}
\end{equation}
It should be expected that functionals of the background geometry would appear in the construction of the canonical variables in the functions $\alpha$, $\beta$, $\eta$ and $\delta$. It seems valid to remark that this matrix is univocally defined up to canonical transformations. We can thus use this fact to choose $\eta$ and $\delta$ as zero; we shall use this choice in order to simplify our analysis.

The Hamiltonian ${\cal H}$ which provides the dynamics of the pair $(Q,P)$ is obtained from the evolution equations of $E$ and $\Sigma$ (\ref{e8}). The condition for the existence of such a Hamiltonian is given by the equation

\begin{eqnarray}
\frac{\dot{\alpha}}{\alpha} + \frac{\dot{\beta}}{\beta} - \xi -\frac{1}{3}\theta +  \frac{2m\xi}{3(1 + \lambda)\rho a^{2}}\left(1 + \frac{3\epsilon}{m}\right) = 0.
\protect\label{h2A}
\end{eqnarray}

It then follows that the Hamiltonian which provides the dynamics of our problem takes the form

\begin{equation}
{\cal H} = \frac{h_{1}}{2} Q^{2} + \frac{h_{2}}{2}P^{2} +2h_{3}PQ,
\protect\label{h3A}
\end{equation}
where $h_{1}$, $h_{2}$ and $h_{3}$ are defined as

\begin{eqnarray}
h_{1} \equiv \frac{\beta}{\alpha} \left[\frac{(1 + \lambda)}{2}\rho + \frac{\xi}{2}\left(\frac{\xi}{2} + \frac{\theta}{3}\right) - \frac{m\xi^{2}}{(1 + \lambda) \rho a^{2}}\left(1 + \frac{3\epsilon}{m}\right) \left(\frac{\lambda}{2} + \frac{1}{3}\right) \right],
\protect\label{h4A}
\end{eqnarray}

\begin{equation}
h_{2} \equiv - \frac{\alpha}{\beta} \left[1 + \frac{2m\lambda}{(1 + \lambda)\rho a^{2}} \left(1 + \frac{3\epsilon}{m}\right)\right],
\protect\label{h5A}
\end{equation}

\begin{equation}
h_{3} \equiv \frac{\theta}{3} - \frac{\dot{\beta}}{\beta} + \frac{\xi}{4} \left[1 + \frac{2m\lambda}{(1 + \lambda)\rho a^{2}} \left(1 + \frac{3\epsilon}{m}\right)\right].
\protect\label{h6A}
\end{equation}

Let us consider the case in which $\xi = 0$, that is, there is no anisotropic pressure. The case where $\xi$ does not vanish presents some interesting peculiarities which will be left to a forthcoming section.

We will choose $\beta = a$ and take $\alpha$ as given by Eq.(\ref{h2A}). We then define the canonical variables $Q$ and $P$ by setting

\begin{equation}
\nonumber
\begin{array}{l}
Q = \Sigma, \\[2ex]
P = a E.
\end{array}
\end{equation}

It then follows that ${\cal H}$ is given by

\begin{equation}
{\cal H} =  - \Delta^{2}(t)P^{2} + \gamma^{2}(t)Q^{2},
\protect\label{h1}
\end{equation}
where $\gamma (t)$ and $\Delta (t)$ are given in terms of the energy density of the background $\rho$, the scale factor $a(t)$ and the wave number $m$ as:

\begin{equation}
\begin{array}{l}
\gamma^{2}(t) \equiv \left[\fracc{(1 + \lambda)}{4}\right] \rho a, \\
\Delta^{2}(t) \equiv \fracc{1}{2a} \left[1 + \fracc{2m\lambda}{(1 + \lambda)\rho a^{2}}\left(1 + \fracc{3\epsilon}{m}\right)\right].
\end{array}
\protect\label{h2}
\end{equation}

Let us make two comments here: first of all, the fact that the system is not conservative (which means $\dot{{\cal H}}$ is not zero) is a consequence of the fact that the ground state of this theory ($Q = P = 0$) corresponds not to
Minkowski flat space-time but to FLRW expanding universe. The second remark is that the same applies to the non-positivity of the Hamiltonian; this is also a consequence of the non-vanishing of the curvature of the fundamental state. The system which we are analyzing is not closed and, so, momentum and energy can be pumped from the background.

We notice that the Hamiltonian structure obtained in terms of the variables $E$ and $\Sigma$ is completely gauge-invariant and, as such, deserves an ulterior analysis, which we will make elsewhere. We would like only to
exhibit an example where this pumping effect can be easily recognized; this will be achieved by applying the Hamiltonian treatment to a static model of the universe.

\subsubsection{Fierz-Lanczos Potential}\protect\label{lanczos}

As it was remarked in a previous section, perturbations of conformally flat spacetimes do not need\footnote{The above quoted gauge problem has been widely discussed in the literature (see \cite{no_sa_ma_re_se3} and references therein).} the complete knowledge of all components of the perturbed metric tensor $\delta g_{\mu\nu}$, although they certainly need to take into account the Weyl conformal tensor, since all the observable information we need is contained in it (namely, $\delta E_{ij}$ and $\delta H_{ij}$).

Let us note at this point that the tensor $W_{\alpha\beta\mu\nu}$ can be expressed in terms of the 3-index Fierz-Lanczos potential tensor---see Fierz (1939) and Lanczos (1962)---that we will denote by $L_{\alpha\beta\mu}$, and which deserves a careful analysis. Indeed, one could consider $\delta L_{\alpha\beta\mu}$ as the good object for studying linear perturbation theory, since, as we shall see, it combines both $\delta\Sigma_{ij}$ and $\delta a_{k}$ (which are alternative variables to describe $\delta E_{ij}$).

Before going into the perturbation-related details, let us summarize here some definitions and properties of $L_{\alpha\beta\mu}$, since the literature has very few papers on this matter\footnote{This tensor was introduced in the 30's to provide, in a similar way as the symmetric tensor $\varphi_{\mu\nu}$ does---in a more used approach---an alternative description of spin-2 field in Minkowski background. In the 60's Lanczos rediscovered it---without recognizing he was dealing with the same object---as a Lagrange multiplier in order to obtain the Bianchi identities in the context of Einstein's General Relativity. However, only recently (cf. Novello, 1992a, b) a complete analysis of Fierz-Lanczos object was undertaken and it was discovered that its generic (Fierz) version describes not only one but two spin-2 fields. The restriction to just a single spin-2 field is usually called the Lanczos tensor. We will limit all our considerations here to this restricted quantity.}.

\underline{{\bf Basic Properties:}} In any 4-dimensional Riemannian geometry there is a 3-index tensor $L_{\alpha\beta\mu}$ which has the following symmetries:

\begin{equation}
L_{\alpha\beta\mu} + L_{\beta\alpha\mu} = 0
\protect\label{b1}
\end{equation}
and
\begin{equation}
L_{\alpha\beta\mu} + L_{\beta\mu\alpha} + L_{\mu\alpha\beta} = 0.
\protect\label{b2}
\end{equation}

With such $L_{\alpha\beta\mu}$ we may write the Weyl tensor in the form of a homogeneous expression in the potential expression, that is

\begin{eqnarray}
W_{\alpha\beta\mu\nu} & = & L_{\alpha\beta [\mu ;\nu ]} +
L_{\mu\nu[\alpha ;\beta ]} + \nonumber \\
& + & \frac{1}{2} [ L_{(\alpha\nu)} g_{\beta\mu} + L_{(\beta\mu)}
g_{\alpha\nu} - L_{(\alpha\mu)} g_{\beta\nu} -
L_{(\beta\nu)} g_{\alpha \mu}] + \nonumber \\
& + & \frac{2}{3}
{L^{\sigma\lambda}}_{  \sigma ;\lambda} \hspace{0.1cm}g_{\alpha\beta\mu\nu},
\protect\label{b3}
\end{eqnarray}
where

\begin{equation}
\nonumber
L_{\alpha\mu} \equiv {{L_{\alpha}}^{\sigma}}_{\mu ;\sigma} -
L_{\alpha ;\mu}
\end{equation}
and

\begin{equation}
\nonumber
L_{\alpha} \equiv {{L_{\alpha}}^{ \sigma}}_{\sigma}.
\end{equation}

Let us point out that, due to the above symmetry properties, Eqs.\ (\ref{b1}) and (\ref{b2}), Lanczos tensor has 20 degrees of freedom. Since Weyl tensor has only 10 independent components, it follows that there is a gauge symmetry involved. This gauge symmetry can be separated into two classes:

\begin{equation}
\Delta^{(1)} L_{\alpha\beta\mu} = M_{\alpha} \hspace{0.1cm}g_{\beta\mu} -
M_{\beta} \hspace{0.1cm}g_{\alpha\mu},
\protect\label{b4}
\end{equation}
and
\begin{eqnarray}
\Delta^{(2)} L_{\alpha\beta\mu} & = & W_{\alpha\beta ;\mu} - \frac{1}{2}
W_{\mu\alpha ;\beta} + \frac{1}{2} W_{\mu\beta ;\alpha} + \nonumber \\
& + & \frac{1}{2} g_{\mu\alpha} \hspace{0.1cm}
{{W_{\beta}}^{\lambda}}_{;\lambda} -
\frac{1}{2} g_{\mu\beta} \hspace{0.1cm}{{W_{\alpha}}^{\lambda}}_{;\lambda},
\protect\label{b5}
\end{eqnarray}
in which the vector $M_{\alpha}$ and the antisymmetric tensor $W_{\alpha\beta}$ are arbitrary quantities.

\underline{{\bf Lanczos Tensor for FLRW Geometry:}} The fact that Friedman-Lemaître-Robertson-Walker geometry is conformally flat implies that the associated Lanczos potential is nothing but a gauge. That is, we can write the Lanczos potential for FRW geometry as

\begin{eqnarray}
L_{\alpha\beta\mu} & = & N_{\alpha} \hspace{0.1cm}g_{\beta\mu} -
N_{\beta} \hspace{0.1cm}g_{\alpha\mu}
+ F_{\alpha\beta ;\mu} - \frac{1}{2} F_{\mu\alpha ;\beta} +\nonumber \\
& + & \frac{1}{2} F_{\mu\beta ;\alpha} + \frac{1}{2} g_{\mu\alpha}
\hspace{0.1cm}{{F_{\beta}}^{\lambda}}_{;\lambda} -
\frac{1}{2} g_{\mu\beta} \hspace{0.1cm}{{F_{\mu}}^{\lambda}}_{;\lambda},
\protect\label{b6}
\end{eqnarray}
for the arbitrary vector $N_{\alpha}$ and the antisymmetric tensor $F_{\alpha\beta}$.

\underline{{\bf Perturbed Fierz-Lanczos Tensor:}} In the case we are examining in this paper (irrotational perturbations) the perturbed Weyl tensor reduces to the form

\begin{equation}
\delta W_{\alpha\beta\mu\nu} = (\eta_{\alpha\beta\gamma\varepsilon}
\hspace{0.1cm} \eta_{\mu\nu\lambda\rho} - g_{\alpha\beta\gamma\varepsilon}
\hspace{0.1cm} g_{\mu\nu\lambda\rho}) V^{\gamma} V^{\lambda} \delta
E^{\varepsilon\rho},
\protect\label{p1}
\end{equation}
since the magnetic part of Weyl tensor remains zero in this case.

It then follows that the perturbed electric tensor is given in terms of Lanczos potential as:

\begin{eqnarray}
- \delta E_{ij} & = & \delta L_{0i[0;j]} + \delta L_{0j[0;i]} -
\frac{1}{2} \delta L_{(00)} \gamma_{ij} \nonumber \\
& - & \frac{1}{2} \delta L_{(ij)} +
\frac{2}{3} \delta {L^{\sigma \lambda}}_{\sigma ;\lambda} \gamma_{ij}.
\protect\label{p2}
\end{eqnarray}

Although the $L_{\alpha\beta\mu}$ tensor is not a unique well defined object (since it has the gauge freedom we discussed above) we can use some theorems---see Novello (1987) and L\'opez Bonilla (1989)---that enable one to write $L_{\alpha\beta\mu}$ in terms of the associated kinematic quantities of a given congruence of curves present in the associated Riemannian manifold. Following these theorems and choosing the case of irrotational perturbed matter, it follows that $\delta L_{\alpha\beta\mu}$ (the perturbed tensor of FLRW background) is given by

\begin{equation}
\delta L_{\alpha\beta\mu} = \delta \sigma_{\mu [\alpha} V_{\beta]} +
F(t) \delta a_{[\alpha} V_{\beta ]} V_{\mu},
\protect\label{p3}
\end{equation}
where

\begin{equation}
F(t) = 1 - \frac{1}{m} \frac{\Sigma}{\Psi} \left(\frac{2}{3} \theta +
\frac{1}{2} \xi\right).
\protect\label{p4}
\end{equation}

In other words, the only non identically zero components of
$\delta L_{\alpha\beta\mu}$ are:

\begin{equation}
\delta L_{0k0} = - F(t) \Psi\hspace{0.1cm}Q_{k},
\protect\label{p5}
\end{equation}
and

\begin{equation}
\delta L_{0ij} = - \Sigma (t) \bar{Q}_{ij},
\protect\label{p6}
\end{equation}
which coincides with the previous results.

From what we have learned in the previous section, we can conclude that this is not a univocal expression, that is, Eqs.\ (\ref{p5}) and (\ref{p6}) are obtained by a specific gauge choice.

Let us apply the above gauge transformation to the present
case. In the first gauge, Eq.\ (\ref{b4}), we decompose vector $M_{\alpha}$ in the spatial harmonics (scalar and vector):

\begin{equation}
M_{0} = M^{(1)}(t) \hspace{0.1cm}Q(x),
\protect\label{p7}
\end{equation}

\begin{equation}
M_{i} = M^{(2)}(t) \hspace{0.1cm}Q_{i}(x),
\protect\label{p8}
\end{equation}
and in the second gauge, Eq.\ (\ref{b5}), we have

\begin{equation}
W_{0i} = W^{(1)}(t) \hspace{0.1cm}Q_{i}(x),
\protect\label{p9}
\end{equation}
and

\begin{equation}
W_{ij} = - \frac{1}{a^{2}} \varepsilon_{ijk} \hspace{0.1cm}
W^{(2)}(t) \hspace{0.1cm}Q^{k}(x).
\protect\label{p10}
\end{equation}

To sum up, asking what the Lanczos tensor is for the perturbed FLRW geometry is one of those questions (like the one about the perturbed tensor $\delta g_{\mu\nu}$) that should be avoided, since this quantity is gauge-dependent. A good question to be asked should be, as we remarked
before: What is the perturbation of Weyl tensor? This was precisely the motivation of the previous section.

\subsection{Friedman Universe: Vectorial Perturbations}\protect\label{genvetorial}

As it has been discussed in the previous section, we will make use of the perturbation formalism of Einstein's theory of gravitation based in gauge-independent and evident physically meaningful quantities, such as the vorticity, shear, electric and magnetic parts of the conformal Weyl tensor and so on.

In the scalar case, the convention was simplified in order to facilitate the resulting system of dynamical equations. For the vectorial and tensorial cases, however, we feel that the convention set in Hawking (1966) is more adequate. Therefore, we will present it here.

The metric of the background is given in the standard Gaussian form, thus defining a class of privileged observers $V^{\alpha} = \delta^{\alpha}_{0}$. The projector $h_{\mu\nu}$ defines, in the 3-dimensional space orthogonal to $V^{\alpha}$, the 3-dimensional quantities with the symbol (\,\,$\widehat{}$\,\,\,). Thus, $\hat{X}_{\alpha} \equiv {h_{\alpha}}^{\beta}\, X_{\beta}$ denotes a projection into the 3-geometry. For the same reasoning, we define the operator $\hat{\nabla}_{\alpha}$ as the covariant derivative in the 3-geometry. The relation between the 3-dimensional Laplacian ($\hat{\nabla}^{2}$) and the 4-dimensional one is given as follows:

\begin{equation}
\nonumber
\hat{\nabla}^{2}\, \hat{X}_{\alpha} =
\left(\frac{\theta}{3}\right)^{2}
\, \hat{X}_{\alpha} + h_{\alpha}^{\beta}\,
\nabla^{2}\, \hat{X}_{\beta}.
\end{equation}

We then introduce the fundamental harmonic basis of the functions projected onto the 3-surface

\begin{equation}
\bigl\{Q(x)\bigr\},\,\bigl\{\hat{P}_{\alpha}(x)\bigr\},\,\bigl\{\hat{U}_{\alpha\beta}(x)\bigr\}.
\end{equation}

In this section, we are interested in the vector basis $\hat{P}_{\alpha}(x)$, which is defined by the following relations:

\begin{equation}
\begin{array}{l}
\hat{P}_{\mu}\, V^{\mu} = 0, \\
\dot{\hat{P}}\mbox{}^{\mu} = 0, \\
\hat{\nabla}^{\alpha}\, \hat{P}_{\alpha} = 0, \\
\hat{\nabla}^{2}\, \hat{P}_{\alpha} = \fracc{m}{a^{2}}\, \hat{P}_{\alpha},
\end{array}
\protect\label{intro2}
\end{equation}
where the eigenvalue (again denoted by $m$, despite the fact that this eigenvalue and the scalar basis one have no relation at all) is given by

\begin{equation}
\protect\label{mP}
m=\left\{
\begin{array}{lll}
q^2+2, & 0<q<\infty,     & \epsilon=+1   \mbox{\rm\ (open)},\\
q,     & 0<q<\infty,     & \epsilon=\ \ 0\mbox{\rm\ (plane) },\\
n^2-2, & n=2,\,3,\ldots, & \epsilon=-1   \mbox{\rm\ (closed)}.
\end{array}
\right.
\end{equation}

From this basis, it is possible to derive a pseudo-vector and a tensor:

\begin{equation}
\begin{array}{l}
\hat{P}^{\ast \alpha} \equiv \eta^{\alpha\beta\mu\nu}\, V_{\beta}\, \hat{P}_{\mu\nu}, \\[2ex]
\hat{P}_{\alpha\beta} \equiv \hat{\nabla}_{\beta}\, \hat{P}_{\alpha}, \\[2ex]
\hat{P}^{\ast}_{\alpha\beta} \equiv \hat{\nabla}_{\beta}\, \hat{P}^{\ast}_{\alpha},
\end{array}
\protect\label{intro3}
\end{equation}
suited to developing pseudo-vectors and tensors.

The following vectorial relations are useful in obtaining the dynamical equations:

\begin{equation}
\begin{array}{l}
\dot{\hat{P}}_{(\alpha\beta )} = - \fracc{1}{3}\,\theta\, \hat{P}_{(\alpha\beta )}, \\[2ex]
\dot{\hat{P}}\mbox{}^{\ast}_{(\alpha\beta )} = - \fracc{2}{3}\, \theta\, \hat{P}^{\ast}_{(\alpha\beta )}, \\[2ex]
\dot{\hat{P}}\mbox{}^{\ast}_{\alpha} =- \fracc{1}{3}\, \theta\,\hat{P}^{\ast}_{\alpha}, \\[2ex]
\hat{\nabla}^{\beta}\, \hat{P}_{(\alpha\beta )} =\fracc{1}{A^{2}}\, (m + 2\epsilon)\, \hat{P}_{\alpha}, \\[2ex]
\hat{\nabla}^{\beta}\, \hat{P}^{\ast}_{(\alpha\beta )} =\fracc{1}{A^{2}}\, (m + 2\epsilon)\, \hat{P}^{\ast}_{\alpha}, \\[2ex]
\hat{\nabla}^{2}\, \hat{P}^{\ast}_{\alpha} =\fracc{m}{a^{2}}\, \hat{P}^{\ast}_{\alpha}, \\[2ex]
\eta^{\alpha\beta\gamma\varepsilon}\, V_{\beta}\,\hat{P}^{\ast}_{\gamma ;\varepsilon} = \fracc{1}{a^{2}}\,(m - 2\epsilon)\, \hat{P}^{\alpha}, \\[2ex]
h^{\mu}_{(\alpha }\, h^{\nu}_{\beta )}\,{\eta_{\mu}}^{\lambda\gamma\tau}\, V_{\tau}\,\hat{\nabla}_{\gamma}\, \hat{P}_{(\nu\lambda )} = h^{\mu}_{(\alpha }\, h^{\nu}_{\beta )}\,\hat{P}^{\ast}_{\mu\nu}, \\[2ex]
h^{\mu}_{(\alpha }\, h^{\nu}_{\beta )}\,{\eta_{\mu}}^{\lambda\gamma\tau}\, V_{\tau}\,\hat{\nabla}_{\gamma}\, \hat{P}_{\left[\nu\lambda\right]} =
- h^{\mu}_{(\alpha }\, h^{\nu}_{\beta )}\,\hat{P}^{\ast}_{\mu\nu}.
\end{array}
\protect\label{intro3.5}
\end{equation}
where we have used the constraint relation below,

\begin{equation}
-\frac{\epsilon}{a^{2}} - \fracc{1}{3}\, \rho + \left(\fracc{\theta}{3}\right)^{2} = 0,
\protect\label{intro7}
\end{equation}
valid in the FLRW background. The following auxiliary relations are also useful:

\begin{equation}
\begin{array}{l}
\dot{\theta} = - \fracc{1}{3}\, \theta^{2} - \fracc{1}{2}\,(\rho + 3p), \\
\dot{\rho} = - \theta\, (\rho + p).
\end{array}
\protect\label{intro8.5}
\end{equation}

With the above basis, we are able to expand any {\it good} perturbed quantity (again denoted by $\delta X$, where $X$ is any quantity associated to the matter content, kinematics and geometry) as

\begin{equation}
\protect\label{decomp}
\begin{array}{lcl}
\delta \omega_{\alpha} &=& \Omega(t)\, \hat{P}^{\ast}_{\alpha},\\[2ex]
\delta q_{\alpha} &=& q(t)\, \hat{P}_{\alpha},\\[2ex]
\delta a_{\alpha} &=& \Psi(t)\, \hat{P}_{\alpha},\\[2ex]
\delta V_{\alpha} &=& V(t)\, \hat{P}_{\alpha},\\[2ex]
\delta\sigma_{\alpha\beta} &=& \Sigma(t)\, \hat{P}_{(\alpha\beta )},\\[2ex]
\delta H_{\alpha\beta} &=& H(t)\, \hat{P}^{\ast}_{(\alpha\beta )},\\[2ex]
\delta E_{\alpha\beta} &=& E(t)\, \hat{P}_{\alpha\beta},\\[2ex]
\delta\Pi_{\alpha\beta} &=& \Pi(t) \, \hat{P}_{(\alpha\beta )}.
\end{array}
\end{equation}

\subsubsection{Dynamics}\protect\label{vetorial}

In order to get simpler equations, we will again approximate the thermodynamic equation,

\begin{equation}
\tau\, \dot{\Pi}_{\alpha\beta} + \Pi_{\alpha\beta} = \xi\, \sigma_{\alpha\beta},
\protect\label{vet1}
\end{equation}
to the limit of small relaxation time $\tau$ (adiabatic approximation) and constant viscosity coefficient $\xi$ to get

\begin{equation}
\delta\Pi_{\alpha\beta} = \xi\, \delta\sigma_{\alpha\beta}\,  \leadsto \, \Pi = \xi\, \Sigma.
\protect\label{vet2}
\end{equation}
The vorticity can be written in terms of the 3-velocity as

\begin{equation}
\delta\omega_{\alpha} = - \frac{1}{2}\,
\delta V_{\alpha}\,  \leadsto \, V = - 2\, \Omega.
\protect\label{vet3}
\end{equation}

We will denote by ($\chi_r$, $\tilde{\Phi}_s$) the fundamental dynamical and constraint equations, respectively. Introducing equation (\ref{decomp}) and equations (\ref{vet2})--(\ref{vet3}) into the perturbed quasi-Maxwellian equations (\ref{apb13})--(\ref{apb24}) and making use of equations (\ref{intro2})--(\ref{intro3.5}), we get

\begin{subequations}
\label{ev_pert_vec_fried}
\begin{eqnarray}
&&\dot{E} - \fracc{1}{2}\xi\, \dot{\Sigma} + \fracc{2}{3}\theta \, E + \fracc{1}{2}(\rho + p)\, \Sigma + \fracc{1}{2a^{2}}\, (m - 2\epsilon)\, H + \fracc{1}{4}q = 0,\label{ev_pert_vec_fried1}\\
&&\dot{\Sigma} + \left(\fracc{\theta}{3} + \fracc{\xi}{2}\right)\, \Sigma + E - \fracc{1}{2}\, \Psi = 0,\label{ev_pert_vec_fried2}\\
&&\dot{\Omega} + \fracc{1}{3}\theta\, \Omega + \fracc{1}{2}\,\Psi = 0,\label{ev_pert_vec_fried3}\\
&&\dot{H} + \fracc{1}{3}\theta\, H - \fracc{1}{2}\, E - \fracc{1}{4}\xi\, \Sigma = 0,\label{ev_pert_vec_fried4}\\
&&\dot{q} + \fracc{4}{3}\theta\, q + \fracc{1}{a^{2}}\, (m + 2\epsilon)\, \xi\, \Sigma + 2\dot{p}\, \Omega + (\rho + p)\,\Psi = 0,\label{ev_pert_vec_fried5}
\end{eqnarray}
\end{subequations}
and

\begin{subequations}
\label{cons_pert_vec_fried}
\begin{eqnarray}
&&\Sigma + \Omega + 2H = 0,\label{cons_pert_vec_fried1}\\
&&\fracc{1}{a^{2}}\, (m + 2\epsilon)\, E - \fracc{1}{2a^{2}}\, (m + 2\epsilon)\, \xi\, \Sigma + \fracc{2}{3}\, \theta\, (\rho + p)\, \Omega - \fracc{1}{3}\theta\, q = 0,\label{cons_pert_vec_fried2}\\
&&\fracc{1}{a^{2}}\, (m + 2\epsilon)\,H - (\rho + p)\, \Omega + \fracc{1}{2}\, q = 0,\label{cons_pert_vec_fried3}\\
&&\frac{1}{a^{2}}\, (m + 2\epsilon)\, \Sigma + \left\{ \fracc{1}{a^{2}}\, (m - 2\epsilon) + 4\, \left(\fracc{\theta}{3}\right)^{2} + \fracc{2}{3}\, (\rho + p)\right\}\, \Omega - q =0.\label{cons_pert_vec_fried4}
\end{eqnarray}
\end{subequations}

By making use of equations (\ref{intro7})--(\ref{intro8.5}), it can be easily shown that constraint\ (\ref{cons_pert_vec_fried4}) is not essential, since it is written in terms of\ (\ref{cons_pert_vec_fried1}) and\ (\ref{cons_pert_vec_fried3}). We also note that we can write constraint\ (\ref{cons_pert_vec_fried2}) in a simpler form as

\begin{equation}
E - \fracc{1}{2}\, \xi\, \Sigma + \fracc{2}{3}\, \theta\, H = 0.
\protect\label{vet7}
\end{equation}

The fundamental differential system is now written as

\begin{subequations}
\label{vet8}
\begin{eqnarray}
&&\dot{E} - \fracc{1}{2}\, \xi\,\dot{\Sigma} + \fracc{2}{3}\, \theta\, E + \fracc{1}{2}\, (\rho + p)\, \Sigma + \fracc{1}{2a^{2}}\, (m - 2\epsilon)\, H + \fracc{1}{4}\, q = 0,\label{vet81}\\
&&\dot{\Sigma} + \left(\fracc{\theta}{3} + \fracc{\xi}{2}\right)\, \Sigma + E - \fracc{1}{2}\, \Psi = 0,\label{vet82}\\
&&\dot{\Omega} + \fracc{1}{3}\,\theta\, \Omega + \fracc{1}{2}\, \Psi = 0,\label{vet83}\\
&&\dot{H} + \fracc{1}{3}\, \theta\,H - \fracc{1}{2}\, E - \fracc{1}{4}\, \xi\, \Sigma = 0,\label{vet84}\\
&&\dot{q} + \fracc{4}{3}\, \theta\,q + \fracc{1}{a^{2}}\, (m + 2\epsilon)\, \xi\, \Sigma - 2\dot{p}\, \Omega + (\rho + p)\, \Psi =0,\label{vet85}
\end{eqnarray}
\end{subequations}
and

\begin{subequations}
\label{vet9}
\begin{eqnarray}
&&\Sigma + \Omega + 2H = 0,\label{vet91}\\
&&E - \fracc{1}{2}\, \xi\, \Sigma +\fracc{2}{3}\, \theta\, H = 0,\label{vet92}\\
&&\fracc{1}{A^{2}}\, (m + 2\epsilon)\, H -(\rho + p)\, \Omega + \fracc{1}{2}\, q = 0.\label{vet93}
\end{eqnarray}
\end{subequations}

It could be argued that the acceleration $\Psi$ should be eliminated from the dynamical system by using the definition

\begin{equation}
\nonumber
a_{\alpha} = \dot{V}_{\alpha} = V_{\alpha ;\beta}\,V^{\beta}.
\end{equation}
If this is done, we obtain

\begin{equation}
\nonumber
\Psi\, \hat{P}_{\alpha} = \left(\dot{V} + \frac{\theta}{3} \, V\right)\, \hat{P}_{\alpha} - \delta\Gamma^{0}_{0\alpha}.
\end{equation}

However, it is easily proven (see Novello, 1995a) that

\begin{equation}
\nonumber
\delta\Gamma^{0}_{0\alpha} = \frac{1}{2}\, (\delta\, g_{00})_{,\alpha} = (\delta\, V_{0})_{,\alpha},
\end{equation}
which is zero in the vector basis. Then, making use of Equation\ (\ref{vet3}), we have the following relation:

\begin{equation}
\nonumber
\Psi = - 2\, \dot{\Omega} - \frac{2}{3}\, \theta\,\Omega,
\end{equation}
which is precisely the dynamical equation\ (\ref{vet83}). The variable $\Psi$ can then be eliminated only by means of losing some degree of freedom. This way we get physically motivated ({\it i.e.,} by observation) algebraic relations between acceleration and other selected variables. We will restrict ourselves here to the three cases that follow.

The first possible choice is to admit a shear-free model for the cosmological perturbation. In such a case, there is no shear, and, from this, the anisotropic pressure vanishes too. Thus we will refer to this case as ``isotropic" throughout the remainder of this section. Therefore\ (\ref{vet82}) becomes

\begin{equation}
\Psi = 2\, E.
\label{h1}
\end{equation}

The second possibility is to admit that no vorticity should be taken into account. As it has been known for long, the presence of a non vanishing vorticity usually brings together along some troubles related to causality violation. So, we motivate this case by eliminating the main source of causality breakdown. In this case, we have

\begin{equation}
\nonumber
\Omega = 0,
\end{equation}
and\ (\ref{vet83}) then results in

\begin{equation}
\Psi = 0.
\label{h2}
\end{equation}

Another possibility is to impose the physical source of curvature to be a Stokesian fluid. This means that the energy flux (heat flux in this case) vanishes. Despite the fact that we can always set this quantity to zero by a suitable choice of observers, it actually represents a true restriction, for our equations are written in such a way that no observer changes can be performed---that is, we have already fixed the observer by imposing the particle flux to vanish. Now\ (\ref{vet85}) yields

\begin{equation}
\Psi = - \fracc{1}{a^{2}}\, (m + 2\epsilon)\, \fracc{\xi}{(\rho + p)}\, \Sigma + 2\, \lambda\, \theta\, \Omega,
\label{h3}
\end{equation}
with

\begin{equation}
\nonumber
(\rho + p) = (1 + \lambda)\, \rho \neq 0 \hspace{0.6cm}
\lambda \equiv const,
\end{equation}
a relation that eliminates $\Psi$ for all but the de Sitter background.  All three possibilities will have their respective dynamics and Hamiltonian treatment investigated in a later section.

\subsubsection{Permanence of Constraints}
\protect\label{vinculos}

Since we obtained a constrained differential system, given by Equations (\ref{vet8}) and (\ref{vet9}), it is useful to consider whether constraints are automatically preserved or not. If one differentiates the expressions (\ref{vet9}) and inserts into the results the relations (\ref{intro7})--(\ref{intro8.5}), one gets directly

\begin{equation}
\begin{array}{l}
\dot{\Phi}_{1} = \chi_{2} + \chi_{3} + 2\chi_{4} - \fracc{\theta}{3}\, \Phi_{1}, \\[2ex]
\dot{\Phi}_{2} = \chi_{1} - \fracc{2}{3}\, \theta\, \chi_{4} - \fracc{1}{2}\, (\rho + p)\, \Phi_{1} - \fracc{\theta}{3}\,
\Phi_{2} + \fracc{1}{2}\, \Phi_{3} - \left[\fracc{-\epsilon}{a^{2}} + \left(\fracc{\theta}{3}\right)^{2} - \fracc{1}{3}\, \rho
\right]\, (\Omega + 2H), \\[2ex]
\dot{\Phi}_{3} = - (\rho + p)\, \chi_{3} + \fracc{1}{a^{2}} \, (m + 2\epsilon)\, \chi_{4} + \fracc{1}{2}\, \chi_{5} - \fracc{1}{2a^{2}}\, (m + 2\epsilon)\, \Phi_{2} - \fracc{2}{3}\, \theta\, \Phi_{3}.
\end{array}
\protect\label{vinc1}
\end{equation}
where $\Phi_{i}$, $i=1,2,3$ are the constraint Eqs.\ (\ref{vet9}) and $\chi_j$, $j=1,...,5$ are the evolution Eqs.\ (\ref{vet8}).
Thus, it follows that no secondary constraint\footnote{Terminology due to Bergmann relative to Dirac's work (1950) on constrained systems.}
(SC) appears in the case of vector perturbations. One should expect it, since this result reflects the fact that our basic (quasi-Maxwellian) equations are dynamically equivalent to Einstein's field equations, which are complete.

\subsubsection{Hamiltonian Treatment of the Vectorial Solution}\protect\label{vecHamilton}

If we keep all degrees of freedom, as we mentioned before, the simplest solution for Equations (\ref{vet8})--(\ref{vet9}) is then to consider $\Psi$ as a small arbitrary function of the  background---{\it i.e.,} $\Psi = \Psi(t)$---which can also be parameterized by the perturbation wavelength $m$.

The constraints can now be used to eliminate three of the five variables, and the most suited pair for this solution is ($\Sigma,\,\Omega$). The resulting free dynamics is

\begin{equation}
\begin{array}{l}
\dot{\Sigma} = - \left(\fracc{2}{3}\, \theta + \xi \right)\, \Sigma - \fracc{\theta}{3}\, \Omega + \fracc{1}{2}\, \Psi, \\[1.5ex]
\dot{\Omega} = - \fracc{\theta}{3}\, \Omega - \fracc{1}{2}\,\Psi,
\end{array}
\protect\label{H1}
\end{equation}
directly integrable as

\begin{equation}
\begin{array}{l}
\Sigma(t) = a^{-2}(t)\, {\rm e}^{-\xi\, t}\, \left\{C_1 + \int\limits^{t}_{(H_{0}^{-1}+c_0)}\, a^2(t')\, {\rm e}^{\xi t'}\, \left[-\fracc{\theta(t')}{3}\, \Omega(t') + \fracc{1}{2}\, \Psi(t')\right]\, dt'\right\}, \\[1.5ex]
\Omega(t) = a^{-1}(t)\, \left\{C_2 - \int\limits^{t}_{(H_{0}^{-1}+c_0)}\, \fracc{1}{2}\, a(t')\, \Psi(t')\, dt'\right\},
\end{array}
\label{H2}
\end{equation}
where $H_{0}$ is the Hubble parameter and $c_0$ a positive integration constant.

Solution (\ref{H2}) can be thought of as a particular case of an arbitrary linear relation\footnote{Linearity is a requirement in order to preserve coherence with our basic assumption of linear perturbations approximation. For the understanding of the physical meaning of such a relation, see the examples given in Section [\ref{vetorial}].} between $\Psi$ and the fundamental variables,

\begin{equation}
\Psi = y(t)\, Q + z(t)\, P + g(t),
\label{H3}
\end{equation}
where

\begin{equation}
\nonumber
\hfill y(t) \colon= \frac{\partial\Psi}{\partial Q},
\hfill z(t) \colon= \frac{\partial\Psi}{\partial P} \hfill
\end{equation}
and ($Q,\,P$) is a pair of canonical variables (as we shall see) that describe the vector perturbations, given by

\begin{equation}
\protect\label{QPpsi}
\begin{pmatrix}
Q\\
P
\end{pmatrix}=
\begin{pmatrix}
\alpha& \beta\\
\gamma& \delta
\end{pmatrix}
\begin{pmatrix}
\Sigma\\
\Omega
\end{pmatrix},\qquad\qquad
\begin{pmatrix}
\Sigma\\
\Omega
\end{pmatrix}=\frac{1}{\Delta}
\begin{pmatrix}
\delta&-\beta\\
-\gamma&\alpha
\end{pmatrix}
\begin{pmatrix}
Q\\
P
\end{pmatrix}
\end{equation}
where $\Delta\equiv \alpha\delta-\beta\gamma\neq 0$.

The choice for the above variables is motivated by traditional results of perturbations assuming a perfect fluid law; within this assumption both the vorticity and the shear are essential variables: none of them may vanish, or else all the system turns out to be trivial (see Goode, 1989). In the more general case, such result does not apply.

Differentiating Equation (\ref{QPpsi}) and making use of the dynamics given above in (\ref{H1}), we get the dynamics written in terms of ($Q$,$P$) as

\begin{equation}
\begin{array}{l}
\dot{Q} = \left[\dot{\alpha} - \left(\fracc{2}{3}\, \theta + \xi\right)\, \alpha\right]\, \Sigma + \left[\dot{\beta} - \fracc{\theta}{3}\, (\alpha + \beta)\right]\, \Omega + \fracc{1}{2}\, (\alpha - \beta)\, \Psi, \\[1.5ex]
\dot{P} = \left[\dot{\gamma} - \left(\fracc{2}{3}\, \theta + \xi\right)\, \gamma\right]\, \Sigma + \left[\dot{\delta} - \fracc{\theta}{3}\, (\gamma + \delta)\right]\, \Omega + \fracc{1}{2}\, (\gamma - \delta)\, \Psi.
\end{array}
\label{H4}
\end{equation}

To ensure that we are actually working with canonically conjugated variables, we write the Hamiltonian constraint

\begin{eqnarray}
\Phi & \colon= & \Delta\, \left(\frac{\partial\dot{Q}}{\partial Q} + \frac{\partial\dot{P}}{\partial P}\right) \nonumber \\
& = & \left[\dot{\alpha} - \left(\fracc{2}{3}\, \theta + \xi\right)\, \alpha\right]\, \delta - \left[\dot{\beta} - \fracc{\theta}{3}\, (\alpha + \beta)\right]\, \gamma - \left[\dot{\gamma} - \left(\fracc{2}{3}\, \theta + \xi\right)\, \gamma\right]\, \beta \nonumber \\
& + & \left[\dot{\delta} - \fracc{\theta}{3}\, (\gamma + \delta)\right]\, \alpha + \fracc{\Delta}{2}\, (\alpha - \beta)\, y + \fracc{\Delta}{2}\,
(\gamma - \delta)\, z \nonumber \\
& = & \dot{\Delta} - (\theta + \xi)\, \Delta + \fracc{1}{2}\, [(\alpha - \beta)\, y + (\gamma - \delta)\, z]\, \Delta,
\label{H5}
\end{eqnarray}
and set the solution of $\Phi = 0$ as

\begin{equation}
\hfill \Delta = \fracc{a^{3}(t)}{a_{0}}\, {\rm e}^{\xi\, t},
\hfill \alpha = \delta = \Delta^{1/2}, \hfill
\label{H5.5}
\end{equation}
with $a_{0} = const$. It will indeed be a solution if

\begin{equation}
\nonumber
(\alpha - \beta)\, y + (\gamma - \delta)\, z = 0
\end{equation}
holds, which leads to the following three possibilities:

\begin{equation}
\protect\label{123}
\begin{array}{llcclr}
i)   & \:y\neq 0,\,\forall z\, & \rightarrow & f \equiv \fracc{z}{y}, & \beta=\alpha(1-f),  & \gamma=0;\\
ii)  & \:y=0,\,z\neq 0 & \rightarrow && \beta=0,       & \gamma=\alpha;\\
iii) & \:y=z=0 & \rightarrow         && \beta=0,       & \gamma=0.
\end{array}
\end{equation}

For the first case, the dynamics results in

\begin{eqnarray}
\dot{Q} & = & \left(\fracc{\dot{\alpha}}{\alpha} - \fracc{2}{3}\, \theta - \xi + \fracc{\alpha}{2}\, z\right)\, Q + \left[- \fracc{\theta}{3}\, f + \xi\, (1 - f) - \dot{f} - \fracc{\alpha}{2}\, f\, z\right]\, P + \fracc{\alpha}{2}\, f\, g, \vspace{0.4cm}\\
\dot{P} & = & - \fracc{\alpha}{2}\, y\, Q + \left(\fracc{\dot{\alpha}}{\alpha} - \fracc{\theta}{3} - \fracc{\alpha}{2}\, z\right)\, P - \fracc{\alpha}{2}\, g,
\label{H6}
\end{eqnarray}
described by the Hamiltonian

\begin{eqnarray}
{\cal H}(Q,P) & = & \fracc{\alpha}{4}\, Q^{2} + \fracc{1}{2}\, \left[-\fracc{\theta}{3}\, f + \xi\, (1 - f) - \dot{f} - \fracc{\alpha}{2}\, f\, z\right]\, P^{2} \nonumber \\
& - & \left(\fracc{\dot{\alpha}}{\alpha} - \fracc{\theta}{3} - \fracc{\alpha}{2}\, z\right\}\, Q\, P + \fracc{\alpha}{2}\, (Q + f\, g\, P).
\label{H7}
\end{eqnarray}

For the second case the dynamics becomes

\begin{equation}
\begin{array}{l}
\dot{Q} = \left(\fracc{\dot{\alpha}}{\alpha} - \fracc{\theta}{3} - \xi\right)\, Q + \left(- \frac{\theta}{3} + \fracc{\alpha}{2}\, z\right)\, P + \fracc{1}{2}\, \alpha\, g, \\[1.5ex]
\dot{P} = \left(\fracc{\dot{\alpha}}{\alpha} - \fracc{2}{3}\, \theta \right)\, P,
\end{array}
\label{H8}
\end{equation}
associated with the Hamiltonian

\begin{equation}
{\cal H}(Q,P) = \fracc{1}{2}\, \left(- \fracc{\theta}{3} + \fracc{\alpha}{2}\, z\right)\, P^{2} - \left(\fracc{\dot{\alpha}}{\alpha} - \fracc{2}{3}\, \theta\right)\, Q\, P + \fracc{\alpha}{2}\, g\, P.
\label{H9}
\end{equation}

The third case is equivalent to the situation given in Equations (\ref{vet8}) with new variables, and can be written as

\begin{equation}
\begin{array}{l}
\dot{Q} = \left(\fracc{\dot{\alpha}}{\alpha} - \fracc{2}{3}\, \theta - \xi + \fracc{\alpha}{2}\, y\right)\, Q + \left(- \fracc{\theta}{3} + \fracc{\alpha}{2}\, y\right)\, P + \fracc{\alpha}{2}\, g, \\[1.5ex]
\dot{P} = - \fracc{\alpha}{2}\, y\, Q + \left(\fracc{\dot{\alpha}}{\alpha} - \fracc{\theta}{3} - \fracc{\alpha}{2}\, y\right)\, P - \fracc{\alpha}{2}\, g.
\end{array}
\label{H10}
\end{equation}
The Hamiltonian associated with this case is then \settowidth{\espaco}{${\cal H}(Q,P) = $}

\begin{equation}
{\cal H}(Q,P) = \fracc{\alpha}{4}\, y\, Q^{2} + \fracc{1}{2}\, \left(- \fracc{\theta}{3} + \fracc{\alpha}{2}\, y\right)\, P^{2} - \left(\fracc{\dot{\alpha}}{\alpha} - \fracc{\theta}{3} - \fracc{\alpha}{2}\, y\right)\, Q\, P + \fracc{\alpha}{2}\, g\,(Q + P).
\label{H11}
\end{equation}

\subsubsection{The Specific Solutions}\protect\label{Particular}

We proceed to study the three particular cases presented in Section [\ref{vetorial}], where the acceleration $\Psi$ was eliminated by an explicit loss of a degree of freedom.

In the first case (the isotropic or shear-free model), we have $\Sigma =0$ and, using Equation (\ref{h1}) in the system (\ref{vet8})--(\ref{vet9}), we obtain the following results:

\begin{equation}
\begin{array}{l}
H(t) = C_1\, a^{-2}(t), \\[1.5ex]
E(t) = - \fracc{2C_1}{3}\, \theta\, a^{-2}(t), \\[1.5ex]
\Omega(t) = -2\, C_1 a^{-2}(t), \\[1.5ex]
\Psi(t) = - \fracc{4C_1}{3}\, \theta\, a^{-2}(t), \\[1.5ex]
q(t) = - 2C_1 a^{-2}(t)\, [(m + 2\epsilon)\, a^{-2}(t) + 2\, (\rho + p)],
\end{array}
\label{P1}
\end{equation}
where $C_1$ is an integration constant. It is worth pointing out that a non-zero heat flux is necessary in order that a shear-free linear perturbation is obtained, since zero shear is a characterizing condition for no perturbation in the perfect fluid case (cf. Goode, 1989).

The second case (irrotational model, $\Omega = 0$) gives, upon substitution of Equation\ (\ref{h2}) in\ (\ref{vet8})--(\ref{vet9}), the results below:

\begin{equation}
\begin{array}{l}
\Sigma(t) = C_2 a^{-2}(t)\, \exp^{-\xi\, t}, \\[1.5ex]
H(t) = - \fracc{C_2}{2}\, a^{-2}(t)\, \exp^{-\xi\, t},\\[1.5ex]
E(t) = C_2\left(\fracc{\theta}{3} + \fracc{\xi}{2}\right)\, a^{-2}(t)\, \exp^{-\xi\, t}, \\[1.5ex]
q(t) = C_2 (m + 2\epsilon)\, a^{-4}(t)\, \exp^{-\xi\, t},
\end{array}
\label{P2}
\end{equation}
where $C_2$ is another integration constant. In this case, the fluid must be non-perfect in order that a linear perturbation with zero vorticity be possible.

Finally, for the third case (Stokesian fluid), with $q = 0$, $p = \lambda\rho$ and Equation\ (\ref{h3}) holding, the system\ (\ref{vet8})--(\ref{vet9}) gives the reduced dynamics

\begin{equation}
\begin{array}{l}
\dot{\Sigma} = - \left[\fracc{2}{3}\, \theta + \xi\left( 1 + \fracc{1}{2a^{2}}\, \fracc{(m + 2\epsilon)}{(\rho + p)}\right) \right]\, \Sigma - (1 - 3\lambda)\, \fracc{\theta}{3}\, \Omega, \\[1.5ex]
\dot{\Omega} = \fracc{1}{2a^{2}}\, \fracc{(m + 2\epsilon)}{(\rho + p)}\, \xi\, \Sigma - (1 + 3\lambda)\, \fracc{\theta}{3}\, \Omega.
\end{array}
\label{P3}
\end{equation}

We again seek a Hamiltonian description with variables ($Q,\,P$), using the same transformation given in Equation\ (\ref{QPpsi}). Differentiating these expressions, we find that Equations\ (\ref{P3}) can be written as

\begin{equation}
\begin{array}{rcl}
\dot{Q} & = & \left\{\dot{\alpha} - \left[\fracc{2}{3}\, \theta + \xi\, \left(1 + \fracc{1}{2a^{2}}\, \fracc{(m + 2\epsilon)}{(\rho + p)}\right)\right]\, \alpha + \fracc{1}{2a^{2}}\, \fracc{(m + 2\epsilon)}{(\rho + p)}\, \xi\, \beta \right\}\, \fracc{1}{\Delta}\, (\delta Q - \beta P) \\[1ex]
& + & \left\{\dot{\beta} - [\alpha\, (1 - 3\lambda) + \beta\, (1 + 3\lambda)]\, \fracc{\theta}{3}\right\}\, \fracc{1}{\Delta}\, (- \gamma\, Q + \alpha\, P), \\[1.5ex]
\dot{P} & = & \left\{\dot{\gamma} - \left[\fracc{2}{3}\, \theta + \xi\, \left(1 + \fracc{1}{2a^{2}}\, \fracc{(m + 2\epsilon)}{(\rho + p)}\right)\right]\, \gamma + \fracc{1}{2a^{2}}\, \fracc{(m + 2\epsilon)}{(\rho + p)}\, \xi\ \delta \right\}\, \fracc{1}{\Delta}\, (\delta Q - \beta P) \\[1ex]
& + & \left\{\dot{\delta} - [\gamma\, (1 - 3\lambda) + \delta\, (1 + 3\lambda)]\, \fracc{\theta}{3}\right\}\, \fracc{1}{\Delta}\,(- \gamma\, Q + \alpha\, P).
\label{P4}
\end{array}
\end{equation}

From Equations\ (\ref{P4}) we read the Hamiltonian constraint

\begin{eqnarray}
\Phi & \equiv & \Delta\, \left( \frac{\partial\dot{Q}}{\partial Q} + \frac{\partial\dot{P}}{\partial P}\right) = \dot{\Delta} - \left[\frac{\theta}{3} + \xi\, \left(1 + \frac{1}{2a^{2}}\, \frac{(m + 2\epsilon)}{(\rho + p)}\right)\right]\, \Delta + \lambda\, \theta\, \Delta =0,
\label{P5}
\end{eqnarray}
whose solution is given by

\begin{equation}
\Delta(t) = a^{(1 - 3\lambda)}(t)\, {\rm e}^{\xi\, \int\limits^{t}_{(H_{0}^{-1}+c_0)}\, \left\{1 + \fracc{1}{2a^{2}}\, \fracc{(m + 2\epsilon)}{(1 + \lambda)\, \rho(t')}\right\}\, dt'},
\label{P6}
\end{equation}
where $c_0$ is again a positive integration constant. We now set the Hamiltonian variables ($Q,\,P$) as given by Equation\ (\ref{QPpsi}) with

\begin{equation}
\nonumber
\begin{array}{l}
\alpha = \delta = \Delta^{1/2}, \\[1.5ex]
\beta = \gamma = 0,
\end{array}
\end{equation}
where $\Delta$ is given by Equation\ (\ref{P6}). Therefore we finally obtain the dynamics

\begin{equation}
\begin{array}{l}
\dot{Q} = \left\{\fracc{\dot{\alpha}}{\alpha} - \fracc{2}{3}\, \theta - \xi\, \left[1 + \fracc{1}{2a^{2}}\, \fracc{(m + 2\epsilon)}{(\rho + p)}\right]\right\}\, Q - \left\{(1 - 3\lambda)\, \fracc{\theta}{3}\right\}\, P, \\[2ex]
\dot{P} = \left[\fracc{1}{2a^{2}}\, \frac{(m + 2\epsilon)}{(\rho + p)}\, \xi\right]\, Q + \left[\fracc{\dot{\alpha}}{\alpha} - (1 - 3\lambda)\, \fracc{\theta}{3}\right]\, P,
\end{array}
\label{P7}
\end{equation}
submitted to the constraint of vanishing heat flux

\begin{equation}
Q = - \left[1 + \frac{2a^{2}}{(m + 2\epsilon)}\, (\rho + p) \right]\, P.
\label{P8}
\end{equation}

The associated Hamiltonian is then given by

\begin{eqnarray}
& &{\cal H}(Q,P) = - \frac{1}{2}\, \left[
\frac{1}{2a^{2}}\, \frac{(m + 2\epsilon)}{(\rho + p)}\, \xi +
\frac{2}{3}\, \frac{(1 + 3\lambda)\, \theta}{1 +
\frac{2a^{2}}{(m + 2\epsilon)}\, (\rho + p)}
\right]\, Q^{2} + \nonumber \\
& & - (1 - 3\lambda)\, \frac{\theta}{3}\, P^{2}
 -  \left\{(1 + \lambda)\, \frac{\theta}{2} +
\frac{\xi}{2}\, \left[1 + \frac{1}{2a^{2}}\,
\frac{(m + 2\epsilon)}{(\rho + p)}\right]\right\}\, Q\, P.
\label{P9}
\end{eqnarray}

As an example, equations of motion\ (\ref{P7}) can be explicitly integrated by taking into account Equation\ (\ref{P8}). Thus the system evolution follows

\begin{equation}
\dot{P} = - \left\{(1 + 9\lambda)\, \frac{\theta}{6} +
\frac{\xi}{2}\, \left[1 + \frac{1}{2a^{2}}\,
\frac{(m + 2\epsilon)}{(\rho + p)}\right]\right\}\, P,
\label{P10}
\end{equation}
which can be readily integrated, and we finally find

\begin{equation}
\begin{array}{l}
Q = - \left[1 + \fracc{2a^{2}}{(m + 2\epsilon)}\, (\rho +
p)\right]\, a^{-\fracc{(1 + 9\lambda)}{2}}\,
\exp\left\{-\fracc{\xi}{2}\,
\int\limits^{t}_{(H_{0}^{-1}+c_0)}\left[
1 + \fracc{1}{2a^{2}(t')}\, \frac{(m + 2\epsilon)}{(1 + \lambda)\,
\rho(t')}\right]\, dt'\right\}, \\[2ex]
P = a^{-\fracc{(1 + 9\lambda)}{2}}\,
\exp\left\{-\fracc{\xi}{2}\,
\int\limits^{t}_{(H_{0}^{-1}+c_0)}\,
\left[1 + \fracc{1}{2a^{2}(t')}\,
\frac{(m + 2\epsilon)}{(1 + \lambda)\, \rho(t')}\right]\, dt'\right\},
\end{array}
\label{P11}
\end{equation}
Returning to the physically relevant variables we particularly find that

\begin{equation}
\Omega(t) = a^{-(1 + 3\lambda)}\,
\exp\left\{-\xi\, \int\limits^{t}_{(H_{0}^{-1}+c_0)}\,
\left[1 + \fracc{1}{2a^{2}}\,
\fracc{(m + 2\epsilon)}{(1 + \lambda)\, \rho(t')}\right]\, dt'\right\}.
\label{P11.5}
\end{equation}
The perturbation in vorticity appears to diverge---thus breaking down our fundamental approach of the linear treatment---for perturbation wavelengths such that

\begin{equation}
m < -2\epsilon - 2\, (1 + \lambda)\, a^{2}\, \rho.
\label{P12}
\end{equation}
However, Equation\ (\ref{mP}) shows that we always have $m > -2\epsilon$, and from this $\Omega$ goes to zero. Such a result could also be expected from the angular momentum conservation law.

Therefore we get the minimal set of observables for the vectorial mode:

\begin{equation}
\nonumber
{\cal M}^{vector}_{A} = \{\Sigma, \Omega, q, \Psi\}.
\end{equation}
However, the system is not closed, since the variable $\Psi$ cannot be written in terms of the other ones. In order to solve this system we have then to eliminate one of the variables involved, thus losing a degree of freedom.

\subsection{Friedman Universe: Tensorial Perturbation}\protect\label{tensorial}

Here we will proceed as in Section\ [\ref{vetorial}] in order to get an ordinary differential system which describes tensorial perturbations in terms of {\it good} variables.

The tensorial basis $\hat{U}_{\alpha\beta}(x)$ is defined by the relations

\begin{equation}
\begin{array}{l}
\dot{\hat{U}}_{\alpha\beta} = 0, \\[2ex]
h^{\mu\nu}\, \hat{U}_{\mu\nu} = 0, \\[2ex]
\hat{\nabla}^{\mu}\, \hat{U}_{\mu} = 0, \\[2ex]
\hat{U}_{\alpha\beta} = \hat{U}_{\beta\alpha}, \\[2ex]
\hat{\nabla}^{2}\, \hat{U}_{\alpha\beta} = \fracc{m}{A^{2}}\, \hat{U}_{\alpha\beta},
\end{array}
\protect\label{intro4}
\end{equation}
where the new eigenvalue $m$ has the following spectrum

\begin{equation}
\protect\label{mU}
m=\left\{
\begin{array}{lll}
q^2+3, & 0<q<\infty,     & \epsilon=+1   \mbox{\rm\ (open)},\\
q,     & 0<q<\infty,     & \epsilon=\ \ 0\mbox{\rm\ (plane)},\\
n^2-3, & n=3,\,4,\ldots, & \epsilon=-1   \mbox{\rm\ (closed)}.
\end{array}
\right.
\end{equation}

Using the tensor basis, we can define the dual tensor

\begin{equation}
\hat{U}^{\ast}_{\mu\nu} \equiv \frac{1}{2}\,h^{\alpha}_{(\mu }\, h^{\beta}_{\nu )} \, {\eta_{\beta}}^{\lambda\varepsilon\gamma} \, V_{\lambda}\,
\hat{\nabla}_{\varepsilon}\, \hat{U}_{\gamma\alpha}.
\protect\label{intro5}
\end{equation}

The tensorial relations below are employed in obtaining the dynamical equations system:

\begin{equation}
\begin{array}{l}
\dot{\hat{U}}^{\ast}_{\alpha\beta} = - \fracc{1}{3}\, \theta\, \hat{U}^{\ast}_{\alpha\beta}, \\
\hat{U}^{\ast\ast}_{\alpha\beta} = \left(\fracc{m}{a^{2}} + \rho - \fracc{1}{3}\, \theta^{2}\right)\, \hat{U}_{\alpha\beta} = \frac{1}{a^{2}}\, (m - 3\epsilon)\, \hat{U}_{\alpha\beta},
\end{array}
\protect\label{intro6}
\end{equation}
We remark that it involves the energy density $\rho$ and the expansion $\theta$, which satisfies:

\begin{equation}
\theta = 3\, \frac{\dot{a}}{a}.
\protect\label{intro8}
\end{equation}
Its inverse is usually referred to as the Hubble parameter, whose present value will be denoted by $H_{0}$.

Now, we expand the {\it good} perturbed quantity in terms of the above basis as

\begin{equation}
\protect\label{decomp_tens}
\begin{array}{lcl}
\delta\sigma_{\alpha\beta} &=&\Sigma(t)\, \hat{U}_{\alpha\beta}\vspace{.2cm}\\
\delta H_{\alpha\beta} &=&H(t)\, \hat{U}_{\alpha\beta}^{\ast}\vspace{.2cm}\\
\delta E_{\alpha\beta} &=&E(t)\, \hat{U}_{\alpha\beta}\vspace{.2cm}\\
\delta\Pi_{\alpha\beta} &=& \Pi(t)\, \hat{U}_{\alpha\beta},
\end{array}
\end{equation}
where the time dependent functions $\Sigma$, $E$, $H$ and $\Pi$ are not related to the vector components of the previous section.

\subsubsection{Dynamics}

Under the properties (\ref{intro4})--(\ref{intro6}) and again making use of relation\ (\ref{vet2}), quasi-Maxwellian equations are written as:

\begin{subequations}
\label{tensor1}
\begin{eqnarray}
&&\dot{E} - \fracc{\xi}{2}\, \dot{\Sigma} + \theta\, E - \fracc{1}{2}\, \left[\fracc{\theta}{3}\, \xi - (\rho + p)\right]\, \Sigma + \fracc{1}{a^{2}}\, (m - 3\epsilon)\,H = 0,\label{tensor11}\\[1.5ex]
&&\dot{H} + \fracc{2}{3}\, \theta\, H + E + \fracc{\xi}{2}\, \Sigma = 0,\label{tensor12}\\[1.5ex]
&&\dot{\Sigma} + \left(\fracc{2}{3}\, \theta + \fracc{\xi}{2}\right)\, \Sigma + E = 0,\label{tensor13}
\end{eqnarray}
\end{subequations}
constrained to

\begin{equation}
\eta\equiv H - \Sigma = 0.
\label{tensor2}
\end{equation}
We also know that $\Phi$ is dynamically preserved as

\begin{equation}
\dot{\eta}= \chi_{2} - \chi_{3} - \fracc{2}{3}\, \theta\, \eta,
\label{tensor3}
\end{equation}
where $\chi_2$ and $\chi_3$ are Eqs.\ (\ref{tensor12}) and (\ref{tensor13}), respectively. From this we are, therefore, authorized to insert it into dynamics. So proceeding we get the unconstrained coupled differential system

\begin{subequations}
\begin{eqnarray}
&&\dot{E} + \left(\theta + \frac{\xi}{2} \right)\, E + \left\{\frac{1}{2}\left[\xi\, \left(\frac{\theta}{3} + \frac{\xi}{2}\right) + (\rho + p)
\right] - \frac{1}{a^{2}}\, (m - 3\epsilon)\right\}\, H = 0,\label{tensor41}\\[1.5ex]
&&\dot{H} + \left(\frac{2}{3}\, \theta + \frac{\xi}{2}\right)\, H + E = 0.\label{tensor42}
\end{eqnarray}
\label{tensor4}
\end{subequations}

It should be remarked that the coefficient $H$ in Eq.\ (\ref{tensor41}) in the de Sitter background yields a positive \footnote{The Hubble constant, here translated to $\theta$, is positive from astronomical observations, despite the fact that its magnitude is not universally agreed upon. Thermodynamical reasoning ensures the nonnegativeness of parameter $\xi$.} constant leading term, for times such that $(1/a^{2}) \simeq 0$.
This feature will be important in Section [\ref{HTensor}].

We also stress that Equations\ (\ref{tensor4}) have no non trivial solution unless both ($E,\,H$) are assumed to be non zero. That is, both variables are essential in describing tensor perturbations---it should be remembered that these variables constitute the electric and magnetic parts of Weyl tensor, allowing one to write ``gravitational waves'' for ``tensor perturbations''.

\subsubsection{Hamiltonian Treatment of the Tensorial Solution}\protect\label{HTensor}

The basic system given by Equations\ (\ref{tensor4}) can be described in the Hamiltonian language, which provides a more elegant interpretation of the dynamical role of our variables. The link between it and perturbation theory has worth on its own. We thus introduce new variables

\begin{equation}
\begin{pmatrix}
Q\\
P
\end{pmatrix}\equiv
\begin{pmatrix}
\alpha & \beta\\
\gamma & \delta
\end{pmatrix}\,
\begin{pmatrix}
E\\
H
\end{pmatrix},
\label{HT1}
\end{equation}
where we suppose

\begin{equation}
\nonumber
\Delta \equiv \det
\begin{pmatrix}
\alpha & \beta \\
\gamma & \delta
\end{pmatrix}=
\alpha\delta - \beta\gamma \neq 0,
\end{equation}
which is proven {\it a posteriori} to be actually correct. Therefore we can use the set ($Q,\,P$) for ($E,\,H$) in order to characterize the gravitational waves. Inserting definitions\ (\ref{HT1}) into Equations\ (\ref{tensor4}) we eventually get

\begin{equation}
\begin{array}{ll}
\dot{Q} = &\left\{\dot{\alpha} - \alpha\, \left(\theta + \fracc{\xi}{2}
\right) - \beta\right\}\, E + \left\{\dot{\beta} - \beta\,
\left(\fracc{2}{3}\, \theta + \fracc{\xi}{2}\right)
- \alpha\, \left(\fracc{1}{2}\, \left[\xi\, \left(\fracc{\theta}{3} +
\fracc{\xi}{2}\right) + (\rho + p)\right] -
\fracc{1}{a^{2}}\, (m - 3\epsilon)\right)\right\}\, H, \\[1.5ex]
\dot{P} = &\left\{\dot{\gamma} - \gamma\, \left(\theta +
\fracc{\xi}{2}\right) - \delta\right\}\, E + \left\{\dot{\delta} - \delta\,
\left(\fracc{2}{3}\, \theta + \fracc{\xi}{2}\right)
- \gamma\, \left(\fracc{1}{2}\, \left[\xi\, \left(\fracc{\theta}{3} +
\fracc{\xi}{2}\right) + (\rho + p)\right] -
\fracc{1}{a^{2}}\, (m - 3\epsilon)\right)\right\}\, H.
\end{array}
\label{HT2}
\end{equation}

We also need to show that our variables are, in fact, canonically conjugated to each other, as suggested by notation. That is, we again make use of the Hamiltonian constraint,

\begin{equation}
\Phi \equiv \Delta\, \left(\frac{\partial\dot{Q}}{\partial Q} + \frac{\partial\dot{P}}{\partial P}\right) = \dot{\Delta} - \left(\frac{5}{3} \theta + \xi\right)\, \Delta = 0.
\label{HT3}
\end{equation}
A particular solution of Equation\ (\ref{HT3}) is

\begin{equation}
\Delta(t) = a^{5}(t)\, {\rm e}^{\xi\, t},
\label{HT4}
\end{equation}
and we then set

\begin{equation}
\begin{array}{l}
\alpha = \Delta^{\omega}, \\[1.5ex]
\delta = \Delta^{(1 - \omega)}, \\[1.5ex]
\beta = \gamma = 0,
\end{array}
\label{HT5}
\end{equation}
where $\omega$ is an arbitrary constant.

With the choice\ (\ref{HT5}), and using solution\ (\ref{HT4}), system\ (\ref{HT2}) becomes

\begin{equation}
\begin{array}{ll}
\dot{P} = &- \left[\left(\fracc{5}{3}\, \omega - 1\right)\theta + \left(\omega-\fracc{1}{2}\right)\, \xi\right]\,P - \Delta^{(1 - 2\omega)}\, Q \\[1.5ex]
\dot{Q} =  &- \left[\fracc{\xi}{2}\, \left(\fracc{\theta}{3} + \fracc{\xi}{2}\right) - \fracc{1}{2}\, (\rho + p) - \fracc{1}{2a^{2}}\, (m - 3\epsilon)\right]\, \Delta^{(2\omega - 1)}\, P +\left[\omega\, \left(\fracc{5}{3}\, \theta + \xi \right) - \left(\theta + \fracc{\xi}{2}\right)\right]\, Q.
\end{array}
\label{HT6}
\end{equation}
From this we directly read the Hamiltonian

\begin{eqnarray}
& {\cal H}(Q,\,P) = & - \frac{1}{2}\, \Delta^{(2\omega-1)}\,
\left[\frac{\xi}{2}\left(\frac{\theta}{3} + \frac{\xi}{2}
\right) + \frac{1}{2}\,(\rho + p) -\frac{1}{a^2}(m - 3\epsilon)
\right]\, P^2 + \hfill \nonumber \\[1.5ex]
& & + \frac{1}{2}\, \Delta{(1-2\omega)}\, Q^2 +
\left[\left(\frac{5}{3}\, \omega - 1\right)\, \theta
+ \left(\omega - \frac{1}{2}\right)\, \xi\right]\, PQ.
\label{HT7}
\end{eqnarray}

This result shows that de Sitter ($\theta=const.$) geometry admits a tensor perturbation Hamiltonian of a typical harmonic oscillator with imaginary mass, which evidences instability.  This is obtained by setting the arbitrary constant parameter

\begin{equation}
\nonumber
\omega=\frac{3}{2}\,
\frac{(2\theta + \xi)}{(5\theta + 3\xi)},
\end{equation}
in the Hamiltonian\ (\ref{HT7}). We thus recover the well known result of instability of de Sitter solution.  The above result also shows, however, that the same remark applies to arbitrary Friedman-like background with no tensorial perturbation in anisotropic pressure tensor, $\xi = 0$. In such cases we set $\omega=3/5$ to get

\begin{equation}
\label{HT8}
{\cal H}(Q,\,P)\Big|_{\xi=0} = - \frac{1}{2}\, \Delta^{1/5}\,
\left[(\rho + p) - \frac{1}{a^2}(m - 3\epsilon)\right]\, P^2
+ \frac{1}{2}\, \Delta^{-1/5}\, Q^2,
\end{equation}
where $(Q,\,P)$ are given by

\begin{equation}
\label{HT9}
\begin{array}{l}
Q = a^3(t)\, {\rm e}^{\frac{3}{5}\xi t}\, E \\[1ex]
P = a^2(t)\, {\rm e}^{\frac{2}{5}\xi t}\, H.
\end{array}
\end{equation}

Summarizing, we have found that there is a complete set of {\it good} perturbed variables for tensorial modes:

\begin{equation}
\nonumber
{\cal M}^{tensor}_{A} = \{E, H\}.
\end{equation}
In this case, the system is closed and completely independent of the other modes, due to the linearity of the harmonic basis.

We have also obtained the Hamiltonian formulation for all modes, according to the previous sections, and we address the possibility to canonically quantize the cosmological perturbations of FLRW universes. This analysis will be discussed with more details in the next section.

\subsection{Friedman Universe: Quantum Treatment of the perturbations}\protect\label{quantumtreat}

In the previous Sections\ [\ref{scalarpert}], [\ref{genvetorial}] and [\ref{tensorial}], we have shown how to treat, in a completely gauge-invariant way, the evolution of the perturbations of FLRW universes. Besides, we have shown that it is possible to select a minimal set of observable quantities to analyze the perturbations of the FLRW universe. We have also shown that the complete dynamical system of the perturbed geometry is described only in terms of two quantities: $E$ and $\Sigma $, (respectively, the electric part of conformal Weyl tensor and the shear) for scalar perturbations; $E$ and $H$, (where the last quantity is the associated magnetic part of Weyl conformal tensor) for tensorial perturbations. For vectorial perturbations, in a more general case, this minimal set should be expanded to include $E$, $\Sigma$, $H$ and the vorticity $\Omega$. Now we have completed the classical treatment of these gauge-independent perturbations; it is rather natural then to go beyond this classical theory. Indeed, the purpose of this section is to treat these perturbations into the quantum framework.

This amounts to using a semi-classical description in which the background geometry is taken in the classical framework and considering the perturbations as quantum variables. There are many ways to perform this task. Here we will follow a very natural way that consists of applying the method of the auxiliary Hamiltonian, which was introduced before. The problem can be stated in the following way: using the quasi-Maxwellian formulation of Einstein's General Relativity, we find out that the complete dynamical system reduces to the form

\begin{equation}
\left[\begin{array}{cc}
\dot{M_{1}} \\
\dot{M_{2}}
\end{array}\right] = {\cal M} \hspace{0.1cm}\left[\begin{array}{cc}
M_{1} \\
M_{2}
\end{array}\right],
\protect\label{1}
\end{equation}
in which ${\cal M}$ is a $2\times2$ matrix containing information that characterizes the background geometry and $M_{j}$ are the ``good" observables which
describe the perturbations\footnote{We are considering
here only the cases in which the minimal closed set of
observables contains only two variables. In the more
general vectorial case, ${\cal M}$ should be a $4\times4$
matrix, as stated above (See Section [\ref{vector}] for
more details).}. The dot ($\,\dot{ }\,$) denotes a temporal
derivative.

For the FLRW background, this set constitutes a non-autonomous dynamical system. A direct inspection of the matrix ${\cal M}$ shows that it is not trace-free; thus, this system does not have a Hamiltonian. Nevertheless, we exhibited a method by which we were able to obtain an auxiliary Hamiltonian, ${\cal H}$, for this system. As we will see later, the linear relation between the associated canonically conjugated variables ($Q$, $P$) and the original physical quantities ($E$, $\Sigma$, $H$ or $\Omega$) is not unique. This is not a drawback of this approach, but merely a consequence of the fact that the set of possible pairs ($Q$, $P$) is related in turn by canonical transformations.

The existence of this Hamiltonian leads us to consider the possibility of employing the canonical method to arrive at the quantum version of the perturbed set. The quantum study of these perturbations in FLRW was done by  Lifshitz (1963), Hawking (1966) and Novello (1983). However, all these previous works deal with variables which either are gauge-dependent or follow the general scheme introduced by Bardeen (1980) and subsequent papers (cf. Ellis, 1989),
which has a difficult physical interpretation. This fact makes the analysis more complex.

Let us remark that the use of our method, in which the gauge problem is inexistent, seems to be really the best way of making the transition to the quantum version. Indeed, alternative methods, for example, the minisuperspaces approach (e.g. Ryan, 1991) suffer not only from the need to fix a gauge, but also from the fact that the order in which this choice is made (before or after the quantization) leads to different theories. Even the schemes that consider gauge-independent variables (as defined by Bardeen initially) have, as a main problem, the fact that these variables do not have an evident physical interpretation---which makes the physical comprehension of the results obtained quite complex. Since the gauge-independent variables in our method are observables, and completely equivalent to Bardeen's variables, the advantages of quantizing the dynamical systems obtained in our procedure are evident.

In order to perform the quantization of our system we will make use of squeezed states of the Quantum Optics, which was first employed in the framework of Cosmology by Grishchuk (1990), Schumaker (1986) and Bialynicha-Birula (1987). However, the use of these approaches suffer from the same difficulties pointed out above. So, the advantages of incorporating to this approach the method of gauge-independent variables are again obvious.

All the definitions and notations that were employed to obtain the results for gauge-invariant, observable perturbations in the FLRW background (scalar, vectorial and tensorial)are equally valid here. So, we write FLRW geometry in the standard Gaussian coordinate system.

For a comoving observer (one with $V^{\alpha} = \delta^{\alpha}_{0}$), we denote $\rho$ as the energy density, $p$ as the isotropic pressure, and $\theta$ as the expansion. The constraint relation

\begin{equation}
-\frac{\epsilon}{a^{2}} - \frac{1}{3}\,\rho + \left(\frac{\theta}
{3}\right)^{2} = 0
\protect\label{constraint}
\end{equation}
holds, as well as the following auxiliar relation,
\begin{equation}
\rho = \rho_{0}\,a^{- 3\,(1 + \lambda)},
\protect\label{relation}
\end{equation}
which comes from Raychaudhuri equation for a fluid with the usual linear state equation $p = \lambda\,\rho$ (with $\lambda$ as a constant defined in the interval $[0,1]$). The parameter $\rho_{0}$ denotes the value of the energy density for $a(t) = 1$.

We will denote the viscosity by $\xi$. As before, it will be taken in the limit of small relaxation times for the adiabatic approximation of the thermodynamic equation (see Novello, 1995a) and, due to this choice as well as to thermodynamic considerations, it will be taken as a negative constant.

As for the conventions, let us point out that we will choose the geometrical units system, $\hbar \equiv k \equiv c = 1$. The constant $m$ will denote the wave number associated to the perturbations in the FLRW background and the arbitrary integration constants $\kappa$ and $b$ will be employed
throughout the whole section, for all three perturbation types. Additionally, we will denote by calligraphic letters the matrixes (such as, for example, the Hamiltonian matrix ${\cal H}$) and by capital letters their corresponding linear counterparts (for example, the Hamiltonian $H$). It should be remarked at this point that, when following the standard quantization procedure, we will drop any extra indication (such as the Hamiltonian $H$ turning into the Hamiltonian operator $\hat{H}$), so as not to make our notation excessively complicated to be read, except the case of creation $\hat a^{\dagger}$ and annihilation $\hat a$ operators where we will maintain the $``\,(\,\,\widehat\,\,\,)\,"$ symbol to avoid confusion with the scale factor $a(t)$.

\subsubsection{The Auxiliary Hamiltonian}\protect\label{auxH}

Let us consider the linear two-dimensional dynamical system for the variables $M_{1}$ and $M_{2}$, which gives the dynamics of the evolution in the FLRW universe background for the minimal closed set of these gauge-independent linear perturbation quantities given by

\begin{equation}
\left[\begin{array}{cc}
\dot{M_{1}} \\
\dot{M_{2}}
\end{array}\right] = {\cal M} \hspace{0.1cm}\left[\begin{array}{cc}
M_{1}\\
M_{2}
\end{array}\right],
\protect\label{1x}
\end{equation}
in which ${\cal M}$ is a $2\times 2$ matrix that may depend on time through the known background quantities and $M_{j}$ are the observables that form the minimal closed set which describe all the perturbations. As it has been pointed out, the variables ($M_{1}$, $M_{2}$) are not canonically conjugated.

We thus define a new set of variables ($Q$, $P$) as:

\begin{equation}
\left[\begin{array}{cc}
Q \\ P
\end{array}\right] = S\, \left[\begin{array}{cc}
M_{1} \\ M_{2}
\end{array}\right] = \left[\begin{array}{ccc}
\alpha & \beta \\ \gamma & \delta
\end{array}\right]\, \left[\begin{array}{cc}
M_{1} \\ M_{2}
\end{array}\right],
\protect\label{2}
\end{equation}
where $\alpha$, $\beta$, $\gamma$ and $\delta$ are functions of time and $S$, the transformation matrix, has a  determinant given by

\begin{equation}
\Delta \equiv det(S) = \alpha\, \delta - \beta\, \gamma
\neq 0,
\protect\label{3}
\end{equation}
and its inverse exists.

As a consequence, the variables ($Q$, $P$) satisfy the following dynamics:

\begin{equation}
\nonumber
\left[\begin{array}{cc}
\dot{Q} \\ \dot{P}
\end{array}\right] = {\cal H}\, \left[\begin{array}{cc}
Q \\ P
\end{array}\right],
\end{equation}
in which ${\cal H}$ depends on time through ${\cal M}$ and the transformation matrix $S$. From Eqs.\ (\ref{1x}) and (\ref{2}) it follows that

\begin{equation}
\nonumber
{\cal H} = S\, {\cal M}\, S^{-1} + \dot{S}\, S^{-1}.
\end{equation}

If we require ($Q$, $P$) to be canonical variables, then the matrix ${\cal H}$ must be traceless:

\begin{equation}
\nonumber
Tr\, {\cal H} = 0.
\end{equation}
From the equation above, it then follows:

\begin{equation}
Tr\, {\cal M} + \frac{\dot{\Delta}}{\Delta} = 0,
\protect\label{4}
\end{equation}
which can be easily integrated.

Thus we have a set ($Q$, $P$) of canonical variables, which has an associated Hamiltonian $H$ that is linearly related to ${\cal H}$.  The above condition ensures that the set ($\alpha$, $\beta$, $\gamma$, $\delta$) has only three independent quantities. These degrees of freedom are fixed by the canonical transformations, as it will be discussed later.

The most general quadratic Hamiltonian for our system can be written as

\begin{equation}
{\cal H} = \frac{h_{1}}{2}\, Q^{2} + \frac{h_{2}}{2}\, P^{2} + 2\,
h_{3}\, P\, Q,
\protect\label{H}
\end{equation}
whose matrix form, ${\cal H}$, is then given as

\begin{equation}
{\cal H} = \left[\begin{array}{cc}
2\, h_{3} & h_{2} \\ - h_{1} & - 2\, h_{3}
\end{array}\right],
\protect\label{Hmatrix}
\end{equation}
where the $h_{i}$ are given in terms of the functions $\alpha$, $\beta$, $\gamma$ and $\delta$, as well as of quantities in the FLRW background. Since ${\cal H}$ is a $2\times 2$ traceless matrix, we can decompose it as ${\cal H} = \vec{\mu}\,.\,\vec{\sigma}$ where $\vec{\mu}$ has the following components:

\begin{equation}
\nonumber
\left(\frac{(h_{2} - h_{1})}{2} \hspace{0.1cm}, \hspace{0.1cm}\frac{i}{2}
\hspace{0.1cm}(h_{1} + h_{2}) \hspace{0.1cm}, \hspace{0.1cm}2h_{3}\right).
\end{equation}
The $h_{k}$ ($k = 1,2,3$) depend indirectly on the parameter $t$, through the known quantities of FLRW background. The vector $\vec{\sigma}$ is built with the Pauli matrices:

\begin{equation}
\nonumber
\sigma_{1} = \left[\begin{array}{ccc}
0 & 1 \\
1 & 0
\end{array}\right]; \hspace{1.0cm}\sigma_{2} = \left[\begin{array}{ccc}
0 & -i \\
i & 0
\end{array}\right]; \hspace{1.0cm}\sigma_{3} = \left[\begin{array}{ccc}
1 & 0 \\
0 & -1
\end{array}\right].
\protect\label{h3}
\end{equation}
Let us remark that the $h_{k}$ (written in terms of background quantities) vary according to the perturbation type. They will be presented in later sections. We then have:

\begin{equation}
\nonumber
\left[\begin{array}{cc}M_{1} \\M_{2}\end{array}\right] =
\left[\begin{array}{cc}\Sigma \\E\end{array}\right]\; ,
\left[\begin{array}{cc}\Sigma \\ \Omega\end{array}\right]\;
, \left[\begin{array}{cc}E \\H\end{array}\right],
\end{equation}
for scalar, vectorial and tensorial perturbations respectively\footnote{The observables for vectorial perturbations give a reduced dynamical system in
the specific case of a Stokesian fluid (for more information on that issue, see Novello (1995b)).}.

Let us make a canonical transformation by changing variables $Q$ and $P$ into $\tilde{Q}$ and $\tilde{P}$:

\begin{equation}
\left[\begin{array}{cc}
\tilde{Q} \\
\tilde{P}
\end{array}\right] = J \hspace{0.1cm}\left[\begin{array}{cc}
Q \\
P
\end{array}\right],
\protect\label{h52}
\end{equation}
We then obtain the new Hamiltonian for the transformed system as a function of the previous one and the transformation matrix $J$, that is,

\begin{equation}
\tilde{{\cal H}} = J \hspace{0.1cm}{\cal H} \hspace{0.1cm}J^{-1} +
\dot{J} \hspace{0.1cm}J^{-1}.
\protect\label{h6}
\end{equation}
The guarantee that the system will keep being described by a Hamiltonian is provided by imposing that $\tilde{{\cal H}}$ be traceless

\begin{equation}
\nonumber
Tr\left(\dot{J} \hspace{0.1cm}J^{-1}\right) = 0,
\end{equation}
that is,

\begin{equation}
det \,J = 1.
\protect\label{h7}
\end{equation}

This is nothing but the well known fact that quadratic Hamiltonians constitute the equivalence class of the harmonic oscillator. The group $SL(2,\rm I\!\rm R)$ describes the canonical transformations on a plane.

\subsubsection{The Scalar Case}\protect\label{scalar}

\underline{{\bf The Auxiliary Hamiltonian:}} In this section we will state the results of the auxiliary Hamiltonian method for the scalar perturbations of FLRW background (cf. Novello, 1995a,b). The resulting dynamical system for scalar perturbations in the general case (with non zero viscosity $\xi$) is given by:

\begin{equation}
\dot{\Sigma} = \xi\, \left[\frac{2m}{(1 + \lambda) \rho a^{2}}\, \left(1 + \frac{3\epsilon}{m}\right)\, \left(\frac{\lambda}{2} + \frac{1}{3}\right) -
\frac{1}{2}\right]\, \Sigma - \left[1 + \frac{2m\lambda}{(1 + \lambda) \rho a^{2}}\, \left(1 + \frac{3\epsilon}{m}\right)\right]\, E,
\protect\label{S1}
\end{equation}
and
\begin{equation}
\dot{E} =\frac{1}{2}\, \left[\frac{2m}{(1 + \lambda)\rho a^{2}}\, \left(1 + \frac{3\epsilon}{m}\right)\, \left(\frac{\lambda}{2} + \frac{1}{3}\right)\, \xi^{2} - \frac{\xi^{2}}{2} - (1 + \lambda)\, \rho - \frac{\theta}{3}\, \xi \right]\, \Sigma  - \left[\frac{m\lambda}{(1 + \lambda)\rho a^{2}}\, \left(1 + \frac{3\epsilon}{m}\right)\, \xi + \frac{\theta}{3} + \frac{\xi}{2}\right]\, E.
\protect\label{S2}
\end{equation}

The components of the Hamiltonian matrix (\ref{Hmatrix}), $h_{k}$, are then written as

\begin{eqnarray}
h_{1} & = & \frac{1}{\Delta}\, \left\{- \dot{\gamma}\, \delta + \gamma\, \dot{\delta} + \delta^{2}\, \frac{(1 + \lambda)}{2}\, \rho - \gamma^{2}\, (1 + \lambda \, L) - \gamma\, \delta\, \frac{\theta}{3}\right. + \nonumber \\
& - & \left.\frac{\xi}{2}\, \left[2\, \gamma\, \delta\, L\, \left(\lambda + \frac{1}{3}\right) - \delta^{2}\, \left(\frac{\xi}{2}\, (1 - \lambda\, L) + \frac{(\theta - \xi\, L)}{3}\right)\right]\right\}, \nonumber \\
h_{2} & = & \frac{1}{\Delta}\, \left\{\alpha\, \dot{\beta} - \dot{\alpha}\, \beta - \alpha^{2}\, (1 + \lambda\, L) -\alpha\, \beta\, \frac{\theta}{3} + \beta^{2}\, \frac{(1 + \lambda)}{2}\, \rho\right. +\nonumber \\
& - & \left. \frac{\xi}{2}\, \left[2\, \alpha\, \beta\, L\, \left(\lambda + \frac{1}{3}\right) - \beta^{2}\, \left(\frac{\xi}{2}\, (1 - \lambda\, L) + \frac{(\theta - \xi\, L)}{3}\right) \right]\right\}  \\
h_{3} & = & \frac{1}{2\Delta}\, \left\{- \alpha\, \dot{\delta} + \beta\, \dot{\gamma} - \beta\, \delta\, \frac{(1 + \lambda)}{2}\, \rho + \alpha\,
\gamma\, (1 + \lambda\, L) + \alpha\, \delta\, \frac{\theta}{3}\right. +\nonumber \\
& - & \frac{\xi}{2}\, \left[\alpha\, \delta\, (1 + \lambda\, L) + \beta\, \gamma\, \left(\frac{2L}{3} - (1 - \lambda\, L)\right)\right. +\nonumber \\
& - & \left.\left. \beta\, \delta\, \left( \frac{\xi}{2}\, (1 - \lambda\, L) + \frac{(\theta - \xi\, L)}{3}\right) \right]\right\} \nonumber .
\protect\label{6}
\end{eqnarray}
The auxiliary quantity $L$, which appears in the above equation, is defined as follows:

\begin{equation}
\nonumber
L \equiv \frac{2m}{(1 + \lambda)\rho a^{2}}\, \left(1 +
\frac{3\epsilon}{m}\right).
\end{equation}
From Eq.\ (\ref{6}), the condition for the existence of the Hamiltonian, Eq.\ (\ref{4}), now reads:

\begin{equation}
\frac{\dot{\Delta}}{\Delta} - \frac{\theta}{3} - \xi\,
\left(1 - \frac{L}{3}\right) = 0.
\protect\label{8}
\end{equation}

We will tackle the simpler case of zero viscosity for scalar perturbations in the FLRW background, which then yields:

\begin{equation}
\Delta (t) = \kappa\, a(t),
\protect\label{detS}
\end{equation}
where $\kappa$ is an integration constant.

\underline{{\bf Canonical Quantization:}} The problem to be analyzed at this point is nothing but a single harmonic oscillator problem with a time-dependent quadratic interaction. This problem appears in many different places, e.g. the equation for quantum test fields in homogeneous and isotropic expanding/contracting universes, Quantum Optics, etc. There are many ways to face this problem; here we will follow the standard procedure of Quantum Optics. The creation and annihilation operators $a$ and $a^{\dagger}$ are defined in the standard way as

\begin{equation}
\begin{array}{ll}
\hat a = \fracc{1}{\sqrt{2}} \hspace{0.1cm}(Q + iP), \\ \\
\hat a^{\dagger} = \fracc{1}{\sqrt{2}} \hspace{0.1cm}(Q - iP), \\ \\
\left[\hat a,\hat a^{\dagger}\right] = 1.
\end{array}
\protect\label{c1}
\end{equation}

Using Eqs.\ (\ref{H}), (\ref{6}) and (\ref{c1}), the Hamiltonian then becomes:

\begin{equation}
{\cal H} = {\cal H}_{0} + {\cal H}_{int},
\protect\label{c2}
\end{equation}
with

\begin{equation}
\begin{array}{l}
{\cal H}_{0} \equiv \omega (t) \hspace{0.1cm}(1 + 2N), \\ \\
\omega (t) \equiv \frac{1}{4}\, (h_{1} + h_{2}),
\end{array}
\protect\label{c3}
\end{equation}
where $N$ is the number of particles (photons): $N \equiv \hat a^{\dagger}\,\hat a.$ The self-interaction Hamiltonian is given by

\begin{equation}
{\cal H}_{int} \equiv \eta (t) \hspace{0.1cm}\hat a^{2} + \eta^{\star}(t) \hspace{0.1cm} (\hat a^{\dagger})^{2},
\protect\label{c5}
\end{equation}
where

\begin{equation}
\eta (t)\equiv \frac{1}{4}\, (h_{1} - h_{2}) - i\hspace{0.1cm}h_{3},
\protect\label{cc5}
\end{equation}
and $\eta^{\star}$ is the complex conjugate of $\eta$.

Schr\"{o}dinger equation is then easily written for the operator ${\cal H}$ as

\begin{equation}
\nonumber
i\, \frac{\partial{\psi(\vec{x},t)}}{\partial t} = {\cal H}\,\psi(\vec{x},t),
\end{equation}
with the wave function $\psi(\vec{x},t)$ given as

\begin{equation}
\nonumber
\psi(\vec{x},t) = U(t,t_{0})\,\psi(\vec{x},t_{0}),
\end{equation}
and $U(t,t_{0})$ is the evolution operator.

Now we will proceed to solve the equation above by employing the formalism of Quantum Optics. It involves writing the time evolution operator as a product of the rotation and the single-mode squeeze operators, along with a phase factor:

\begin{equation}
\nonumber
U(t,t_{0}) = e^{i\phi}\,S_{(r,\varphi)}\,R_{(\Gamma)},
\end{equation}
where $\phi$ is a time-dependent phase and the rotation operator $R_{(\Gamma)}$ and the single-mode squeeze operator $S_{(r,\varphi)}$ are defined respectively as:

\begin{equation}
\nonumber
\begin{array}{ll}
R_{(\Gamma)} \equiv e^{- i\Gamma\,\hat a\,\hat a^{\dagger}}, \\ \\
S_{(r,\varphi)} \equiv \exp\left\{\fracc{r}{2}\,\left[ e^{- 2i\varphi}\,\hat a^{2} - e^{2i\varphi}\,(\hat a^{\dagger})^{2} \right]\right\},
\end{array}
\end{equation}
where $\Gamma$, $r$ and $\varphi$ depend of time through the known quantities in the FLRW background and are defined as the rotation angle, the squeeze factor and the squeeze angle, respectively. It should be remarked that all these quantities are real. For further details and explanations the reader should see Novello et al. (1996).

Schr\"{o}dinger equation thus reduces, after a direct but somewhat long calculation, to the following first order coupled differential system:

\begin{equation}
\begin{array}{ll}
\phi(t) = \fracc{1}{2}\,\theta(t), \\[2ex]
\dot{\Gamma} = \fracc{2\omega}{\cosh(2r)}, \\[2ex]
\dot{r} = \fracc{1}{2}\, (h_{1} - h_{2})\, \sin{(2\varphi)} - 2\, h_{3}\, \cos{(2\varphi)}, \\[2ex]
r\, \dot{\varphi} = \fracc{1}{4}\, (h_{1} - h_{2})\, \cos{(2\varphi)} + h_{3}\, \sin{(2\varphi)} - \omega (t)\,\tanh(2r),
\end{array}
\protect\label{c8}
\end{equation}
where $\omega (t)$ is defined by Eq. (\ref{c3}).

We proceed now to solve the above system, Eq. (\ref{c8}). The last two equations are coupled, which makes their integration very much involved. However, this task can be easily accomplished if we perform a transformation in the Hamiltonian matrix ${\cal H}$, by means of a canonical transformation matrix ${\cal J}$, given by Eqs. (\ref{h6}) and (\ref{h7}), such that the transformed Hamiltonian has now:

\begin{equation}
\tilde{\omega} \equiv \frac{1}{4}\, (\tilde{h}\mbox{}_{1} + \tilde{h}\mbox{}_{2}) = 0,
\protect\label{c09}
\end{equation}
and we will drop the ($\,\tilde{ }\,$) symbol from now on, to keep our notation simpler. This choice can always be made, since it amounts only to making the right choice of the original functions $\alpha$, $\beta$, $\gamma$ and $\delta$. The first step taken is then to choose

\begin{equation}
{\cal S} \equiv
\begin{pmatrix}
\alpha & \beta \\
\gamma & \delta
\end{pmatrix}=
\begin{pmatrix}
\alpha & 0 \\
0 & \kappa\, a/\alpha
\end{pmatrix},
\protect\label{C1}
\end{equation}
where $\kappa$ is an integration constant that comes from the condition of existence of ${\cal H}$, Eq. (\ref{8}).

The condition (\ref{c09}) may then be written as

\begin{equation}
\nonumber
\frac{(1 + \lambda)}{2}\, \rho\,
\left(\frac{\kappa a}{\alpha}\right)^{2} - (1 + \lambda L)\,
\alpha^{2} = 0,
\end{equation}
which then gives

\begin{equation}
\alpha^{4} (t) =
\frac{\kappa^{2}(1 + \lambda)\rho_{0}}{2a^{(1 + 3\lambda)}}
\, \left\{1 + \frac{2\lambda}{(1 + \lambda)\rho_{0}}\,
(m + 3\epsilon)\, a^{(1 + 3\lambda)}\right\}^{-1},
\protect\label{C2}
\end{equation}
where $\rho_{0}$ is the matter density when $a(t) = 1$. The above quantity is always positive (since $m > -3\epsilon$ in all circumstances). In this case we have the following results:

\begin{equation}
\begin{array}{ll}
h_{1} = \fracc{\kappa^{2}(1 + \lambda)\rho}{2a}\, \alpha^{-2} = - h_{2}, \\ \\
h_{3} = \fracc{1}{2}\, \fracc{\dot{\alpha}}{\alpha}, \\ \\
\eta (t) \equiv \fracc{1}{4}({h_{1}} - {h_{2}}) - ih_{3} = \fracc{\kappa^{2}(1 + \lambda)\rho}{4a\alpha^{2}} - \fracc{i}{2}\, \fracc{\dot{\alpha}}{\alpha}.
\end{array}
\protect\label{C3}
\end{equation}

Proceeding now to integrate the differential equations in $r$ and $\varphi$, we now obtain:

\begin{equation}
\nonumber
r\, e^{-2i\varphi} = - 2i\, \int \eta(t)\, dt.
\end{equation}
To make the integration above simpler we will choose the case of 3-curvature zero and $\lambda = 1/3$---radiation era, the regime at the quantum phase of the Universe (cf. Novello, 1995b) ---, to find

\begin{eqnarray}
\int \eta(t)\, dt & = & - \frac{\kappa}{2}\, \left\{\sqrt{\frac{m}{\rho_{0}a^{2}}  + \frac{2}{a^{4}}} + \frac{\sqrt{2}}{4}\, \frac{m}{\rho_{0}}
\, \ln\left[\sqrt{\frac{2\rho_{0}}{ma^{2}}} + \sqrt{1 + \frac{2\rho_{0}}{ma^{2}}} \right]\right\} + \nonumber \\
& + & \frac{i}{4}\, \left\{\ln\, a(t) + \frac{1}{2}\, \ln\left(a^{2} + \frac{2\rho_{0}}{m}\right)\right\}.
\protect\label{C4}
\end{eqnarray}

With the above result it is easy to decouple the differential equations to find

\begin{eqnarray}
r\, \sin{(2\varphi)} & = & - \sqrt{2}\,\kappa\,\left\{ \frac{1}{a}\,\sqrt{\frac{m}{2\rho_{0}}\,a^{2} + 1} + \frac{m}{4\rho_{0}}\, \ln\left(\sqrt{
\frac{2\rho_{0}}{ma^{2}}} + \sqrt{\frac{2\rho_{0}}{ma^{2}} + 1}\right)\right\} \\ \nonumber \\
r\, \cos{(2\varphi)} & = & \frac{1}{4}\, \ln\left[ a^{4}\,\left(1 + \frac{2\rho_{0}}{m}\right)\right]\nonumber. \protect\label{C5}
\end{eqnarray}

\underline{{\bf Observables:}} From the above construction it follows that the observables of the theory are written in terms of the corresponding creation and annihilation operators in the same modes.

For the shear $\Sigma$ and for the electric part of Weyl tensor $E$ we have

\begin{equation}
\begin{array}{c}
\Sigma = \chi (t)\hspace{0.1cm}\hat a + \chi^{\star}(t)\hspace{0.1cm}\hat a^{\dagger},\\ \\
E = \Psi (t) \hspace{0.1cm}\hat a + \Psi^{\star}(t)\hspace{0.1cm}\hat a^{\dagger},
\end{array}
\protect\label{c13}
\end{equation}
where $\chi(t)$ and $\Psi(t)$ are defined as

\begin{equation}
\begin{array}{c}
\chi (t) \equiv \fracc{1}{\sqrt{2}\Delta} \hspace{0.1cm}( \delta + i\beta ), \\ \\
\Psi (t) \equiv \fracc{1}{\sqrt{2}\Delta} \hspace{0.1cm}(-\gamma + i\alpha),
\end{array}
\protect\label{c14}
\end{equation}
with $\alpha,\,\beta,\,\gamma,\,\delta$ the same quantities defined on Equation (\ref{2}). In terms of the same solution, Eqs. (\ref{C1}), (\ref{C2}), we easily find that

\begin{equation}
\begin{array}{ll}
\chi (t) = \fracc{1}{\sqrt{2}\alpha} = \chi^{\star}(t), \\ \\
\Psi (t) = i\,\fracc{\alpha}{\sqrt{2}\kappa a} = -
\Psi^{\star}(t).
\end{array}
\protect\label{c15}
\end{equation}

The commuter is then easily calculated to give

\begin{equation}
\left[\Sigma,E\right] = - i\, \frac{1}{\kappa a(t)},
\protect\label{c16}
\end{equation}
if the choice $\hbar \equiv c \equiv k = 1$ holds\footnote{In the general case we have:
\begin{equation}
\nonumber
\begin{array}{ll}
Q = \left(\frac{\hbar^{6}k^{7}}{c^{5}}\right)^{1/4}\,
\alpha (t)\, \Sigma, \\ \\
P = \left(\hbar k^{2}c^{7}\right)^{-1/4}\, \frac{\kappa
a(t)}{\alpha (t)}\, E,
\end{array}
\end{equation}
in order that ($Q$,$P$) be adimensional. The commutator is then written as
\begin{equation}
\nonumber
\left[\Sigma,E\right] = i\, \frac{c^{3}}{\left(\hbar k
\right)^{-5/4}}\, \frac{1}{\kappa a(t)}.
\end{equation}}.

The total noise of the observables $\Sigma$ and $E$ can be calculated as\footnote{This quantity is defined as the mean square uncertainty in the annihilation operator $a$. The total noise of a Gaussian Pure State (GPS) is conserved even if the total number of photons is not, and, it therefore
is more useful to describe the quantum wave functions that are obtained from Schr\"{o}dinger equation. See Appendix for more details on this issue.}

\begin{equation}
\begin{array}{ll}
\big <\psi||\Delta\Sigma|^{2}|\psi\big > = \cosh{(2r)}\, \big <0||\Delta\Sigma|^{2}|0\big >, \\ \\
\big <\psi||\Delta E|^{2}|\psi\big > = \cosh{(2r)}\, \big <0||\Delta E|^{2}|0\big >,
\end{array}
\protect\label{c17}
\end{equation}
where $\big <\psi||\Delta X|^{2}|\psi\big >$ is the total noise calculated in the time $t$, $\big <0||\Delta X|^{2}|0\big >$ is the total noise in the vacuum state and $r(t)$ is given by Eq. (\ref{C5}).

\subsubsection{The Vectorial Case}\protect\label{vector}

In this case we find that the resulting dynamical system is not closed: it depends on a choice of the perturbation in the acceleration $\Psi$. Three different reduced dynamics were studied:
\begin{itemize}
 \item{{\it Stokesian Fluid:} $q = 0$; $p = \lambda\rho$;}
 \item{{\it Shear-Free model;}}
 \item{{\it Vorticity-Free model.}}
\end{itemize}

The second and third models give very simple, directly integrable results. The Stokesian Fluid model implies, for the perturbed acceleration $\Psi$, that $\Psi = 2\lambda\, \theta\, \Omega$ and we obtain a closed reduced dynamics for the observables $\Sigma$ and $\Omega$ (respectively the
shear and vorticity). This is the case that will be quantized here.

\underline{{\bf Auxiliary Hamiltonian:}} As for the scalar case, the results of the auxiliary Hamiltonian method will be given here.

The reduced closed dynamical system for the special case of Stokesian fluid with non zero viscosity and zero heat flux $q$ is:

\begin{equation}
\begin{array}{ll}
\dot{\Sigma} = - \left\{\fracc{2}{3}\,\theta + \xi\,
\left[1 + \fracc{(m + 2\epsilon)}{2(1 + \lambda)\rho a^{2}}\right]
\right\}\,\Sigma - (1 - 3\lambda)\,\fracc{\theta}{3}\,
\Omega, \\[2ex]
\dot{\Omega} = \xi\,\fracc{(m + 2\epsilon)}{2(1 + \lambda)\rho
a^{2}}\,\Sigma - (1 + 3\lambda)\,\fracc{\theta}{3}\,\Omega,
\end{array}
\protect\label{V1}
\end{equation}
where we must have $q = 0$, $p = \lambda\rho$ and $\Psi = 2\lambda\,\theta\,\Omega$. The condition for the existence of a Hamiltonian is then written as

\begin{equation}
\Delta (t) = \kappa\, [a(t)]^{3(1+\lambda)}\, e^{\xi\,M(t)},
\protect\label{V2}
\end{equation}
where $\kappa$ is again an integration constant and $M(t)$ is an auxiliary quantity, defined by

\begin{equation}
\nonumber
M(t) = \int_{t}\,\left[1 + \frac{(m + 2\epsilon)}
{2(1 + \lambda)\rho a^{2}(t)}\right]\; dt.
\end{equation}

We can now proceed to the same quantization formalism described by Eqs.\ (\ref{c1})-(\ref{c8}). Making the choice

\begin{equation}
\begin{array}{ll}
\beta \equiv \gamma = 0, \\ \\
\delta = \fracc{\Delta (t)}{\alpha}, \\ \\
\alpha^{4} = - \fracc{3\xi}{2}\, \fracc{(m + 2\epsilon)}{(1 + \lambda )(1 - 3\lambda)}\, \fracc{\kappa\,a^{(7 + 9\lambda )}\,e^{2\xi M(t)}}{\rho_{0}\theta},
\end{array}
\protect\label{V3}
\end{equation}
we obtain the Hamiltonian coefficients $h_{j}$, ($j = 1,2,3$) as:

\begin{equation}
\begin{array}{ll}
h_{1}(t) = - h_{2}(t) = - \sqrt{- \fracc{\xi}{6}\,
\fracc{(m + 2\epsilon)\theta}{\rho_{0}}\,a^{(1 + 3\lambda)}\,
\left(\fracc{1 - 3\lambda}{1 + \lambda}\right)}, \\ \\
h_{3}(t) = \fracc{1}{2}\,\left(\fracc{\dot{\alpha}}{\alpha}
- \fracc{2}{3}\,\theta\right).
\end{array}
\protect\label{V4}
\end{equation}
It should be noted that, since the viscosity $\xi$ is negative due to thermodynamic considerations, and the fact that $m > -2\epsilon$ in all cases (see Novello, 1995b), the function $\alpha (t)$ is real.

We are then able to decouple the differential system that results from Schr\"{o}dinger equation, Eq.(\ref{c8}), to find:

\begin{equation}
\begin{array}{ll}
r\,\cos{(2\varphi)} = \ln\left(b\,\fracc{a^{2}}{\alpha}
\right), \\ \\
r\,\sin{(2\varphi)} = \fracc{3}{\rho_{0}}\,\sqrt{\xi\,
\fracc{(3\lambda - 1)(m + 2\epsilon)}{6(1 + \lambda )}}\,
a^{(1 + 3\lambda )},
\end{array}
\protect\label{V5}
\end{equation}
with $\alpha$ given in Eq.(\ref{V3}) and $b$ being an integration constant.

\underline{{\bf Observables:}} The same method applied for the scalar observables can be employed here in the vectorial case, giving (for the choice $\hbar \equiv k \equiv c = 1$)\footnote{For the general case, the perturbations should be written as
\begin{equation}
\nonumber
\begin{array}{ll}
\Sigma = \left(\fracc{c^{9}}{\hbar^{5}\,k^{6}}\right)^{1/4}\,
\fracc{1}{\sqrt{2}\alpha}\,(\hat a + \hat a^{\dagger}), \\ \\
\Omega = - i\,\fracc{1}{k^{2}}\left(\fracc{c^{9}}{\hbar^{5}}
\right)^{1/4}\,\fracc{\alpha\,e^{\xi M(t)}}{\sqrt{2}\kappa
a^{3(1 + \lambda )}}\,(\hat a - \hat a^{\dagger}),
\end{array}
\end{equation}
and the commutator $[\Sigma , \Omega]$ should then be accordingly rewritten.}:

\begin{equation}
\begin{array}{ll}
\Sigma = \fracc{1}{\alpha}\,Q = \fracc{1}{\sqrt{2}\,\alpha}\,
(\hat a + \hat a^{\dagger}), \\ \\
\Omega = \fracc{\alpha\,e^{-\xi M(t)}}{\sqrt{2}\kappa
a^{3(1 + \lambda )}}\,P = - i \fracc{\alpha\,e^{-\xi M(t)}}
{\sqrt{2}\kappa a^{3(1 + \lambda )}}\,(\hat a - \hat a^{\dagger}).
\end{array}
\protect\label{V6}
\end{equation}
and, in this case, the commutator between the perturbed variables gives the following result:

\begin{equation}
\nonumber
[\Sigma,\Omega] = i\,\frac{e^{-\xi M(t)}}{\kappa
a^{3(1 + \lambda)}(t)}.
\end{equation}

Finally, the relation between the total noises of $\Sigma$ and $\Omega$ in a time $t$ and their respective values of total noises in the vacuum state is given by

\begin{equation}
\nonumber
\begin{array}{ll}
\big <\Psi ||\Delta\Sigma |^{2}|\Psi\big > = \cosh{(2r)}\,\big <0||
\Delta\Sigma |^{2}|0\big >, \\ \\
\big <\Psi ||\Delta\Omega |^{2}|\Psi\big > = \cosh{(2r)}\,\big <0||
\Delta\Omega |^{2}|0\big >,
\end{array}
\end{equation}
where $r$ is given by Eq.(\ref{V5}).

\underline{{\bf The Case $\xi = 0$:}} If we consider the special case of vectorial perturbations in a Stokesian fluid with zero viscosity ($\xi = 0$), we can follow the same steps detailed in previous subsections to find the Hamiltonian coefficients as

\begin{equation}
\begin{array}{ll}
h_{1} = - h_{2} = 0, \\ \\
h_{3} = \fracc{1}{2}\,\left(\fracc{\dot{\alpha}}{\alpha} -
 \fracc{2}{3}\theta\right),
\end{array}
\protect\label{V8}
\end{equation}
where we have the choice

\begin{equation}
\nonumber
\begin{array}{ll}
\alpha \equiv \mbox{arbitrary function of time}, \\ \\
\beta \equiv \gamma = 0, \\ \\
\delta = \fracc{\kappa}{\alpha}\,a(t)^{4},
\end{array}
\end{equation}
and the following condition must hold:

\begin{equation}
\lambda = \frac{1}{3},
\protect\label{V9}
\end{equation}
so that $\omega (t)$ is zero and the system which arises from Schr\"{o}dinger equation, Eq.(\ref{c8}), is easily decoupled to give:

\begin{equation}
\begin{array}{ll}
r = \ln\left(b\,\fracc{a^{2}(t)}{\alpha (t)}\right), \\ \\
\varphi = 0, \qquad\qquad \mbox{or} \qquad\qquad \varphi = - \pi,
\end{array}
\protect\label{V10}
\end{equation}
with $b \equiv const.$ again. Let us note that the condition on $\lambda$, Eq.(\ref{V9}), ensures that our model of vectorial perturbations in a Stokesian fluid with zero viscosity applies only to the radiation era, as expected.

The observables $\Sigma$ and $\Omega$ are then written as

\begin{equation}
\begin{array}{ll}
\Sigma = \fracc{1}{\alpha (t)}\,Q = \fracc{1}{\sqrt{2}
\alpha (t)}\,(\hat a + \hat a^{\dagger}), \\ \\
\Omega = \fracc{\alpha (t)}{\kappa a^{4}}\,P = - i\,
\fracc{\alpha (t)}{\sqrt{2}\kappa a^{4}}\,(\hat a - \hat a^{\dagger}),
\end{array}
\protect\label{V11}
\end{equation}
if $\hbar \equiv k \equiv c = 1$\footnote{In the general case we have
\begin{equation}
\nonumber
\begin{array}{ll}
\Sigma = \left(\fracc{c^{9}}{\hbar^{5}k^{6}}\right)^{1/4}\,
\fracc{1}{\sqrt{2}\alpha (t)}\,(\hat a +\hat a^{\dagger}), \\ \\
\Omega = - i\,k(\hbar\,c^{3})^{-1/4}\,\fracc{\alpha (t)}
{\sqrt{2}\kappa a^{4}}\,(\hat a -\hat a^{\dagger}),
\end{array}
\end{equation}
with the commutator $[\Sigma, \Omega]$ modified accordingly.}.
The commutator between $\Sigma$ and $\Omega$ will then be

\begin{equation}
[\Sigma,\Omega] = - i\,\fracc{1}{\kappa a^{4}},
\protect\label{V13}
\end{equation}
and the total noises of the observables in terms of their respective values for the vacuum state will be written as

\begin{equation}
\begin{array}{ll}
\big <\Psi ||\Delta\Sigma |^{2}|\Psi\big > =
\cosh{\left(2\,\ln\left(b\,\fracc{a^{2}(t)}{\alpha (t)}\right)
\right)}\,\big <0||\Delta\Sigma |^{2}|0\big >, \\ \\
\big <\Psi ||\Delta\Omega |^{2}|\Psi\big > =
\cosh{\left(2\,\ln\left(b\,\fracc{a^{2}(t)}{\alpha (t)}\right)
\right)}\,\big <0||\Delta\Omega |^{2}|0\big >.
\end{array}
\protect\label{V14}
\end{equation}

\subsubsection{The Tensorial Case}\protect\label{tensor}

In this case we obtain a new dynamical closed system on the observables $E$ and $H$ (respectively the electric and magnetic parts of Weyl tensor), which follows as:

\begin{equation}
\begin{array}{ll}
\dot{E} = - \left(\theta + \fracc{\xi}{2}\right)\,E -
\left\{\fracc{1}{2}\,\left[\xi\left(\fracc{\theta}{3} +
\fracc{\xi}{2}\right) + (\rho + p)\right] - \fracc{1}{a^{2}}\,
(m - 3\epsilon)\right\}\,H, \\ \\
\dot{H} = - E - \left(\fracc{2}{3}\theta + \fracc{\xi}{2}
\right)\,H.
\end{array}
\protect\label{SA1}
\end{equation}
The transformation for the variables ($Q$,$P$) is made in the same way as in the scalar and vectorial cases, Eqs.(\ref{2})-(\ref{Hmatrix}). The condition for the existence of a Hamiltonian is then written as

\begin{equation}
\Delta (t) = \kappa\,a^{5}(t)\,e^{\xi\,t},
\protect\label{SA2}
\end{equation}
where $\kappa$ is again an arbitrary integration constant and $H$ is again given in the form of Eq.(\ref{H}), with the following coefficients:

\begin{equation}
\protect\label{S3}
\begin{array}{ll}
h_{1} =& \fracc{1}{\Delta}\,\left\{- \dot{\gamma}\, \delta + \delta^{2} + \fracc{\theta}{3}\, \gamma\,\delta + \gamma\,\dot{\delta} - \gamma^{2}\,\left[\fracc{\xi}{2}\,\left( \fracc{\theta}{3} + \fracc{\xi}{2}\right) + \fracc{1}{2}\,(\rho + p) - \fracc{(m - 3\epsilon)}{a^{2}}\right]\right\},\\[2ex]
h_{2} =& \fracc{1}{\Delta}\,\left\{- \dot{\alpha}\,\beta + \beta^{2} + \fracc{\theta}{3}\,\alpha\,\beta + \alpha\, \dot{\beta}- \alpha^{2}\,\left[\fracc{\xi}{2}\,\left( \fracc{\theta}{3} + \fracc{\xi}{2}\right) + \fracc{1}{2}\, (\rho + p) - \fracc{(m - 3\epsilon)}{a^{2}}\right]\right\},\\[2ex]
h_{3} =& \fracc{1}{2\Delta}\, \left\{- \dot{\alpha}\, \beta - \dot{\beta}\,\gamma - \beta\,\delta - \alpha\, \delta\,\left(\theta + \fracc{\xi}{2}\right) + \beta\,\gamma\,\left(\fracc{2}{3}\,\theta + \fracc{\xi}{2}\right) +\right.\\[2ex]
&\left.+\alpha\,\gamma\,\left[\fracc{\xi}{2} \,\left(\fracc{\theta}{3} + \fracc{\xi}{2}\right) +\fracc{1}{2}\,(\rho + p) - \fracc{(m - 3\epsilon)}{a^{2}}\right]\right\}.
\end{array}
\end{equation}

We are then able to perform the quantization by employing the standard method described by Eqs.\ (\ref{c1})--(\ref{c8}) and by making the same choice $\omega (t) = 0$ in order to decouple the first order differential system, Eq.\ (\ref{c8}). We then find that

\begin{equation}
\begin{array}{ll}
\alpha^{4}(t) = \kappa^{2}\,a^{10}\,e^{2\xi t}\,\left[
\fracc{\xi}{2}\,\left(\fracc{\theta}{3} + \fracc{\xi}{2}
\right) + \fracc{1}{2}\,(\rho + p) - \fracc{(m - 3\epsilon)}{a^{2}}
\right]^{-1}, \\ \\
\beta (t) \equiv \gamma (t) = 0, \\ \\
\delta (t) = \fracc{\kappa}{\alpha (t)}\,a(t)^{5}\,e^{\xi t}.
\end{array}
\protect\label{S4}
\end{equation}
Therefore, the coefficients of the Hamiltonian for the choice given by Eq.\ (\ref{S4}) are:

\begin{equation}
\begin{array}{ll}
h_{1}(t) = \fracc{\kappa}{\alpha^{2}(t)}\,a(t)^{5}\, e^{\xi t} = - h_{2}(t), \\ \\
h_{3}(t) = \fracc{1}{2}\,\left(\fracc{\dot{\alpha}}{\alpha} -\theta - \fracc{\xi}{2}\right).
\end{array}
\protect\label{S5}
\end{equation}

The decoupled first order differential system then gives the following results:

\begin{equation}
\begin{array}{ll}
r\,\cos{(2\varphi )} = \ln\left(b\,\fracc{a^{3}}{\alpha}\, e^{\frac{\xi}{2}t}\right), \\ \\
r\,\sin{(2\varphi )} = - \int_{t}\,\left[\fracc{\xi}{2}\, \left(\fracc{\theta}{3} + \fracc{\xi}{2}\right) + \fracc{1}{2}\,(\rho + p) - \fracc{(m - 3\epsilon)}{a^{2}}\right]^{1/2}\,
dt^{\prime},
\end{array}
\protect\label{S6}
\end{equation}
where $b$ is an integration constant. We then have\footnote{In the general case, Eq.\ (\ref{S7}) can be written as
\begin{equation}
\nonumber
\begin{array}{ll}
E = \fracc{1}{k^{2}}\left(\fracc{c^{15}}{\hbar^{7}}\right)^{1/4}, \\ \\
H = - i\,\hbar k^{3/4}\,\fracc{\alpha\,e^{-\xi t}}{\sqrt{2}\kappa a^{5}}\,(\hat a - \hat a^{\dagger}),
\end{array}
\end{equation}
and the commutator $[E,H]$ is suitably altered.}:

\begin{equation}
\begin{array}{ll}
E = \fracc{1}{\alpha}\,Q = \fracc{1}{\sqrt{2}\alpha}\,(\hat a +
\hat a^{\dagger}), \\ \\
H = \fracc{\alpha}{\kappa}\,a(t)^{- 5}\,e^{- \xi t}\,P =- i\,\fracc{\alpha}{\sqrt{2}}\,a(t)^{- 5}\,e^{- \xi t}\,(\hat a -
\hat a^{\dagger}).
\end{array}
\protect\label{S7}
\end{equation}
and therefore the commutator between the above observables is given by

\begin{equation}
[E,H] = i\,\frac{e^{-\xi\,t}}{\kappa}\,a(t)^{- 5}.
\protect\label{S9}
\end{equation}
Finally, the total noises for $E$ and $H$ in terms of their respective total noise values for the vacuum state are written in the same way as before:

\begin{equation}
\begin{array}{ll}
\big <\Psi ||\Delta E|^{2}|\Psi\big > = \cosh{(2r)}\,\big <0||\Delta E|^
{2}|0\big >, \\ \\
\big <\Psi ||\Delta H|^{2}|\Psi \big > = \cosh{(2r)}\,\big <0||\Delta H|^
{2}|0\big >,
\end{array}
\protect\label{S10}
\end{equation}
where the function $r$ is given by Eq.(\ref{S6}). From the above results it follows that the total noise in a time $t$ is always greater than its corresponding vacuum value, increasing with $r$.

\subsection{Milne Background}

There is a particular class of FLRW geometries dealt with in this present section that seems worth to be explicitly examined. This is the case analyzed by Milne and contains a portion of Minkowski geometry. The metric is then FLRW-type, where the radius of the universe, the 3-curvature and the expansion are given respectively by:

\begin{equation}
\begin{array}{ll}
a(t) = t, \\ \\
\epsilon = +1, \\ \\
\theta = \fracc{3}{t}.
\end{array}
\protect\label{apdx1}
\end{equation}
We will present only the results which follow:

\subsubsection{Scalar Perturbations}

If we consider the case of scalar perturbations, the vorticity should vanish, which implies that the magnetic part of Weyl conformal tensor will also be zero; thus we have:

\begin{equation}
\begin{array}{ll}
\delta\omega_{ij} = 0, \\ \\
\delta H_{ij} = 0.
\end{array}
\protect\label{apdx2}
\end{equation}

Following the notations used before, the other perturbed quantities are listed below:

\underline{{\bf Geometric Quantity:}}

\begin{equation}
\nonumber
\delta E_{ij} = E(t) \hspace{0.1cm}\hat{Q}_{ij}(\vec{x}).
\end{equation}

\underline{{\bf Kinematic Quantities:}}

\begin{equation}
\nonumber
\begin{array}{ll}
\delta V_{0} = -\delta V^{0} = \fracc{1}{2} \delta g_{00} = \fracc{1}{2} \beta (t) Q(\vec{x}) + \fracc{1}{2} Y(t), \\ \\
\delta V_{k} = V(t) \hspace{0.1cm}Q_{k}(\vec{x}), \\ \\
\delta a_{k} = \Psi (t) \hspace{0.1cm}Q_{k}(\vec{x}), \\ \\
\delta\sigma_{ij} = \Sigma (t) \hspace{0.1cm}\hat{Q}_{ij}
(\vec{x}), \\ \\
\delta\theta = B(t)\hspace{0.1cm}Q(\vec{x}) + Z(t).
\end{array}
\end{equation}

\underline{{\bf Matter Quantities:}}
\begin{equation}
\nonumber
\begin{array}{ll}
\delta\rho = N(t)\hspace{0.1cm}Q(\vec{x}) + L(t), \\ \\
\delta\Pi_{ij} = \xi\hspace{0.1cm}\delta\sigma_{ij} =
\xi\hspace{0.1cm}\Sigma (t)\hspace{0.1cm}\hat{Q}_{ij}
(\vec{x}), \\ \\
\delta p = \lambda \hspace{0.1cm}\delta\rho, \\ \\
\delta q_{k} = q(t)\hspace{0.1cm}Q_{k}(\vec{x}),
\end{array}
\end{equation}
where we have used again the proportionality relation between the perturbed anisotropic pressure and the shear; we also consider the standard formulation in which the perturbed pressure is proportional to the density. The quantity $\beta (t)$ is gauge-dependent and $Y(t)$, $Z(t)$ and $L(t)$ are homogeneous terms.

Making use of the quasi-Maxwellian equations we obtain the system for the above quantities as:

\begin{equation}
\dot{E} = -\frac{\xi}{2} + \frac{\theta}{3} E + \frac{\xi\theta}{6}\Sigma + \frac{m}{2} q = 0,
\protect\label{apdx3}
\end{equation}

\begin{equation}
\frac{2\theta^{2}}{3} \left(\frac{1}{3} + \frac{\epsilon}{m}\right) \left[E - \frac{\xi}{2} \Sigma\right] + N + \theta q = 0,
\protect\label{apdx4}
\end{equation}

\begin{equation}
\dot{B} + \frac{2\theta}{3} B + \frac{\theta ^{2}}{6} \beta (t) +\frac{\theta ^{2}}{9} m \Psi + \frac{(1 + 3\lambda )}{2} N = 0,
\protect\label{apdx5}
\end{equation}

\begin{equation}
\dot{\Sigma} + E + \frac{\xi}{2} \Sigma - m\hspace{0.1cm}\Psi = 0,
\protect\label{apdx6}
\end{equation}

\begin{equation}
V = \left(\frac{1}{3} + \frac{\epsilon}{m}\right) \Sigma - \frac{3}{\theta ^{2}} B - \frac{9}{2\theta ^{2}} q,
\protect\label{apdx7}
\end{equation}

\begin{equation}
\dot{N} + (1 + \lambda )\theta N - \frac{\theta ^{2}}{9} q = 0,
\protect\label{apdx8}
\end{equation}

\begin{equation}
\dot{q} + \theta q - \lambda N - \frac{2\xi\theta ^{2}}{9} \left(\frac{1}{3} + \frac{\epsilon}{m}\right)\Sigma = 0.
\protect\label{apdx9}
\end{equation}
The dynamical equations on the homogeneous terms $Z(t)$ and $L(t)$ are written as:

\begin{equation}
\dot{Z} + \frac{2\theta}{3} Z + \frac{(1 + 3\lambda )}{2} L + \frac{\theta ^{2}}{6} Y = 0,
\protect\label{apdx10}
\end{equation}

\begin{equation}
\dot{L} + (1 + \lambda ) \theta L = 0.
\protect\label{apdx11}
\end{equation}

Let us solve this system for the special simple case where $q = 0$. We have then (from Eq.\ (\ref{apdx9}), the dynamical equation for $q$),

\begin{equation}
-\lambda N - \frac{2\xi\theta ^{2}}{9} \left(\frac{1}{3} + \frac{\epsilon}{m}\right) \Sigma = 0.
\protect\label{apdx12}
\end{equation}

Equations (\ref{apdx1}) and (\ref{apdx8}) give

\begin{equation}
N(t) = N_{0} t^{-3(1 + \lambda )},
\protect\label{apdx13}
\end{equation}
where $N_{0}$ is a constant. From (\ref{apdx12}) and (\ref{apdx13}) we obtain

\begin{equation}
\Sigma (t) = - \frac{\lambda N_{0}}{2\xi}\left(\frac{1}{3} + \frac{\epsilon}{m} \right)^{-1}t^{-(1 + 3\lambda )}.
\protect\label{apdx14}
\end{equation}
These results applied in equation (\ref{apdx4}) give

\begin{equation}
E(t) = - \frac{N_{0}}{6} \left(1 + \frac{3\lambda}{2}\right) \left(\frac{1}{3} + \frac{\epsilon}{m}\right)^{-1}t^{-(1 + 3\lambda )}.
\protect\label{apdx15}
\end{equation}
Equation (\ref{apdx3}) is automatically valid if we make use of the above results for $N(t)$, $\Sigma (t)$ and $E(t)$. Equation (\ref{apdx6}) then gives $\Psi (t)$ as:

\begin{equation}
\Psi (t) = \frac{N_{0}}{2m} \left(\frac{1}{3} + \frac{\epsilon}{m}\right)^{-1} \left[\frac{\lambda (1 + 3\lambda )}{\xi} t^{-1} - \frac{(2 + 9\lambda )}{6}\right] t^{- (1 + 3\lambda )}.
\protect\label{apdx16}
\end{equation}
It must be noted that the constant $N_{0}$ cannot be zero, since this would give a trivial result. Equations (\ref{apdx5}) and (\ref{apdx7}) give the
quantities $B(t)$ and $V(t)$ in terms of $N(t)$, $\Psi (t)$ and $\Sigma (t)$, $B(t)$ respectively. Both quantities may be obtained if the gauge-dependent function $\beta (t)$ is chosen. They are therefore \lq \lq bad\rq \rq quantities to analyze. The minimal
closed set of quantities for perturbations in Milne universe is

\begin{equation}
\nonumber
{\cal M}^{scalar}_{[A]} = \{E, \Sigma , N, \Psi \}.
\end{equation}

The homogeneous part of $(\delta\rho )$, $L(t)$, is directly determined by equation (\ref{apdx10}):

\begin{equation}
L(t) = L_{0}\hspace{0.1cm}t^{- 3(1 + \lambda )},
\protect\label{apdx17}
\end{equation}
where again $L_{0}$ denotes a constant. The function $Z(t)$, whose dynamics is given by equation (\ref{apdx10}), can only be integrated by choosing another homogeneous term ($Y(t)$). That completes the solution for the case $q = 0$.

We can analyze the behavior of the solution above for different values of $\lambda $. The results are as follows:

\begin{enumerate}
 \item{$\lambda > - \frac{1}{3}$: \vspace{.3cm}\\
$E$, $\Sigma $, $N$ and $\Psi $ go to zero when $t \longrightarrow \infty $};
 \item{$\lambda = - \frac{1}{3}$: \vspace{.3cm}\\
$E$, $\Sigma $ and $\Psi $ are constant; $N$ goes to zero when $t \longrightarrow \infty $};
 \item{$- 1 < \lambda < - \frac{1}{3}$: \vspace{.3cm}\\
$E$, $\Sigma $ and $\Psi $ diverge when $t \longrightarrow \infty $ and $N$ goes to zero};
 \item{$\lambda = - 1$ (vacuum $\Lambda $): \vspace{.3cm}\\
$E$, $\Sigma $ and $\Psi $ diverge when $t \longrightarrow \infty $ and $N$ is constant};
 \item{$\lambda < - 1$ (unphysical situation): \vspace{.3cm}\\
$E$, $\Sigma $, $N$ and $\Psi $ diverge when $t \longrightarrow \infty $.}
\end{enumerate}

\subsubsection{Vector Perturbations}

In this case the original dynamical system, Equations\ (\ref{vet8})--(\ref{vet9}) give

\begin{equation}
\begin{array}{l}
\dot{E} - \fracc{\xi}{2}\, \dot{\Sigma} + \fracc{2}{3}\, \theta\, E + \fracc{1}{2a^{2}}\, (m - 2\epsilon)\, H + \fracc{1}{4}\, q = 0, \\[1.5ex]
\dot{\Sigma} + \left(\fracc{\theta}{3} + \fracc{\xi}{2}\right)\, \Sigma + E - \fracc{1}{2}\, \Psi = 0, \\[1.5ex]
\dot{\Omega} + \fracc{\theta}{3}\, \Omega + \fracc{1}{2}\, \Psi = 0, \\[1.5ex]
\dot{H} + \fracc{\theta}{3}\, H - \fracc{1}{2}\, E - \fracc{\xi}{4}\, \Sigma = 0, \\[1.5ex]
\dot{q} + \fracc{4}{3}\, \theta\, q + \fracc{1}{a^{2}}\, (m + 2\epsilon)\, \xi\, \Sigma = 0,
\end{array}
\label{apdx17.5}
\end{equation}
and

\begin{equation}
\begin{array}{l}
\Sigma + \Omega + 2\, H = 0, \\[1.5ex]
E - \fracc{\xi}{2}\, \Sigma + \fracc{2}{3}\, \theta\, H = 0, \\[1.5ex]
\fracc{1}{a^{2}}\, (m + 2\epsilon)\, H + \fracc{1}{2}\, q = 0.
\end{array}
\label{apdx18}
\end{equation}
We will present here only the three cases dealt with in Section\ (\ref{Particular}): isotropic, irrotational and Stokesian fluid. The results are as follows:

\underline{{\bf Isotropic Model:}} For $\Sigma = 0$, we obtain

\begin{equation}
\begin{array}{l}
E(t) = \mu\, t^{-2} \\[1.5ex]
H(t) = \fracc{\mu}{2}\, t^{-1} \\[1.5ex]
\Psi(t) = 2\, \mu\, t^{-2} \\[1.5ex]
q(t) = - (m + 2)\, \mu\, t^{-3},
\end{array}
\label{apdx19}
\end{equation}
where $\mu$ is an integration constant and we used \mbox{$\epsilon = + 1$}. These functions of $t$ diverge when $t$ goes to zero and become null for infinite values of $t$.

\underline{{\bf Irrotational Model:}} In the case $\Omega = 0$, the acceleration $\Psi$ is also zero and

\begin{equation}
\begin{array}{l}
\Sigma(t) = \nu\, t^{-2}\, \exp^{-\xi\, t}, \\[1.5ex]
E(t) = \nu\, \exp^{-\xi\, t}\, t^{-2}\, \left(\fracc{1}{t} + \fracc{\xi}{2}\right), \\[1.5ex]
H(t) = - \fracc{\nu}{2}\, t^{-2}\, \exp^{-\xi\, t}, \\[1.5ex]
q(t) = \nu\, (m + 2)\, t^{-4}\, \exp^{-\xi\, t}.
\end{array}
\label{apdx20}
\end{equation}
These functions also diverge when $t$ goes to zero and become zero when $t$ goes to infinity.

\underline{{\bf Stokesian fluid:}} If we consider $q = 0$, the only possible solution is trivially zero. We conclude therefore that vector perturbations in Milne universes must have a non zero heat flux.

\subsubsection{Tensor Perturbations}

The original Equations\ (\ref{tensor1}) give a closed dynamical system in the variables ($E,\,\Sigma$):

\begin{equation}
\begin{array}{l}
\dot{E} + \left(\theta + \fracc{\xi}{2}\right)\, E + \left[\fracc{\xi}{2}\, \left(\fracc{\theta}{3} + \fracc{\xi}{2}\right) - \fracc{1}{a^{2}}\, (m - 3\epsilon)\right]\, \Sigma = 0, \\[1.5ex]
\dot{\Sigma} + \left(\fracc{2}{3}\, \theta + \fracc{\xi}{2} \right)\, \Sigma + E = 0,
\end{array}
\label{apdx21}
\end{equation}
where $H$ is given by the constraint

\begin{equation}
\nonumber
\Sigma = H.
\end{equation}
We have then the following set of {\it good} quantities for tensorial perturbations in Milne background:

\begin{equation}
\nonumber
{\cal M}_{A} = \{E, H\}.
\end{equation}

\subsection{WIST Model: Scalar Perturbations}

This section deals with the dynamical system of the perturbed quantities that are relevant for the complete knowledge of the system. The equations completely describe the perturbation evolution, according to the quasi-Maxwellian equations of gravitation.

In this case, we will assume that the background of the model can be characterized by a source consisting of a scalar field minimally coupled to the gravitational field. The energy-momentum tensor for a minimal coupling scalar field is represented by a perfect fluid and it can be demonstrated that the general linear perturbations of this fluid also behaves as a perfect fluid. This property of the source simplifies the equations and the minimal set of observables that determine the scalar linear perturbations of the model can be obtained from the following equations:

\begin{equation}
\label{qm_bounce}
(E^{\alpha\beta})\dot{}+ E^{\alpha\beta} - \fracc{3}{2} E^{\mu(\alpha}\sigma^{\beta)}{}_{\mu}+h^{\alpha\beta} = -\fracc{1}{2}(p + \rho)\sigma^{\alpha\beta},
\end{equation}

\begin{equation}
\label{shear_bounce}
(\delta\sigma_{\mu\nu})\dot{} + \fracc{1}{3}h_{\mu\nu}(\delta a^{\alpha})_{;\alpha} - \fracc{1}{2}\delta a_{(\mu;\nu)} + \fracc{2}{3}\theta\delta\sigma_{\mu\nu} = -\delta E_{\mu\nu},
\end{equation}

\begin{equation}
\label{accel_bounce}
(p + \rho)\delta a_{\mu} = (\delta p)_{,\beta}h^{\beta}_{\mu} - p_{,\beta}\delta(h^{\beta}_{\mu}).
\end{equation}

The equations are considered to be linear in the perturbations; so, to solve them we will split the perturbations in terms of the scalar spherical harmonics basis defined by Eq.\ (\ref{d13}) in terms of the conformal scale factor $a(\eta)$. Since the model we are investigating has an open three-section, the eigenvalue $m$ can assume the following values:

\begin{equation}
\label{k_bounce_rel}
m = q^2+1,\hspace{1cm} 0\leq q\leq\infty.
\end{equation}

With the scalar function $Q$, we can construct vector $\hat Q_{\alpha}$ and tensor $\hat Q_{\mu\nu}$ quantities. Using this base we can expand the perturbations as follows:

\begin{equation}
\label{exp_e_boun}
\delta E_{\mu\nu} =\sum_qE_{(q)}{}^{(q)}\hat Q_{\mu\nu},
\end{equation}

\begin{equation}
\label{exp_shear_boun}
\delta\sigma_{\mu\nu} =\sum_q\sigma_{(q)}{}^{(q)}\hat Q_{\mu\nu},
\end{equation}

\begin{equation}
\label{exp_accel_boun}
\delta a_{\mu} =\sum_q\psi_{(q)}{}^{(q)}\hat Q_{\mu},
\end{equation}

\begin{equation}
\label{exp_v_boun}
\delta V_{\mu} =\sum_qV_{(q)}{}^{(q)}\hat Q_{\mu}.
\end{equation}

\subsubsection{Dynamics}

From now on we will suppress the indices $q$ to simplify the notation. Substituting this decomposition in Eqs.\ (\ref{qm_bounce}) and\ (\ref{shear_bounce}), after simple algebraic calculations (see Novello, 1995a), we obtain the following dynamical system for each mode of the variables $E$ and $\Sigma$:

\begin{equation}
\label{ev_sig_boun}
\Sigma'=\left[\fracc{1}{\rho a}(3 + m) - a\right]E,
\end{equation}

\begin{equation}
\label{ev_sig_boun}
E'= -\fracc{a'}{a}E - \rho a\Sigma.
\end{equation}
The prime means covariant derivative projected on $V^{\alpha}$.

In terms of matrix, this dynamical system can be written in compact form as

\begin{equation}
\label{dyn_sys_e_shear_boun}
\begin{pmatrix}
\Sigma\\
E
\end{pmatrix}'={\cal M}
\begin{pmatrix}
\Sigma\\
E
\end{pmatrix},
\end{equation}
The components of the matrix ${\cal M}$ are: ${\cal M}_{11}=0$; ${\cal M}_{12}=-a+(3+m)/(\rho a)$; ${\cal M}_{21}=-\rho a$; ${\cal M}_{22}=-a'/a$.

\subsubsection{Hamiltonian Treatment}

The examination of the perturbations of Robertson-Walker geometries, using the variables associated to the perturbed metric tensor $\delta g_{\mu\nu}$, admits a Hamiltonian formulation. In this vein, it was shown in detail by Novello (1995b) that the present formulation using variables $E$ and $\Sigma$ also admits a Hamiltonian formulation. The usual way to do this is introducing auxiliary field variables as

\begin{equation}
\label{can_trans_e_shear_boun}
\begin{pmatrix}
P\\
Q
\end{pmatrix}=
\begin{pmatrix}
\alpha & \beta\\
\gamma & \delta
\end{pmatrix}
\begin{pmatrix}
\Sigma\\
E
\end{pmatrix}.
\end{equation}

The matrix $S$ with components $\alpha$, $\beta$, $\gamma$ and $\delta$ is univocally defined up to canonical transformations.

In terms of the auxiliary variables, the dynamical system in matrix notation becomes

\begin{equation}
\label{dyn_sys_e_shear_boun}
\begin{pmatrix}
P\\
Q
\end{pmatrix}'=\Xi
\begin{pmatrix}
P\\
Q
\end{pmatrix},
\end{equation}
where

\begin{equation}
\label{trans_matr}
\Xi = S{\cal M}S^{-1}- S'S^{-1}.
\end{equation}
The requirement that $(Q, P)$ are canonical variables implies the necessary and sufficient condition that

\begin{equation}
\label{can_trans_req}
tr\Xi = tr{\cal M}+\fracc{\Delta'}{\Delta} = 0,
\end{equation}
where $\Delta$ is the determinant of $S$. The Hamiltonian in our case is given by

\begin{equation}
\label{ham_boun}
H = \fracc{\Xi_{21}}{2}P^2-\fracc{\Xi_{12}}{2}Q^2-\fracc{\Xi_{11}}{2}(PQ + QP).
\end{equation}
It seems worthwhile to remark that the matrix ${\cal M}$ and $S$ univocally determine $H$ up to canonical transformations. This freedom can be used to simplify our analysis in each particular case.

The background model we will investigate is asymptotic flat in the limits of the conformal time $\eta\rightarrow\pm\infty$. Convenient canonical transformations can be used to fix the functions of the matrix $S$, in order that in the limit $\eta\rightarrow-\infty$ we obtain, for each mode $m$, the Hamiltonian of an harmonic oscillator in a Minkowski space–time. With this choice, the Hamiltonian that describes the system is given by

\begin{equation}
\label{sp_ham_boun}
H =\fracc{1}{2}P^2+\fracc{1}{2}\left[8\tanh(2\eta) + (m-7) + \fracc{\Sigma_0}{a^4}\right]Q^2+ [\tanh(2\eta) + 1](PQ + QP).
\end{equation}
Using Hamilton's equations, after a simple calculation, we obtain

\begin{equation}
\label{ev_q_boun}
Q''_{m}(\eta) + [q^2 - 3(\tanh(2\eta)2 - 1)] Q_m(\eta) = 0 ,
\end{equation}

\begin{equation}
\label{eff_pot_q_boun}
w^2_m(\eta) = q^2 - V_{eff}.
\end{equation}
The exact solution to Eq.\ (\ref{ev_q_boun}) is given by

\begin{equation}
\label{sol_ev_q_boun}
Q_m(\eta) = AF(a_1, b_1, c_1, z) + B \sinh(2\eta)F(a_2, b_2, c_2, z)],
\end{equation}
where $F(a_1, b_1, c_1, z)$ and $F(a_2, b_2, c_2, z)$ are hypergeometric functions with parameters

\begin{equation}
\label{par_hyp_boun}
\begin{array}{ll}
a_1 = \fracc{3}{4}+\fracc{I}{4}\sqrt{m-1},& a_2 = \fracc{5}{4}+\fracc{I}{4}\sqrt{m-1},\\[2ex]
b_1 = \fracc{3}{4}-\fracc{I}{4}\sqrt{m-1},& b_2 = \fracc{5}{4}-\fracc{I}{4}\sqrt{m-1},\\[2ex]
c_1 = \fracc{1}{2},& c_2 = \fracc{3}{2}.
\end{array}
\end{equation}
The variable $z$ is given by

\begin{equation}
\nonumber
z = -\sinh^2(2\eta).
\end{equation}
The asymptotic limits $\eta\rightarrow\pm\infty$ are given by

\begin{equation}
\label{asymp_q_in}
Q^{in}_{m}(\eta) = \lim_{\eta\rightarrow-\infty}Q_m(\eta) = \fracc{1}{\sqrt{w_m}}e^{-iq\eta},
\end{equation}

\begin{equation}
\label{asymp_q_out}
Q^{out}_{m}(\eta) = \lim_{\eta\rightarrow\infty}Q_m(\eta) = d_1(m)e^{-iq\eta}+ d_2(m)e^{iq\eta}.
\end{equation}
The amplitudes $d_1(m)$ and $d_2(m)$ are expressed in terms of trigonometric and gamma functions as

\begin{equation}
\label{amp_d_1_2}
\begin{array}{l}
d_1 = \fracc{1}{8}\fracc{\sqrt{q}(q^2 + 1)\Gamma\left(\fracc{iq}{2}\right)^2\sinh(\pi q/2)}{\Gamma\left(\fracc{3}{2}+\fracc{iq}{2}\right)^2\left[\sin(\pi(1+iq)/4)^2-\sin(\pi(1-iq)/4)^2\right]},\\[2ex]
d_2=\fracc{-2+\cos(\pi(5+iq)/4)^2+\cos(\pi(3+iq)/4)^2}{\sqrt{q}\left[\cos(\pi(5+iq)/4)^2-\cos(\pi(3+iq)/4)^2\right]}
\end{array}
\end{equation}

Using the scalar base $l_m(x)$ the conjugate variables $Q$ and $P$, describing our dynamical system, can be expanded in terms of traveling waves according to

\begin{equation}
\label{four_q_boun}
Q({\bf x},\eta) = \fracc{1}{\sqrt{2}}\int \fracc{d^3{\bf m}}{(2\pi)^{3/2}}[Q^*_m(\eta)l^*_m({\bf x}) + Q_m(\eta)l_m({\bf x})],
\end{equation}
and
\begin{equation}
\label{four_p_boun}
P({\bf x},\eta) = \fracc{1}{\sqrt{2}}\int \fracc{d^3{\bf m}}{(2\pi)^{3/2}}[P^*_m(\eta)l^*_m({\bf x}) + P_m(\eta)l_m({\bf x})].
\end{equation}

\subsubsection{Quantum Treatment of the Perturbations}

The conjugate variables $Q(x)$ and $P(x)$ can be quantized following standard procedures, by transforming the mode functions into operators (cf. Birrel, 1982)

\begin{equation}
\label{pq_quant_boun}
\begin{array}{l}
\hat Q({\bf x},\eta) = \fracc{1}{\sqrt{2}}\int \fracc{d^3{\bf m}}{(2\pi)^{3/2}}[\hat a^{-}_mQ^*_ml^*_m({\bf x}) + \hat a^{+}_mQ_ml_m({\bf x})]\\[2ex]
\hat P({\bf x},\eta) = \fracc{1}{\sqrt{2}}\int \fracc{d^3{\bf m}}{(2\pi)^{3/2}}[\hat a^{-}_mP^*_ml^*_m({\bf x}) + \hat a^{+}_mP_ml_m({\bf x})].
\end{array}
\end{equation}
The classical canonical variables are replaced by operators $\hat P$ and $\hat Q$ which satisfy the commutation relations $[\hat Q_{\alpha},\hat P_{\beta}] = i\hbar\delta_{\alpha\beta}$ and $[\hat Q_{\alpha},\hat Q_{\beta}] = [\hat P_{\alpha},\hat P_{\beta}] = 0$, and through this we can define the operators of creation and annihilation as

\begin{equation}
\label{def_a_boun}
\hat a^{\pm}_{\alpha} =\sqrt{\fracc{\omega_{\alpha}}{2}}(\hat Q_{\alpha}(t)\mp\hat P_{\alpha}),
\end{equation}
and from these operators the new commutation relations $[\hat a^{-}_{\alpha},\hat a^{+}_{\beta}] = \delta_{\alpha\beta}$ and $[\hat a^{-}_{\alpha},\hat a^{-}_{\beta}] = [\hat a^{+}_{\alpha},\hat a^{+}_{\beta}] = 0$.

Through the expansion\ (\ref{four_q_boun}), we can write two different sets of functions corresponding to in-going and out-going waves, respectively given by

\begin{equation}
\label{four_q_in_boun}
\hat Q^{in}({\bf x},\eta) = \fracc{1}{\sqrt{2}}\int \fracc{d^3{\bf m}}{(2\pi)^{3/2}}[\hat a^{-}_mv^*_ml^*_m({\bf x}) + \hat a^{+}_mv_ml_m({\bf x})]
\end{equation}
and
\begin{equation}
\label{four_q_out_boun}
\hat Q^{out}({\bf x},\eta) = \fracc{1}{\sqrt{2}}\int \fracc{d^3{\bf m}}{(2\pi)^{3/2}}[\hat b^{-}_mu^*_ml^*_m({\bf x}) + \hat b^{+}_mu_ml_m({\bf x})]
\end{equation}

The operators $\hat a_m$ and $\hat b_m$ are related by Bogolyubov coefficients $\alpha_m$ and $\beta_m$,

\begin{equation}
\label{bog_rel_boun}
\hat b_m = \alpha_m\hat a_m + \beta*_m\hat a_{-m},
\end{equation}
satisfying the normalization condition $|\alpha_m|^2 - |\beta_m|^2 = 1$. The two sets of isotropic mode functions $u_m$ and $v_m$ form a base of {\it in} and {\it out} functions. We can write this base, according with\ (\ref{asymp_q_in}) and\ (\ref{asymp_q_out}), as follows:

\begin{equation}
\label{v_boun}
v_q(\eta) = \fracc{1}{\sqrt{w_m}}e^{-Iw_m\eta},
\end{equation}

\begin{equation}
\label{u_boun}
u_q(\eta) = d_1(q)e^{-iq\eta}+ d_2(q)e^{iq\eta}.
\end{equation}

The ``in" vacuum is described by the standard Minkowski mode function and the mode corresponding to ``out" vacuum can be written as a linear combination of the basis $v_q$:

\begin{equation}
\label{bas_rel_boun}
u_q = \alpha_qv_q - \beta^*{}_qv^*{}_q.
\end{equation}

\subsection{Nonlinear Electrodynamics: Scalar Perturbations}

The existence of singularities appears to be a property inherent to most of the physically relevant solutions of Einstein equations, in particular to all known up-to-date black hole and conventional cosmological solutions (see Hawking and Ellis, 1977). In the case of black holes, to avoid the singularity some models have been proposed---cf. Bardeen (1968), Barrabes and Frolov (1996), Cabo and Ayon-Beato (1997) and, Mars et al. (1996). These models nonetheless are not exact solutions of Einstein equations since there are no physical sources associated to them. Many attempts try to solve this problem by modifying general relativity, for instance, Cvetic (1993), Tseytlin (1995), and Horne and Horowitz (1992). More recently it has been shown that in the framework of standard general relativity it is possible to find spherically symmetric singularity-free solutions of the Einstein field equations that describe a regular black hole. The source of these solutions are generated by suitable nonlinear vector field Lagrangians, which in the weak field approximation become the usual linear Maxwell theory demonstrated by Ayon-Beato and Garcia (1998, 1999a-b). Similarly in Cosmology many non-singular cosmological models with bounce were constructed where the energy conditions or the validity of Einstein gravity were violated (see Sec.\ [\ref{non_sol_non_elec}] for details).

In 2002, de Lorenci et al. investigated a cosmological model with a source produced by a nonlinear generalization of electrodynamics and succeeded to obtain a regular cosmological model. The Lagrangian of such model is a function of the field invariants up to second order---Eq. (\ref{gen_lag_max}). This modification is expected to be relevant when the fields reach large values, as is the case in the primeval era of our universe. The model is in the framework of the Einstein field equations and the bounce is possible because the singularity theorems are circumvented by the appearance of a negative pressure (although the energy density is positive definite). Recently some papers started a detailed investigation of the transition from contraction to expansion in the bounce of several models (see Cartier et al., 2003). In particular, in Einstein general relativity, models with stress-energy sources constituted by a collection of perfect fluids and FLRW like geometry were examined by Peter and Pinto-Neto (2001). The claim in that paper is that a generic result about the behavior of scalar adiabatic perturbations was obtained. The result is the following: scalar adiabatic perturbations can grow without limit in two situations represented by the points where the scale factor attains its minimum value and where $\rho+p=0$. The first point corresponds to the moment in which the Universe passes through the bounce; the second corresponds to the transition from the region where the Null Energy Condition (NEC) is violated to the region where it is not. However, these instabilities are not an intrinsic property of generic models with bounce but a consequence of the existence of a divergence already appearing in the background solution, when the source is described as a perfect fluid. We show a specific example of a model with bounce, generated by a source representing two non-interacting perfect fluids, that has regular perturbations.

\subsubsection{The Model}
We set the fundamental line element as being an FLRW metric. According to the definitions presented in Sec. [\ref{non_sol_non_elec}], we rewrite the Einstein equations and the equation of energy conservation written for this metric below

\begin{equation}
\label{nlm_fried1}
\left(\fracc{\dot a}{a}\right)^2+\fracc{\epsilon}{a^2}-\fracc{1}{3}\rho_{\gamma}=0,
\end{equation}

\begin{equation}
\label{nlm_fried2}
2\fracc{\ddot a}{a}+\left(\fracc{\dot a}{a}\right)^2+\fracc{\epsilon}{a^2}+p_{\gamma}=0,
\end{equation}

\begin{equation}
\label{nlm_fried3}
\dot\rho_{\gamma} + 3(\rho_{\gamma} + p_{\gamma})\fracc{\dot a}{a}=0.
\end{equation}
Inserting Eqs.\ (\ref{rho_h0}) and (\ref{p_h0}) in (\ref{nlm_fried3}), it yields Eq.\ (\ref{h_sca_fac}) for the magnetic field, where $H_o$ is an arbitrary constant. With this result the equation (\ref{nlm_fried1}) can be integrated. For the case $\epsilon=0$ the solution is given by Eq.\ (\ref{eucl_sc_fac}).

The interpretation of the source as a one component perfect fluid in an adiabatic regime has some difficulties (see Peter and Pinto-Neto (2001). The sound velocity of the fluid in this case is given by

\begin{equation}
\label{spe_sound}
\left(\fracc{\partial p_{\gamma}}{\partial \rho_{\gamma}}\right)=\fracc{\dot p_{\gamma}}{\dot\rho_{\gamma}}= - \fracc{\dot p_{\gamma}}{\theta(\rho_{\gamma} + p_{\gamma})}.
\end{equation}

This expression, involving only the background, is not defined at those points where the energy density attains an extremum given by $\theta=0$ or $\rho_{\gamma}+p_{\gamma}=0$. In terms of the cosmological time, they are determined by $t=0$ and $\pm tc=12\alpha/kc^2$. These points are well-behaved regular points of the geometry indicating that the description of the source is not appropriate. This difficulty can be circumvented if one adopts another description for the source of the model. This can be achieved by separating the part of the source related to Maxwell dynamics from the additional non-linear term dependent on $a$ on the Lagrangian. By doing this, the source automatically splits in two noninteracting perfect fluids:

\begin{equation}
\label{en_mom_12}
T_{\mu\nu}=T^1_{\mu\nu}+T^2_{\mu\nu},
\end{equation}
where

\begin{equation}
\label{en_mom_1}
T^1_{\mu\nu}=(\rho_1+p_1)V_{\mu}V_{\nu}-p_1g_{\mu\nu},
\end{equation}

\begin{equation}
\label{en_mom_2}
T^2_{\mu\nu}=(\rho_2+p_2)V_{\mu}V_{\nu}-p_2g_{\mu\nu},
\end{equation}
and

\begin{equation}
\label{rho_en_mom_1}
\rho_1=\fracc{1}{2}H^2,
\end{equation}

\begin{equation}
\label{p_en_mom_1}
p_1=\fracc{1}{6}H^2,
\end{equation}

\begin{equation}
\label{rho_en_mom_2}
\rho_2=-4\alpha H^4,
\end{equation}

\begin{equation}
\label{p_en_mom_2}
p_2=-\fracc{20}{3}\alpha H^4.
\end{equation}
Using the above decomposition, it follows that each one of the two components of the fluid independently satisfies equation (\ref{nlm_fried3}). This indicates that the source can be described by two non-interacting perfect fluids with equation of states $p_1=\rho_1/3$ and $p_2=5\rho_2/3$. The equation of state for the second fluid should be understood only formally as a mathematical device to allow for a fluid description.

\subsubsection{Gauge Invariant Treatment of Perturbation}

The source of the background geometry is represented by two-fluids, each one having an independent equation of state relating the pressure and the energy density. Following the standard procedure we consider arbitrary perturbations that preserve each equation of state. Thus the general form of the perturbed energy-momentum tensor is written as

\begin{equation}
\label{en_mom_pert_nlm}
\delta T^i_{\mu\nu}=(1+\lambda_i)\delta(\rho_iV_{\mu}V_{\nu})-\lambda_i\delta(\rho_ig_{\mu\nu}).
\end{equation}
The background geometry is conformally flat. Thus any perturbation of the Weyl tensor is a true perturbation of the gravitational field. It is convenient to represent the Weyl tensor $W_{\alpha\beta\mu\nu}$ in terms of its corresponding electric $E_{\mu\nu}$ and magnetic $H_{\mu\nu}$ parts, because these variables have the advantage that since they are null in the background, their perturbations are gauge invariant quantities (Hawking, 1966).

The equations of motion for the first order perturbations are linear so it is useful to develop all perturbed quantities in the spherical harmonics basis. Here, we will limit our analysis to perturbations represented in the scalar base. This basis, we also take from Sec. [\ref{scalarpert}].

In the case of scalar perturbations the fundamental set of equations determining the dynamics of the perturbations are:
\begin{equation}
\label{apb132}
(\delta E_{1}^{\mu\nu})^{\bullet}{h_{\mu}}^{\alpha}{h_{\nu}}^{\beta} + ( \delta E_{2}^{\mu\nu})^{\bullet}{h_{\mu}}^{\alpha}{h_{\nu}}^{\beta} + \theta (\delta E_{1}^{\alpha\beta}+\delta E_{2}^{\alpha\beta}) = - \frac{1}{2}(\rho_{1} + p_{1})\, \delta \sigma_{1}^{\alpha\beta} - \frac{1}{2}(\rho_{2} + p_{2}) \, \delta \sigma_{2}^{\alpha\beta},
\end{equation}

\begin{equation}
\label{apb16}
(\delta E^{1}_{\alpha\mu} + \delta E^{2}_{\alpha\mu})_{;\nu} h^{\alpha\varepsilon} h^{\mu\nu} = \frac{1}{3} (\delta\rho_{1}+\delta\rho_{2})_{,\alpha} h^{\alpha\varepsilon} - \frac{1}{3} \dot{\rho}_{1}\delta v_{1}^{\varepsilon} -\frac{1}{3} \dot{\rho}_{2} \delta v_{2}^{\varepsilon},
\end{equation}

\begin{equation}
\label{apb18_2}
\begin{array}{l}
\left(\delta \sigma^{1}_{\mu\nu}\right)^{\bullet} + \left( \delta\sigma^{2}_{\mu\nu}\right)^{\bullet} + \frac{1}{3}h_{\mu\nu} (\delta {a_{1}^{\alpha}}+\delta a_{2}^{\alpha})_{;\alpha} - \frac{1}{2}(\delta a^{1}_{\alpha;\beta}+\delta a^{2}_{\alpha;\beta}) {h_{(\mu}}^{\alpha} {h_{\nu)}}^{\beta} +  \frac{2}{3}\theta (\delta\sigma^{1}_{\mu\nu}+\delta\sigma^{2}_{\mu\nu})=\\[2ex]
= - \delta E^{1}_{\mu\nu}- \delta E^{2}_{\mu\nu},
\end{array}
\end{equation}

\begin{equation}
\label{apb242}
- \lambda_{1}\delta \left(\rho_{,\beta} {h^{\beta}}_{\mu}\right) + (1 +\lambda_{1})\rho \delta a^{1}_{\mu}=0,
\end{equation}

\begin{equation}
\label{apb25}
- \lambda_{2}\delta \left(\rho_{,\beta} {h^{\beta}}_{\mu}\right) + (1 +\lambda_{2})\rho \delta a^{2}_{\mu}= 0.
\end{equation}
The acceleration $a^{\mu}$, the expansion $\theta$ and the shear $\sigma_{\mu\nu}$ that appear in the above equations are parts of the irreducible components of the covariant derivative of the velocity field.

The expansion of the perturbations in terms of the spherical harmonic basis is\footnote{Note that once we are dealing with a linear process each mode can be analyzed separately.}

\begin{equation}
\delta\rho= N(t)Q,
\end{equation}

\begin{equation}
\delta V^{\mu}= V(t)h^{\mu\alpha}Q_{,\alpha},
\end{equation}

\begin{equation}
\delta a^{\mu}= \dot{V} \, h^{\mu\alpha}Q_{,\alpha},
\end{equation}

\begin{equation}
\delta E^{\mu\nu}=E(t)P^{\mu\nu},
\end{equation}

\begin{equation}
\delta \sigma^{\mu\nu}= \Sigma(t)P^{\mu\nu},
\end{equation}

\subsubsection{Dynamics}

After presenting the necessary formalism we shall start the analysis of the perturbations of the bouncing cosmological model displayed in a previous section. Using the above expansion into the equations\ (\ref{apb132})-(\ref{apb25}) we obtain:

\begin{equation}
E_{1}+E_{2}=\frac{a^2}{6\epsilon+k^2}\left(N_{1}+ N_{2}-{\dot{\rho}}_{1}V_{1}-{\dot{\rho}}_{2}V_{2}\right),
\end{equation}

\begin{equation}
\dot{E}_{1}+ \dot{E}_{2} + \frac{1}{3} \theta\left(E_{1}+E_{2}\right) = -\left(\frac{1+\lambda_{1}}{2}\right)\rho_{1}\Sigma_{1}- \left(\frac{1+\lambda_{1}}{2}\right)\rho_{2}\Sigma_{2},
\end{equation}

\begin{equation}
\dot{\Sigma_{1}}+\dot{\Sigma_{2}}-\dot{V}_{1}-\dot{V}_{2}=-E_{1}-E_{2},
\end{equation}

\begin{equation}
-\lambda_{1}\left(N_{1}-\dot{\rho}_{1}V_{1}\right)+ \left(1+\lambda_{1}\right)\rho_{1}\dot{V}_{1}=0,
\end{equation}

\begin{equation}
-\lambda_{2}\left(N_{2}-\dot{\rho}_{2}V_{2}\right)+ \left(1+\lambda_{2}\right)\rho_{2}\dot{V}_{2}=0.
\end{equation}
These equations can be rewritten in a more convenient way as:

\begin{equation}
\dot{\Sigma_{1}}= - \left(\frac{2\lambda_{1}(3\epsilon+k^2)}{a^2(1+\lambda_{1})\rho_{1}}+1\right)E_{1},
\end{equation}

\begin{equation}
\dot{\Sigma_{2}}= - \left(\frac{2\lambda_{1}(3\epsilon+k^2)}{a^2(1+\lambda_{2})\rho_{2}}+1\right)E_{2},
\end{equation}

\begin{equation}
\dot{E_{1}}+\frac{1}{3}\theta E_{1}=-\frac{1}{2}\left(1+\lambda_{1}\right)\rho_{1}\Sigma_{1},
\end{equation}

\begin{equation}
\dot{E_{2}}+\frac{1}{3}\theta E_{2}=-\frac{1}{2}\left(1+\lambda_{2}\right)\rho_{2}\Sigma_{2},
\end{equation}
The whole set of scalar perturbations can be expressed in terms of the two basic variables:  $E_{i}$ and $\Sigma_{i}$. The corresponding equations can be decoupled. The result in terms of variables $E_{i}$ is the following:

\begin{equation}
\ddot{E}_{i} + \frac{4+3\lambda_{i}}{3}\theta \dot{E}_{i}+ \left[\frac{2+3\lambda_{i}}{9}\theta^2 - \left(\frac{2}{3}+\lambda_{i}\right)\rho_{i} - \frac{1}{6}(1+3\lambda_{j})\rho_{j} -\frac{(3\epsilon+k^2)\lambda_{i}}{a^2} \right] E_{i}=0.
\end{equation}

Note that there is no sum in indices and $j\neq i$ in this expression. In our case the values of $\lambda_{i}$ are $\lambda_{i}=\left(\frac{1}{3}, \frac{5}{3}\right)$. In the first case the equation for the variable $E_{1}$ became:

\begin{equation}
\ddot{E}_{1}+\frac{5}{3} \theta \dot{E}_{1}+\left[\frac{1}{3}\theta^2-\rho_{1}-\rho_{2} -\frac{5 k^2}{3 a^2}\right]E_{(1)}=0.
\end{equation}

We should analyze the behavior of these perturbations in the neighborhood of the points where the energy density attains an extremum. This means not only the bouncing point but also the point in which $\rho + p$ vanishes. Let us start by examining at the bouncing  point $t=0$.

If we consider up to second order terms in the perturbation theory, the expansion of the equation of $E_1$ in the neighborhood of the bouncing is given by:

\begin{equation}
\label{equa1}
\ddot{E_1}+a_ct\dot{E_1}+(b+b_1t^2)E_1=0.
\end{equation}
The constant $a_c$ and the parameters $b$ and $b_1$ are defined as follows

\begin{equation}
a_c=\frac{5}{2t_c^2},
\end{equation}

\begin{equation}
b=-\frac{m^2}{\sqrt{6} H_0 t_c},
\end{equation}
and
\begin{equation}
b_1=-\frac{b}{2t_c^2}-\frac{3}{4t_c^4}.
\end{equation}
Defining a new variable f as

\begin{equation}
f(t)=E_1(t)\exp{\left(+\frac{a_c}{4}-\frac{i}{2}\sqrt{b_1-\frac{a_c^2}{4}}\right)t^2},
\end{equation}
and doing a coordinate transformation for time as indicated bellow

\begin{equation}
\xi=-i ( \sqrt{b_1-\frac{a_c^2}{4}}) t^2,
\end{equation}
we obtain the following confluent hypergeometric equation (cf. Abramowitz and Stegun, 1974)

\begin{equation}
\xi \ddot{f}+(1/2-\xi)\dot{f}+ e f=0,
\end{equation}
where

\begin{equation}
e=\frac{i(b-a_c/2)}{4(b_1-a_c^2/4)^{1/2}}-\frac{1}{2}.
\end{equation}

The solution of this equation is given by:

\begin{equation}
f(t)= f_o \; {\cal  M}\left[d,1/2,-i ( \sqrt{4b_1-a_c^2})\fracc{t^2}{2}\right],
\end{equation}
where $f_o$ is an arbitrary constant and ${\cal M}(d,1/2,\xi)$ is a confluent hypergeometric function. The confluent hypergeometric function is well behaved in this neighborhood and so is the perturbation $E_{1}(t)$ given by:

\begin{equation}
E_1(t)=\Re\left[ f_o \; {\cal M}\left(d,1/2,-i ( \sqrt{4b_1-a_c^2})\fracc{t^2}{2}\right)\exp{\left(-\frac{a_c}{4}+\frac{i}{2}\sqrt{b_1-\frac{a_c^2}{4}}\right)t^2}\right].
\end{equation}

The perturbation $E_2$ at this same neighborhood, after the same procedure we did before, results in the following equation:

\begin{equation}
\label{equa2}
\ddot{E_2}+a_ct\dot{E_2}+(b+b_1t^2)E_2=0.
\end{equation}
This is the same equation we obtained for $E_1$, they differ only by the values of the parameters $a_c, b$ and $b_1$ that in this case are:

\begin{equation}
a_c=\frac{9}{2t_c^2},
\end{equation}

\begin{equation}
b= \frac{3}{2t^2_c}- 5\frac{m^2}{\sqrt{6} H_0 t_c},
\end{equation}
and
\begin{equation}
b_1=-\frac{5 m^2}{t_c^3 H_0\sqrt{6}}-\frac{5}{t_c^4}.
\end{equation}
Then the solution in this case is:

\begin{equation}
E_2(t)=\Re\left[ f_o\,{\cal M}\left(d,1/2,-i ( \sqrt{4b_1-a_c^2}\right)\fracc{t^2}{2}) \exp{-\left(\frac{a_c}{4}-\frac{i}{2}\sqrt{b_1-\frac{a_c^2}{4}}\right)t^2}\right].
\end{equation}
Again the confluent hypergeometric function is well behaved in this neighborhood and so is the perturbation $E_{2}(t)$. At the neighborhood of the point $t=t_c$ the equation for the perturbation $E_1$ is given by

\begin{equation}
\label{equa3}
\ddot{E}_1 + a_c\dot{E}_1 + \left(b+b_1 t\right)E_1=0,
\end{equation}
where the parameters $a_c$, $b$ and $b_1$ in this case are given by

\begin{equation}
a_c=\frac{5}{4t_c},
\end{equation}

\begin{equation}
b_=-\frac{3}{4t_c^2}-\frac{\sqrt{3} m^2}{6 H_0 t_c},
\end{equation}
and
\begin{equation}
b_1=\frac{\sqrt{3}}{4t_c^2}\left(\frac{m^2}{3H_0}-\frac{3}{2t_c}\right).
\end{equation}

We would like to remark that this equation is different from the equations (\ref{equa1}) and (\ref{equa2}) obtained in the neighborhood of $t=0$. We proceed doing the following variable transformation:

\begin{equation}
E_1(t)= w(t)\exp{\left(-\frac{a_c t}{2}\right)}.
\end{equation}
The differential equation for this new variable is

\begin{equation}
\ddot{w}+\left( b-(a_c/2)^2+b_1t\right)w=0.
\end{equation}

The solution for this equation is

\begin{equation}
w(t)= \left[w_0\,\mbox{AiryAi}\left(-\frac{b-(a_c/2)^2 + b_1t}{b^{2\\/3}}\right)\right].
\end{equation}

The AiryAi are regular well behaved functions in this neighborhood and also the perturbations $E_1$. Finally, we look for the equation of $E_2$ at the neighborhood of $t=t_0$, and it becomes

\begin{eqnarray}
\ddot{E}_2 + a_c\dot{E}_2 + \left(b+b_1 t\right)E_2=0,
\end{eqnarray}
where the parameters $a_c$, $b$ and $b_1$ in this case are given by

\begin{equation}
a=\frac{9}{4t_c},
\end{equation}

\begin{equation}
b=\frac{5}{t_c}\left(\frac{5}{4t_c}-\frac{\sqrt{3} m^2}{6 H_0}\right),
\end{equation}
and
\begin{equation}
b_1=\frac{5\sqrt{3}}{2t_c^2}\left(\frac{1}{t_c}-\frac{m^2}{6H_0}\right).
\end{equation}
This equation differs from Eq.(\ref{equa3}) only by the numerical values of the parameters $a_c$, $b$ and $b_1$, so we obtain the same regular solution

\begin{equation}
E_2= \Re\left[ \exp{\left(-\frac{a_ct}{2}\right)}\,w_0 \,\mbox{AiryAi}\left(-\frac{b-(a_c/2)^2+ b_1 t}{b_1^{2/3}}\right) \right].
\end{equation}

Summarizing, recently there has been renewed interest on nonsingular cosmology. As a direct consequence of this, some authors have argued against these models based on instability reasons. Peter and Pinto-Neto (2001) have argued that a rather general analysis shows that there are instabilities
associated to some special points of the geometrical configuration. They correspond to the points of bouncing of the model and  maxima of the energy
density, where the description of the matter content in terms of a single perfect fluid does not apply. In the present paper we have shown, by a direct analysis of a specific nonsingular universe, that the result claimed in the quoted paper does not apply to our model. We took the example from a paper, De Lorenci et al. (2002), in which the avoidance of the singularity comes from a nonlinear electrodynamic theory. We used the quasi Maxwellian equations of motion---cf. Novello et al., 1995a-b, 1996---in order to undertake the analysis of the perturbed set of Einstein equations of motion. We showed that in the neighborhood of the special points in which a change of regime occurs, all independent perturbed quantities are well behaved. Consequently the model does not present any difficulty concerning its instability. This paves the way to investigate models with bounce in more detail and to consider them as good candidates to describe the evolution of the Universe.

\section{On the role of initial conditions to the equivalence theorem}\label{init_cond}

In this section we briefly discuss the question that appears explicitly in perturbation theory: how do Einstein equations determine the Weyl tensor if they contain information only of the traces of the curvature tensor?

Einstein equations relate the energy-momentum tensor with the traces of the curvature tensor (local quantities), leaving the remaining components of curvature tensor, which correspond to the Weyl tensor (nonlocal quantity), undetermined. However, this indetermination is apparent, because Bianchi's identities relate the traces of the curvature tensor with the Weyl tensor through equation (\ref{bianchi_R}). Nevertheless, remark that this relation involves partial derivative. Thus, substituting the Einstein equations into (\ref{bianchi_R}), we obtain the set of equations that involves the energy-momentum and Weyl tensors, which leaves the traces of curvature out---see equation (\ref{bianchi_T}). These equations are direct consequence of Einstein equations, but they are not equivalent in principle. In order to get such equivalence, it is necessary to impose an appropriate initial conditions---see details about Lichnerowicz's theorems in \cite{lichnerowicz}.

From the mathematical point of view, the problem is automatically solved, however it does not occurs from the physical viewpoint, because it involves the determination of initial conditions from empirical data on a Cauchy surface. The empirical data that we get they determine the initial conditions of the physical system that we want to describe taking into account the curvature and energy-momentum tensors. Actually, these empirical data are precisely the curvature tensor of space-time and the energy-momentum tensor. Once the curvature tensor has $20$ independent components, then we face the question: how can we determine the Cauchy data specifying the metric components $g_{ij}$ and first time derivatives $g_{ij,0}$ ($i,j=1,2,3$)? The Einstein equations are differential equations for the metric potential $g_{\mu\nu}$, they do not give any information about the curvature tensor as initial condition, and then, this situation imposes a serious ambiguity on the determination of boundary conditions. This ambiguity clearly appears in gravitational waves process and can be consistently shown even in simple cases of perturbation theory, where the gravitational theory is linearized. Indeed, considering an exact solution with a perfect fluid as source of curvature, the perturbed equations describing gravitational waves reduce to

\begin{equation}
\label{var_ricci}
\delta R_{\mu\nu}=0.
\end{equation}
In the QM formalism, yields

\begin{equation}
\label{var_weyl}
\delta W^{\alpha\beta\mu\nu}{}_{;\nu}=-\fracc{1}{2}(\rho+p)\delta V_{\mu;[\alpha}V_{\beta]}.
\end{equation}
The equations (\ref{var_weyl}) are not precisely equivalent to (\ref{var_ricci}) because they consider distinct initial conditions. Besides, Eq.\ (\ref{var_weyl}) has a kind of ``source" for the gravitational field (in this case, due to coupling terms with shear tensor) absent in the Lifshitz perturbation theory. To fade away this ambiguity with clear examples, we have to resort to two recent papers on tensorial perturbations of isotropic metrics using the QM formalism \cite{hogan,oshea}. In these works, the authors use null coordinates to analyze tensorial perturbations in Robertson-Walker metrics and they encountered a gravitational wave type of solution different from those one obtained through the Bardeen method of gauge-invariant perturbations \cite{Bardeen,Ellis et al}. The definition of gravitational waves in \cite{hogan,oshea} leads to no restriction on the equation of state of the fluid on the isotropic background, opposite to what is obtained by \cite{wain,bicak}, where the speed of sound is equal to the speed of light. According to what we said before, this difference can be associated to different boundary conditions given in each approach, which are fundamental to consistently define gravitational waves in each case. We will not enter the details of their works, but we would like to emphasize the question: in a nonlinear regime of gravitational theory, how can we translate the information contained in the nonnull Weyl tensor in terms of the perturbed potential $\delta g_{ij}$?

In other words, GR does not contemplate in its boundary conditions any kind of information about intrinsic accelerations of a given configuration on the manifold, which would lead to the development of a nonlocal description of the space-time as suggested by \cite{mash08,mash,chic,blome}. This type of information shall explicitly be given to the QM formalism, because these equations have partial derivatives of the energy-momentum tensor and contain higher order derivatives of the metric tensor in the dynamics.


\section{Concluding Remarks}

During the whole history of the quasi-Maxwellian equations of general relativity, they have been just considered as an alternative (and, eventually more complicated) approach equivalent to the Einstein equations. However, even Prof. Jordan and collaborators knew the importance of this issue, as we quoted: \textit{With this paper we start an enterprise which, considering the present state of the scientific development of this field, is of some urgency. We want to collect and describe all those exact solutions which are dispersed in the literature and thus probably, in their completeness,
known to very few authors only.} \cite{jek_eq} After all these years of their seminal work, we summarize here all those works which have been done in this area showing that this formalism is actually as powerful as the standard one, once the well-known solutions of GR are easily regained from QM-formalism.

In the case of conformally flat universes, singular \cite{no_sa_ma_re_se} or not \cite{nov_sal_06}, this approach is more convenient to treat all types of perturbations (scalar, vectorial and tensorial). It reduces the dynamics to a pair of equations in $\Sigma$ and $E$, which are variables physically meaningful, providing a dynamical planar system and a reparametrization of these variables allows to establish a gauge-invariant Hamiltonian treatment for the perturbation. In particular, some years ago, there has been a renewed interest on nonsingular cosmology and we review that these models are completely stable based on results obtained from the QM-approach. In the case of Schwarzschild metric \cite{renato} as well as the Kasner case \cite{martha}, it is also possible to use the Stewart Lemma \cite{Stewart} for applying such method and get significant results.

There are many areas of research on this topic. Some of them concern the development of deep analogies between QM-equations and the Maxwell theory \cite{erico,erico2}. Another topic concerns modifications on the dynamics of general relativity suggested by Bianchi's identities for the Weyl tensor \cite{nov_75}. Most of these issues are still under development.


\section{{\bf APPENDIX:} An example of further developments in QM-formalism}

This appendix presents an example in which the QM-formalism is modified  and the analogy with the electromagnetism is strengthened. Modifications of this kind indicate possible manners to solve problems that appear in the general relativity theory in a more intuitive way. We present an enlightening example next.

Electrodynamics is the paradigm of field theory. Its theoretical and experimental properties have been simulated and sought for in many other theories, and in particular in the analysis of gravitational phenomena. Many works have been done along this line and discussed the resemblance between electrodynamics and gravidynamics\footnote{This resemblance is far from being accepted by all physics community. Indeed, in the final session of the 1972 Copenhagen International Conference on Gravitation and Relativity, A. Trautman argued that perhaps many of the difficulties of gravitational theory may be due to the extension of this similarity to all aspects of both fields.}. However, it seems possible to further improve this similarity, as we shall show.

In this vein, we review here a modification of Einstein's theory of general relativity under certain special states of the geometry of the space-time. Since the original proposal of Einstein's geometrization of gravitational processes, many physicists have discussed alternative models of gravitation. The kind of theory we shall analyze here is given by means of the metric properties---represented by a symmetrical metric tensor $g_{\mu \nu}(x)$---and by two other functions, $\varepsilon (x)$ and $\mu (x)$, which are independent from the metric\footnote{This hypothesis is made here only for simplicity. It is an over-simplification under certain drastic situations, such as, for instance, for very strong gravitational fields.} but have intimate connection to the space-time.

For pedagogical reasons, we think it is convenient to limit our considerations in this presentation to the case in which both $\varepsilon$ and $\mu$  are constants. The meaning one should attribute to these two constants is obtained by a direct analogy with the dielectric and permeability constants of a given medium in electrodynamics.

We shall simplify the model by merely stating that  $\varepsilon$  and $\mu$ can be provisionally identified with the characteristics of certain states of tensions, in free space-time, due to an average procedure on (quantum) properties of gravitation.

In other words, $\varepsilon$  and $\mu$  are interpreted as the result---in a macroscopic level---of some sort of averaging microscopic field fluctuation\footnote{We call the reader's attention to the fact that we are, perhaps, in a situation similar to that experienced by Maxwell, a century ago.
His theory described the electromagnetic fields in the interior of substances by means of the same type of fields in vacuum  and by characterizing the distortion produced by the matter on the fields, as given by macroscopic quantities: the dielectric constant $\varepsilon_{\mbox{Max}}$ and the
permeability $\mu_{\mbox{Max}}$ (by ``Max" we mean Maxwell). It took many years before Lorentz - having at the disposal the atomic theory of matter -- made Maxwell's theory rigorously understood be averaging properties of microscopic fields in a macroscopic scale.}. This is perhaps not  difficult to assume if we can say exactly how the equations of motion of gravity phenomena must be modified by them, as we shall do later. We remark that we are not supporting this interpretation but merely suggesting it as a provisional {\it sursis} of the model.

We shall describe gravitational interaction\footnote{We remark that in the present paper we shall limit ourselves to the sourceless case, i.e.  the so-called vacuum gravitational fields. A further generalization to include matter is straightforward and does not present difficulties.} by means of a fourth-rank tensor $Q_{\alpha \beta \mu \nu}$. We shall set up its algebraic properties and give its dynamics. It is possible to separate this tensor, for an observer moving with four-velocity $V^{\mu}$, into four second-order symmetric trace-free tensors $E_{\alpha \beta}$, $B_{\alpha \beta}$, $D_{\alpha \beta}$ and $H_{\alpha \beta}$. The principal result is then obtained by showing that it is possible to select a class of observers with velocity $\ell^{\mu}$ in such a way as to have equations of motion for $Q_{\alpha \beta \mu \nu}$ similar to the Maxwell equations for the Electrodynamics. That is, for $E_{\alpha \beta}$, $D_{\alpha \beta}$, $B_{\alpha \beta}$ and   $H_{\alpha \beta}$ separated into two groups: one containing only $E_{\alpha \beta}$  and $B_{\alpha \beta}$ (and their derivatives) and the other containing only $H_{\alpha \beta}$ and $D_{\alpha \beta}$ (and their derivatives). These equations have the same formal structure of Maxwell's equations in a given general medium. So, we arrived at the conclusion that in the present theory there is a class of privileged observers in which gravitational field equations admit this simple separated form. Any other observer, which is in motion with respect to $\ell^{\mu}$, mixes the terms $E_{\alpha \beta}$, $D_{\alpha \beta}$, $H_{\alpha \beta}$ and $B_{\alpha \beta}$ into the equations. This situation could be thought of as defining a new type of ether. However, unlike the ether of the pre-Einstein epoch, our ether is not a substance, but it is only a preferred frame of observation.

In the remainder of this presentation we discuss in some detail a very particular situation of these tensors, that is, the case in which they can be reduced to two tensors plus two constants: they are $\varepsilon$ and $\mu$. Then we show that Einstein's theory is obtained from this for a
particular set of values of $\varepsilon$  and $\mu$, that is, $\varepsilon =\mu =1$. It is in this sense that we can call this theory a
generalization of Einstein's gravidynamics.

\subsection{The Q-Field}

\subsubsection{Definitions}

Let us define in a four-dimensional Riemannian manifold a fourth-rank tensor $Q_{\alpha \beta \mu \nu}$  described by an observer $V^{\mu}$ in terms of four second-order tensors $E_{\alpha \beta}$, $D_{\alpha \beta}$, $H_{\alpha \beta}$ and $B_{\alpha \beta}$. We set, by analogy with the irreducible decomposition of the Weyl tensor, that

\begin{equation}
\label{qfield}
Q_{\alpha\beta}{}^{\mu\nu}=V_{[\alpha}D_{\beta ]}{}^{[\mu} V^{\nu]}+V_{[\alpha}E_{\beta ]}{}^{[\mu} V^{\nu]}+\delta^{[\mu}_{[\alpha} E^{\nu ]}_{\beta ]}-\eta_{\alpha\beta\rho\sigma}V^{\rho}B^{\sigma[\mu}V^{\nu]}-\eta^{\mu\nu\rho\sigma}V_{\rho}H_{\sigma[\alpha}V_{\beta]}.\end{equation}

The tensors $E_{\alpha\beta}$, $D_{\alpha\beta}$, $B_{\alpha\beta}$ and $H_{\alpha\beta}$, represented below by $X_{\alpha\beta}$, satisfy the following properties:

\begin{subequations}
\begin{eqnarray}
&&X^{\alpha}{}_{\alpha}=0,\\
&&X^{\alpha \beta}V_{\alpha}=0,\\
&&X_{\alpha \beta}=X_{\beta \alpha}.
\end{eqnarray}
\end{subequations}

We can write $D_{\alpha \beta}$, $E_{\alpha \beta}$ etc., in terms of $Q_{\alpha \beta \mu \nu}$ and projections on $V^{\mu}$ like, for instance

\begin{equation}
\nonumber
D_{\alpha \beta}=-Q_{\varepsilon \alpha \mu \beta}V^{\varepsilon}V^{\mu}
\end{equation}
and so on.

\subsubsection{Algebraic properties}

From definition (\ref{qfield}) of $Q_{\alpha \beta \mu \nu}$  it follows directly the properties:

\begin{eqnarray}
&& Q_{\alpha \beta}{}^{\mu\nu}=-Q_{\beta\alpha}{}^{\mu\nu},\\
&& Q_{\alpha \beta}{}^{\mu \nu}=-Q_{\alpha \beta}{}^{\nu \mu},\\
&&Q^{\alpha}_{\ \ \beta \alpha \nu}=E_{\beta \nu}-D_{\beta \nu},\\
&&Q^{\alpha}{}_{ \alpha}=0
\end{eqnarray}

\subsubsection{Dynamics}

By analogy with Einstein's equations in vacuum we impose on $Q_{\alpha \beta \mu \nu}$   the equation of
motion\footnote{Indeed, as we shall see, in the case in which $E_{\alpha \beta} =D_{\alpha \beta}$ and $B_{\alpha \beta}=H_{\alpha \beta}$, $Q_{\alpha \beta \nu \mu}$ can be identified to Weyl's tensor and Eqs.~(10) reduces to Einstein's equation in the vacuum.}
\begin{equation}
\label{qfield_dyn}
Q^{\alpha \beta \mu \nu}{}_{;\nu}=0
\end{equation}

Now, we shall use the above properties to project the system of Eq.\ (\ref{qfield_dyn}) parallel and orthogonal to the rest frame of a selected observer with $4-$velocity $\ell^{\mu}$ from the whole class of $V^{\mu}$. We impose that the congruence generated by $\ell^{\mu}$ satisfy the properties.

\begin{subequations}
\label{gen_kin_quant}
\begin{eqnarray}
&& \ell^{\mu}\ell_{\mu}=+1,\\
&&w_{\alpha \beta}=\frac{1}{2}\ h_{[\alpha}{\ }^{\lambda}\ h_{\beta ]}{\ }^{\varepsilon}\ell_{\lambda;\varepsilon }=0,\\
&&\theta_{\alpha \beta}=\frac{1}{2}\ h_{(\alpha}{\ }^{\lambda}\ h_{\beta )}{\ }^{\varepsilon}\ell_{\lambda;\varepsilon }=0,\\
&& \dot\ell^{\mu}=\ell^{\mu}{}_{;\nu}\ell^{\nu}=0.\\
\end{eqnarray}
\end{subequations}
where $h_{\mu \nu}$ is the projector in the plane orthogonal to $\ell^{\mu}$, that is
\begin{equation}
h_{\mu \nu}=g_{\mu \nu}-\ell_{\mu}\ell_{\nu}
\end{equation}
So, the congruence generated by  $\ell_{\mu}$  is geodesic, irrotational, non-expanding and shear-free. The reason for selecting such a particular class of observers will appear clear later on. Then, Eq.\ (\ref{qfield_dyn}) assumes the form:
\begin{subequations}
\label{qfield_max}
\begin{eqnarray}
&& D_{\alpha \mu;\nu}h^{\mu \nu}h^{\alpha}_{\ \ \varepsilon}=0\ , \\
&&\dot D_{\alpha \mu}h^{\alpha}_{(\sigma}\ h{\ }^{\mu}_{\varepsilon )}+h^{\alpha}_{(\sigma}\ \eta_{\varepsilon )}
{\ }^{\nu \rho \tau}\ell_{\rho}H_{\tau \alpha ;\nu}=0,\\
&&B_{\alpha \mu ;\nu}h^{\mu \nu}h^{\alpha}_{\varepsilon}=0,\\
&&\dot{B}^{\mu \nu}h_{\mu (\sigma}\ h{\ }_{\lambda )\nu}-h_{(\sigma}^{\ \alpha}\ \eta {\ }_{\lambda )}{}^{\nu \rho \tau}\ell_{\rho}E_{\tau \alpha; \nu}=0 \ ,
\end{eqnarray}
\end{subequations}
in which a parenthesis means symmetrization.

This set of equations has a striking resemblance with Maxwell's macroscopic equations of electrodynamics. Indeed, we can formally understand the  above set as being \cite{jek_eq}
\begin{subequations}
\label{qfield_max_way}
\begin{eqnarray}
&&\nabla \cdot \vec D=0\ , \\
&&\dot{\vec{D}}-\nabla\times \vec{H}=0\ ,\\
&&\nabla \cdot  \vec{B}=0\ ,\\
&&\dot{\vec{B}}+\nabla\times \vec{E}=0\ ,
\end{eqnarray}
\end{subequations}
where the symbol $\rightarrow$ is put over $D$, $E$ etc. only to represent its tensorial character; the $\nabla$ operator represents the generalizations of the usual well-known differential operators.
		
So, we can understand the reason for selecting the above privileged set of observers, given by the tangential
vector $\ell^{\mu}$. Eq.\ (\ref{qfield_dyn}) takes the form\ (\ref{qfield_max}) only for this class of frame. Any other observer which is in motion with respect to $\ell^{\mu}$ will mix into the equations of motion of the set of tensors $(E_{\alpha \beta},B_{\alpha \beta})$ with the set of tensors $(D_{\alpha \beta},H_{\alpha \beta})$. So, it is in this sense that there is a natural selection of observers, with respect to the equation of motion satisfied by $Q_{\alpha \beta \mu \nu}$.

\subsubsection{$\varepsilon$ and $\mu$ states of tension}

A particular class of states of space-time occurs in the case in which there is a linear function relating the tensors $B_{\alpha \beta}$ with $H_{\alpha \beta}$ and $E_{\alpha \beta}$  with  $D_{\alpha \beta}$  by the intermediary of two constants, $\varepsilon$ and $\mu$.

We set
\begin{subequations}
\label{qm_dielectric}
\begin{eqnarray}
&&B_{\alpha \lambda}=\mu H_{\alpha \lambda}, \\
&&D_{\alpha \lambda}=\varepsilon H_{\alpha \lambda}.
\end{eqnarray}
\end{subequations}
If we put expressions\ (\ref{qm_dielectric}) into definition\ (\ref{qfield}) of $Q_{\alpha \beta \mu \nu}$, a straightforward calculation shows that it is possible to write  $Q_{\alpha \beta \mu \nu}$ in terms of the Weyl tensor $C_{\alpha \beta \mu \nu}$, and its "electric" and "magnetic" parts ${\cal E}_{\alpha \beta}$ and ${\cal H}_{\alpha \beta}$, if we identify the tensor  $E_{\alpha \beta}$ with ${\cal E}_{\alpha \beta}$  and $H_{\alpha \beta}$ with ${\cal H}_{\alpha \beta}$. Then we can write
\begin{eqnarray}
Q_{\alpha \beta} &=& C_{\alpha \beta}^{\ \ \ \mu \nu}+(\varepsilon -1)\ell_{[\alpha}\ {\cal E}_{\beta ]}^{[\mu }\ \ell^{\nu ]}+ (1-\mu )\eta_{\alpha \beta \rho \sigma}{\cal H}^{\sigma [\mu}\ \ell^{\nu ]}\ell^{\rho}\ ,
\end{eqnarray}
where
\begin{eqnarray}
C_{\alpha \beta}^{\ \ \ \mu \nu} &=& 2\ell_{[\alpha}\ {\cal E}_{\beta ]}{}^{[\mu}\ell^{\nu]}+\delta^{[\mu}_{[\alpha}\
{\cal E}^{\nu ]}_{\beta ]}-\eta_{\alpha \beta \lambda \sigma}\ell^{\lambda}{\cal H}^{\sigma [\mu}\ell^{\nu ]}-
\eta^{\mu \nu \rho \sigma}\ell_{\rho}{\cal H}_{\sigma [\alpha}\ \ell_{\beta ]}
\end{eqnarray}
and, consequently,
\begin{equation}
{\cal E}_{\alpha \beta}=-C_{\alpha \mu \beta \nu}\ell^{mu}\ell^{\nu}\ ,
\end{equation}
\begin{equation}
{\cal H}_{\alpha \beta}=\frac{1}{2}\ \eta_{\alpha \mu}{}^{\rho\sigma}C_{\rho \sigma \beta \lambda}\ell^{\mu}\ell^{\lambda}
\end{equation}
The resulting equations of motion (13)-(14) turn into the set:
\begin{subequations}
\label{qm_dielectric_dyn}
\begin{eqnarray}
&&{\cal E}_{\alpha \mu ||\nu}h^{\mu \nu}h^{\alpha}_{\ \ \varepsilon}=0\ ,\\
&&\varepsilon\dot{\cal E}_{\alpha \mu}h^{\alpha}_{(\sigma}\ h^{\mu}_{\varepsilon )}+h^{\alpha}_{(\sigma}\ \eta_{\varepsilon )}^{\nu \rho \tau}\ell_{\rho} {\cal H}_{\tau \alpha || \nu}=0\ ,\\
&&{\cal H}_{\alpha \mu ||\nu}h^{\mu \nu}h^{\alpha}_{\varepsilon}=\ ,\\
&&\mu \dot{\cal H}_{\alpha \mu}h^{\alpha}_{(\sigma }\ h^{\mu}_{\varepsilon )}-h^{\alpha}_{(\sigma}\ \eta_{\varepsilon )}^{\nu \rho \tau}\ell_{\rho}{\cal E}_{\tau \alpha || \nu}=0\ .
\end{eqnarray}
\end{subequations}
By the same argument that guided us to Eqs.\ (\ref{qfield_max}) we see from the above set that we can identify $\varepsilon$ as being the gravitational analogue of the dielectric constant of electrodynamics, and $\mu$ as being the permeability of space-time.

Now, we recognize in Eqs.\ (\ref{qm_dielectric_dyn}) Einstein's equations for the free gravitational field for the particular case in which  $\varepsilon =\mu =1$.\footnote{It seems worthwhile to observe that this equivalence is complete only if we impose as initial date the set  $R_{\mu \nu}=0$ on a given space-like hypersurface.}

So, it seems natural to interpret Eqs.\ (\ref{qm_dielectric_dyn}) for the general case ($\varepsilon$, $\mu$ different from unity) as the equations for the gravitational field on states of space-time that are macroscopically characterized (in the sense discussed in the introduction) by the two constants $\varepsilon$ and $\mu$

\subsubsection{Conformal behavior of $Q_{\alpha \beta \mu \nu}$}

A conformal transformation of the metric $g_{\alpha \mu}$ is given by the map
\begin{eqnarray}
&&g_{\mu \nu}(x) \longrightarrow \tilde{g}_{\mu \nu}(x)=\Omega^2(x)g_{\mu \nu}(x)\nonumber \\
&&g^{\mu \nu}(x) \longrightarrow \tilde{g}^{\mu \nu}(x)=\Omega^{-2}(x)g^{\mu \nu}(x)
\end{eqnarray}
How we can set $\eta_{\alpha \beta}^{\ \ \ \mu \nu}$ as independent from the conformal transformation:
\[
\tilde{\eta}_{\alpha \beta}^{\ \ \ \mu \nu}=\eta_{\alpha \beta}^{\ \ \ \mu \nu}
\]
Then, it is easy to see that the electric and magnetic parts of Weyl tensor remain unchanged,
\begin{subequations}
\begin{eqnarray}
&&\tilde{{\cal E}}_{\mu \nu}=-\tilde{C}_{\mu \rho \nu \sigma}\tilde{\ell}^{\rho}\tilde{\ell}^{\dot{\sigma}}=
{\cal E}_{\mu\nu}\ ,\\
&&\tilde{{\cal H}}_{\mu \nu}=-\frac{1}{2}\ \tilde{\eta}_{\mu \alpha}^{\ \ \ \rho \sigma}
\tilde{C}_{\rho \sigma \nu \lambda}\tilde{\ell}^{\alpha}\tilde{\ell}^{\lambda}={\cal H}_{\mu \nu}\ ,
\end{eqnarray}
\end{subequations}
where we have used conformal transformation of the velocity $\ell^{\mu}$  as usual,
\begin{equation}
\tilde{\ell}^{\mu}=\Omega^{-1}\ell^{\mu}
\end{equation}
As a consequence of the above mapping, $Q_{\alpha \beta \mu \nu}$ behaves, under the conformal transformation, as the Weyl tensor,
\begin{equation}
\tilde{Q}_{\alpha \beta \mu \nu}(x)=\Omega^2(x)Q_{\alpha \beta \mu \nu}(x)\ .
\end{equation}

\subsection{Gravitational Energy in an $\varepsilon -\mu$ State of Tension}

There have been many discussions, since Einstein's 1916 paper, concerning the definition of the energy of a given gravitational field. We do not intend to discuss this subject here but we shall limit ourselves to consider one reasonably successful suggestion of Bel \cite{bel} for the form of the energy-momentum tensor of gravitational radiation.

The point of departure, see \cite{bel}, comes from the similitude of the equation of motion of gravity and electrodynamics. He defines a fourth-rank tensor $T^{\alpha \beta \mu \nu}$ given in terms of quadratic components of the field (identified with the Riemann tensor) and written in terms of the Weyl tensor $C^{\alpha \beta \mu \nu}$.

Bel's super-energy tensor takes the form:
\begin{equation}
T^{\alpha \beta \mu \nu}=\frac{1}{2}\left\{C^{\alpha \rho \mu \sigma}C^{\beta}{}_{\rho}{}^{\nu}{}_{\sigma}+
C^{\ast \alpha \rho \mu \sigma} C^{\ast \beta}{}_{\rho}{}^{\nu}{}_{\sigma}\right\}\ ,
\end{equation}
where $^{\ast}$ is the dual operator. Note that the symmetry of the Weyl tensor ($C^{\,\ast}_{\alpha \beta \mu \nu} = C^{\,\,\,\,\,\,\,\,\,\ast}_{\alpha \beta\mu \nu} = \,^{\ast}C_{\alpha \beta \mu \nu}$) does not hold for $Q_{\alpha \beta \mu \nu}$. This is related to the lack of $Q_{\alpha \beta \mu \nu}\neq Q_{\mu \nu \alpha \beta}$ symmetry. Indeed, we have
\begin{eqnarray}
&&Q^{\,\ast}_{\alpha\beta}{}^{\mu \nu}=C^{\ast}_{\alpha \beta}{}^{\mu \nu}+\frac{1}{2}(\varepsilon -1)
\eta_{\alpha \beta \rho \sigma}\ell^{[\rho}\ {\cal E}^{\sigma ]}{}^{[\mu}\ \ell^{\nu ]}+\frac{1}{2}\ (1-\mu )\eta_{\alpha \beta \rho \sigma}\eta^{\rho \sigma \varepsilon \sigma}\ell_{\varepsilon}
{\cal H}_{\tau}^{[\mu}\ \ell^{\nu ]}
\nonumber
\end{eqnarray}
and
\begin{eqnarray}
&&Q_{\alpha\beta}{}^{\mu^{\ast} \nu}=C^{\,\ast}_{\alpha \beta}{}^{\mu \nu}+\frac{1}{2}(\varepsilon -1)
\eta^{\mu \nu}_{\ \ \rho \sigma}\ell^{[\alpha}\ {\cal E}^{\beta ]}{}^{[\rho}\ \ell^{\sigma ]}+\frac{1}{2}\ (1-\mu )\eta^{\mu \nu}{}_{\rho \sigma}\eta_{\alpha \beta \varepsilon \tau}\ell^{\varepsilon}
{\cal H}^{\tau}{}^{[\rho}\ \ell^{\sigma ]}
\nonumber
\end{eqnarray}
Then we have
\begin{eqnarray}
&&Q^{\,\,\,\,\,\,\,\,\,\ast}_{\alpha \beta \mu\nu}\ell^{\beta}\ell^{\nu}=C^{\,\ast}_{\alpha \beta \mu \nu}\ell^{\beta}\ell^{\nu}={\cal H}_{\alpha \mu}\ ,\nonumber \\
&&Q^{\,\ast}_{\alpha\beta \varepsilon \sigma}\ell^{\beta}\ell^{\sigma}=\mu\ {\cal H}_{\alpha \varepsilon}\ .\nonumber
\end{eqnarray}
This $T^{\alpha \beta \mu \nu}$ tensor has properties very similar indeed to the Minkowski energy-momentum tensor of electrodynamics.

The scalar constructed with $T^{\alpha \beta \mu \nu}$  and the tangent vector $\ell^{\nu}$, for instance, takes the form
\begin{equation}
U_{(T)}=T^{\alpha \beta \mu \nu}\ell_{\alpha}\ell_{\beta}\ell_{\mu}\ell_{\nu}
\end{equation}
and gives the ``energy" of the field
\begin{equation}
U_{(T)}=\frac{1}{2}\left({\cal E}^2+{\cal H}^2\right)\ ,
\end{equation}
where
\begin{subequations}
\begin{eqnarray}
&&{\cal E}^2={\cal E}_{\alpha \beta}{\cal E}^{\alpha \beta}\ .\\
&&{\cal H}^2={\cal H}_{\alpha \beta}{\cal H}^{\alpha \beta}\ \ .
\end{eqnarray}
\end{subequations}
In the context of the present extended theory, for a space-time in the state $\varepsilon -\mu$ of tension, we are led to modify $T^{\alpha \beta \mu \nu}$ into $\Theta^{\alpha \beta \mu \nu}$ defined in an analogous manner by
\begin{equation}
\Theta^{\alpha \beta \mu \nu} = \frac{1}{2}\left\{Q^{\alpha \rho \mu \sigma}C^{\beta}{}_{\rho}{}^{\nu}{}_{\sigma} + Q^{\alpha^{\ast}\rho\mu\sigma} {}^{\ast}C^{\beta}{}_{\rho}{}^{\nu}{}_{\sigma}\right\}\ .
\end{equation}
Then, the energy $U_{(\varepsilon ,\mu )}$ as viewed by an observer $\ell^{\mu}$  will be given by
\begin{equation}
U_{(\varepsilon ,\mu )}=\theta^{\alpha \beta \mu \nu}\ell_{\alpha}\ell_{\beta}\ell_{\mu}\ell_{\nu}=\frac{1}{2}
\left(\varepsilon{\cal E}^2+\mu {\cal H}^2\right)
\end{equation}
in complete analogy with the electrodynamics case in a general medium.

We would like to make an additional remark by presenting two special properties of $\Theta^{\alpha \beta \mu \nu}$
\begin{subequations}
\begin{eqnarray}
&&\Theta^{\alpha}_{\ \ \beta \alpha \mu}=\frac{1}{2}(1-\varepsilon )\varepsilon^{\rho \sigma}C_{\beta \rho \mu \sigma}\ ,\label{th_tr1}\\
&&\Theta = \Theta^{\alpha \mu}{}_{\alpha \mu}=0 \ .\label{th_tr2}
\end{eqnarray}
\end{subequations}
Property\ (\ref{th_tr1}) states that not all traces of $\Theta^{\alpha \beta \mu \nu}$  are null for a general state of tension of space-time and that the non-null parts of the contracted tensor are independent of the ``permeability" $\mu$. The second property\ (\ref{th_tr2}) states that the scalar obtained by taking the trace of $\Theta^{\alpha \beta \mu \nu}$ twice is null, independent from the state of tension of the space-time.

\subsection{The Velocity of Propagation of Gravitational Disturbances in States of Tension}

In order to know the velocity of gravitational waves in  $\varepsilon -\mu$ states of space-time, let us perturb the set of equations\ (\ref{qfield_max}). The perturbation will be represented by the map:
\begin{subequations}
\begin{eqnarray}
&&{\cal E}_{\mu \nu}\longrightarrow {\cal E}_{\mu \nu}+\delta {\cal E}_{\mu \nu}\ ,\\
&&{\cal H}_{\mu \nu}\longrightarrow {\cal H}_{\mu \nu}+\delta {\cal H}_{\mu \nu} \,
\end{eqnarray}
\end{subequations}
in which $\delta {\cal E}_{\mu \nu}$, $\delta {\cal H}_{\mu \nu}$ are null quantities. Then, Eqs.\ (\ref{qfield_max}) go into the perturbed set of equations.
\begin{subequations}
\label{qfield_max_pert}
\begin{eqnarray}
&&\delta {\cal E}_{\alpha}{}^{\beta}{}_{;\beta}\approx 0 \ ,\label{qfield_max_pert1}\\
&&\varepsilon \delta \dot{{\cal E}}_{\alpha \mu}+\frac{1}{2}\ h_{(\alpha}^{\lambda}\eta_{\mu )}{}^{\rho\sigma \tau}\ell_{\rho}\,\delta {\cal H}_{\tau \lambda;\rho}\approx 0 \ ,\label{qfield_max_pert2}\\
&&\delta {\cal H}_{\alpha}{}^{\beta}{}_{;\beta}\approx 0 \ ,\label{qfield_max_pert3}\\
&&\mu \ \delta \dot{{\cal H}}-\frac{1}{2}\ h_{(\alpha}^{\lambda}\eta_{\mu )}{}^{\rho \sigma \tau}{\cal E}_{\tau \lambda;\rho}\approx 0\ ,\label{qfield_max_pert4}
\end{eqnarray}
\end{subequations}
where the covariant derivative is taken in the background -– and we limit ourselves to the linear terms of the perturbation.

Now, let us specialize the background to be the flat Minkowski space-time\footnote{It seems worthwhile to remark that we are certainly not considering a usual Minkowski space-time. Here we are considering a more complex structure that takes into account the fluctuations of  space-time (as in quantum gravidynamics). These fluctuations are assumed to be represented on an average by the macroscopic quantities $\varepsilon$ and $\mu$, as stated in the introduction.}. In this case the covariant derivatives are the usual derivation and we can use commutative property to write:

\begin{equation}
\label{qfield_max_pert_e}
\varepsilon \delta \ddot{{\cal E}}_{\alpha \beta}+\frac{1}{2}\ h_{(\alpha}^{\lambda}\eta_{\beta )}{}^{\rho \sigma \tau}
\ell_{\sigma}\delta \dot{\cal H}_{\tau \lambda,\rho}\approx 0 \ ,
\end{equation}
by taking the derivative of Eq.\ (\ref{qfield_max_pert2}) projected in the privileged direction $\ell^{\mu}$. Multiplying Eq.\ (\ref{qfield_max_pert4}) by the factor

\begin{equation}
\nonumber
\frac{1}{2 \mu}\ h_{(\alpha}^{\nu} \eta_{\beta )}{}^{\gamma\sigma\tau}\ell_{\tau}\ \frac{\partial}{\partial x^{\sigma}}\ ,
\end{equation}

we find
\begin{eqnarray}
\label{qfield_max_pert_h}
&&\frac{1}{2}h_{(\alpha }^{\nu}\ \eta_{\beta )}{}^{\gamma\sigma \tau}\ell_{\tau}\,\delta \dot{{\cal H}}_{\gamma \nu, \sigma}-\frac{1}{4\mu}\ h_{(\alpha }^{\nu}\ \eta_{\beta )}{}^{\sigma \tau \gamma}\ \ell_{\tau}\ell_{\rho}h^{\varepsilon}{}_{(\,\gamma}\ \eta_{\nu)}{}^{\psi \rho \phi}\delta \varepsilon_{\psi \varepsilon ,\psi ,\sigma}\approx
0 \ .
\end{eqnarray}
Substituting Eq.\ (\ref{qfield_max_pert_e}) into\ (\ref{qfield_max_pert_e}) we finally find

\begin{equation}
\label{qfield_wave_pert}
\delta \ddot{{\cal E}}_{\alpha \mu}-\frac{1}{\varepsilon \mu}\nabla^2\delta{\cal E}_{\alpha \mu}=0 \ ,
\end{equation}
where $\nabla^2$ is the Laplacian operator defined in the three-dimensional space orthogonal to $\ell^{\mu}$.

In the same way an analogous wave equation can be obtained for ${\cal H}_{\alpha \mu}$. From Eq.\ (\ref{qfield_wave_pert}) we obtain the expected result: the velocity of propagation of gravitational waves in $\varepsilon$, $\mu$  states of tension of space-time is equal to $1/\sqrt{\varepsilon \mu}$.

The set of privileged observers that we dealt with here may be enlarged by somehow weakening the defining conditions (see Eqs.\ (\ref{gen_kin_quant})). This point deserves further investigation.

\section*{Acknowledgments}
The authors would like to thank all the participants of the Pequeno Semin\'ario at CBPF for valuable  contribution on some parts of this review. We also thank CNPq, FAPERJ and FINEP for their financial support. MN and EB acknowledge the ICRANet-Pescara staff for the hospitality during some stages of this work.


\begin{thebibliography}{200}
\bibitem{abramowitz}
M. Abramowitz and I. A. Stegun, \textit{Handbook of Mathematical Functions}, Dover Publications, New York, p.504 (1974).
\bibitem{acacio}
J. Acacio de Barros, N. Pinto-Neto and M. A. Sagioro-Leal, {\em Phys.\ Lett.\ A} {\bf 241} 229 (1998).
\bibitem{adler}
R. Adler, M. Bazin, M. Schiffer, \textit{Introduction to General Relativity}, McGraw-Hill, New York, Chap. 6 (1975).
\bibitem{LopezBonilla}
G. Ares de Parga, O. Chavoya, J.L. L\'{o}pez Bonilla, {\em J.\ Math.\ Phys.} {\bf 30} 1294 (1989).
\bibitem{ayon1}
E. Ayon-Beato and A. Garcia, {\em Phys.\ Rev.\ Lett.} {\bf 80} 5056 (1998).
\bibitem{ayon2}
E. Ayon-Beato and A. Garcia, {\em Gen.\ Rel.\ Grav.} {\bf 31} 629 (1999a).
\bibitem{ayon3}
E. Ayon-Beato and A. Garcia, {\em Phys.\ Lett.\ B} {\bf 464} 25 (1999b).
\bibitem{Bardeen2}
J. Bardeen, \textit{in Proceedings of GR5}, Tbilisi, U.S.S.R, 174 (1968).
\bibitem{Bardeen}
J. Bardeen, {\em Phys.\ Rev.\ D} {\bf 22} 1882 (1980).
\bibitem{barrabes}
C. Barrabes and V. P. Frolov, {\em Phys.\ Rev.\ D} {\bf 53} 3215 (1996).
\bibitem{bateman}
H. Bateman, \textit{The Mathematical Analysis of Electrical and Optical Wave Motion}, Dover, New York (1955).
\bibitem{bel}
L. Bel, in Colloques Internationaux du Centre National de la Recherche Scientifique, Paris, p. 119 (1962).
\bibitem{bkl}
V. A. Belinski, I. M. Khalatnikov and E. M. Lifshtz, {\em Usp.\ Fiz.\ Nauk} {\bf 102} 463 (1970); V. A. Belinski, I. M. Khalatnikov and E. M. Lifshtz, {\em Adv.\ Phys.} {\bf 19} 525 (1880).
\bibitem{gasp12}
I. Ben-Dayan, M. Gasperini, G. Marozzi, F. Nugier and G. Veneziano, arXiv:astro-ph.CO 1202.1247v2 (2012).
\bibitem{Poloneses}
Z. Bialynicka-Birula, I. Bialynicka-Birula, {\em J.\ Opt.\ Soc.\ Am.\ B} {\bf 4} 10 (1987).
\bibitem{bicak}
J. Bicak and J. Griffiths, {\em Ann.\ Phys.\ (N.Y.)} {\bf 252} 180 (1996).
\bibitem{birrel}
N. D. Birrel, P. C. W. Davis, \textit{Quantum Fields in Curved Space}, Cambridge University Press, Cambridge (1982).
\bibitem{eduardo}
E. Bittencourt, Master's Thesis, CBPF/Rio de Janeiro (2009).
\bibitem{blome}
H. J. Blome, C. Chicone, F. W. Hehl and B. Mashhoon, arXiv:gr-qc/1002.1425 (2010).
\bibitem{brady}
P. R. Brady, S. Droz, W. Israel and S. M. Morsink, {\em Class.\ Quantum\ Grav.} {\bf  13} 2211 (1996).
\bibitem{rob_brand}
R. Brandenberger, {\em Proceedings\ of\ the\ VIII\ Brazilian\ School\ of\ Cosmology\ and Gravitation}, edited by M. Novello, Editions Fronti\'eres, Singapore, (1996).
\bibitem{BrandenbergerKhan}
R. Brandenberger, R. Khan and W. H. Press, {\em Phys.\ Rev.\ D} {\bf 28} 1809 (1983).
\bibitem{brand_muk_sorn}
R. Brandenberger, V. Mukhanov, and A. Sornborger, {\em Phys.\ Rev.\ D} {\bf 48} 1629 (1993).
\bibitem{bruni}
M. Bruni, P. K. S. Dunsby and G. F. R. Ellis, {\em Astrophys.\ J.} {\bf 395} 34 (1992).
\bibitem{mat_95}
M. Bruni, S. Matarrese, O. Pantano, {\em Phys.\ Rev.\ Lett.} {\bf 74} 1916 (1995).
\bibitem{cabo}
A. Cabo and E. Ayon-Beato, arXiv:gr-qc/9704073 (1997).
\bibitem{cald_dare_stein}
R. R. Caldwell, R. Dare, and P. J. Steinhardt, Phys. Rev. Lett. 80, 1582 (1998).
\bibitem{marcela}
M. Campista, Ph. D. Thesis, CBPF/Rio de Janeiro (2010).
\bibitem{art_marcela}
M. Campista, M. Novello, J. M. Salim, {\em Int.\ J.\ Mod.\ Phys.\ A} {\bf 25} 3095 (2010).
\bibitem{campos}
A. Campos and B. L. Hu, {\em Phys.\ Rev.\ D} {\bf 58} 125021 (1998).
\bibitem{cartier}
C. Cartier, R. Durrer, E. J. Copeland, {\em Phys.\ Rev.\ D} {\bf 67} 103517 (2003).
\bibitem{chic}
C. Chicone, B. Mashhoon, arXiv:gr-qc/1111.4702 (2008).
\bibitem{choquet_bruhat}
Y. Choquet-Bruhat, R. Geroch, {\em Comm.\ Math.\ Phys.} {\bf 14} 329 (1969).
\bibitem{choq_book}
Y. Choquet-Bruhat, \textit{General Relativity and Einstein's Equations}, Oxford University Press, New York (2009).
\bibitem{coq}
R. Coqueraux and A. Grossmann, {\em Ann.\ Phys.\ (N.Y.)} {\bf 143} 296 (1982).
\bibitem{cvetic}
M. Cvetic, {\em Phys.\ Rev.\ Lett.} {\bf 71} 815 (1993).
\bibitem{dabro_stel}
M. Dabrowski and J. Stelmach, {\em Astron.\ J.} {\bf 92} 1272 (1986).
\bibitem{lorenci}
V. A. De Lorenci, R. Klippert, M. Novello and J. M. Salim, {\em Phys.\ Rev.\ D} {\bf 65} 063501 (2002); arXiv:gr-qc/9806076v2 (1998).
\bibitem{Debever}
R. Debever, {\em Cahiers de Physique} {\bf 168} (1964).
\bibitem{Dirac}
P. A. M. Dirac, {\em  Can.\ J.\ Math.} {\bf 2} 129 (1950).
\bibitem{dunne_hall}
G. Dunne and T. Hall, {\em Phys.\ Rev.\ D} {\bf 58} 105022 (1998).
\bibitem{dunne}
G. Dunne, {\em Int.\ J.\ Mod.\ Phys.\ A} {\bf 12} 1143 (1997).
\bibitem{dunsky}
P. K. S. Dunsby, B. A. C. C. Bassett and G. F. R. Ellis, {\em Class.\ Quantum\ Grav.} {\bf 14} 1215 (1997).
\bibitem{ed}
D. Edwards, {\em Astrophys.\ Space\ Sci.} {\bf 24} 563 (1973).
\bibitem{ehlers}
J. Ehlers, {\em Abh.\ Akad.\ Wiss.\ Lit.\ Mainz.\ Mat.\ Nat.\ Kl.} {\bf 1} (1961); J. Ehlers, {\em Gen.\ Rel.\ Grav.} {\bf 25} 1225 (1993).
\bibitem{einstein}
A. Einstein, {\em Ann.\ Phys.} {\bf 354} 769 (1916); A. Einstein, {\em Rev.\ Mod.\ Phys} {\bf 20} 35 (1048).
\bibitem{Ellis}
G. F. R. Ellis, {\em Proc.\ Int.\ School of Physics, Enrico Fermi Course XLVII}, London Academic 104 (1971).
\bibitem{ellis_carg}
G.F.R. Ellis, {\em Carg\`ese Lectures in Physics Vol. VI}, edited by E. Schatzmann, Gordon and Breach, London (1971).
\bibitem{Ellis et al}
G. F. R. Ellis and M. Bruni, {\em Phys.\ Rev.\ D} {\bf 40} 1804 (1989).
\bibitem{fab_sal_sau}
J. C. Fabris, J. M. Salim and S. L. Sautu, {\em Mod.\ Phys.\ Lett.\ A} {\bf 13} 953 (1998).
\bibitem{Fierz}
M. Fierz, {\em Helv.\ Phys.\ Acta} {\bf 12} 379 (1939).
\bibitem{garc_breton}
R. Garcia-Salcedo and N. Breton, {\em Int.\ J.\ Mod.\ Phys.\ A} {\bf 15} 4341 (2000).
\bibitem{gasp11}
M. Gasperini, G. Marozzi, F. Nugier and G. Veneziano, {\em JCAP} {\bf 07} 008 (2011).
\bibitem{gasp}
M. Gasperini and G. Veneziano, {\em Phys.\ Rep.} {\bf 373} 1 (2003); arXiv:hep-th/0207130 (2002).
\bibitem{giov_shap}
M. Giovannini and M. Shaposhnikov, {\em Phys.\ Rev.\ D} {\bf 57} 2186 (1998).
\bibitem{Goode}
S. W. Goode, {\em Phys.\ Rev.\ D} {\bf 39} 2882 (1989).
\bibitem{erico}
E. Goulart, Ph. D. Thesis, CBPF/Rio de Janeiro (2007).
\bibitem{erico2}
E. Goulart and M. Novello, {\em Grav.\ Cosmology} {\bf 14} 321 (2008).
\bibitem{grad}
I. S. Gradshteyn and I. M. Ryzhik, \textit{Table of Integrals, Series, and Products}, Academic, London, (1965).
\bibitem{Grishchuk}
L. P. Grishchuk and Yu V. Sidorov, {\em Phys.\ Rev.\ D} {\bf 42} 3413 (1990).
\bibitem{hawking}
S. Hawking, G. F. R. Ellis, \textit{The Large Scale Structure of Space-Time}, Cambridge University Press, Cambridge, England, (1977).
\bibitem{hawking_66}
S. Hawking, {\em Ap.\ J.} {\bf 145} 544 (1966).
\bibitem{heis_euler}
W. Heisenberg and H. Euler, {\em Z.\ Phys} {\bf 98} 714 (1936).
\bibitem{hind_ever}
M. Hindmarsh and A. Everett, {\em Phys.\ Rev.\ D} {\bf 58} 103505 (1998).
\bibitem{hogan88}
P. A. Hogan, {\em Astrophys.\ J.} {\bf 333} 64 (1988).
\bibitem{hogan_ellis}
P. A. Hogan and G. F. R. Ellis, {\em Class.\ Quantum\ Grav.} {\bf 14} A171 (1997).
\bibitem{hogan}
P. A. Hogan and E. M. O'Shea, {\em Phys.\ Rev.\ D} {\bf 65} 124017 (2002).
\bibitem{hogan_traut}
P. A. Hogan and A. Trautman, \textit{in Gravitation and Geometry: Papers in Honour of I. Robinson}, edited by W. Rindler and A. Trautman, Bibliopolis, Naples, (1987).
\bibitem{horne}
J. H. Horne and G. T. Horowitz, {\em Nucl.\ Phys.} {\bf B368} 444 (1992).
\bibitem{Vishniac}
J. C. Hwang and E. T. Vishniac, {\em Ap.\ J.} {\bf 353} 1 (1990).
\bibitem{ibanez}
J. Ib\'a\~nez, R. J. van den Hoogen, A. A. Coley, {\em Phys.\ Rev.\ D} {\bf 51} 928 (1995).
\bibitem{Israel}
W. Israel, {\em An.\ Phys.} {\bf 100} 310 (1976).
\bibitem{jedam}
K. Jedamzik, V. Jatalini\'c, and A. V. Olinto, {\em Phys.\ Rev.\ D} {\bf 57} 3264 (1998).
\bibitem{Jones}
B. J. T. Jones, {\em Rev.\ Mod.\ Phys.} {\bf 48} 107 (1976).
\bibitem{Jordan et al}
P. Jordan, J. Ehlers and R. Sachs, {\em Akad.\ Wiss.\ Lt.\ Meinz Abh.\ Math.\ Nat.\ Kl.} {\bf 1} 3 (1961).
\bibitem{jek_eq}
P. Jordan, J. Ehlers, W. Kundt, {\em Akad.\ Wiss.\ Lit.\ (Mainz) Abhandl.\ Math.\-Nat.\ Kl.\ 2} {\bf 21} (1960); P. Jordan, J. Ehlers, W. Kundt, {\em Gen.\ Rel.\ Grav.} {\bf 41} 2191 (2009).
\bibitem{joyce}
M. Joyce and M. Shaposhnikov, {\em Phys.\ Rev.\ Lett.} {\bf 79} 1193 (1997).
\bibitem{kalnins}
E. G. Kalnins, W. Miller, Jr., {\em J.\ Math.\ Phys.} {\bf 32} 698 (1991).
\bibitem{renato}
R. Klippert, Ph. D. Thesis, CBPF/Rio de janeiro (1998).
\bibitem{klip_lor_nov_sal}
R. Klippert, V. A. De Lorenci, M. Novello, and J. M. Salim, {\em Phys.\ Lett.\ B} {\bf 472} 27 (2000).
\bibitem{kof_lind_star}
L. Kofman, A. Linde, and A. A. Starobinsky, {\em Phys.\ Rev.\ D} {\bf 56} 3258 (1997).
\bibitem{Lanczos}
C. Lanczos, {\em Rev.\ Mod.\ Phys.} {\bf 34} 379 (1962).
\bibitem{landau}
L. D. Landau and E. M. Lifshitz, \textit{Fluid Mechanics}, Pergamon Press, N.Y. (1982).
\bibitem{lichnerowicz}
A. Lichnerowicz, {\em Ann.\ Mat.\ Pura\ ed\ Appl.} {\bf 50} 1 (1960).
\bibitem{lifshitz}
E. M. Lifshitz, I. M. Khalatnikov, {\em Adv.\ Phys.} {\bf 12} 185 (1963).
\bibitem{marklund}
M. Marklund, P. K. S. Dunsby and G. Brodin, {\em Phys.\ Rev.\ D} {\bf 62} 101501 (2000).
\bibitem{mars}
M. Mars, M. M. Martin-Prats, and J. M. M. Senovila, {\em Clas.\ Quantum\ Grav.} {\bf 13} L51 (1996).
\bibitem{mash08}
B. Mashhoon, {\em Ann.\ Phys.} {\bf 17} 705 (2008); arXiv:gr-qc/0805.2926 (2008).
\bibitem{mash}
B. Mashhoon, arXiv:gr-qc/1101.3752 (2011).
\bibitem{Matacz}
A. L. Matacz, {\em Phys.\ Rev.\ D} {\bf 49} 788 (1994).
\bibitem{mat_93}
S. Matarrese, O. Pantano, D. Saez, {\em Phys.\ Rev.\ D} {\bf 47} 1311 (1993).
\bibitem{mat_94}
S. Matarrese, O. Pantano, D. Saez, {\em Phys.\ Rev.\ Lett.} {\bf 72} 320 (1994).
\bibitem{mied}
P. G. Miedema, {\em Phys.\ Rev.\ D} {\bf 50} 2431 (1994).
\bibitem{moes_trod}
R. Moessner and M. Trodden, {\em ibid} {\bf 51} 2801 (1995).
\bibitem{martha}
M. C. Motta da Silva, Ph. D. Thesis, CBPF/rio de Janeiro (1998).
\bibitem{muk_brand}
V. Mukhanov and R. Brandenberger, {\em Phys.\ Rev.\ Lett.} {\bf 68} 1969 (1992).
\bibitem{MukhanovBrand}
V. F. Mukhanov, H. A. Feldmann and R. H. Brandenberger, {\em Phys.\ Rep.} {\bf 215} (1992).
\bibitem{mukh_winit}
V. F. Mukhanov, S. Winitzki, \textit{Introduction to Quantum Effects in Gravity}, Cambridge University Press, Cambridge (2007).
\bibitem{murphy}
G. L. Murphy, {\em Phys.\ Rev.\ D} {\bf 8} 4231 (1973).
\bibitem{mutoh}
H. Mutoh, T. Hirai, K. Maeda, {\em Phys.\ Rev.\ D} {\bf 55} 3276 (1997).
\bibitem{new_penr}
E. T. Newman, R. Penrose, {\em J.\ Math.\ Phys.} {\bf 3} 566 (1962).
\bibitem{noh}
H. Noh, J. Hwang, {\em Phys.\ Rev.\ D} {\bf 52} 1970 (1995).
\bibitem{nov_75}
M. Novello, {\em Inter.\ Atomic\ Energy\ Agency\ Ic} {\bf 75} 61 1 (1975);
\bibitem{nov_80}
M. Novello, \textit{in Proceedings of the II Brazilian School of Cosmology and Gravitation}, edited by M. Novello, J. Sansom \& Cia., Rio de Janeiro, (1980) in Portuguese.
\bibitem{novello}
M. Novello, \textit{Cosmology and Gravitation}, Frontieres Editions, Chap. 23-24, Singapore (1994); \textit{Minimal Programm on Cosmology}, edited by M. Novello, N. Pinto-Neto and S. E. P. Bergliaffa, (2010).
\bibitem{nov_bit}
M. Novello and E. Bittencourt, {\em Gravitation\ and\ Cosmology} {\bf 17} 3 230 (2011).
\bibitem{nov_sant}
M. Novello and S. E. P. Bergliaffa, {\em Phys.\ Rep.} {\bf 463} 127 (2008);	 arXiv:astro-ph/0802.1634v1 (2008).
\bibitem{nov_duque}
M. Novello, S. L. S. Duque, {\em Physica A} {\bf 168} 1073 (1990).
\bibitem{nov_luc}
M. Novello, L. R. de Freitas, {\em Int.\ J.\ Mod.\ Phys.\ A} {\bf 13} 363 (1998).
\bibitem{NovelloNelson}
M. Novello and N. Pinto-Neto, {\em Fortschrift der Physik} {\bf 40} 173 (1992a).
\bibitem{NovelloNelson2}
M. Novello and N. Pinto-Neto, {\em Fortschrift der Physik} {\bf 40} 195 (1992b).
\bibitem{nov_oli_sal_elb}
M. Novello, L. Oliveira, J. M. Salim and E. Elbaz, {\em Int.\ J.\ Mod.\ Phys.\ D} {\bf 1} 641 (1993).
\bibitem{nov_sal_79}
M. Novello and J. M. Salim, {\em Phys.\ Rev.\ D} {\bf 20} 377 (1979);
\bibitem{Novello e Salim}
M. Novello and J. M. Salim, {\em Fund.\ of Cosmic\ Phys.} {\bf 8} 201 (1983).
\bibitem{nov_sal}
M. Novello and J. M. Salim, CBPF-NF-021/85, Rio de Janeiro, (1985).
\bibitem{nov_sal_06}
M. Novello and J. M. Salim, {\em The\ Tenth\ Marcel\ Grossmann\ Meeting}, edited by M. Novello S. E. P. Bergliaffa and R. Ruffini, Cingapura: World Scientific, (2006).
\bibitem{NovelloSalim}
M. Novello, J. M. Salim and H. Heintzmann {\it in J. Leite Lopes Festschrift}, Edited by S. Joffily, (1988).
\bibitem{no_sa_ma_re_se}
M. Novello, J. M. Salim, M. C. Motta da Silva, S. E. Jor\'as, and R. Klippert, {\em Phys.\ Rev.\ D} {\bf 51} 450 (1995a).
\bibitem{no_sa_ma_re_se2}
M. Novello, J. M. Salim, M. C. Motta da Silva, S. E. Jor\'as, and R. Klippert, {\em Phys.\ Rev.\ D} {\bf 52} 730 (1995b).
\bibitem{no_sa_ma_re_se3}
M. Novello, J. M. Salim, M. C. Motta da Silva, R. Klippert, {\em Phys.\ Rev.\ D} {\bf 54} 2578 (1996).
\bibitem{no_sa_ma_re_se4}
M. Novello, J. M. Salim, M. C. Motta da Silva, R. Klippert, {\em Phys.\ Rev.\ D} {\bf 61} 124025 (2000).
\bibitem{NovelloVeloso}
M. Novello and A. L. Velloso, {\em Gen.\ Rel.\ Grav.} {\bf 19} 1251 (1987).
\bibitem{oli_sal_sau}
H. P. de Oliveira, J. M. Salim and S. L. Sautu, {\em Class.\ Quantum Grav.} {\bf 14} 2833 (1997).
\bibitem{Olson}
D. W. Olson, {\em Phys. Rev. D} {\bf 14} 327 (1976).
\bibitem{papapetrou_regina}
A. Papapetrou, \textit{Lectures on General Relativity}, D. Reidel Publishing Company, Dordrecht, Holland (1974).
\bibitem{peebles}
P. J. E. Peebles, \textit{The Large-Scale Structure of the Universe}, Princeton University Press, Princeton, (1980).
\bibitem{peter}
P. Peter and N. Pinto-Neto, {\em Phys.\ Rev.\ D} {\bf 65} 023513 (2001).
\bibitem{pirani}
F. A. E. Pirani, {\em Phys.\ Rev.} {\bf 105} 1089 (1957).
\bibitem{oshea}
E. M. O'Shea, {\em Phys.\ Rev.\ D} {\bf 69} 064038 (2004); arXiv:gr-qc/0401106 (2004).
\bibitem{reg_wheel}
T. Regge, J. A. Wheeler, {\em Phys.\ Rev.} {\bf 108} 1063 (1957).
\bibitem{rob}
H. P. Robertson, {\em Rev.\ Mod.\ Phys.} {\bf 5} 62 (1933).
\bibitem{robinson}
I. Robinson, {\em J.\ Math.\ Phys.} {\bf 2} 290 (1961).
\bibitem{rob_traut}
I. Robinson and A. Trautman, {\em J.\ Math.\ Phys.} {\bf 24} 1425 (1983).
\bibitem{Ryan}
M. P. Ryan Jr., {\em lecture on VI Brazilian School of Cosmology and Gravitation} (1991).
\bibitem{Schumaker}
B. L. Schumaker. {\em Phys.\ Rep.} {\bf 135} 317 (1986).
\bibitem{salim}
J. M. Salim, Ph. D. thesis, CBPF/Rio de Janeiro (1982).
\bibitem{salim_oliv}
J. M. Salim and H. P. de Olivera, {\em Acta\ Phys.\ Pol.\ B} {\bf 19} 649 (1988).
\bibitem{sal_sau}
J. M. Salim and S. L. Sautu, {\em Class.\ Quantum\ Grav.} {\bf 13} 353 (1996).
\bibitem{saa}
A. Saa, E. Gunzig, L. Brenig, V. Faraoni, T. M. Rocha Filho, and A. Figueiredo, arXiv:gr-qc/0012105 (2000).
\bibitem{sagnotti}
A. Sagnotti, B. Zwiebach, {\em Phys.\ Rev.\ D} {\bf 24} 305 (1981).
\bibitem{desitter}
W. de Sitter, {\em Proc.\ K.\ Ned.\ Akad.\ Wet.} {\bf 19} 1217 (1917).
\bibitem{Stewart2}
J. M. Stewart, {\em Class.\ Quantum\ Grav.} {\bf 7} 1169 (1990).
\bibitem{Stewart}
J. M. Stewart and M. Walker, {\em Proc.\ R.\ Soc.\ London\ A} {\bf 341} 49 (1974).
\bibitem{subra}
K. Subramanian and J. D. Barrow, {\em ibid} {\bf 58} 883502 (1998).
\bibitem{taj}
T. Tajima, S. Cable, K. Shibata, and R. M. Kulsrud, {\em Astrophys.\ J.} {\bf 390} 309 (1992).
\bibitem{thom_blaes}
C. Thompson and O. Blaes, {\em Phys.\ Rev.\ D} {\bf 57} 3219 (1998).
\bibitem{tol_ehren}
R. C. Tolman and P. Ehrenfest, {\em Phys.\ Rev.} {\bf 36} 1791 (1930).
\bibitem{traut_book}
A. Trautman, \textit{in Recent Developments in General Relativity}, PWN, Warsaw (1962).
\bibitem{tseytlin}
A. A. Tseytlin, {\em Phys.\ Lett.\ B} {\bf 363} 223 (1995).
\bibitem{ven}
G. Veneziano, arXiv:hep-th/0002094 (2000).
\bibitem{vishves}
C. V. Vishveshwara, {\em Phys.\ Rev.\ D} {\bf 1} 2870 (1970).
\bibitem{wain}
J. Wainwright and B. J. Marshman, {\em Phys.\ Lett.} {\bf 72A} 275 (1979).
\bibitem{wald}
R. M. Wald, \textit{General Relativity}, Univ. Chicago Press, Chicago, USA (1984).
\bibitem{weyl}
H. Weyl, \textit{Space, Time and Matter}, Dover Publications, INC., USA (1922).
\bibitem{wilt11}
D. L. Wiltshire, {\em Class.\ Quantum Grav.} {\bf 28} 164006 (2011).
\bibitem{wilt12}
D. L. Wiltshire,  P. R. Smale, T. Mattsson and R. Watkins, arXiv:astro-ph.CO/1201.5371v1 (2012).
\bibitem{zakharov}
V. D. Zakharov, \textit{Gravitational Waves in Einstein's Theory}, Halsted Press, Jerusalem, Israel (1973).
\end{thebibliography}
\end{document}